\documentclass[twocolumn]{aastex63}
\pdfoutput=1 
\usepackage{amsmath,amstext,amssymb}
\usepackage[T1]{fontenc}
\usepackage{apjfonts} 
\usepackage[figure,figure*]{hypcap}
\usepackage{multirow}
\usepackage{graphicx}
\usepackage{enumerate}
\usepackage[caption=false]{subfig}
\usepackage{comment}
\usepackage{textcomp}
\usepackage{gensymb}
\usepackage{subfig}

\usepackage{leftidx}

\newcommand{\jyb}{Jy beam$^{-1}$}
\newcommand{\kms}{km s$^{-1}$}
\newcommand{\formai}{$\rm H_{2}CO(3_{0,3}-2_{0,2})$} 
\newcommand{\formaii}{$\rm H_{2}CO(3_{2,1}-2_{2,0})$} 
\newcommand{\soi}{$\rm SO(6_{5}-5_{4})$} 
\newcommand{\forma}{$\rm H_{2}CO$}
\newcommand{\so}{$\rm SO$}

\shorttitle{Hypercompact HII regions in W51 A}
\shortauthors{Rivera-Soto et al.}

\begin{document}

\title{Recombination Lines and Molecular Gas from Hypercompact HII regions in W51 A}

\author[0000-0002-2162-8441]{Rudy Rivera-Soto}
\affiliation{Instituto de Radioastronom\'ia y Astrof\'isica (IRyA), UNAM, Apdo. Postal 72-3 (Xangari), Morelia, Michoac\'an 58089, Mexico.}
\author[0000-0003-1480-4643]{Roberto Galv\'an-Madrid}
\affiliation{Instituto de Radioastronom\'ia y Astrof\'isica (IRyA), UNAM, Apdo. Postal 72-3 (Xangari), Morelia, Michoac\'an 58089, Mexico.}
\author[0000-0001-6431-9633]{Adam Ginsburg}
\affiliation{Jansky Fellow. National Radio Astronomy Observatory, 1003 Lopezville Rd, Socorro, NM 87801, USA.}
\affiliation{Department of Astronomy, University of Florida, PO Box 112055, USA.}
\author{Stan Kurtz}
\affiliation{Instituto de Radioastronom\'ia y Astrof\'isica (IRyA), UNAM, Apdo. Postal 72-3 (Xangari), Morelia, Michoac\'an 58089, Mexico.}

\correspondingauthor{Roberto Galv\'an-Madrid}
\email{r.galvan@irya.unam.mx}

\accepted{July 18, 2020}

\begin{abstract}
We present a detailed characterization of the population of compact radio-continuum sources in W51 A using subarcsecond VLA and ALMA observations. 
We analyzed their 2-cm continuum, the  recombination lines (RL's) H77$\alpha$ and H30$\alpha$, and the lines of \formai~, \formaii~, and \soi.  
We derive diameters for 10/20 sources in the range $D \sim 10^{-3}$ to $\sim 10^{-2}$ pc, thus placing them in the regime of hypercompact HII regions (HC HII's). Their continuum-derived electron densities are in the range $n_{\rm e} \sim 10^4$ to $10^5$ cm$^{-3}$, lower than typically considered for HC HII's.  
We combined the RL measurements and independently derived $n_{\rm e}$, finding the same range of values but significant offsets for individual measurements between the two methods. 
We found that most of the sources in our sample are ionized by early B-type stars, and a comparison of $n_{\rm e}$ vs $D$ shows that they follow the inverse relation previously derived for ultracompact (UC) and compact HII's. 
When determined, the ionized-gas kinematics is  always (7/7) indicative of outflow. Similarly, 5 and 3 out of the 8 HC HII's still embedded in a compact core show evidence for expansion and infall motions in the molecular gas, respectively. 
We hypothesize that there could be two different types of \textit{hypercompact} ($D< 0.05$ pc) HII regions: those that essentially are smaller,  expanding UC HII's; and those that are also \textit{hyperdense} ($n_{\rm e} > 10^6$ cm$^{-3}$), probably associated with O-type stars in a specific stage of their formation or early life. 
\end{abstract}

\keywords{ISM: individual: W51 --- H II regions --- stars: formation}

\section{Introduction}
Massive stars form in the densest clumps within Giant Molecular Clouds (GMCs), mostly in a clustered manner \citep[for a recent review see][]{Motte18_ARAA}. They gain mass via accretion and eventually reach the Zero-Age Main Sequence (ZAMS). The mass at which this happens is in the range $M_\mathrm{ZAMS} \sim 10 - 30~M_\odot$, depending on geometry and rate of (proto)stellar accretion \citep{Hosokawa10}. 
When the ZAMS is reached, the hot stellar atmospheres emit enough extreme ultraviolet (EUV) photons to ionize their own accretion flows \citep{Keto03,Keto07,Peters10,TT16}.
HII regions smaller than about a pc are categorized by their ever decreasing sizes and increasing densities as compact, ultracompact, and hypercompact \citep[see][]{Kurtz05}. 
The simplest interpretation of the relation between hypercompact (HC) and ultracompact (UC) HII regions is that they are successive stages in the early evolution of ionization soon after massive stars reach the ZAMS \citep{WC89,Kurtz94,Hoare07}. However, models and simulations show that some HC HII's are associated with active accretion \citep{Keto07,Peters10,Peters10b}, and that early HII region evolution can be a time-variable process where the size is not necessarily a predictor of absolute age \citep{GM11,DePree14}. 

HC HII's are typically faint and deeply embedded. Therefore, most of our knowledge about them is based on their cm continuum properties \citep{SanchezMonge11,Ginsburg16,Rosero16,Yang19}. Further characterization of their Hydrogen recombination lines (RLs) is helpful to assess their densities and kinematics, which in turn gives additional insight into their physical nature \citep[e.g.,][]{Sewilo04,Guzman14}. Earlier interferometric studies were able to detect (sub)mm RLs only in the brightest HC HII's \citep[e.g.,][]{KZK08,GM09,Shi10a}. With ALMA it is now possible to go much deeper \citep[e.g.,][]{Peters12,Klaassen18,ZhangY19} and characterize larger samples.   

In this paper we use the Very Large Array (VLA) and the Atacama Large Millimeter  Array (ALMA) to investigate the nature of compact centimeter continuum sources in the high-mass star cluster formation region W51 A. This region has been studied in detail in the past, mostly in the cm radio \citep[e.g.,][]{Gaume93,Mehringer94,Ginsburg16}, (sub)mm \citep[e.g.,][]{ZhangQ98,Tang13,Ginsburg17}, infrared \citep[e.g.,][]{Kang09,Saral17}, and X-rays \citep[e.g.,][]{Townsley14}.

The paper is ordered as follows. In \S \ref{sec:data} we present the observational data and source selection criteria.
In \S \ref{sec:cont_rls} and \S  \ref{sec:physical_props} we describe our results on the cm continuum and recombination lines. In \S \ref{sec:molec_lines} we give our results on the molecular lines and a comparison to the recombination lines, when detected. In \S \ref{sec:disc} we discuss our findings. Finally, in \S \ref{sec:concl} we give our conclusions. 

\section{Data} \label{sec:data}

\begin{deluxetable}{c|c}[!t]
\tablecaption{Source catalogs \label{tab:catalogs}}
\tablehead{
\colhead{catalog Name} &  \colhead{Sources} \\
\colhead{(\# of sources)} &
}
\startdata
A    & d2, d4e, d4w, d5, d6, d7, e1, e2, e3, e4, \\ 
(29) & e5, e6, e8n, e8s, e9, e10, e11, e12, e13, e14, \\
     & e15, e16, e17, e18, e19, e20, e21, e22, e23 \\
\hline
B & d2, d6, d7, e1, e2, e3, e4, e5, e6, e8n, \\
(20) & e8s, e9, e10, e12, e13, e14, e20, e21, e22, e23 \\
\hline
B-H30 & d2, e1, e2, e3, e4, e5, e6, e8n \\
(8)   & \\
B-H77 & d2, e1, e2, e3, e4, e5, e6, e9 \\
(8)   & \\
B-H30-H77 & d2, e1, e2, e3, e4, e5, e6 \\
(7)   & \\
\enddata 
\end{deluxetable}

\subsection{Observational Data} \label{sec:obs_data}

The VLA Ku-band observations were executed for a total time of 1 hour in D-configuration on March 02, 2013, plus 5 hours in B-configuration on October 01, 2013. The observations were made under program 13A-064 and were originally reported in \citet{Ginsburg16}.
We use the 2-cm continuum and Hydrogen $77\alpha$ RL images and refer the reader to their paper for details on the observations and data reduction. 
The continuum image has a central frequency $\nu_{\rm 0, cm} = 13.436$ GHz (2.2 cm), an \textit{rms} noise of about $50$ $\mu$\jyb, and a synthesized beam FWHM $0.34\arcsec \times 0.33\arcsec$, $\mathrm{P.A.}=14.8 \degree$. The H77$\alpha$ cube ($\nu_{\rm 0, ~H77} = 14.129$ GHz) was created with uniform weighting, has a channel width of 1.33 \kms, a velocity range from -207.9 \kms to 288.0 \kms, and a synthesized beam FWHM of $0.39\arcsec \times 0.34\arcsec$, $\mathrm{P.A.}=75.6 \degree$. The typical \textit{rms} noise in channels with bright emission is 0.46 m\jyb, whereas in channels free of emission it is about 0.32 m\jyb.

The ALMA observations were executed as part of Cycle 2 project 2013.1.00308.S in two 12m-array configurations. We refer to \citet{Ginsburg17} for details on the data reduction. 
The H$30\alpha$ ($\nu_{\rm 0, H30} = 231.901$ GHz) cube has a channel width of $1.2$ \kms, a velocity range from 25.0 \kms to 93.4 \kms, and a synthesized beam FWHM of $0.32\arcsec \times 0.31\arcsec$, $\mathrm{P.A.} = 50.7 \degree$. The typical \textit{rms} noise in all channels is about $3.53$ m\jyb.

\begin{figure*}[t!]
   \centering
   \includegraphics[scale=0.75]{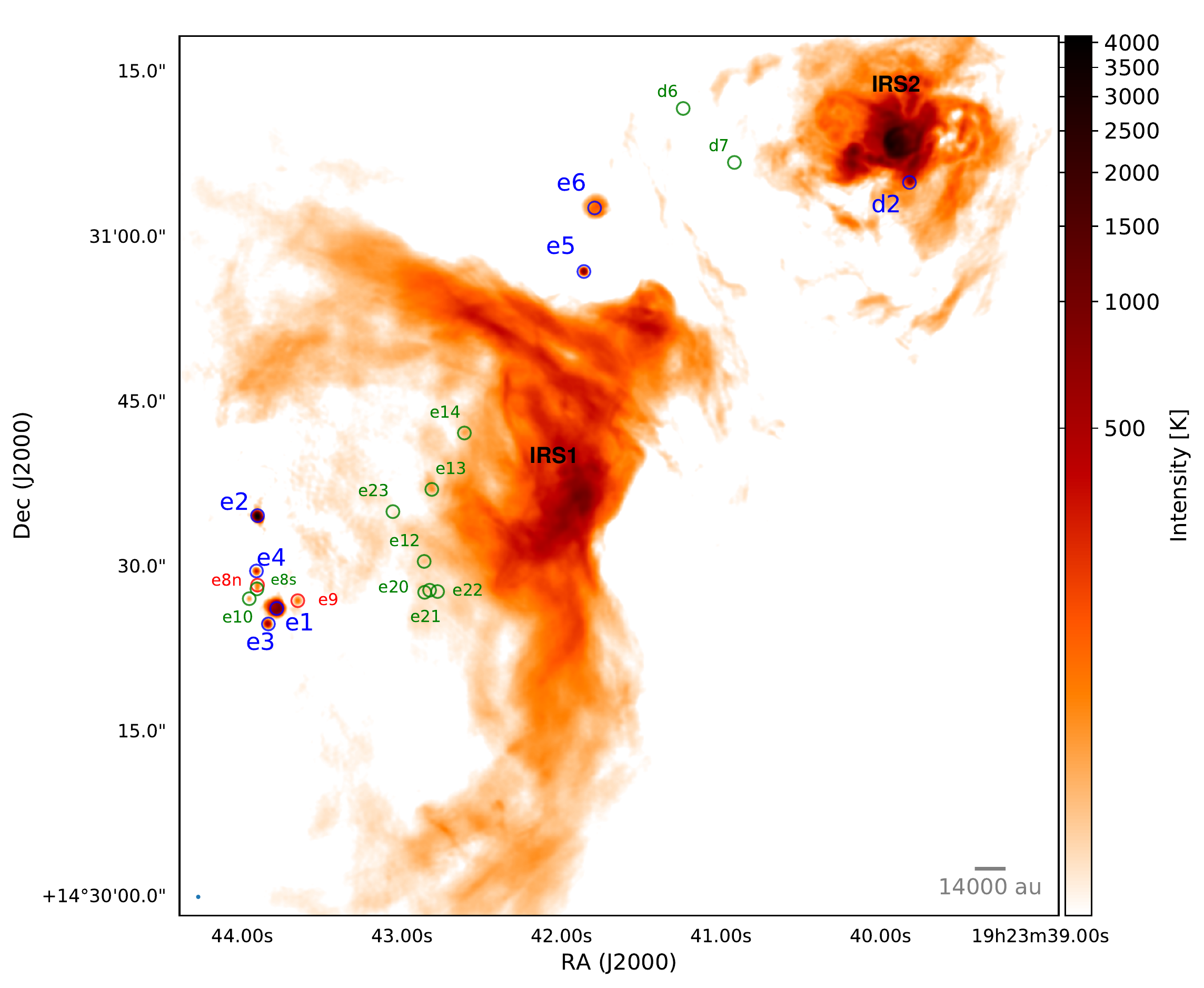} 
    \caption{VLA image of the 2-cm continuum in brightness temperature [K] units. All labeled sources are compact objects in catalog $B$. Sources labeled in \textit{blue} are also detected in both RLs. Sources in \textit{red} are detected only in a single RL. Sources in \textit{green} have no RL detection. The HPBW is $0.40 \arcsec$ and the \textit{rms} noise is about $4.8$ K in locations away from bright emission. The large regions of bright emission known as IRS1 and IRS2 are labeled.} \label{fig:ContImage}
\end{figure*}

To allow for a uniform comparison between tracers of ionized gas, we regridded both the 2-cm continuum and H77$\alpha$ images to pixel dimensions of $0.05 \arcsec$, matching those of the ALMA images, and afterwards convolved the 2-cm continuum, H$77\alpha$, and H$30\alpha$ images to a common circular beam with HPBW$=0.40\arcsec$. We use the convolved images in the analysis of these tracers unless otherwise specified. 

We also use the molecular-line cubes from the data release of \citet{Ginsburg17}. The lines of interest are: 
\formai~ at $\nu_{0} = 218.22219$ GHz, 
\formaii~ at $\nu_{0} = 218.76007$ GHz, 
and \soi~ at $\nu_{0} = 219.94944$ GHz. 
The \formai~ cube has a channel width of 0.17 \kms, a velocity range from 20 \kms to 89 \kms, and a synthesized beam FWHM of $0.68\arcsec \times 0.53\arcsec$, $\mathrm{P.A.}=-66.0 \degree$. The \formaii~ cube has a channel width of 0.67 \kms and the same velocity range and beam size as the previous formaldehyde cube. The \soi~ cube has a channel width of 0.67 \kms, a velocity range from -0.1 \kms to 130.1 \kms, and a synthesized beam FWHM of $1.23\arcsec \times 1.00\arcsec$, $\mathrm{P.A.}=-70.2\degree$. We retrieved atomic and molecular parameters from  \textit{Splatalogue}\footnote{https://www.cv.nrao.edu/php/splat/}, mainly taking the laboratory measurements from the Cologne Database for Molecular Spectroscopy (CDMS)  \citep{Muller05}.

\subsection{Source Selection}

We produce a sensitivity-limited sample of 
compact cm continuum sources in W51 A. We first impose a size cut to the catalog of \cite{Ginsburg16} and take only the cm-continuum sources with radii $<1\arcsec$, which corresponds to 0.026 pc 
at a distance of $d = 5.4$ kpc \citep{Sato10}. This size limit roughly corresponds to the threshold definition of HC HII's \citep{Kurtz05}. The resulting catalog {\it A} is composed of 29 objects (see table \ref{tab:catalogs}). 
Since we need both cm and mm RLs for part of our analysis, we select from catalog {\it A}
those sources that lie within the smaller ALMA field of view to define a new catalog {\it B}. Then we extracted the 
RL spectra from sources in {\it B} and group the H$77\alpha$ and H$30\alpha$ detections in catalogs {\it B-H77} and {\it B-H30}, respectively. 
The intersection of the previous two catalogs is defined as {\it B-H30-H77}, to which we limit our main RL analysis.

\section{cm Continuum and Recombination Lines} \label{sec:cont_rls}

\begin{deluxetable*}{c|c|c|c|c|c|c}[t!]
\tablecaption{Positions and sizes for sources in catalog B \label{tab:Obs_Params}}
\tablehead{
\colhead{Sources} & \colhead{RA} & \colhead{Dec} & \colhead{$\leftidx{^{\theta_{\rm maj}}_{\theta_{\rm min}}~}{\Theta_{\rm conv}} $} & \colhead{$N_{\rm B, conv}$} & \colhead{$\leftidx{^{\theta_{\rm maj}}_{\theta_{\rm min}}~}{\Theta_{\rm deconv}} $} & \colhead{$D$}\\
\colhead{} & \colhead{[h:m:s]} & \colhead{[\degree : \arcmin : \arcsec]} & \colhead{$[\arcsec]$} & \colhead{} & \colhead{$[\arcsec]$} & \colhead{[pc]}
}
\startdata
d2 & 19:23:39.821 $\pm$ 0.002 & $+$ 14:31:05.03 $\pm$ 0.04 & $\leftidx{^{0.41}_{0.41}~}{0.41 \pm 0.01}$ & 1.06 $\pm$ 0.01 & $\leftidx{^{0.1}_{0.1}~}{0.1\pm 0.01}$ & 0.0026 $\pm$ 0.0001\\
d6 & 19:23:41.238 $\pm$ 0.001 & $+$ 14:31:11.58 $\pm$ 0.01 & -- & -- & -- & -- \\
d7 & 19:23:40.919 $\pm$ 0.001 & $+$ 14:31:06.58 $\pm$ 0.02 & $\leftidx{^{0.85}_{0.53}~}{0.67 \pm 0.02}$ & 2.83 $\pm$ 0.20 & $\leftidx{^{0.75}_{0.35}~}{0.51\pm 0.03}$ & 0.0135 $\pm$ 0.0009\\
e1 & 19:23:43.785 $\pm$ 0.002 & $+$ 14:30:26.11 $\pm$ 0.03 & $\leftidx{^{1.04}_{1.03}~}{1.03 \pm 0.04}$ & 6.69 $\pm$ 0.55 & $\leftidx{^{0.96}_{0.94}~}{0.95\pm 0.05}$ & 0.0250 $\pm$ 0.0012\\
e2 & 19:23:43.906 $\pm$ 0.0002 & $+$ 14:30:34.48 $\pm$ 0.003 & $\leftidx{^{0.56}_{0.47}~}{0.51 \pm 0.01}$ & 1.65 $\pm$ 0.03 & $\leftidx{^{0.40}_{0.24}~}{0.31\pm 0.01}$ & 0.0082 $\pm$ 0.0002\\
e3 & 19:23:43.842 $\pm$ 0.0001 & $+$ 14:30:24.72 $\pm$ 0.002 & $\leftidx{^{0.51}_{0.47}~}{0.49 \pm 0.01}$ & 1.52 $\pm$ 0.03 & $\leftidx{^{0.32}_{0.25}~}{0.29\pm 0.01}$ & 0.0075 $\pm$ 0.0002\\
e4 & 19:23:43.913 $\pm$ 0.0001 & $+$ 14::30:29.49 $\pm$ 0.002 & $\leftidx{^{0.43}_{0.42}~}{0.43 \pm 0.01}$ & 1.14 $\pm$ 0.02 & $\leftidx{^{0.17}_{0.13}~}{0.15\pm 0.01}$ & 0.0039 $\pm$ 0.0003\\
e5 & 19:23:41.863 $\pm$ 0.0001 & $+$ 14:30:56.73 $\pm$ 0.001 & $\leftidx{^{0.44}_{0.43}~}{0.43 \pm 0.01}$ & 1.18 $\pm$ 0.01 & $\leftidx{^{0.19}_{0.15}~}{0.17 \pm 0.01}$ & 0.0044 $\pm$ 0.0001\\
e6 & 19:23:41.785 $\pm$ 0.002 & $+$ 14:31:02.56 $\pm$ 0.03 & $\leftidx{^{1.55}_{1.46}~}{1.50 \pm 0.05}$ & 14.12 $\pm$ 0.89 & $\leftidx{^{1.50}_{1.40}~}{1.45\pm 0.05}$ & 0.0380 $\pm$ 0.0013\\
e8n & 19:23:43.906 $\pm$ 0.001 & $+$ 14:30:28.17 $\pm$ 0.02 & -- & -- & -- & -- \\
e8s & 19:23:43.907 $\pm$ 0.0003 & $+$ 14:30:27.91 $\pm$ 0.01 & -- & -- & -- & -- \\
e9 & 19:23:43.654 $\pm$ 0.001 & $+$ 14:30:26.81 $\pm$ 0.02 & $\leftidx{^{0.68}_{0.61}~}{0.64 \pm 0.02}$ & 2.58 $\pm$ 0.15 & $\leftidx{^{0.54}_{0.46}~}{0.50 \pm 0.02}$  & 0.0132 $\pm$ 0.0006\\
e10 & 19:23:43.958 $\pm$ 0.0003 & $+$ 14:30:26.98 $\pm$ 0.005 & $\leftidx{^{0.45}_{0.43}~}{0.44 \pm 0.01}$ & 1.23 $\pm$ 0.04 & $\leftidx{^{0.21}_{0.17}~}{0.19 \pm 0.02}$  & 0.0049 $\pm$ 0.0005\\
e12 & 19:23:42.861 $\pm$ 0.001 & $+$ 14:30:30.41 $\pm$ 0.01 & -- & -- & -- & -- \\
e13 & 19:23:42.819 $\pm$ 0.002 & $+$ 14:30:37.11 $\pm$ 0.03 & -- & -- & -- & -- \\
e14 & 19:23:42.605 $\pm$ 0.001 & $+$ 14:30:42.11 $\pm$ 0.02 & -- & -- & -- & -- \\
e20 & 19:23:42.857 $\pm$ 0.001 & $+$ 14:30:27.72 $\pm$ 0.01 & -- & -- & -- & -- \\
e21 & 19:23:42.848 $\pm$ 0.001 & $+$ 14:30:27.69 $\pm$ 0.01 & -- & -- & -- & -- \\
e22 & 19:23:42.781 $\pm$ 0.0004 & $+$ 14:30:27.67 $\pm$ 0.01 & -- & -- & -- & -- \\
e23 & 19:23:43.058 $\pm$ 0.0004 & $+$ 14:30:34.92 $\pm$ 0.01 & -- & -- & -- & -- \\
\enddata
\tablecomments{
Observational parameters obtained from Gaussian fitting of the 2-cm sources. These are the position coordinates in J2000 RA and DEC, the convolved $\Theta_{\rm conv}$ and deconvolved $\Theta_{\rm deconv}$ FWHM sizes  -- with the  corresponding major $\theta_{\rm maj}$ and minor $\theta_{\rm min}$ axis components as a super- or subscript, the convolved size in beam units $N_{\rm B, conv}$, and the deconvolved diameter $D$. The ten objects marked with dashes are those without valid fits because they are  significantly mixed with their background.}
\end{deluxetable*}

\begin{deluxetable*}{c|c|c|c|c|c|c|c}[t!]
\tablecaption{Physical parameters from 2-cm continuum}
\tablehead{
\colhead{Sources} & \colhead{$I_{\rm pk}$} & \colhead{$S$} & \colhead{$S_{\rm -bg}$} & \colhead{$\tau_{\rm c}$} & \colhead{$\rm EM$} & \colhead{$n_{\rm e, c}$} & \colhead{$\log_{10}(L_{\rm c})$} \\
\colhead{} & \colhead{[m\jyb]} & \colhead{[mJy]} & \colhead{[mJy]} & \colhead{} & \colhead{[$10^{8}$ pc cm$^{-6}$]} & \colhead{[$10^{5}$ cm$^{-3}$]} & \colhead{$\rm [s^{-1}]$}
}
\startdata
d2$^a$ & 19.50 $\pm$ 1.00 & 19.50 $\pm$ 1.00 & 14.73 $\pm$ 1.00 & $>$ 1.0 & $>$ 4.84 & $>$ 4.30 & $>$ 46.01 \\
d6 & 0.26 $\pm$ 0.05 & 0.26 $\pm$ 0.05 & 0.23 $\pm$ 0.07 & -- & -- & -- & -- \\
d7 & 0.43 $\pm$ 0.02 & 1.22 $\pm$ 0.08 & 0.99 $\pm$ 0.16 & $<$ 0.01 & 0.02 & 0.11 $\pm$ 0.01 & 44.96 $\pm$ 0.11\\
e1 & 26.90 $\pm$ 1.60 & 180.00 $\pm$ 12.00 & 178.10 $\pm$ 12.01 & 0.19 & 0.94 $\pm$ 0.12 & 0.61 $\pm$ 0.04 & 47.25 $\pm$ 0.07\\
e2$^a$ & 108.00 $\pm$ 1.20 & 178.20 $\pm$ 3.00 & 177.78 $\pm$ 3.00 & $>$ 1.0 & $>$ 4.84 & $>$ 2.43 & $>$ 46.99 \\
e3 & 15.44 $\pm$ 0.21 & 23.46 $\pm$ 0.48 & 23.11 $\pm$ 0.49 & 0.30 & 1.43 $\pm$ 0.10 & 1.39 $\pm$ 0.05 & 46.39 $\pm$ 0.04\\
e4 & 8.78 $\pm$ 0.10 & 9.99 $\pm$ 0.19 & 9.96 $\pm$ 0.20 & 0.54 & 2.62 $\pm$ 0.55 & 2.61 $\pm$ 0.29 & 46.08 $\pm$ 0.11\\
e5$^a$ & 26.27 $\pm$ 0.15 & 31.05 $\pm$ 0.29 & 31.17 $\pm$ 0.30 & $>$ 1.0 & $>$ 4.84 & $>$ 3.31 & $>$ 46.46 \\
e6 & 4.50 $\pm$ 0.20 & 63.6 $\pm$ 3.0 & 63.00 $\pm$ 3.08 & 0.03 & 0.13 $\pm$ 0.02 & 0.19 $\pm$ 0.01 & 46.77 $\pm$ 0.05\\
e8n& 2.68 $\pm$ 0.05 & 2.68 $\pm$ 0.05 & 2.55 $\pm$ 0.07 & -- & -- & -- & -- \\
e8s & 2.38 $\pm$ 0.05 & 2.38 $\pm$ 0.05 & 2.25 $\pm$ 0.07 & -- & -- & -- & -- \\
e9 & 2.86 $\pm$ 0.12 & 7.39 $\pm$ 0.41 & 6.23 $\pm$ 0.43 & 0.02 & 0.11 $\pm$ 0.01 & 0.29 $\pm$ 0.02 & 45.76 $\pm$ 0.07\\
e10 & 2.02 $\pm$ 0.04 & 2.48 $\pm$ 0.09 & 2.47 $\pm$ 0.11 & 0.07 & 0.31 $\pm$ 0.06 & 0.80 $\pm$ 0.09 & 45.37 $\pm$ 0.12\\
e12 & 1.21 $\pm$ 0.05 & 1.21 $\pm$ 0.05 & 0.35 $\pm$ 0.07 & -- & -- & -- & -- \\
e13 & 2.19 $\pm$ 0.05 & 2.19 $\pm$ 0.05 & 1.46 $\pm$ 0.07 & -- & -- & -- & -- \\
e14 & 1.82 $\pm$ 0.05 & 1.82 $\pm$ 0.05 & 0.75 $\pm$ 0.07 & -- & -- & -- & -- \\
e20 & 1.00 $\pm$ 0.05 & 1.00 $\pm$ 0.05 & 0.32 $\pm$ 0.07 & -- & -- & -- & -- \\
e21 & 0.95 $\pm$ 0.05 & 0.95 $\pm$ 0.05 & 0.27 $\pm$ 0.07 & -- & -- & -- & -- \\
e22 & 0.75 $\pm$ 0.05 & 0.75 $\pm$ 0.05 & 0.10 $\pm$ 0.07 & -- & -- & -- & -- \\
e23 & 0.60 $\pm$ 0.05 & 0.60 $\pm$ 0.05 & 0.08 $\pm$ 0.07 & -- & -- & -- & -- \\
\enddata
\tablecomments{
Values of 2-cm peak intensity $I_{\rm pk}$, flux density $S$ prior to background subtraction, flux density $S_{\rm -bg}$ after background subtraction, optical depth $\tau_{\rm c}$, emission measure $\rm EM$, continuum derived electron density $n_{\rm e, c}$, and Lyman photon rate $\log_{10}(L_{\rm c})$. $^a$ d2, e2 , and e5 have $T_{\rm B}$ > $T_{\rm e} = 7500$ K, so we set their $\tau_{\rm c} > 1$. 
The ten objects marked with dashes are those without valid fits because they are  significantly mixed with their background.}
\label{tab:Cont_Params}
\end{deluxetable*}

\begin{figure*}[t!]
   \setlength{\lineskip}{0pt}
   \centering
   \includegraphics[scale=0.46]{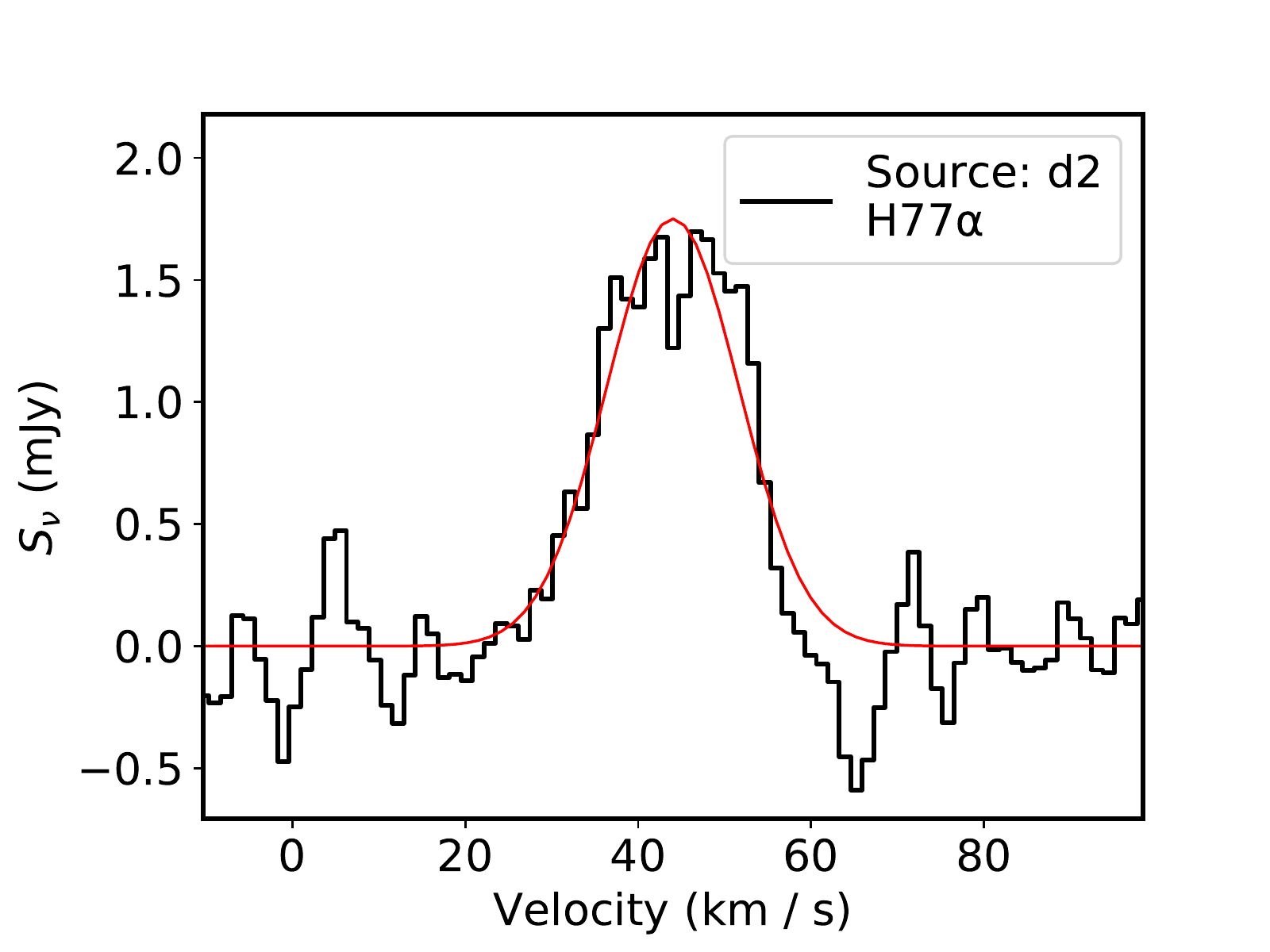}
   \includegraphics[scale=0.46]{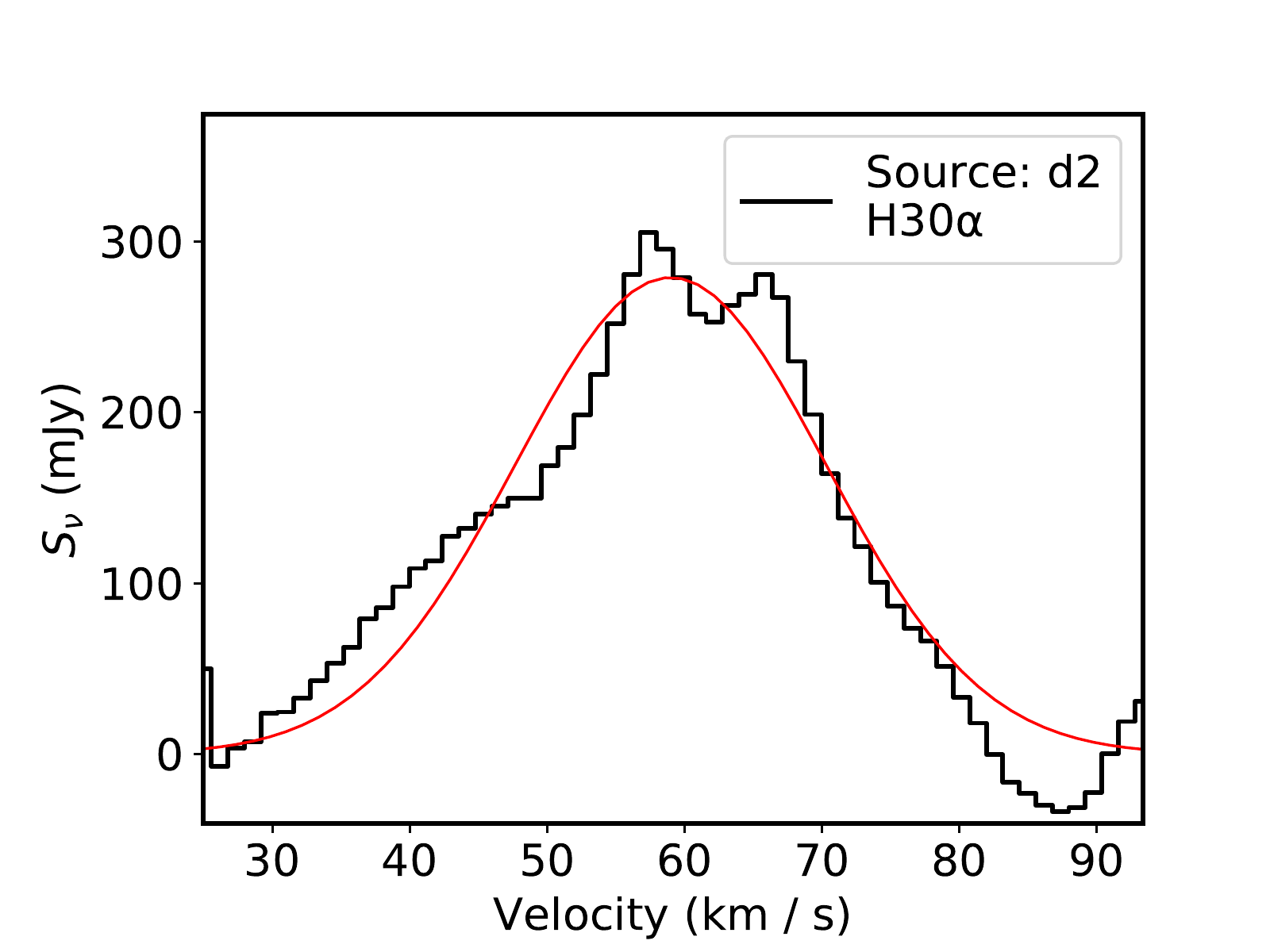}
   \includegraphics[scale=0.46]{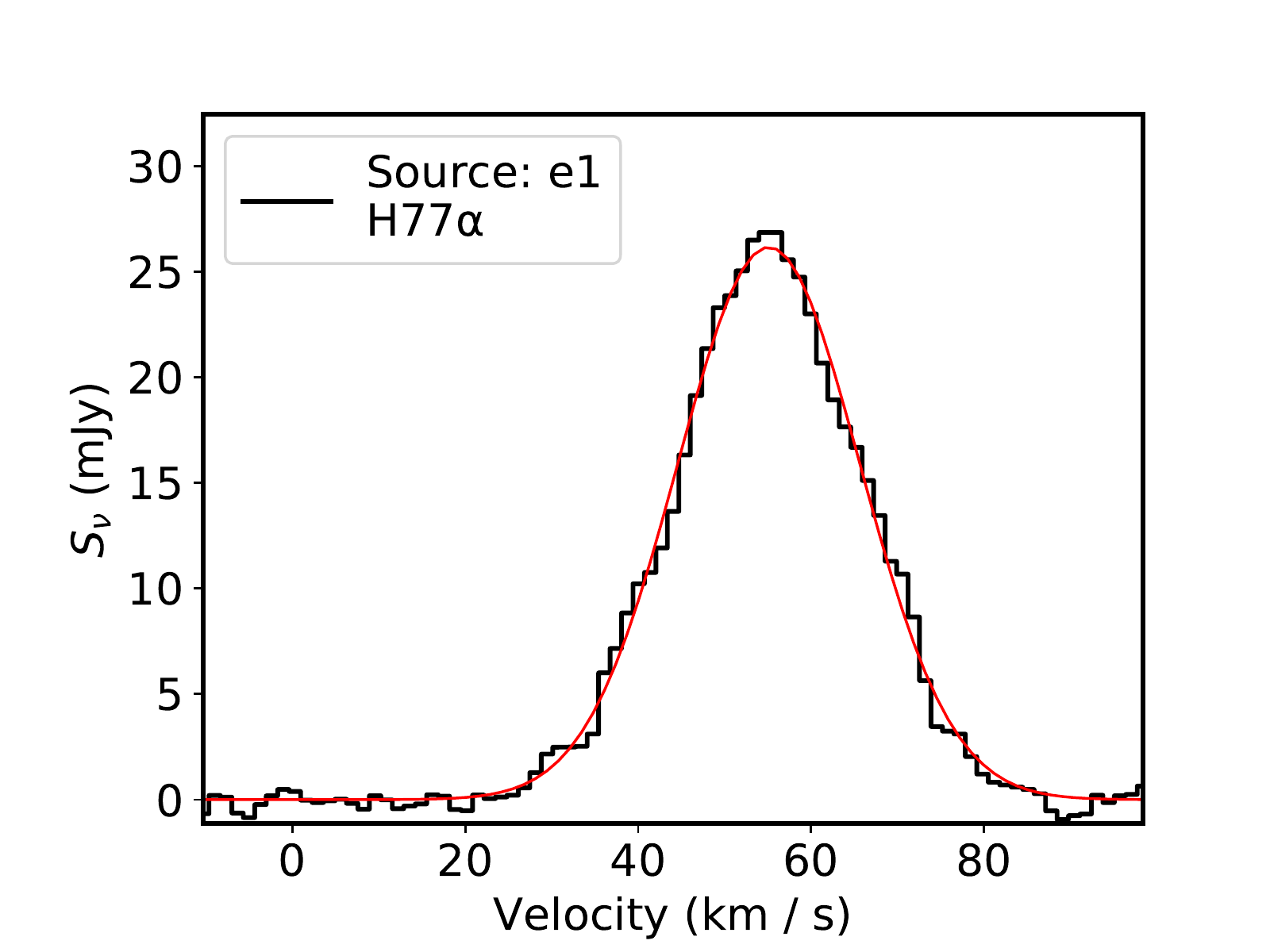} 
   \includegraphics[scale=0.46]{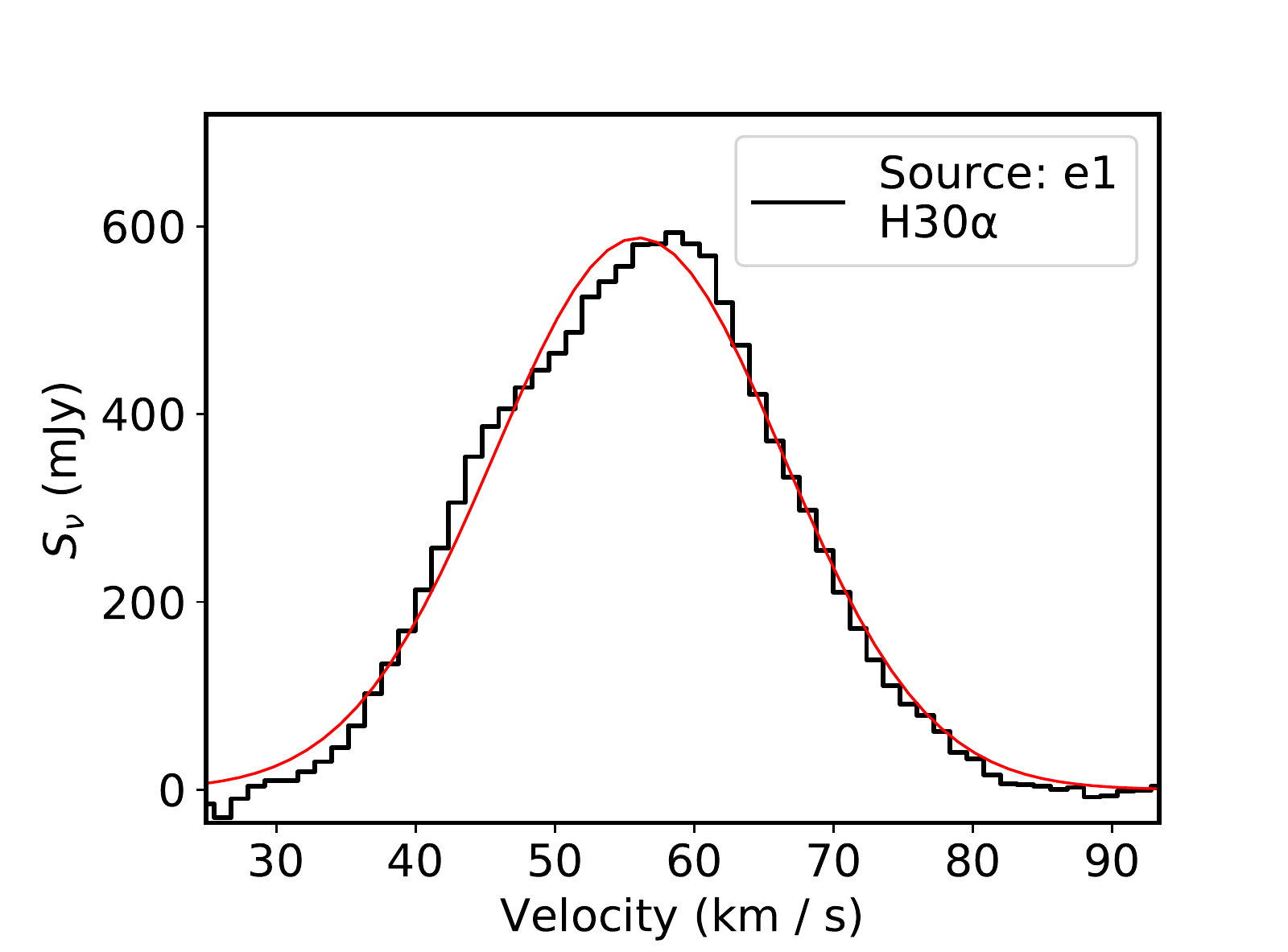} 
   \includegraphics[scale=0.46]{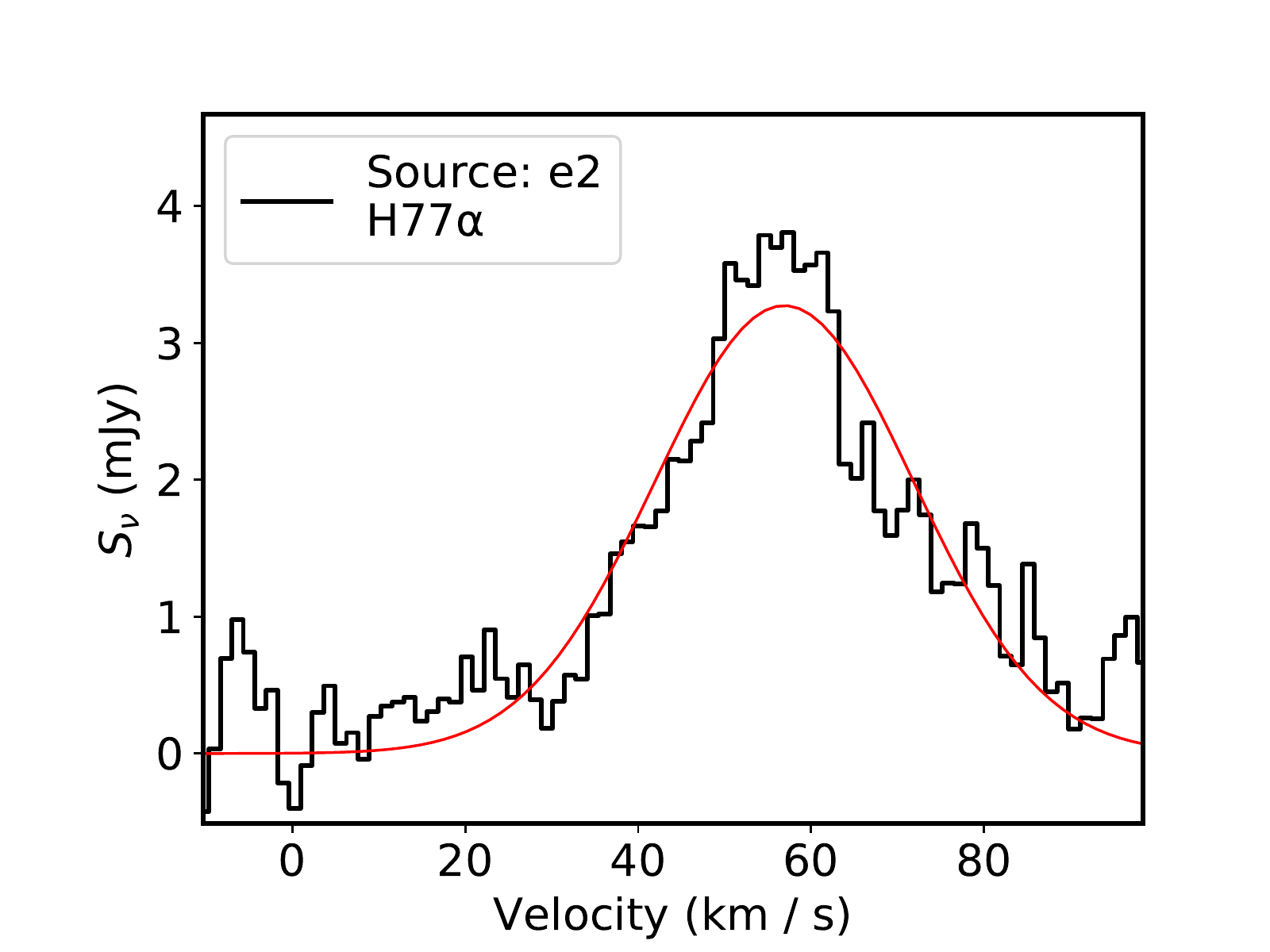}
   \includegraphics[scale=0.46]{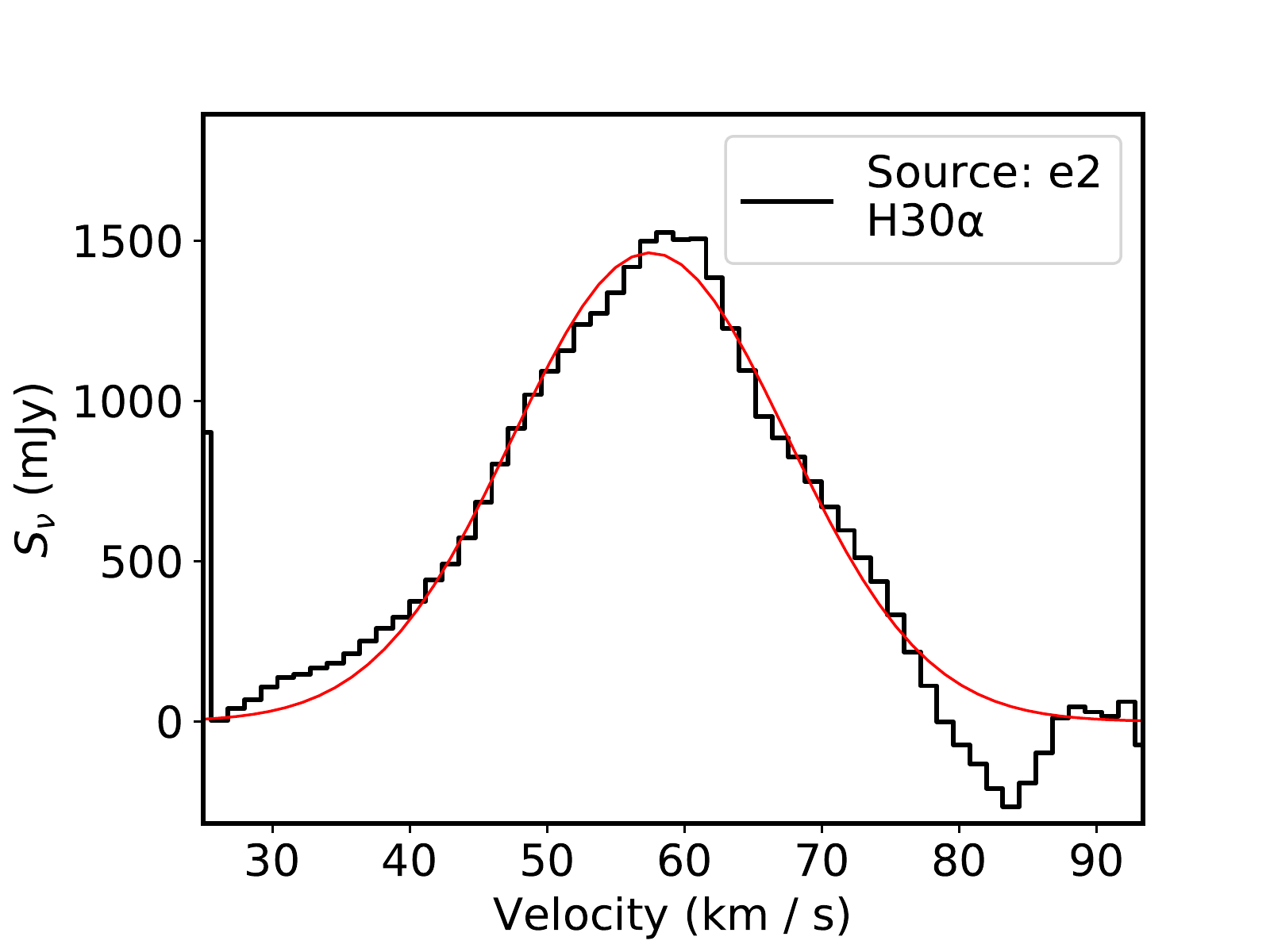}
   \includegraphics[scale=0.46]{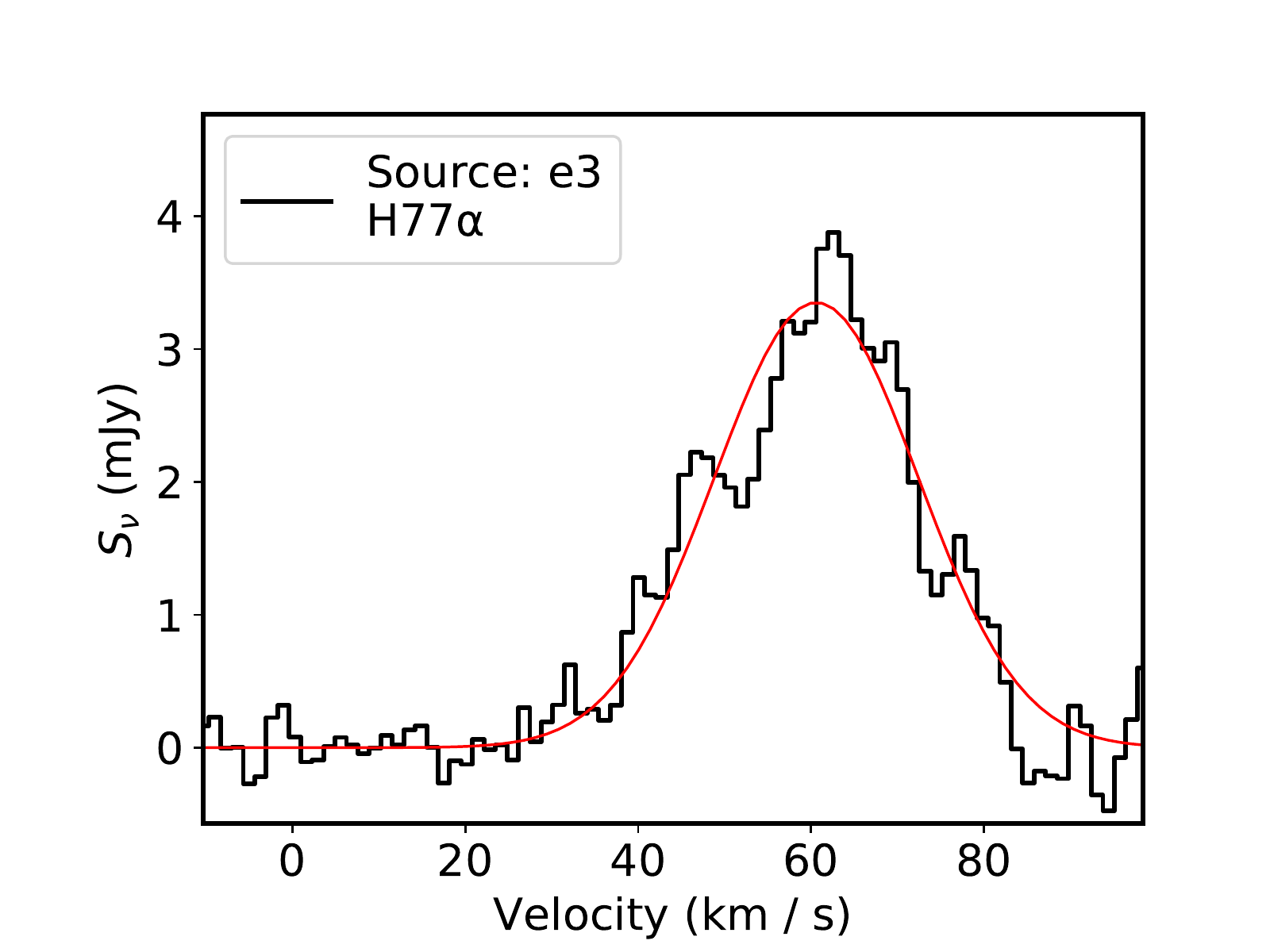}
   \includegraphics[scale=0.46]{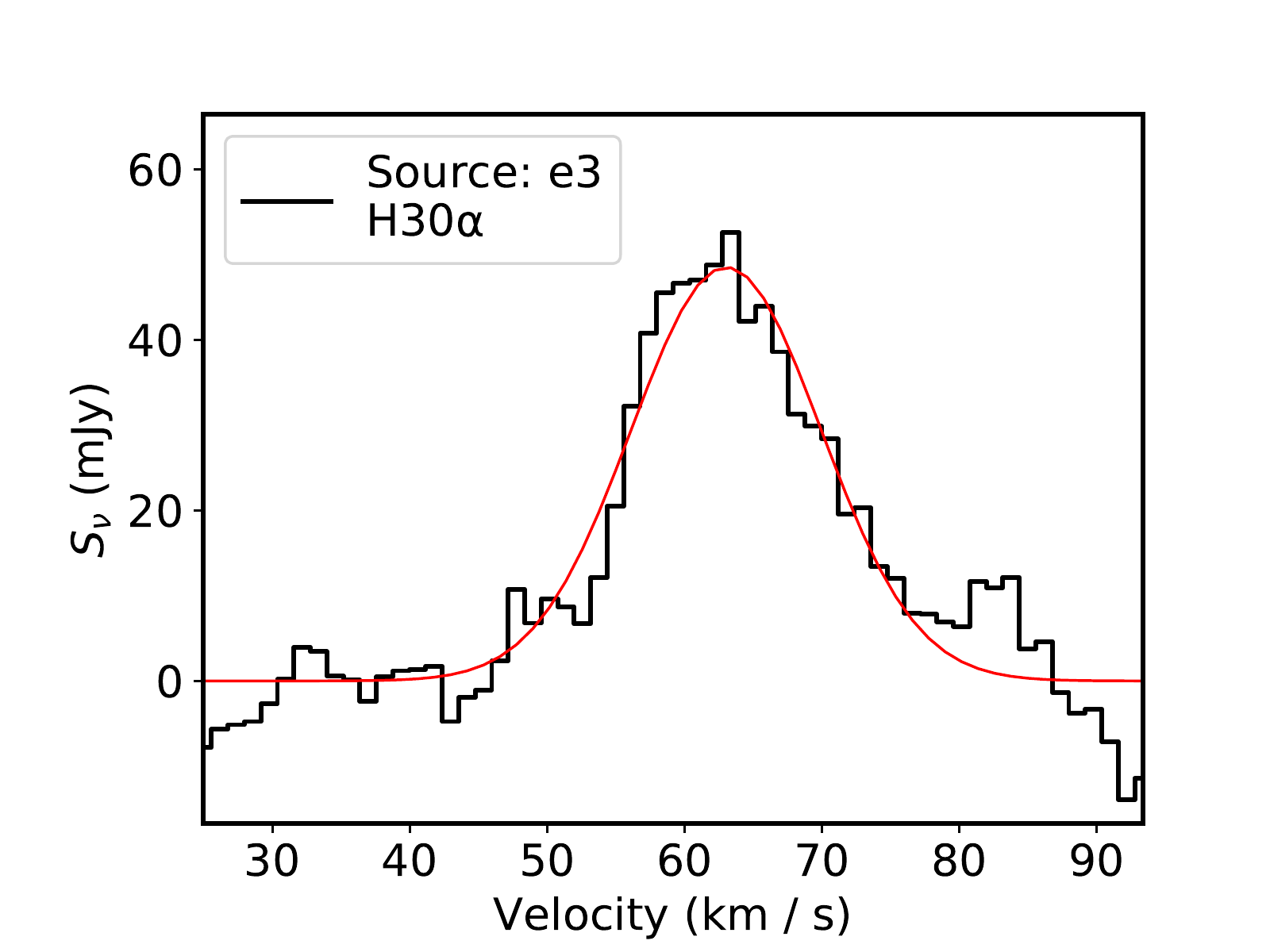}
    \caption{H77$\alpha$ (left) and H30$\alpha$ (right) RL spectra for sources d2, e1, e2, e3.  Gaussian fits are indicated by the red line.} \label{fig:RL_detections1}
\end{figure*}

\begin{figure*}[t!]
   \ContinuedFloat
   \setlength{\lineskip}{0pt}
   \centering
   \includegraphics[scale=0.46]{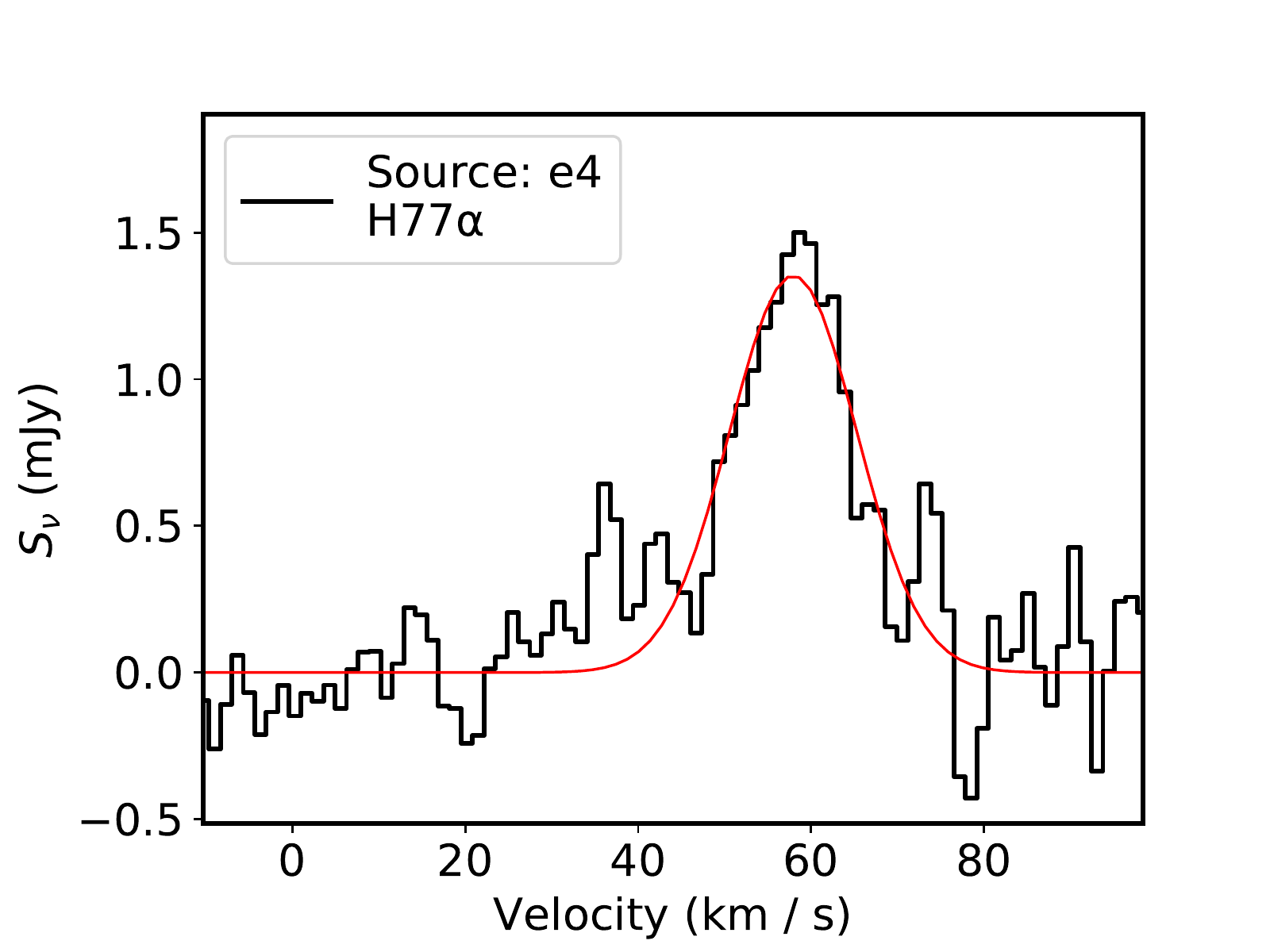}
   \includegraphics[scale=0.46]{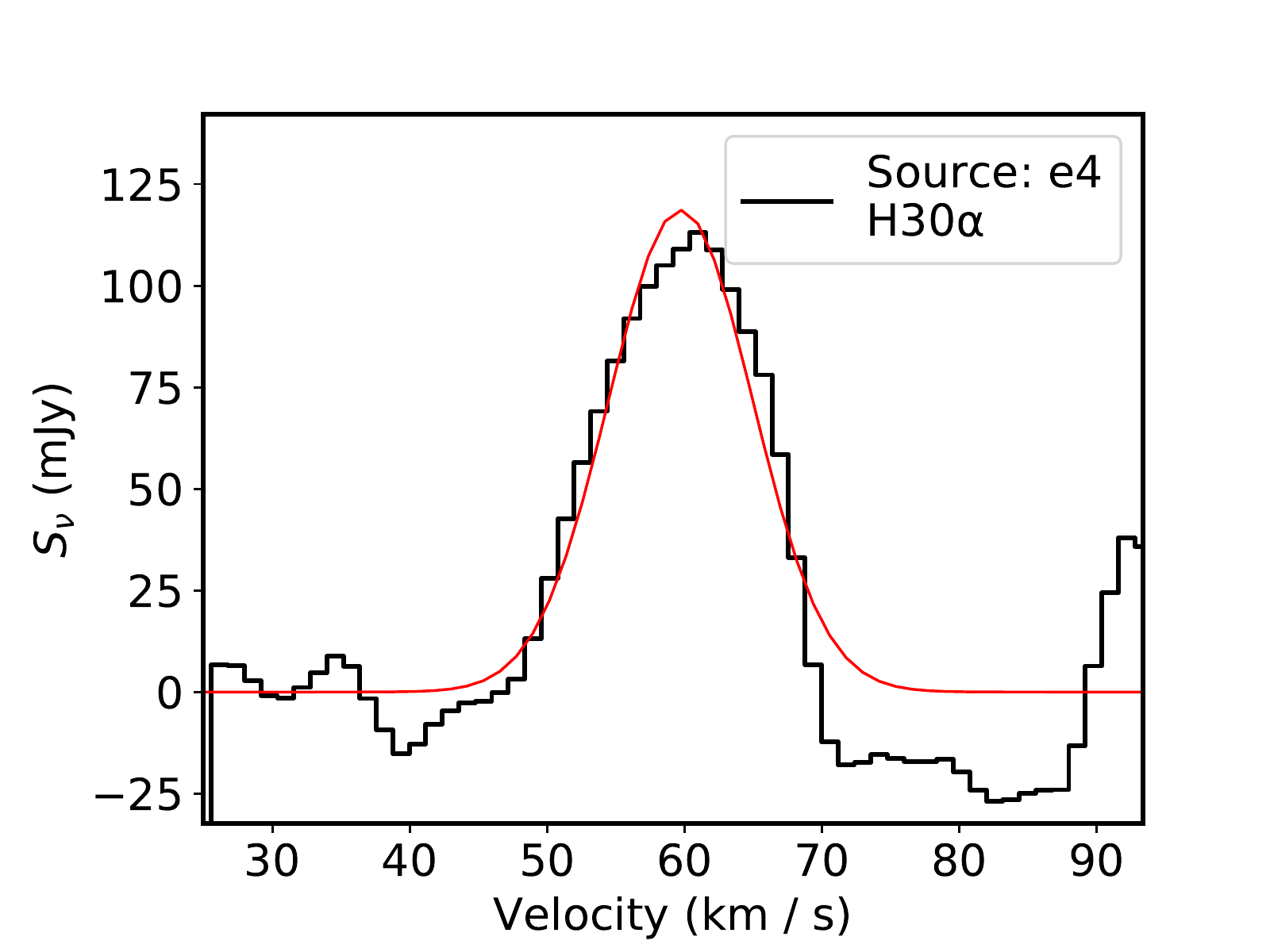}
   \includegraphics[scale=0.46]{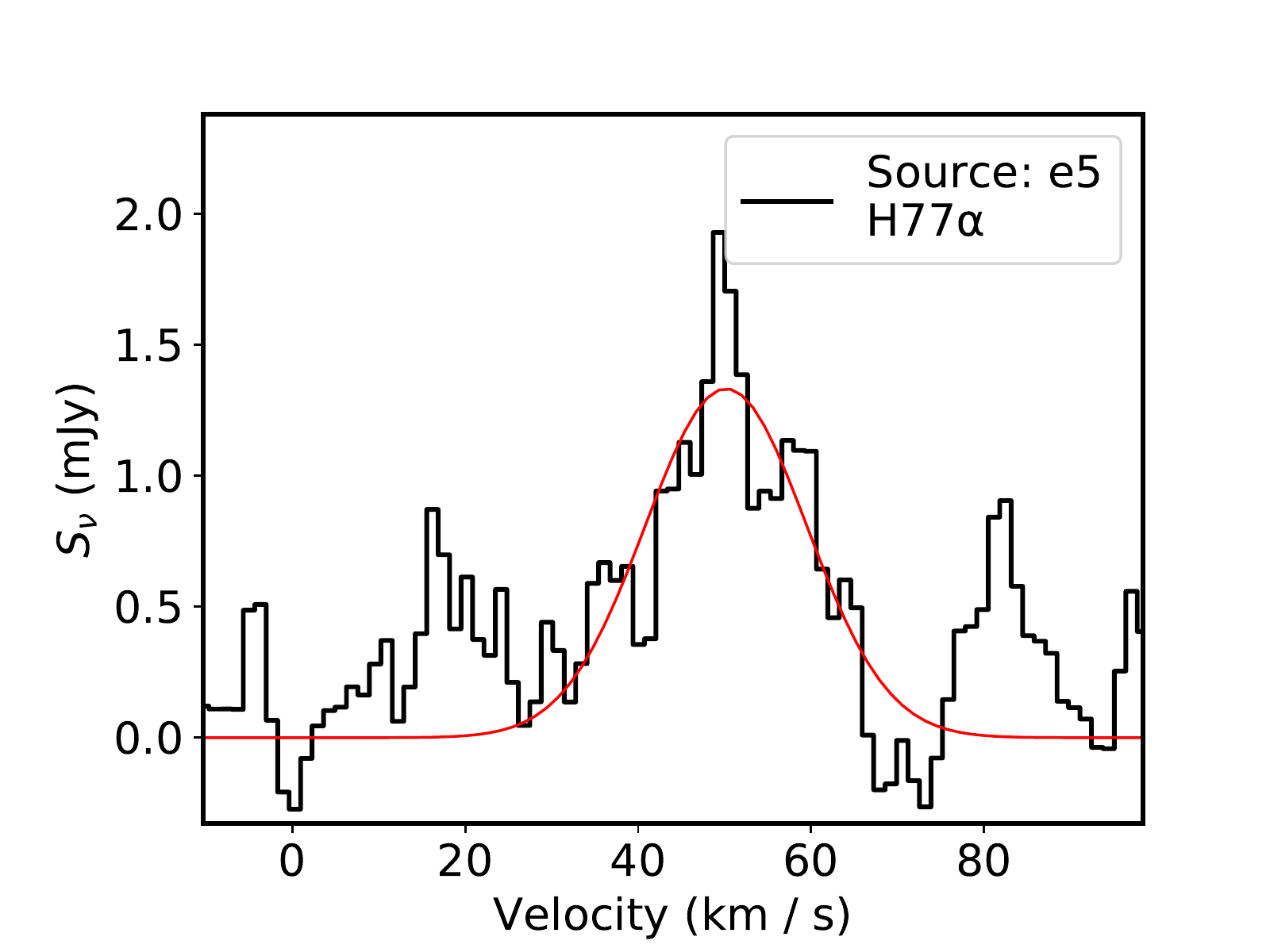}
   \includegraphics[scale=0.46]{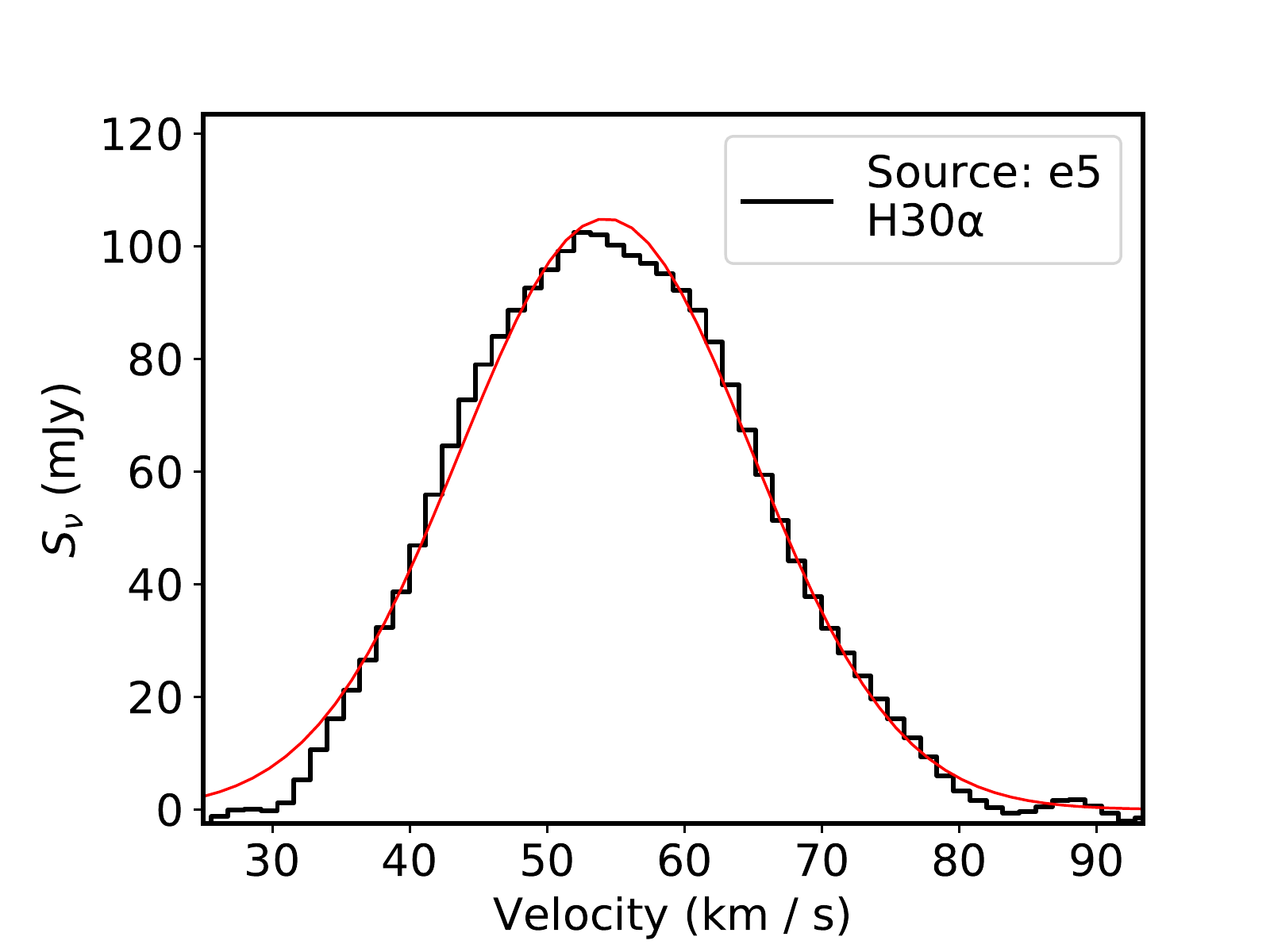}
   \includegraphics[scale=0.46]{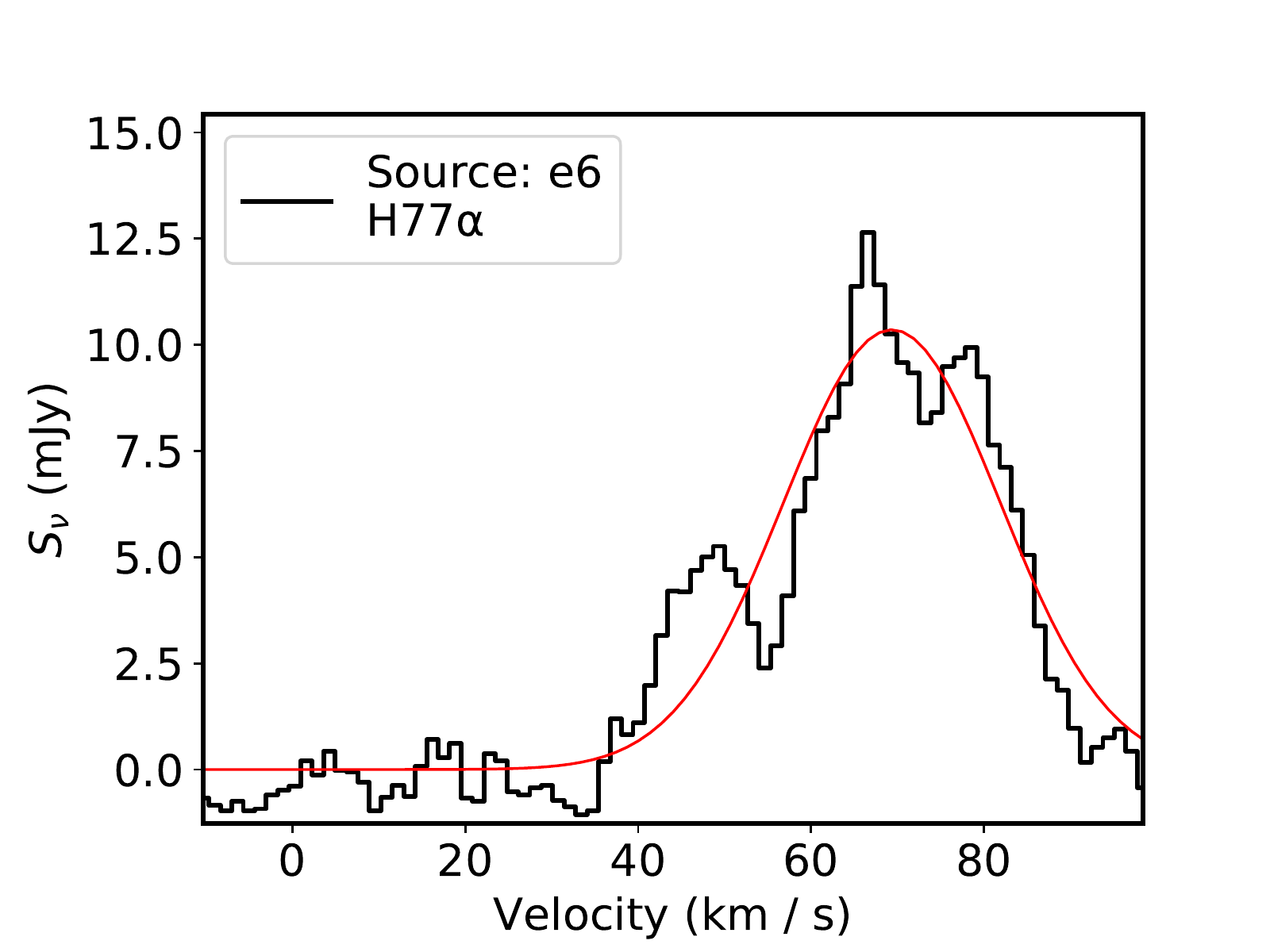}
   \includegraphics[scale=0.46]{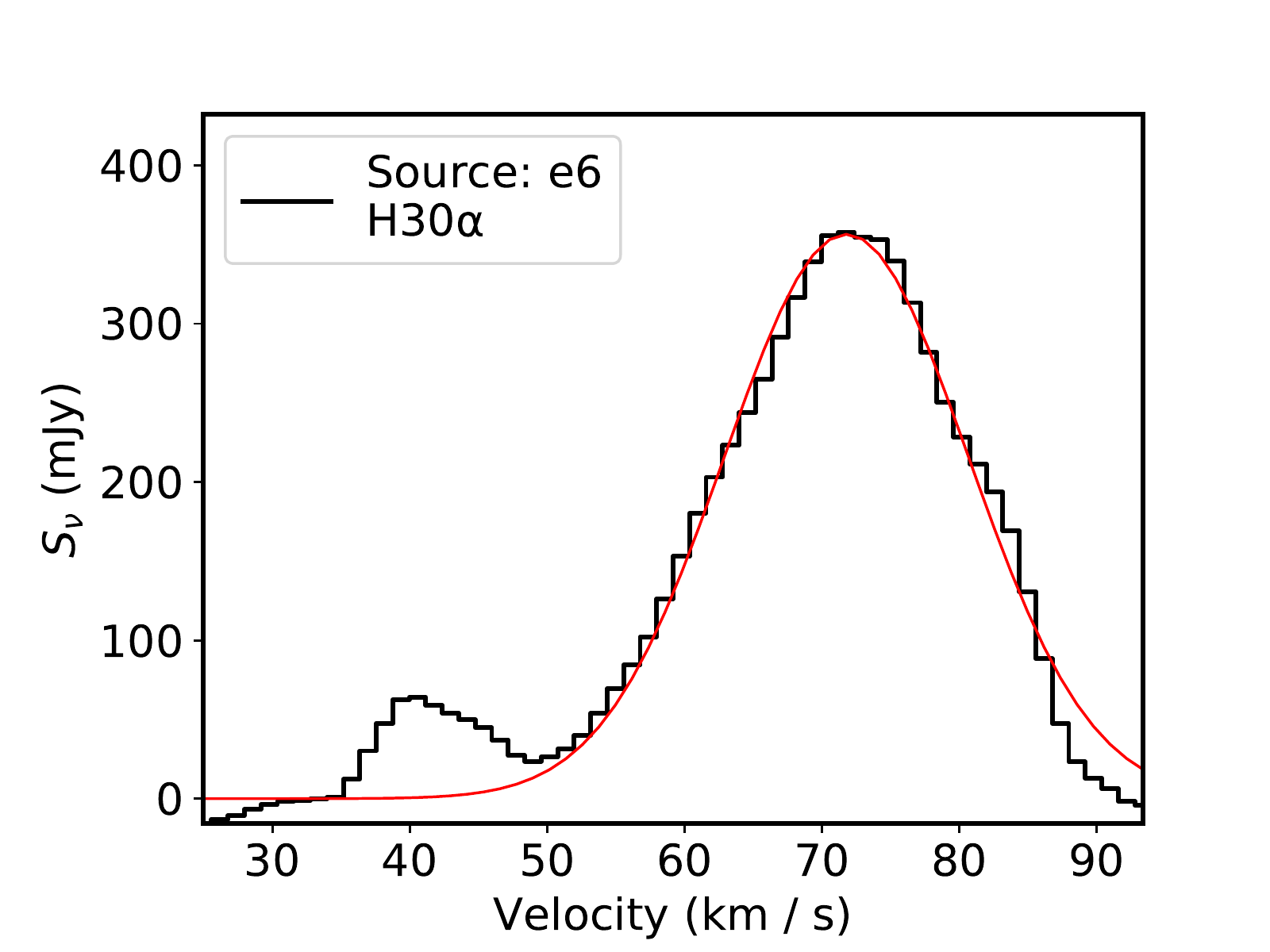}
   \includegraphics[scale=0.46]{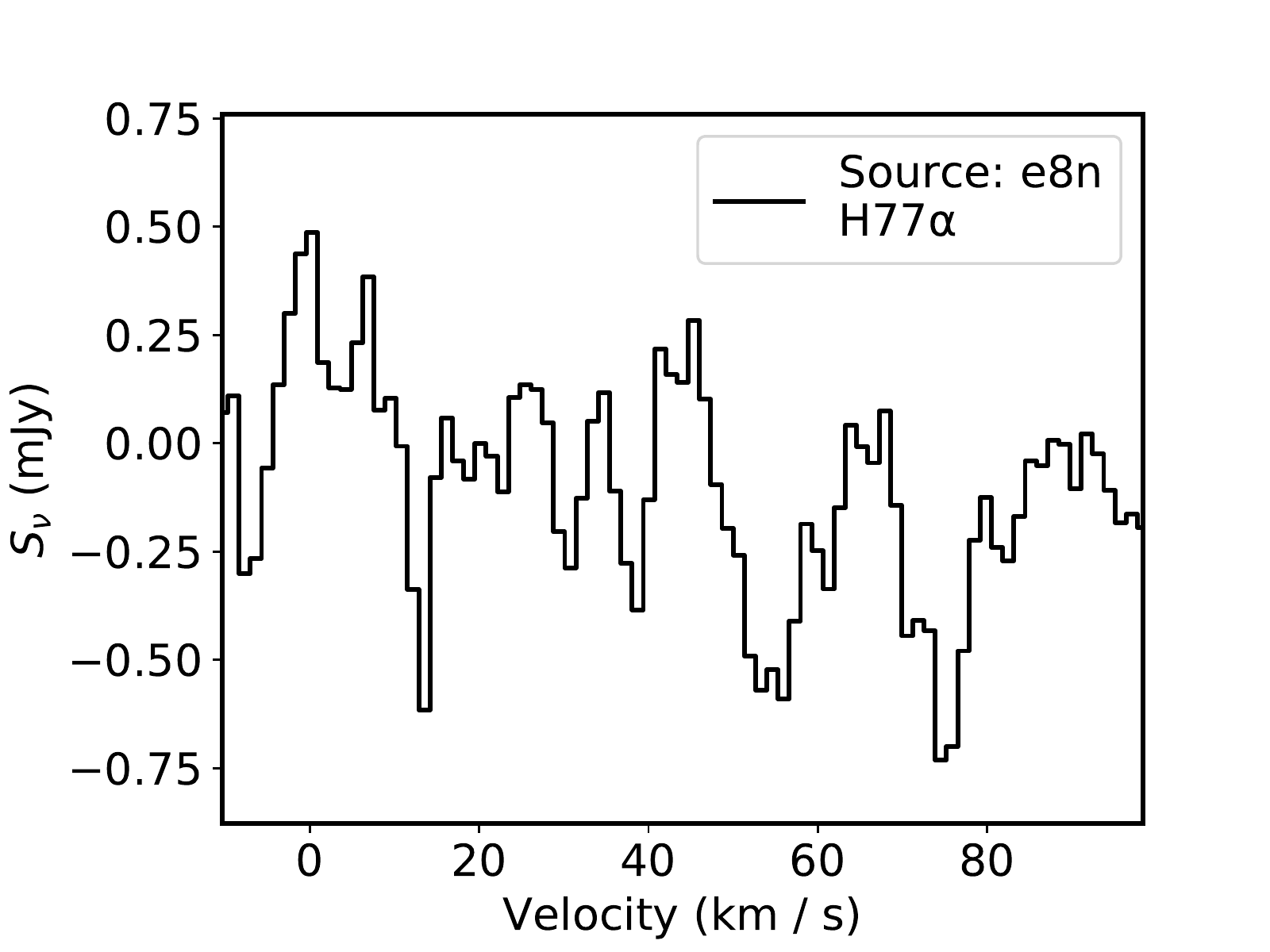}
   \includegraphics[scale=0.46]{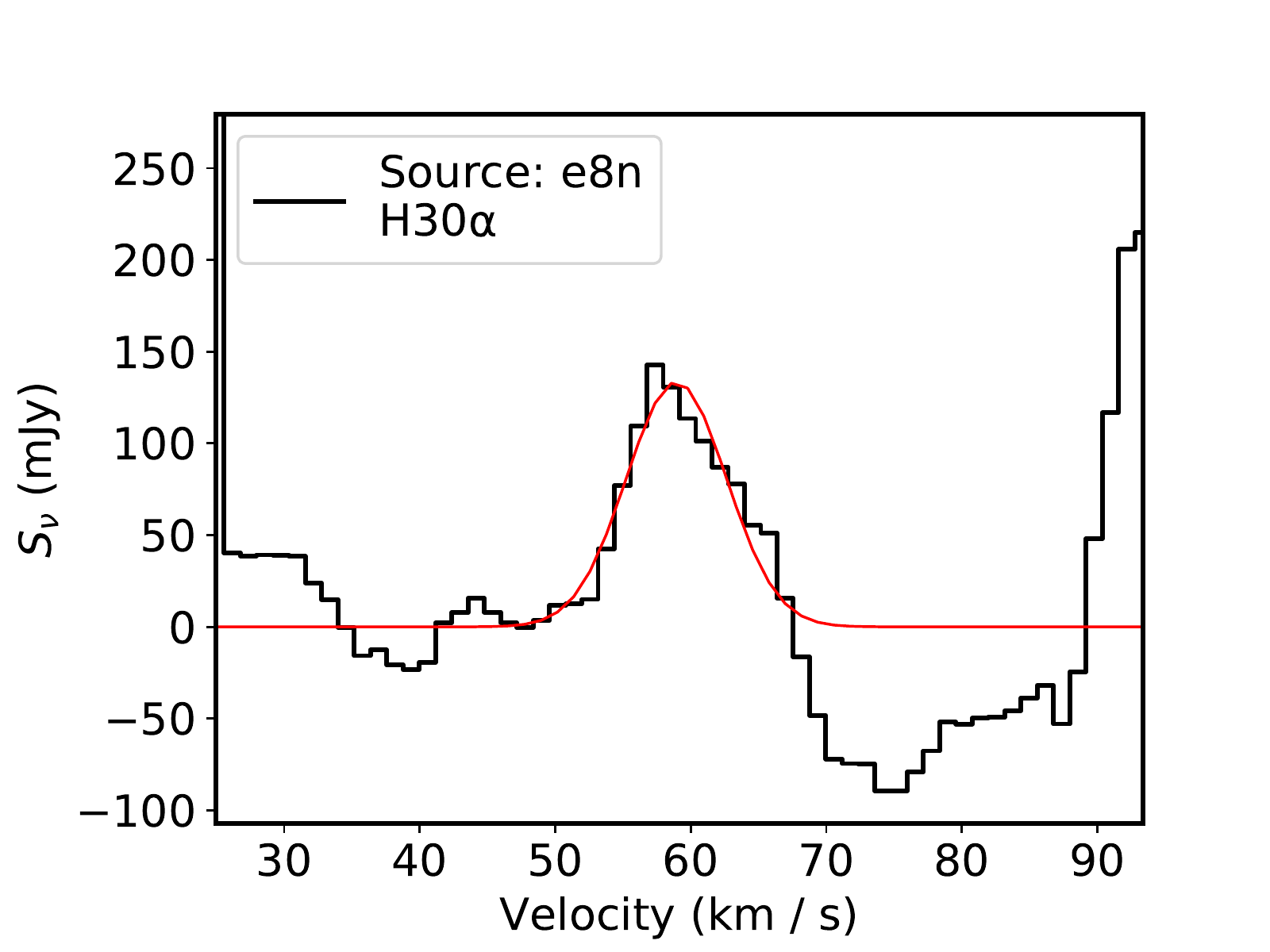}
    \caption{\textit{contd.} H77$\alpha$ (left) and H30$\alpha$ (right) RL spectra for sources e4, e5, e6, and e8n. Gaussian fits are indicated by the red line.}
\end{figure*}

\begin{figure*}[ht!]
   \ContinuedFloat
   \centering
   \includegraphics[scale=0.48]{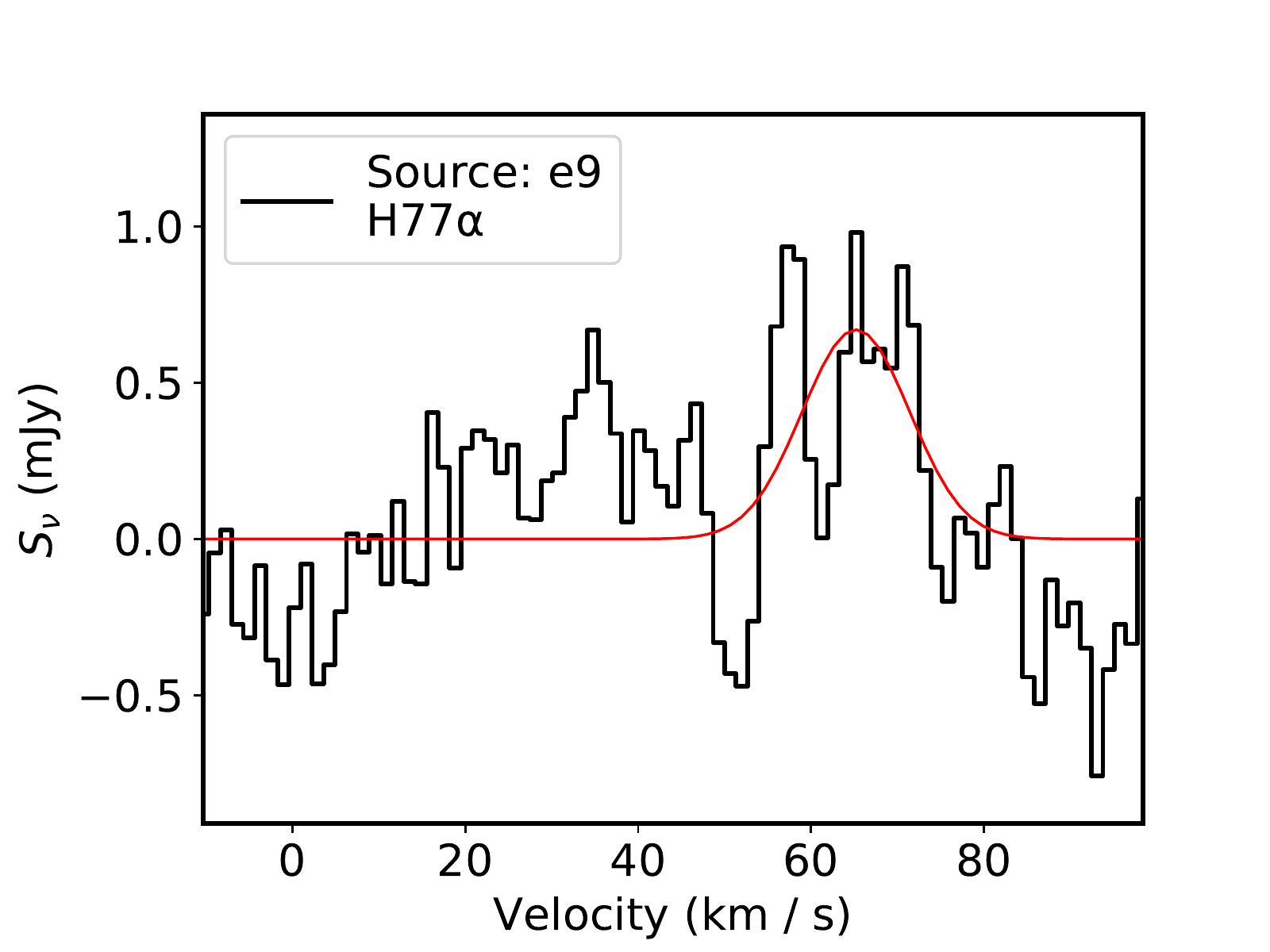}
   \includegraphics[scale=0.48]{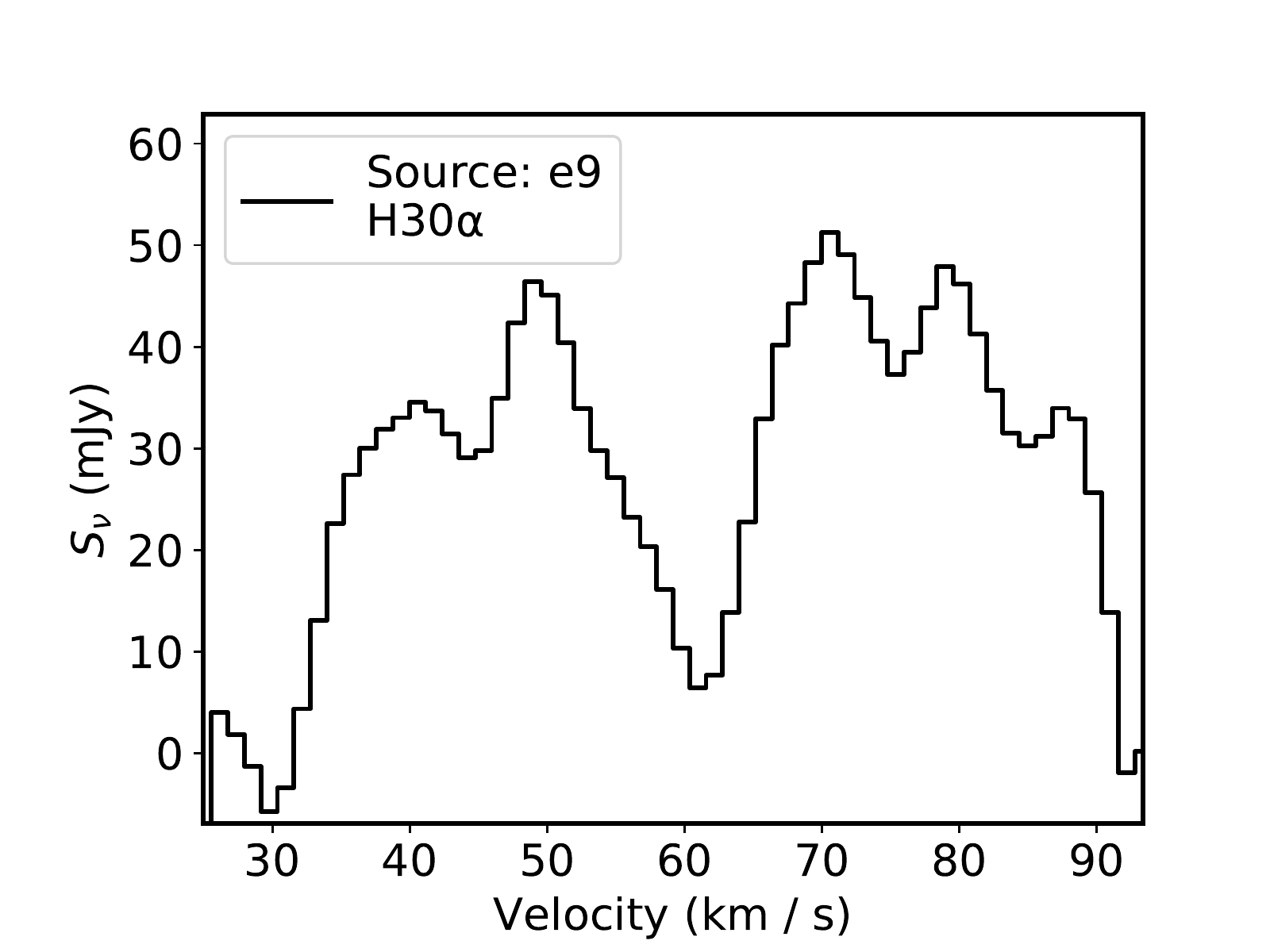}
    \caption{\textit{contd.} H77$\alpha$ (left) and H30$\alpha$ (right) RL spectra for source e9. Gaussian fit is indicated by the red line.}
\end{figure*}

\subsection{2-\textit{cm} Continuum}
The positional and size parameters obtained for the sources in catalog \textit{B} are detailed in table \ref{tab:Obs_Params}, while the properties derived from the cm continuum are given in table \ref{tab:Cont_Params}. We performed an interactive 2D Gaussian fitting in CASA v.5.3.0-143 \citep{McMullin07} on each source to obtain their FWHM convolved and deconvolved size components along their major $\theta_{\rm maj}$ and minor $\theta_{\rm min}$ axes. From these we define our convolved and deconvolved sizes as the respective geometric mean of the  components (Eq. \eqref{FWHM}). Then we calculate the convolved source size in beam units $N_{\rm B, conv}$ (Eq. \eqref{Nbeams_conv}) and the deconvolved physical $D$ size in pc (Eq. \eqref{Diameter}): 

\begin{subequations}
\begin{equation}
\label{FWHM}
    \Theta = \sqrt{\theta_{\rm maj} \theta_{\rm min}}, 
\end{equation}
\begin{equation}
\label{Nbeams_conv}
    N_{\rm B, conv} = \Bigl ( \frac{\Theta_{\rm conv}}{\Theta_{\rm beam}} \Bigr )^{2},
\end{equation}
\begin{equation}
\label{Diameter}
    D \approx 4.84 \times 10^{-6} (d \Theta_{\rm deconv}),
\end{equation}
\end{subequations}

\noindent
where d = 5400 pc and $\Theta_{\rm deconv}$ is in arcsec. 
To obtain the peak intensities $I_{\rm pk}$ and flux densities $S$ we subtract the local background emission by estimating the average local background intensity $\langle I_{\rm bg} \rangle$ in \jyb ~as the average intensity in a ring around the source. Then, the background-subtracted flux density is estimated as $S_{\rm -bg} = S - \langle I_{\rm bg} \rangle N_{\rm B, conv}$, where $S$ is the source flux before subtraction. 

The sources in tables \ref{tab:Obs_Params} and \ref{tab:Cont_Params} marked with dashes (--) are significantly mixed with background emission,
such that the results of Gaussian fitting are unreliable. We treat those sources as intrinsically point like, thus their peak intensity and flux density are equal. After subtraction of their locally-measured background, a few of these sources are marginal detections\footnote{Yet they are robust identifications from multiwavelength observations in \citet{Ginsburg16}.} (see the  $S_\mathrm{-bg}$ column in table \ref{tab:Cont_Params} for d6, e22, e23). The analysis presented in this paper only uses those high S/N cm sources with valid Gaussian fits.

We define our sources to be resolved if $\Theta_{\rm deconv} > 2\Theta_{\rm beam}$ (e1, e6), marginally resolved if $\Theta_{\rm beam} < \Theta_{\rm deconv} < 2\Theta_{\rm beam}$ (d7, e9), and unresolved if $\Theta_{\rm deconv} < \Theta_{\rm beam}$ (d2, e2, e3, e4, e5, e10).  
For the sources for which we retrieved a valid $\Theta_{\rm deconv}$, their deconvolved solid angle $\Omega_{\rm deconv}$ in steradian units is given by equation \eqref{deconvsize}. Therefore, their background-subtracted 2-cm brightness temperatures $T_{\rm B, c}$ and optical depths $\tau_\mathrm{c}$ are as in equations \eqref{Tb} and \eqref{tau_cont}:  
\begin{subequations}
\begin{equation}
\label{deconvsize}
    \Omega_{\rm deconv} = \Bigl (\frac{\pi}{4\ln{2}} \Bigr ) (4.84 \times 10^{-6}  \Theta_{\rm deconv} )^2, 
\end{equation}
\begin{equation}
\label{Tb}
    T_{\rm B, c} = \Bigl ( \frac{c^{2}}{2k_{\rm B}\nu_{\rm 0, cm}^{2}} \Bigr ) \Bigl (\frac{S_{\rm -bg}}{\Omega_{\rm deconv}} \Bigr ),
\end{equation}
\begin{equation} 
\label{tau_cont}
    \tau_{\rm c} = \ln{\left[\Bigl (1 - \frac{T_{\rm B, c}}{T_{\rm e}} \Bigr )^{-1} \right]}, 
\end{equation}
\end{subequations}

\noindent
where $c$ is the speed of light, $k_\mathrm{B}$ is the Boltzmann constant, and $T_e$ is the electron temperature. 
We find that when assuming $\rm T_{\rm e} = 7500K$ as the typical temperature for an ionized plasma \citep{Osterbrock06} sources d2, e2, and e5 have brightness temperatures larger than $T_{\rm e}$, so we set their lower-limit $\tau_{\rm c}$ to unity. The rest of the sources are approximately optically thin. Optical depths are shown in table \ref{tab:Cont_Params}.

We also determined emission measures and electron densities from continuum parameters for these sources, using 
\citep[see equation 10.35 of][]{ToolsRA}:  
\begin{subequations}
\begin{equation} \label{continuumEM}
    EM = 12.143 \tau_{\rm c} \biggl ( \frac{\nu_{\rm 0, cm}}{[{\rm GHz}]} \biggr ) ^{2.1} \biggl ( \frac{T_{\rm e}}{[{\rm K}]} \biggr )^{1.35},
\end{equation}
\begin{equation} \label{continuumdensity}
    n_{\rm e, c}  =  \sqrt{EM / D}.
\end{equation}
\end{subequations}

where $D$ is the deconvolved source diameter in pc. The calculated $EM$ values are $\sim 10^7$ to $10^8$ pc cm$^{-6}$, while electron densities are of order $n_{\rm e, c} \sim 10^4$ to $10^5$ cm$^{-3}$. The values for d2, e2, and e5 are lower limits. 

Further, we can use the electron densities and diameters of a given source to determine the amount of Lyman-continuum photons per second required for it to produce the observed emission, assuming ionization-recombination balance \citep{Osterbrock06}: 

\begin{equation} \label{LYMAN}
    L_{\rm c} = \frac{\pi}{6} \alpha_{B} n_{\rm e, c}^{2} D^{3},
\end{equation}

\noindent
where $\alpha_{B}$ is the Hydrogen recombination coefficient excluding transitions to the ground state, equal to $2 \times 10^{-13}$ cm$^{3}$s$^{-1}$. All of the calculated values are shown in table \ref{tab:Cont_Params}.

\subsection{Recombination Lines}
Recombination line width has thermal and `pressure' -- collisional -- contributions on top of the dynamical width. Pressure broadening increases as a steep power law of the principal quantum number \citep{GS02}. Thus, for HC HII's pressure broadening can be significant in cm RL's such as H$77\alpha$, but is negligible ($<0.6$ \kms) in the mm H30$\alpha$ line for all electronic densities $n_e < 10^7$ cm$^{-3}$ \citep{GM12}. In this paper we perform a RL analysis similar to that presented in \citet{KZK08} for brighter RL sources. Thanks to the sensitivity of ALMA, now we can perform such analysis in fainter HC HII's.  
 
We extracted the RL spectra of sources in catalog  $B$ using an aperture of radius equal to the $2\sigma$ level of each sources' convolved model Gaussian $R_{\rm apert} = 2 \times 0.42465 \Theta_{\rm conv}$, which gives the flux within two standard deviations $\sigma$ from the emission centroid. The spectra are then scaled up by a 1/0.95 factor to account for the flux outside the aperture. Figure \ref{fig:RL_detections1} shows plots of the RL spectra and Gaussian fits for those sources with at least one detection (catalogs $B-H77$ and $B-H30$ in table \ref{tab:catalogs}). Eight objects were detected in H$77\alpha$ and eight were detected in H$30\alpha$ as well. Seven sources -- d2, e1, e2, e3, e4, e5, e6 -- were detected in both lines (catalog \textit{B-H30-H77}).
The Gaussian fitting and baseline removal was done using {\it PySpecKit}  \citep{GinsburgMirocha11}. From the fitting results we derived the flux density, centroid velocity, and FWHM width of the line emission as listed in table \ref{tab:RL_params}. Then we proceeded to derive the thermal, dynamical and pressure broadening components following the prescriptions in \citet{KZK08} and \citet{GM12}. Those results are summarized in table \ref{tab:Broadening_Components}. 

A few special considerations were made in some sources. 
For e8n H$30\alpha$ was detected but 
the intrinsically fainter H$77\alpha$ line was not. This is not unexpected. 
However, for e9 H$77\alpha$ was detected but H$30\alpha$ was not. After inspection of the images we concluded that in the H$30\alpha$ cube source e9 lies at the position of a sidelobe caused by nearby bright emission, thus preventing detection. Sources d2, e2, e4, and e8n have a negative bowl in their H$30\alpha$ spectra toward redshifted velocities caused by sidelobes of nearby bright emission. For source d2 these artefacts were partially alleviated by taking the spectrum from a single pixel at the peak rather than in an aperture. To be self-consistent the same was done for the H$77\alpha$ extraction in this source. 

\begin{deluxetable*}{c|ccc|ccc|c}[ht!]
\tablecaption{Fits to recombination line spectra}
\tablehead{
\colhead{} & \colhead{} & H77$\alpha$ & \colhead{} & \colhead{} & H30$\alpha$ & \colhead{}  & \colhead{}\\
\colhead{Sources} & \colhead{Peak Flux} & \colhead{Centroid} & \colhead{FWHM} & \colhead{Peak Flux} & \colhead{Centroid} & \colhead{FWHM} & \colhead{$\Delta \nu_{77-30}$}\\
\colhead{} & \colhead{[mJy]} & \colhead{[\kms]} & \colhead{[\kms]} & \colhead{[mJy]} & \colhead{[\kms]} & \colhead{[\kms]} & \colhead{\kms}
}
\startdata
d2 & $1.76\pm0.18$ & $44.00\pm0.95$ & $18.08\pm2.23$ & $278.88\pm3.50$ & $58.98\pm0.16$ & $26.67\pm0.39$ & $-14.97\pm1.11$\\
e1 & $26.16\pm0.13$ & $55.11\pm0.06$ & $24.87\pm0.15$ & $588.87\pm2.75$ & $56.01\pm0.06$ & $24.47\pm0.13$ & $-0.90\pm0.12$\\
e2 & $3.26\pm0.10$ & $56.84\pm0.53$ & $34.42\pm1.25$ & $1365.84\pm3.11$ & $58.31\pm0.02$ & $20.90\pm0.06$ & $-1.47\pm0.55$\\
e3 & $3.36\pm0.13$ & $60.59\pm0.53$ & $27.94\pm1.24$ & $48.50\pm2.92$ & $63.03\pm0.48$ & $16.24\pm1.13$ & $-2.44\pm1.01$\\
e4 & $1.35\pm0.18$ & $57.96\pm1.09$ & $16.95\pm2.57$ & $118.89\pm3.52$ & $59.70\pm0.18$ & $12.37\pm0.42$ & $-1.73\pm1.27$\\
e5 & $1.32\pm0.14$ & $50.31\pm1.07$ & $20.59\pm2.51$ & $105.28\pm1.67$ & $54.27\pm0.20$ & $25.14\pm0.46$ & $-3.96\pm1.26$\\
e6 & $10.54\pm0.56$ & $69.25\pm0.80$ & $30.89\pm1.88$ & $325.80\pm10.34$ & $71.65\pm0.29$ & $18.61\pm0.69$ & $-2.41\pm1.09$\\
e8n & -- & -- & -- & $105.52\pm16.40$ & $59.06\pm0.53$ & $6.92\pm1.24$ & -- \\
e9 & $0.67\pm0.17$ & $65.19\pm1.87$ & $14.68\pm4.40$ & -- & -- & -- & -- \\
\enddata
\tablecomments{Parameters derived from Gaussian fits to the H77$\alpha$ and H30$\alpha$ spectra, along with velocity centroid differences between the RLs. Source e8n lacks an H77$\alpha$ detection, and source e9 lacks an H30$\alpha$ detection. The d2 spectra was extracted from the central pixel rather than from an aperture due to image sidelobes in the H$77\alpha$ image caused by caused IRS2. This spectra could still be biased, so $\Delta v_{77-30}$ could be artificially exaggerated for d2.}
\label{tab:RL_params}
\end{deluxetable*} 

\smallskip

The contribution to line broadening due to the thermal velocity distribution of the gas particles has a Gaussian shape. Neglecting microturbulence, its FWHM is given by:
\begin{equation}
    \Delta \upsilon_{\rm th} = \biggl (8 \ln{2}\frac{k_{\rm B}T_{\rm e}}{m_{\rm H}}\biggr)^{1/2}, 
\end{equation}

\noindent
where $m_{\rm H}$ is the Hydrogen mass.

We exploit the fact that the H30$\alpha$ line is free of pressure broadening to infer the dynamical component from bulk motions. Assuming it is Gaussian, its FWHM is given by:  
\begin{equation} \label{dynamic}
    \Delta \upsilon_{\rm dy} = \sqrt{\Delta \upsilon_{\rm H30}^{2} - \Delta \upsilon^{2}_{\rm th}}.
\end{equation}

Considering that the profile of a pressure broadened line is a Lorentzian, whereas thermal and dynamic widths are taken as Gaussians, the H77$\alpha$ FWHM linewidth is given by a Voigt profile which can be expressed as \citep{KZK08}: 
\begin{equation} \label{eq:pressure_br}
    \Delta \upsilon_{\rm H77} = 0.534 \Delta \upsilon_{\rm pr} + (\Delta \upsilon^{2}_{\rm dy} + \Delta \upsilon^{2}_{\rm th} + 0.217 \Delta \upsilon^{2}_{\rm pr})^{1/2}.
\end{equation} 

Source d2 is anomalous in the sense that its H$77\alpha$ line appears to be narrower and blueshifted compared to H$30\alpha$. We concluded that this could be due to the above mentioned image artifacts, in particular sidelobes from the bright, nearby IRS2. 
Thus, in table \ref{tab:Broadening_Components} we do not report pressure broadening values for d2. The H30$\alpha$ line for sources e3 and e4 is narrower than $\Delta \upsilon_{\rm th}(T_{\rm e} = 7500 K) = 18.52$ \kms, so for these we assume that there is no dynamical contribution to the broadening and derive the upper limit $T_{\rm e} = T_{\rm e, upper}$ necessary to account for their H30$\alpha$ linewidth. The resulting temperatures, listed in table \ref{tab:Broadening_Components}, are still consistent with fully ionized gas, but indeed lower than the typically assumed value.
The difference of the H$77\alpha$ and H$30\alpha$ linewidths for e5 is consistent with zero within $2\sigma$. 
Then, for e5 we obtain an upper limit to the pressure broadening $\Delta \upsilon_{\rm pr}$ and electron density $n_e$ using equation \ref{eq:pressure_br}, and then substituting $\Delta \upsilon_{\rm H77}$ by $\Delta \upsilon_{\rm H30} + \sigma_{\Delta \upsilon_{\rm H77}}$ and 
$\Delta \upsilon^{2}_{\rm dy} + \Delta \upsilon^{2}_{\rm th}$ by 
$(\Delta \upsilon_{\rm H30}-\sigma_{\Delta \upsilon_{\rm H30}})^2$.

\begin{deluxetable}{c|c|c|c|c|c}[ht!]
\tablecaption{Broadening components and electron densities}
\tablehead{
\colhead{Sources} & \colhead{$\Delta \upsilon_{\rm th}$} & \colhead{$\Delta \upsilon_{\rm dy}$} & \colhead{$\Delta \upsilon_{\rm pr}$} & \colhead{$T_{\rm e}$} & \colhead{$n_{\rm e, RL}$} \\
\colhead{} & \colhead{[km/s]} & \colhead{[km/s]} & \colhead{[km/s]} & \colhead{[K]} & \colhead{[$10^{5}$ cm$^{-3}$]}
}
\startdata
~d2$^a$ & 18.52 & 19.20 & -- & 7500 & -- \\
e1 & 18.52 & 16.00 & 0.73 $\pm$ 0.36 & 7500 & 0.13 $\pm$ 0.06\\
~e2$^a$ & 18.52 & 9.68 & 21.18 $\pm$ 1.70 & 7500 & 3.62 $\pm$ 0.29\\
~e3$^b$ & 16.24 & 0.00 & 18.06 $\pm$ 2.13 & 5767 & 3.08 $\pm$ 0.36\\
~e4$^b$ & 12.37 & 0.00 & 7.64 $\pm$ 3.92 & 3345 & 1.31 $\pm$ 0.67\\
~ ~e5${^{a,}}{^c}$ & 18.52 & 16.99 & $\leq$ 5.34 & 7500 & $\leq$ 0.913\\
e6 & 18.52 & 1.84 & 19.18 $\pm$ 2.69 & 7500 & 3.28 $\pm$ 0.46\\
\enddata
\tablecomments{$^a$: d2 does not have a derived $\Delta \upsilon_{\rm pr}$ nor $n_{\rm e, RL}$ since its $\Delta \upsilon_{\rm H30} > \Delta \upsilon_{\rm H77}$. $^b$: e3 and e4 have derived $n_{\rm e}$ using upper limit $T_{\rm e, upper}$ found from assuming H30$\alpha$ width is entirely thermal ($\Delta \upsilon_{\rm dy} = 0$). $^c$: for e5 derived $\Delta \upsilon_{\rm pr}$ and $n_{\rm e}$ are upper limits as described in the text.}
\label{tab:Broadening_Components}
\end{deluxetable}

For sources e1, e2, e3, e4, e5, and e6 we calculate electron densities $n_{\rm e, RL}$ from the derived pressure broadening components via equation (A.4) of \citet{GM12}, for quantum number $\rm n$:
\begin{equation} \label{RLdensity}
    n_{\rm e, RL} = \Delta \upsilon_{\rm pr} \Bigl ( \frac{8.2 \nu_{\rm 0, H77}^{2}}{c} \Bigr ) \Bigl (\frac{\rm n + 1}{100} \Bigr )^{-4.5} \Bigl (1 + \frac{2.25 \Delta \rm n}{\rm n + 1} \Bigr )^{-1},
\end{equation}

\noindent
where $n$ is the quantum number and $\Delta n = 1$ for $\alpha$ transitions. 

The electron densities obtained from this RL analysis are $n_{\rm e, RL} \sim 10^4$ to $10^5$ cm$^{-3}$ as shown in table \ref{tab:Broadening_Components}.
Although the range of derived electron densities is the same using both the continuum and RL methods, the individual values are typically offset by a factor of a few, and even an order of magnitude for e6. These differences are larger than the error bars resulting from measurement uncertainty propagation (see Fig. \ref{fig:NeVsD}), which suggests that some assumptions such as LTE line emission or that both tracers arise from the same gas might not hold. 

\smallskip

More information on the internal kinematics of the ionized gas can be extracted from the RL velocity centroids. Given that the H$30\alpha$ line traces on average higher densities than H$77\alpha$, and under the assumption of a density profile decreasing with increasing radius, a blue- or redshifted H$77\alpha$ centroid with respect to H$30\alpha$ can be interpreted as outflowing or inflowing ionized gas, respectively \citep[e.g.,][]{KZK08}. From table \ref{tab:RL_params} it is seen that the nominal centroid values for H$77\alpha$ are {\it all blueshifted} with respect to  H$30\alpha$.
The velocity differences range from $\Delta v_{77-30} \approx - 1 $ to $- 4$ \kms ~ for all sources except for d2, which we deem to be artificially
exaggerated (see above). Our main conclusion from this analysis is that the bulk of ionized gas at beam scales ($\sim 0.01$ pc) in all these HC HII's seems to be going outwards, not inwards. 

\section{Physical Properties of the HII Regions} \label{sec:physical_props}

Interestingly, the $n_{\rm e}$ and $EM$ values we find through both methods are significantly smaller than those often-quoted for HC HII's \citep[$n_{\rm e} \geq 10^6$ cm$^{-3}$, $EM \geq 10^{10}$ cm$^{-6}$ pc,][]{Kurtz05}. The source diameters, however, are  all in the range $D \sim 10^{-3}$ to $10^{-2}$ pc, which clearly puts them in the HC HII range. 

\begin{figure}[!h]
\centering 
\includegraphics[scale=0.57]{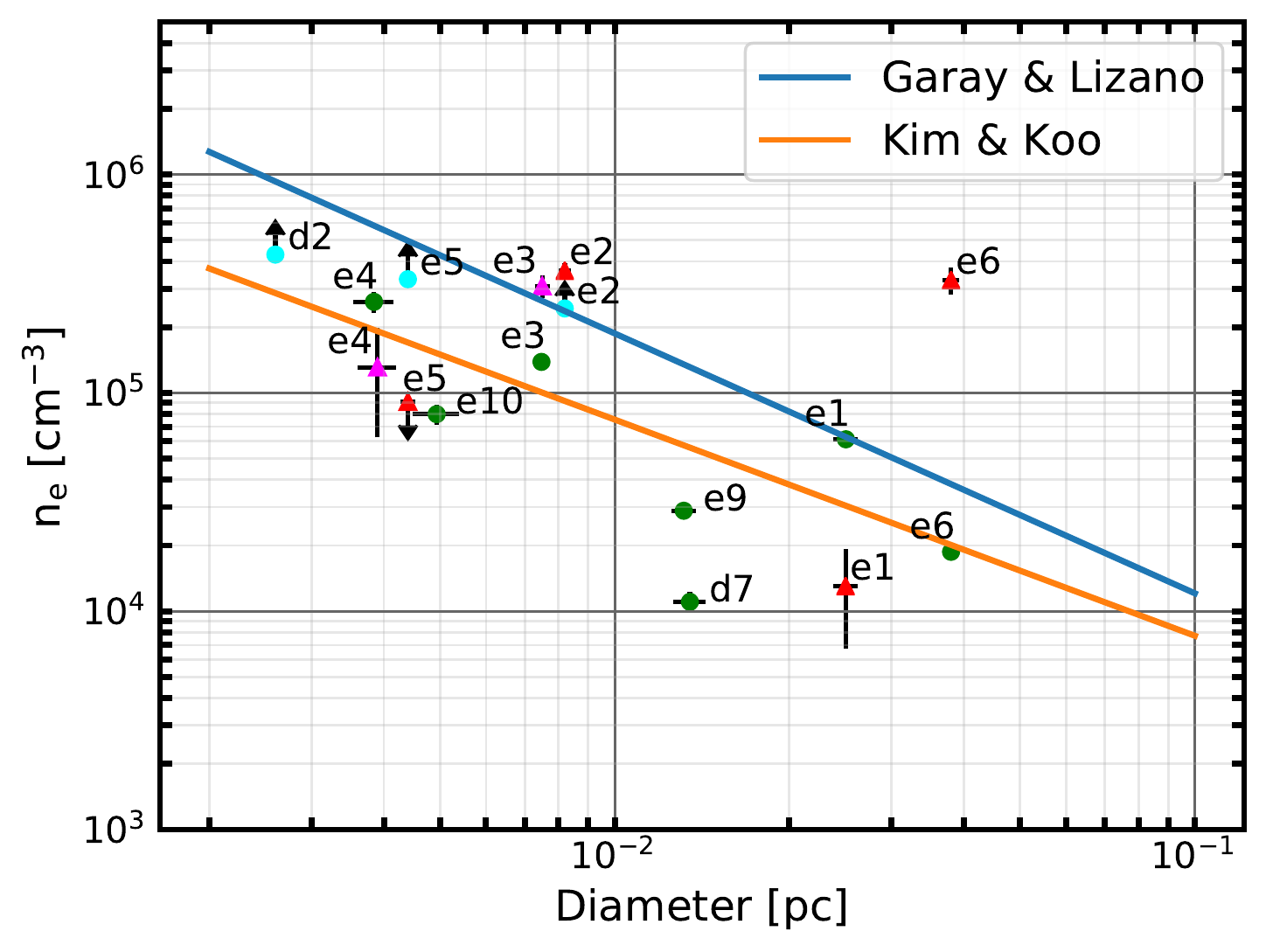}
\caption{$n_e$ vs $D$ for sources in catalog $B$ with a valid diameter measurement. {\it Triangles} correspond to RL-derived densities, color coded in \textit{red} for those calculated with $T_{\rm e} = 7500$ K and \textit{magenta} for those calculated with a $T_{\rm e} = T_{\rm e, upper}$. {\it Circles} correspond to continuum-derived values, color coded in  \textit{green} for optically thin sources, and \textit{cyan} for optically thick sources assuming $\tau_{\rm c} = 1$.
Arrows indicate lower or upper limits. The \textit{blue} and \textit{orange} lines show the relations previously derived by \citet{GarayLizano99}, $n_{\rm e} = 780 \times D^{-1.19}$, and \citet{KimKoo}, $n_{\rm e} = 790 \times D^{-0.99}$, where $n_e$ is in $cm^{-3}$ and $D$ in pc.}
\label{fig:NeVsD}
\end{figure}

We now show that our values are consistent with an extrapolation of UC HII's to smaller sizes. Figure \ref{fig:NeVsD} shows $n_{\rm e}$ versus $D$ for our sources along with the relations previously found in the literature for surveys of compact and UC HII regions at high angular resolution \citep{Garay1993,GarayLizano99,KimKoo}. We find that our 
objects do follow these relations. Although radio surveys of UC HII's can be considered to be comprehensive, HC HII's are more difficult to detect, and it appears that most of them exist in crowded environments such as W51 A \citep{Ginsburg20}. 
It is possible that previous detailed characterizations of HC HII's are biased toward landmark objects, which happen to be relatively isolated and satisfy more specific selection criteria, such as also being \textit{hyperdense} ($n_{\rm e} > 10^6$ cm$^{-3}$) or having broad (FWHM $> 50$ km s$^{-1}$) millimeter RL's \citep[e.g.,][]{Sewilo04,KZK08,SanchezMonge11}.  

\begin{figure}[!h]
\centering 
\includegraphics[scale=0.58]{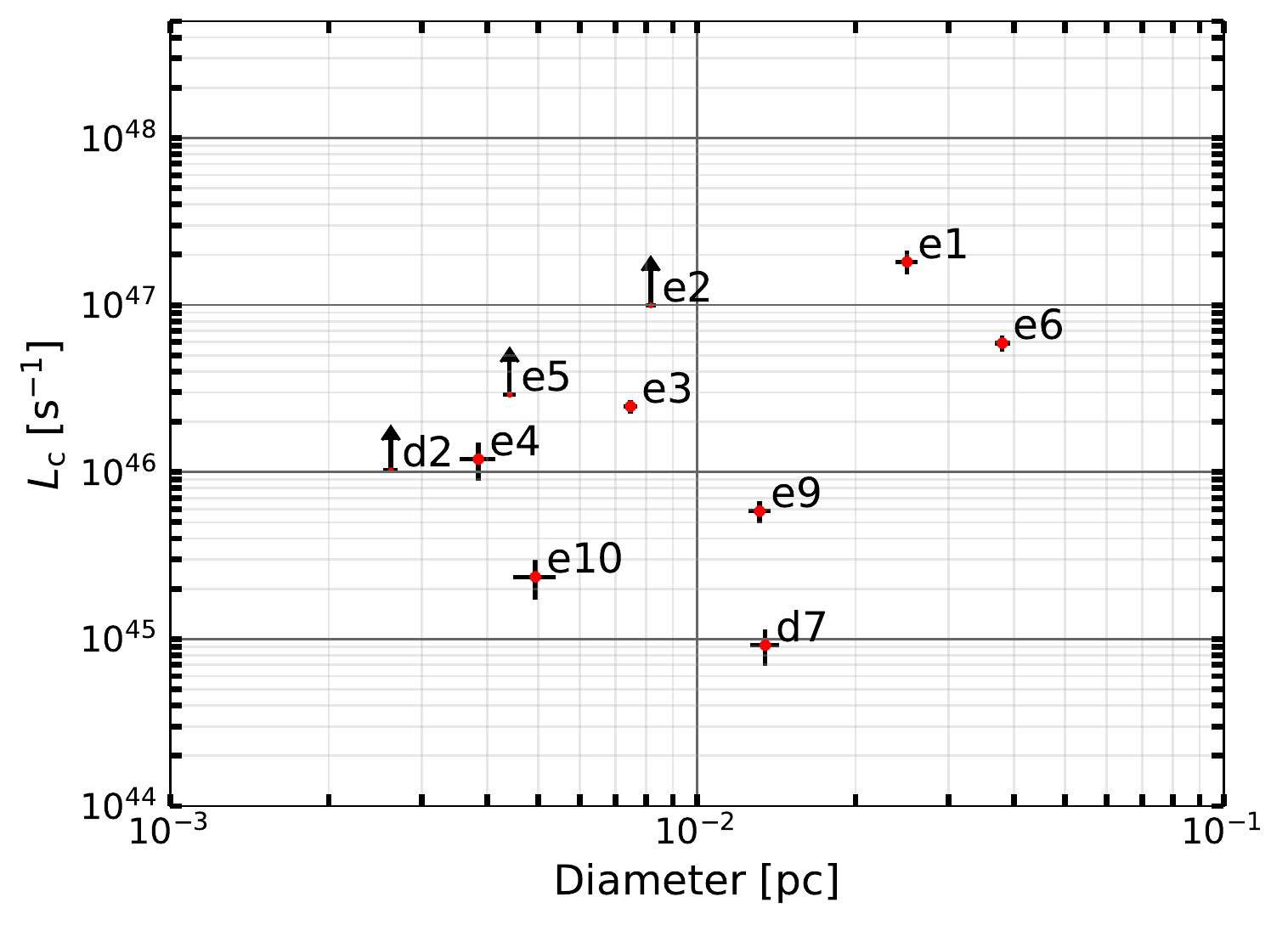}
\caption{$L_{\rm c}$ vs $D$ for sources in catalog B with a valid diameter measurement. The arrows represent lower limits due to high optical depths.}
\label{fig:LymanvsDiameter}
\end{figure}

More can be learned from the determination of the spectral type of the stars ionizing our HII regions. The derived ionizing photon rates are shown in table \ref{tab:Cont_Params} and figure \ref{fig:LymanvsDiameter}. Only 2 (e1 and e2) out of 10 sources have measurements or lower limits above the threshold between O-type and B-type ZAMS stars $L_\mathrm{c} \approx 10^{47}$ s$^{-1}$ \citep{Panagia73}. e6 is close at $L_\mathrm{c} \approx 46.8$ s$^{-1}$, and formally the lower limits for d2 and e5 could be above this threshold too. Therefore, 5 to 8 out of 10 sources are ionized by early B-type stars, not O-type stars. There is also a hint of a positive correlation between $L_\mathrm{c}$ and $D$ that could be further explored with larger samples. 

\smallskip

\begin{deluxetable}{c|c|c|c|c|c}[ht!]
\tablecaption{Molecular line properties \label{tab:Molec_Props}}
\tablehead{
\colhead{Molecule} & \colhead{$\rm E_{\rm l}$} & \colhead{$\rm E_{\rm u}$} & \colhead{$A_{\rm ul}$} & \colhead{$B_{\rm ul}$} & \colhead{$n_{\rm crit}^{\rm 100K}$}\\
\colhead{} & \colhead{[K]} & \colhead{[K]} & \colhead{[$\rm 10^{-4} s^{-1}$]} & \colhead{[$\rm 10^{-11} cm^{3} s^{-1}$]} & \colhead{[$\rm 10^{6} cm^{-3}$]}
}
\startdata
\formai & 10.5 & 21.0 & 2.8 & 8.4 & 3.4 \\
\formaii & 57.6 & 68.1 & 1.6 & 4.9 & 3.2 \\
\soi & 24.4 & 35.0 & 1.3 & 5.8 & 2.3 \\
\enddata
\tablecomments{Lower and upper level energies for each molecular transition, Einstein coefficients, and critical densities at 100 K.}
\end{deluxetable}

\section{Molecular Lines Toward Radio Continuum Sources} \label{sec:molec_lines}

Using the ALMA data release from \citet{Ginsburg17}, we investigate the dense molecular gas content of our HC HII region sample (catalog $B$). Table \ref{tab:Molec_Props} lists the physical properties of the available line cubes covering the entire mosaic field of view:  formaldehyde \formai,  \formaii, and of sulfur monoxide \soi. The critical densities ($n_{\rm cr}\sim 10^6$ cm$^{-3}$) and upper-level energies ($E_U = 21$ to 68 K) are similar among the three lines, but the chemistry of \forma~and \so~is expected to be different: \forma~is considered a standard tracer of dense gas in molecular clouds \citep[e.g., ][]{Henkel83, Ginsburg16}, whereas \so~and other sulphur-bearing molecules are thought to trace shocked regions \citep{Pineau93,Guzman18}. 


\begin{figure*}[!t]
    \centering
    \includegraphics[scale=0.38]{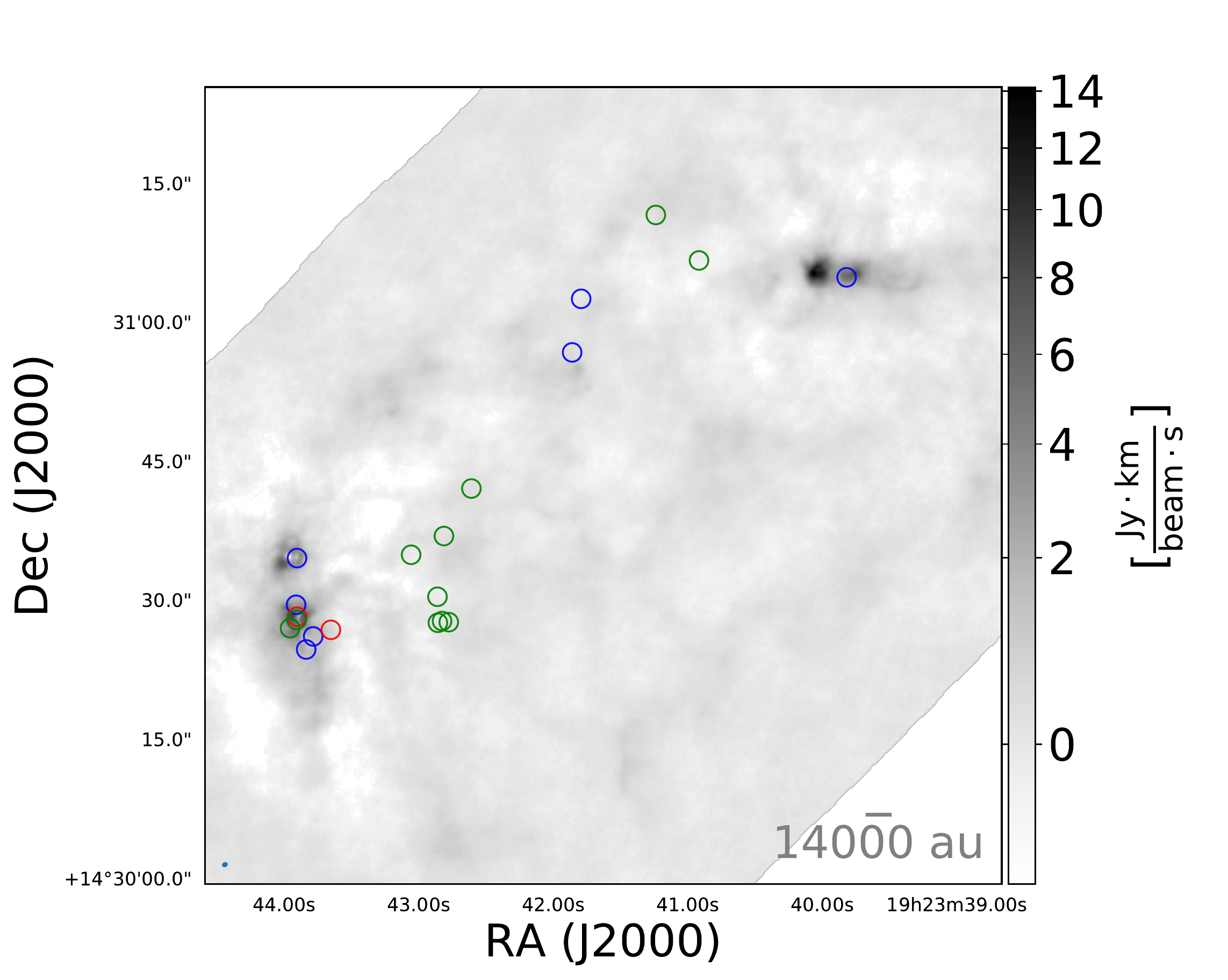}
    \includegraphics[scale=0.38]{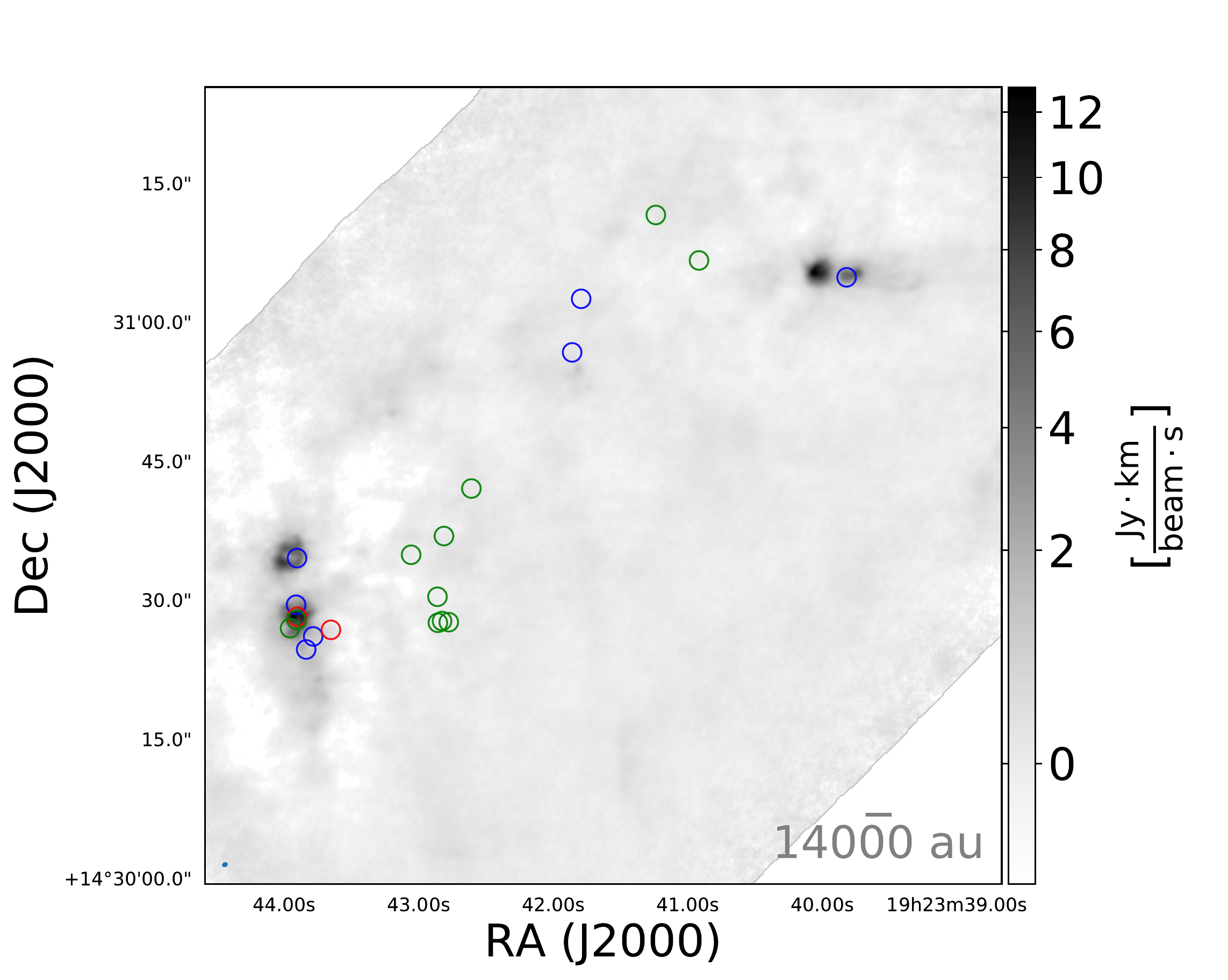}
    \includegraphics[scale=0.38]{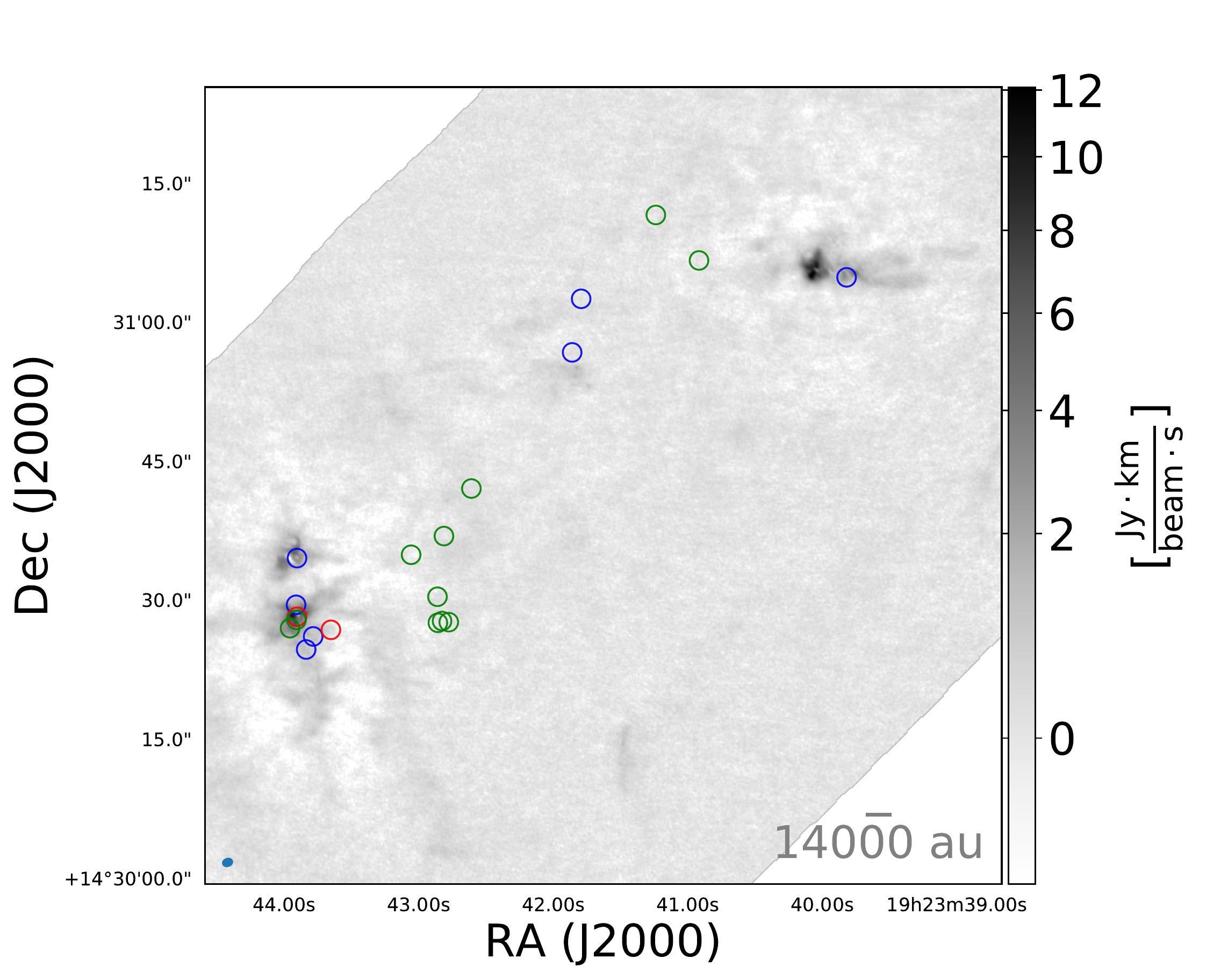}
    \caption{Velocity-integrated (moment 0) images of \formai~(top-left, $\sigma_\mathrm{rms} \sim 0.15$ Jy beam$^{-1}$ km s$^{-1}$), \formaii~(top-right, $\sigma_\mathrm{rms} \sim 0.10$ Jy beam$^{-1}$ km s$^{-1}$), and \so~(bottom, $\sigma_\mathrm{rms} \sim 0.14$ Jy beam$^{-1}$ km s$^{-1}$). 
    The color stretch is square-root from $-3\sigma_\mathrm{rms}$ to the absolute maximum in the respective image. Velocity integration was performed from 35 \kms to 75 \kms. 
    Circles mark the position of 2-cm sources as labeled in figure \ref{fig:ContImage}.} 
    \label{fig:Moment0}
    \end{figure*} 

\subsection{Molecular morphology and profiles} \label{sec:molec_profiles}

Figure \ref{fig:Moment0} shows the velocity-integrated intensity (moment 0) maps for the three molecular lines. As is known, \citep[e.g.,][]{Ginsburg17}, the total emission is dominated by IRS2 and the e2/e8 subcluster. However, it is seen that the compact radio  sources in catalog $B$ inhabit within structures of molecular gas. \forma~emission is detected at the position of 20 out of 20 cm sources and the \so~line is detected in 14 out of 20.  
Figure \ref{fig:mom0_zoom} in the appendix shows zoomed-in regions of the moment 0 maps, where it is appreciated that some HC HII's (d2, e1, e2, e3, e4, e8n, e8s, e10) are still embedded in their compact hot molecular cores (HMC's), whereas the rest (d6, d7, e5, e6, e9, e12, e13, e14, e20, e21, e22, e23) are embedded in molecular structures that are somewhat more extended. This distinction is not sharp since some HC HII's in W51 A are known to be surrounded by a common envelope \citep[e.g., those around e8n/e8s,][]{Ginsburg16,Koch18}

We extracted the molecular-line spectra of all the cm sources in catalog $B$ over the same apertures as for the RL's. In the following we compare the line profiles with respect to each other.
The profiles of 8 objects could not be analyzed because most of the fainter sources (d6, d7, e12, e20, e21, e22, e23) and source e6 were affected by image sidelobes from nearby, bright emission. These uncharacterized spectra are shown in figure \ref{fig:profiles_sidelobes} in the appendix. 
For the sources that were characterized, figure \ref{fig:StellarVelocities1} shows the molecular spectra of sources with H$30\alpha$ detection (catalog {\it B-H30}, except e6), which tend to be among the brightest in molecular emission too. The complementary figure \ref{fig:profiles_noH30a} in the appendix shows the molecular spectra of the analyzed sources without H$30\alpha$.

The molecular profiles have a variety of shapes, but all of them have FWHM widths from $\sim 5$ to 20 \kms, clearly signaling bulk motions far larger than the thermal width $\sim 0.3$ to 0.4 \kms~ for \so~ and \forma~ gas at $T\sim 100$ K. These bulk motions could be due to expansion, outflow, infall, rotation, or shear motions, both locally in the surrounding core or within more extended structures. When localized, the nature of these  motions could be figured out from the shape of the molecular spectra under reasonable assumptions for the temperature, density, and velocity profiles \citep[e.g.,][]{Myers93}.

All of the 8 sources still embedded in compact molecular emission (see above) also satisfy the requirement of having spectra without artifacts at the position of the cm continuum. For these  we  qualitatively group their molecular spectra (see Figs.  \ref{fig:StellarVelocities1} and \ref{fig:profiles_noH30a}) in three categories based on their profiles: 1) a single, almost symmetrical emission component, 2) the presence of significant asymmetries or multiple emission components, and 3) the presence of a prominent absorption component, often accompanied by emission features. The first group is consistent with molecular gas that is optically thin. The second group has asymmetries which we label as ``blue-'' or ``red-peaked'' according to the relative position of the maximum in the spectra. This group could also have dips in the profile. In the third group we only have source e2 with an inverse P-Cygni profile. In this source the continuum background from the hot HC HII is relevant, which results in the observed prominent absorption. Finally, regardless of category, in some sources we additionally identify high-velocity line wings, which indicate the presence of faster outflows.

Our interpretation of blue-peaked profiles follows the widely used models of inside-out collapse with radially decreasing temperature and velocity profiles \citep[e.g.,][]{Shu1977, Myers93, Churchwell}, in which the decreased brightness at redshifted velocities is caused by self-absorption from the cooler, outer layers of the object in the observer's side. 
Hence, we denote those sources with blue-peaked profiles as \textit{infall} candidates. Red-peaked profiles have been described in YSO surveys \citep[e.g.,][]{Mardones1997} and have been associated with bulk expansion of molecular envelopes \citep[e.g.,][]{Thompson2004,Velusamy2008}. Therefore, we denote such sources as \textit{expansion} candidates.  
The inverse P-Cygni profile in e2 is a signature of infall. 
We do not consider more complex kinematics as alternate mechanisms for producing profile asymmetries, such as bipolar outflows, rotation, or multiplicity 
\citep[e.g.,][]{Cabrit1986,Izquierdo18}.
Table \ref{tab:MolecularKinematics} summarizes our assessment of the molecular profiles for the
 aforementioned 8 sources.
We now comment on their individual line profiles. 

- d2 lines are single-peaked. The \forma~ lines are slightly red-peaked  and \so~ is symmetrical. There is a blue high-velocity wing in all three lines. Expansion candidate.

- e1 is red-peaked in the \formai ~and \so ~lines. The \so ~ also has a broad, redshifted wing extending up to $\sim 40$ \kms~ from the absorption dip, suggesting the presence of a fast outflow. Expansion candidate. 

- e2 has two emission peaks on each side of the prominent absorption peak in all spectra. The line profiles are similar to inverse P-Cygni. Infall candidate. 

- e3 is blue-peaked in all spectra. The red emission shoulder in \formai~ becomes a secondary peak in \formaii~ and \so. A second, fainter shoulder is also seen in \forma~ lines. Infall candidate. 

- e4 is symmetrical in \formaii~ and blue-peaked with a prominent redshifted shoulder in \formai~ and \so. Blue- and redshifted wings are seen in all lines. Infall candidate. 

- e8n spectra are red-peaked with a prominent blue shoulder in all lines. \so ~shows self-absorption at velocities slightly blueshifted from the main peak. Expansion candidate. 

- The spectra of e8s are very similar to its neighbour e8n. Expansion candidate as well. These two sources are separated by 0.13\arcsec, so their spectra are likely mixed. This it at odds with the presence of $> 100~M_\odot$ of molecular gas within a radius of 0.5\arcsec ~ \citep{Ginsburg17}. 

- The spectra in e10 are very broad. \formai~appears to have several dips, \formaii~is red-peaked, and \so~appears almost symmetrical with prominent shoulders. In all lines the main emission extends from $\sim 40$ to 80 \kms. We consider e10 as a fast expansion candidate. 

\begin{deluxetable}{c|c|c|c|c}[t!]
\tablecaption{Inferred core molecular kinematics}
\tablehead{
\colhead{Sources} & \colhead{\formai} & \colhead{\formaii} & \colhead{\soi} & \colhead{Candidate}
}
\startdata
d2 & Red+BW & Red+BW & Sym+BW & \textit{Exp} \\
e1 & Red & Sym & Red+RW & \textit{Exp} \\
e2 & inv. P-Cyg & inv. P-Cyg & inv. P-Cyg & \textit{Inf} \\
e3 & Blue & Blue & Blue & \textit{Inf} \\
e4 & Blue+BRW & Sym+BRW & Blue+BRW & \textit{Inf} \\
e8n & Red & Red & Red & \textit{Exp} \\
e8s & Red & Red & Red & \textit{Exp} \\
e10 & Sym+BRW & Red+BRW & Sym+BRW & \textit{Exp} \\
\enddata
\tablecomments{2$^{nd}$, 3$^{rd}$, 4$^{th}$ columns: Red/Blue: line profile with \textit{red} or \textit{blue} absolute peak, respectively. Sym: symmetric profile. inv. P-Cyg: inverse P-Cygni profile. +RW/+BW: additional red/blue high-velocity wings in line profile. 5$^{th}$ column: \textit{Exp} for expansion candidates, \textit{Inf} for infall candidates.}
\label{tab:MolecularKinematics}
\end{deluxetable}

\smallskip
In summary, we have 3 out of 8 HC HII's (e2, e3, e4) whose local molecular gas appears to have bulk infall motions, whereas the other 5 (d2, e1, e8n, e8s, e10) appear to have bulk expansion. Four sources (1 with infall and 3 with expansion) have high-velocity line wings signaling additional, faster outflows. The 3 infall candidates are in the e2/e8 subcluster, which is expected as this is one the most active sites of current star formation. 

\subsection{Comparison to stellar velocities} \label{sec:comparison_stellar}

Given that the material closest to the massive (proto)stars within HC HII's is ionized, and that the H30$\alpha$ is a better kinematical tracer of the denser ionized gas compared to lower-frequency RLs, it is  reasonable to take the H30$\alpha$ velocity centroid as a proxy for the stellar velocity \citep[e.g.,][]{KZK08,ZhangY19}.
A comparison of the molecular spectra with the H30$\alpha$ velocity for catalog \textit{B-H30} (except e6, see above) is shown in figure \ref{fig:StellarVelocities1}.

From the 8 sources that are embedded within a local core as determined in \S \ref{sec:molec_profiles}, 6 have an H$30\alpha$ detection too. Any large 
velocity differences between the molecular and ionized tracers would indicate significant offsets between the bulk motions of the respective HMC and the HC HII region.
All these 6 sources have the main features of their molecular spectra (emission peaks and absorption dips) within $\approx 1$ to 5 \kms~ of the H$30\alpha$ velocity centroid. In particular, sources d2, e1, and e8n  have the molecular and ionized tracers aligned within 2 \kms. The slightly larger offsets for sources e2, e3, and e4 seem to have the H$30\alpha$ systematically redshifted (see Fig. \ref{fig:StellarVelocities1}). We tentatively interpret this as a similar velocity offset between these HC HII's and their parent molecular cores, which is plausible given that e2, e3, and e4 form part of the same (sub)cluster. 
Velocity offsets between molecular and ionized tracers have been observed in other UC and HC HII's \citep[e.g.,][]{Liu12,Klaassen18}. They are expected due to the complex interaction between these gas phases in a dynamical scenario for massive star formation \citep[e.g.,][]{Peters10,Peters10b}. 

Finally, for most sources the \so ~line is systematically blueshifted with respect to the \forma ~lines. As mentioned above, a plausible explanation is that \so ~is tracing a shocked layer more external than \forma, with kinematics more often influenced by outflow motions. Indeed, the \so~ line tends be more asymmetric and to have more prominent line wings (Figs. \ref{fig:StellarVelocities1} and \ref{fig:profiles_noH30a}).   

\begin{figure*}[htb!]
   \centering
   \includegraphics[scale=0.38]{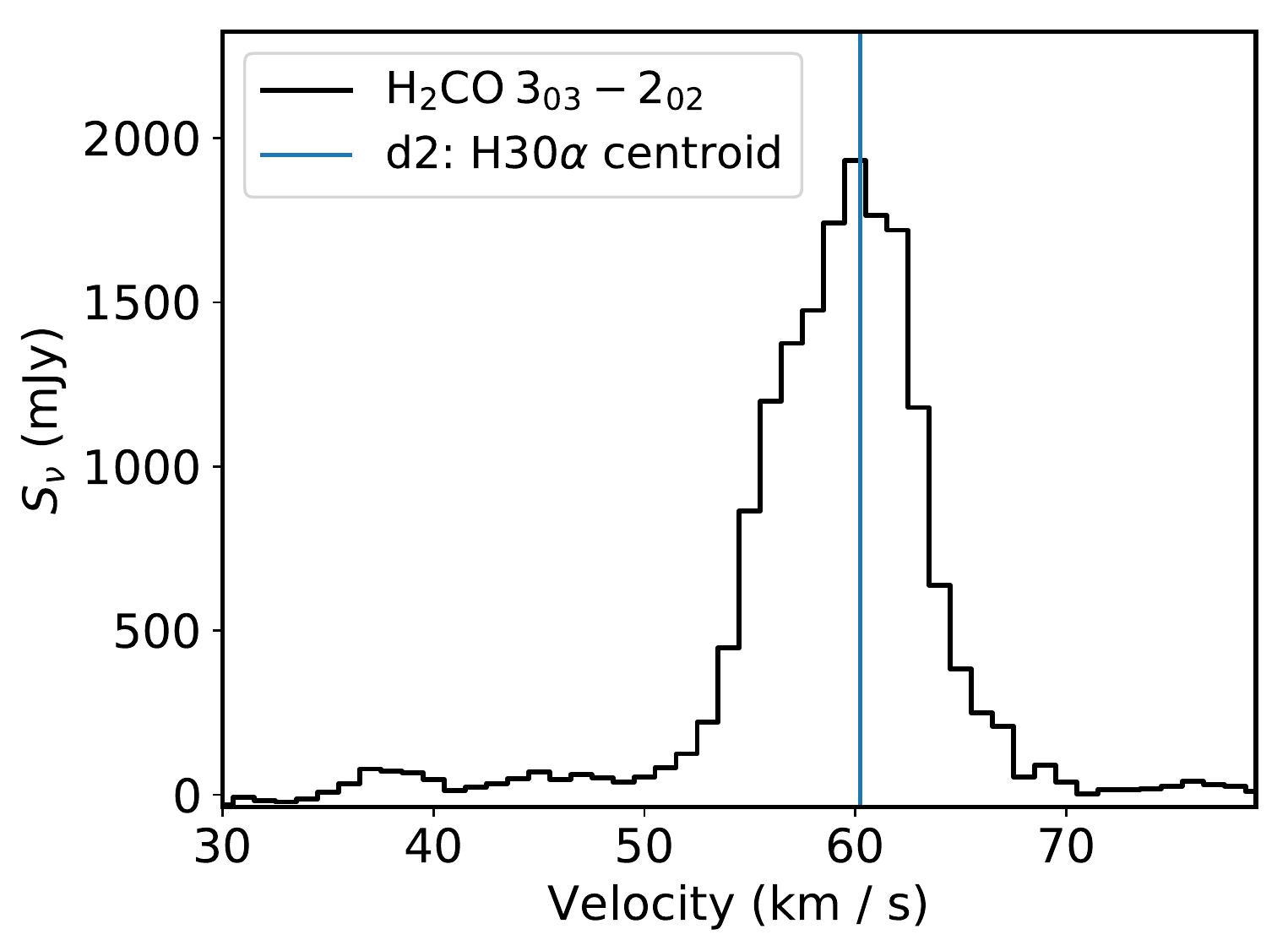}
   \includegraphics[scale=0.38]{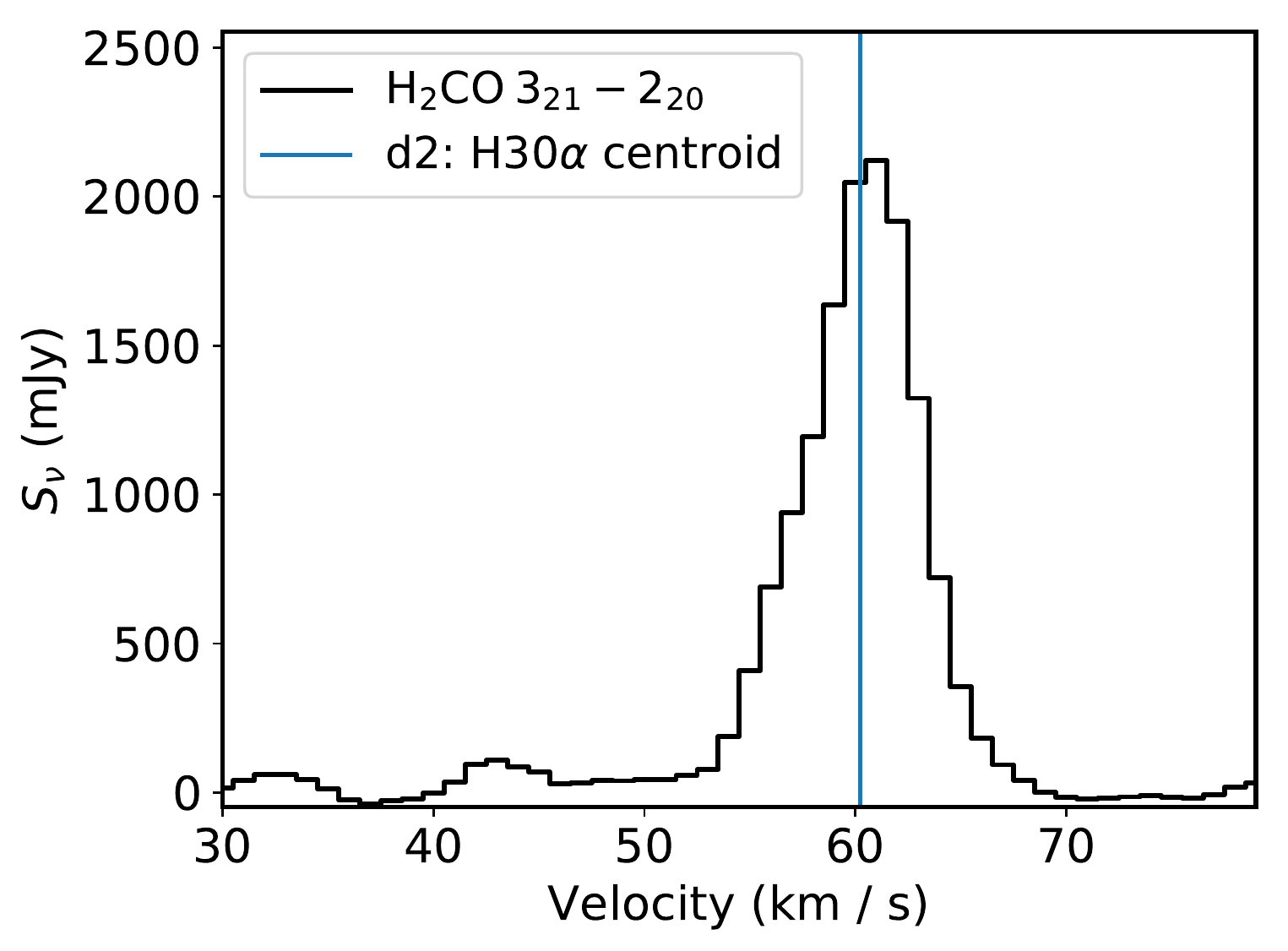}
   \includegraphics[scale=0.38]{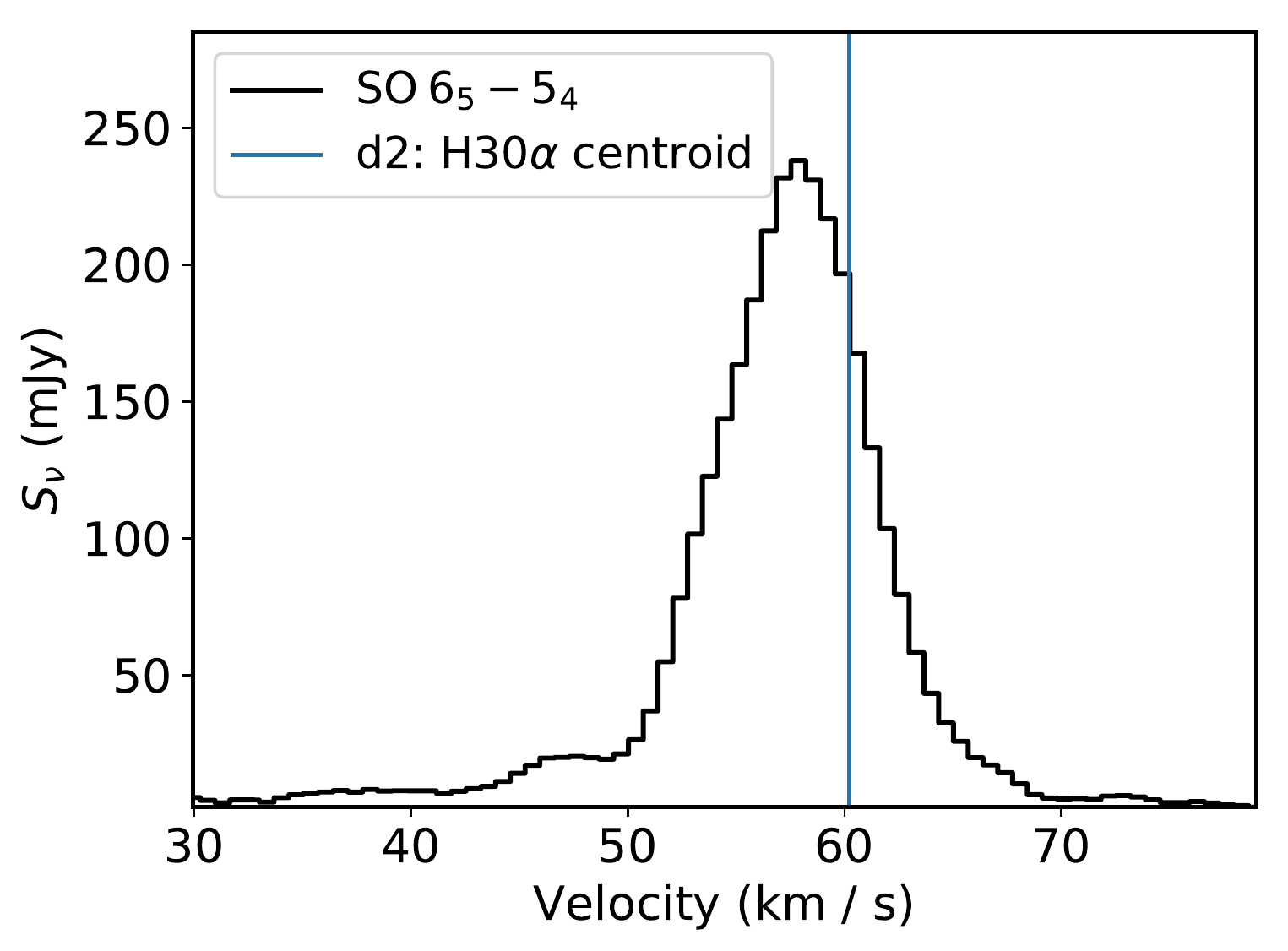}
   \includegraphics[scale=0.38]{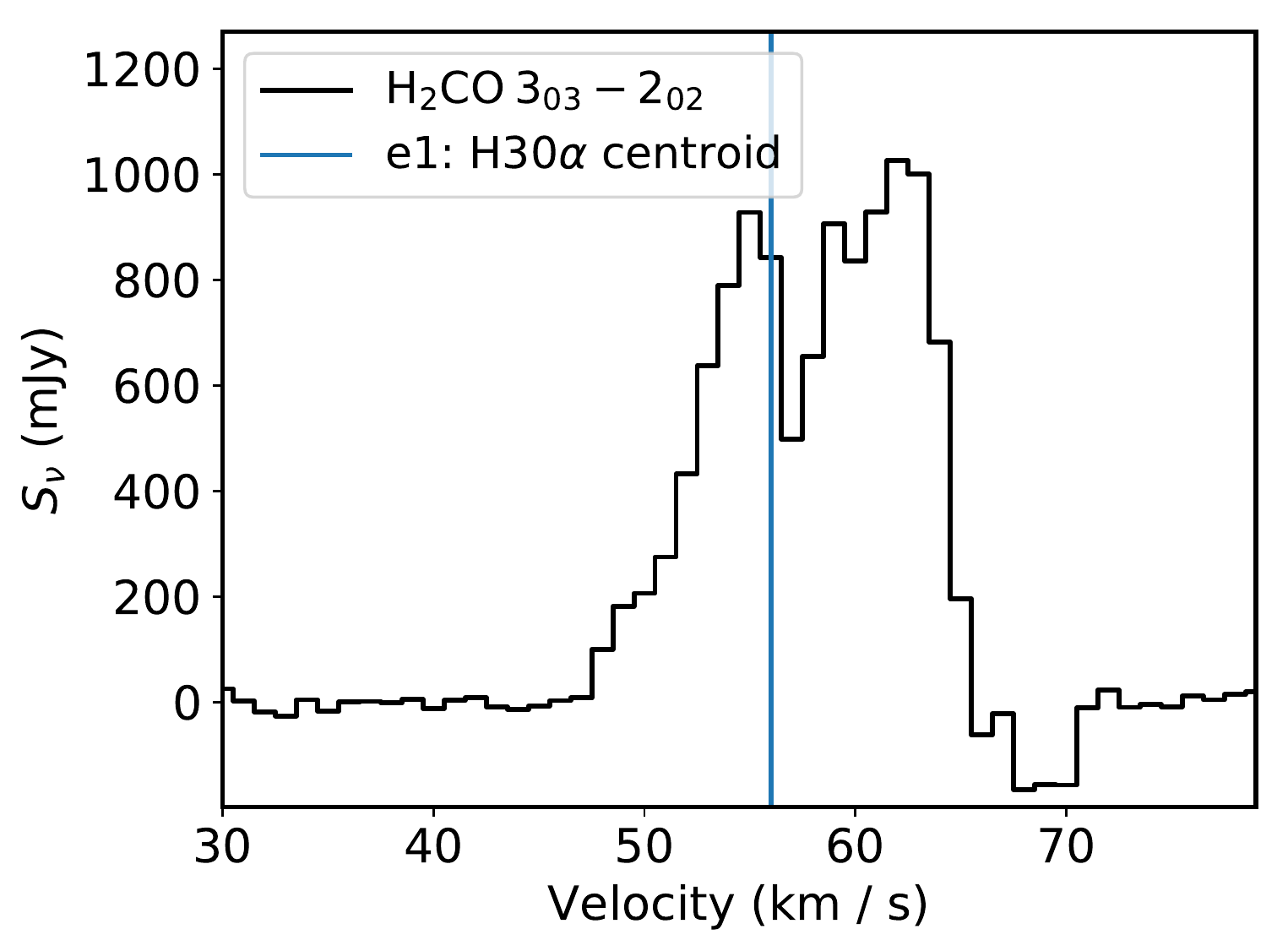}
   \includegraphics[scale=0.38]{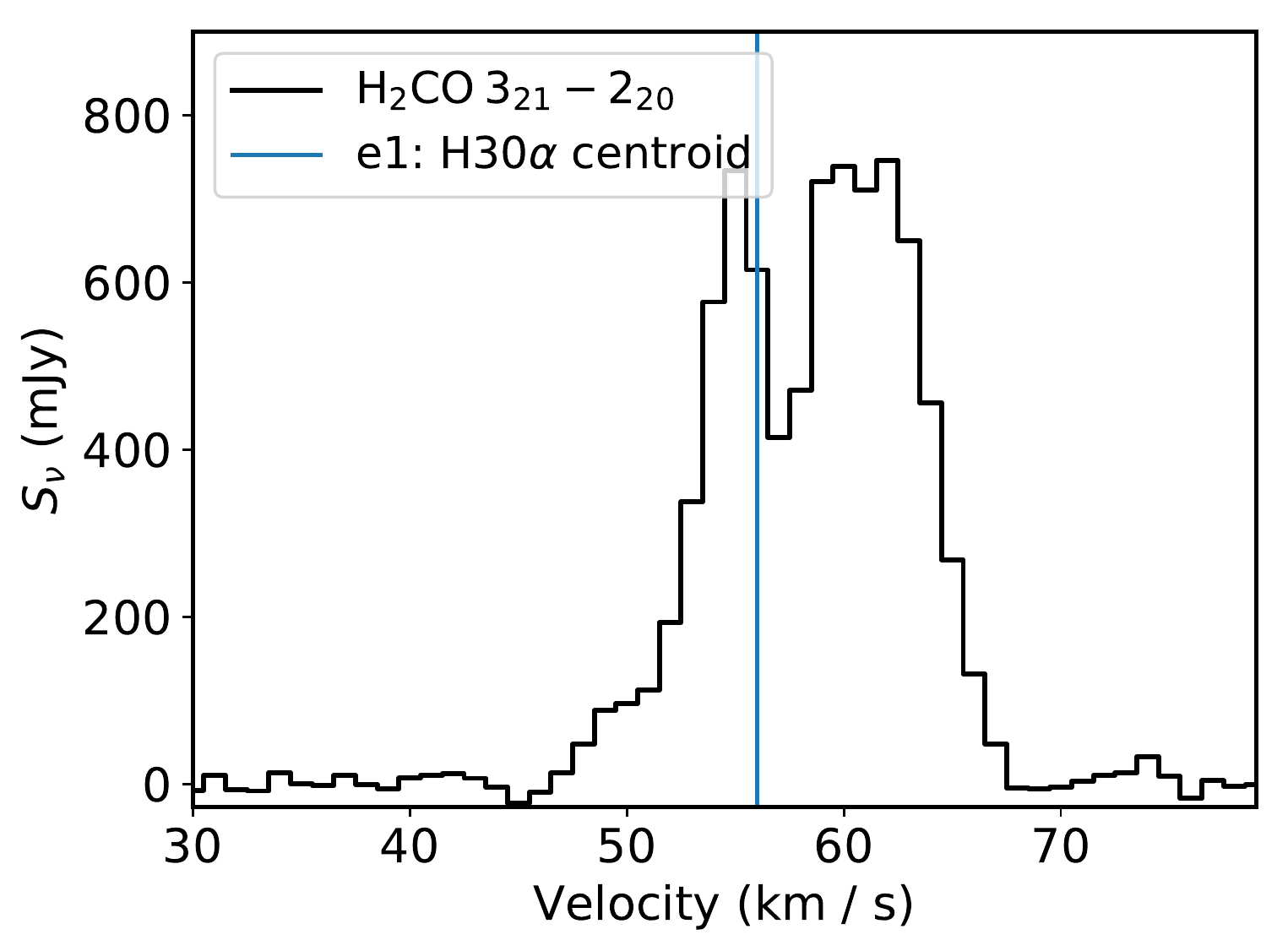}
   \includegraphics[scale=0.38]{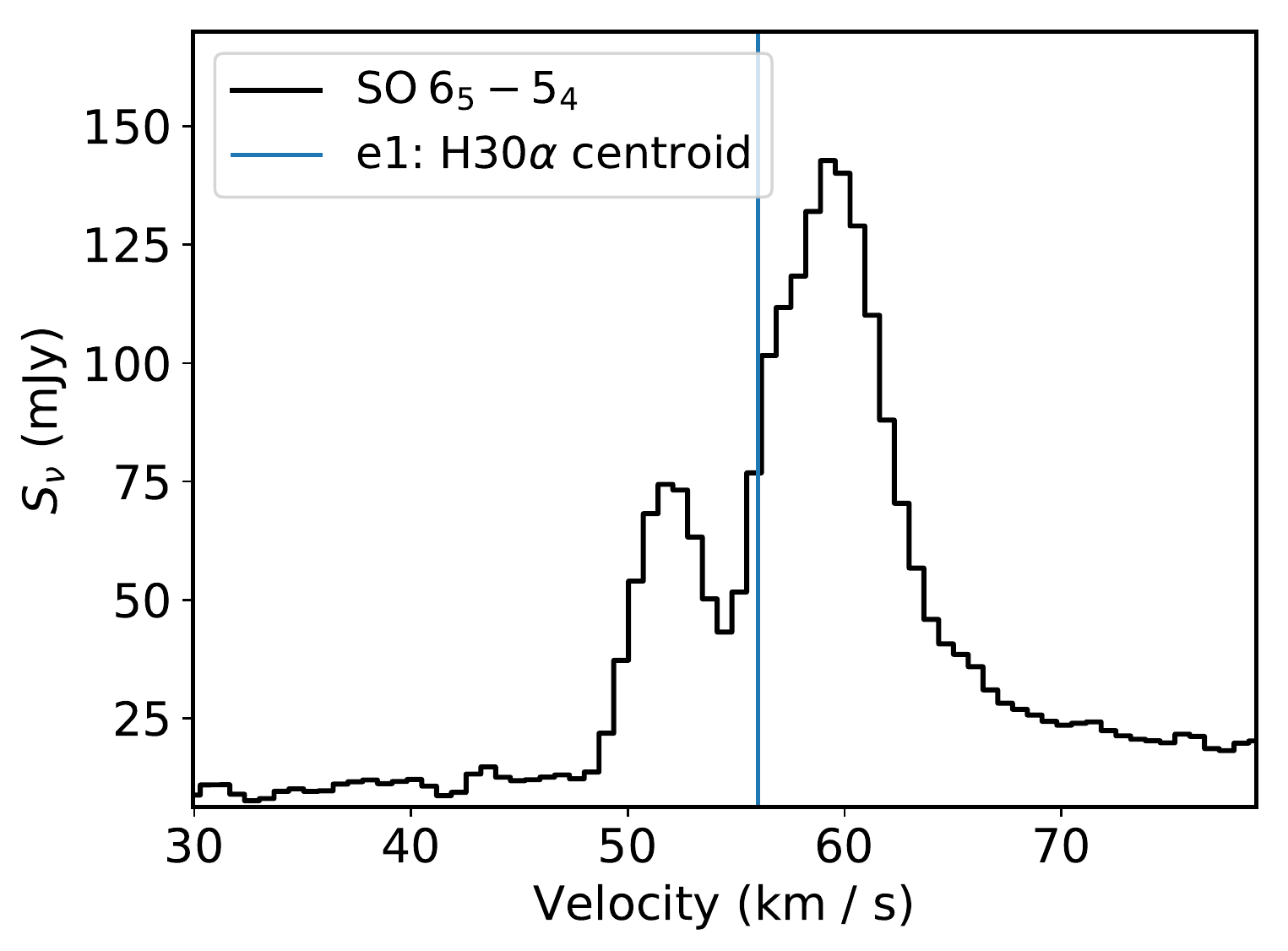}
   \includegraphics[scale=0.38]{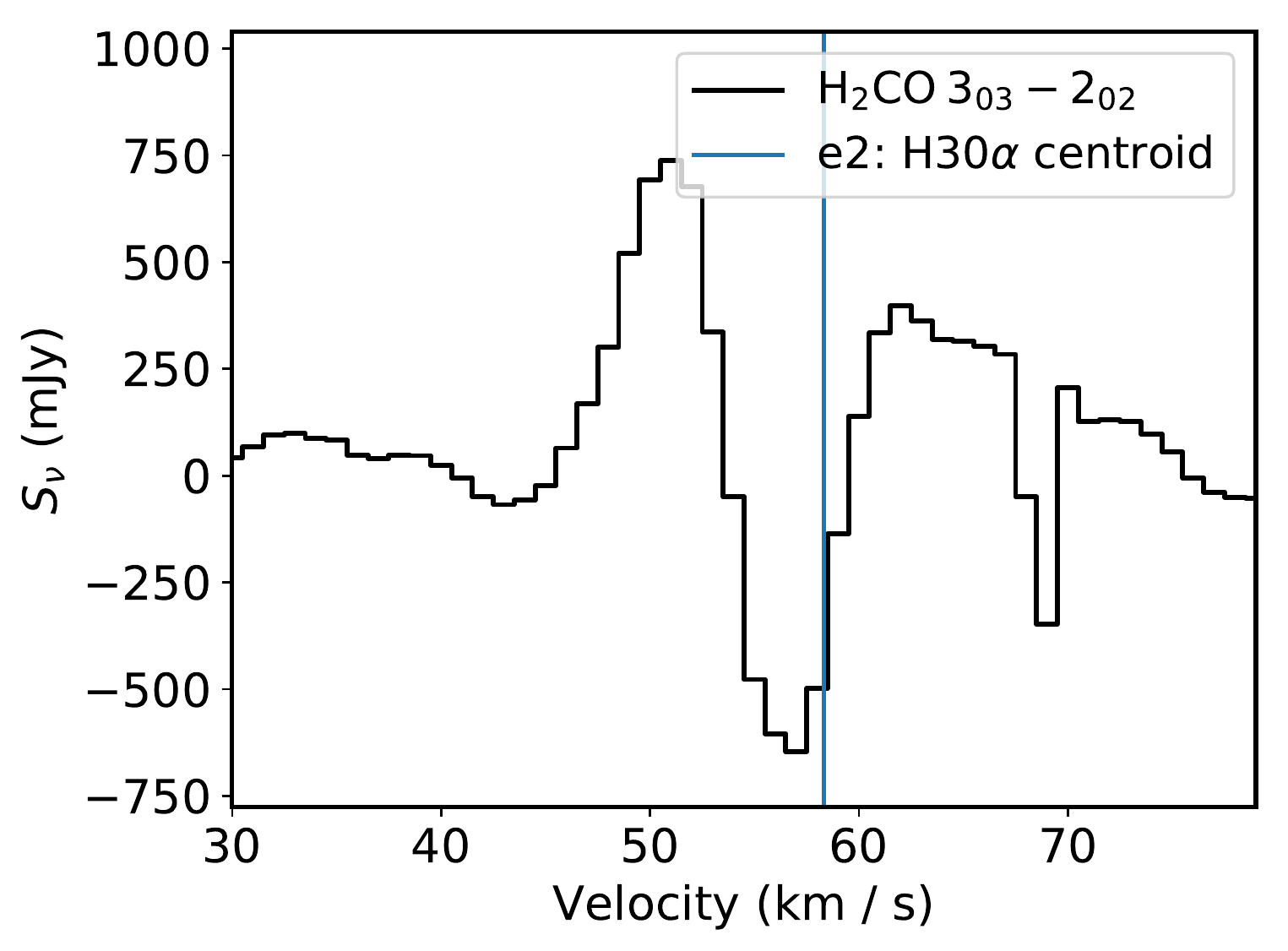}
   \includegraphics[scale=0.38]{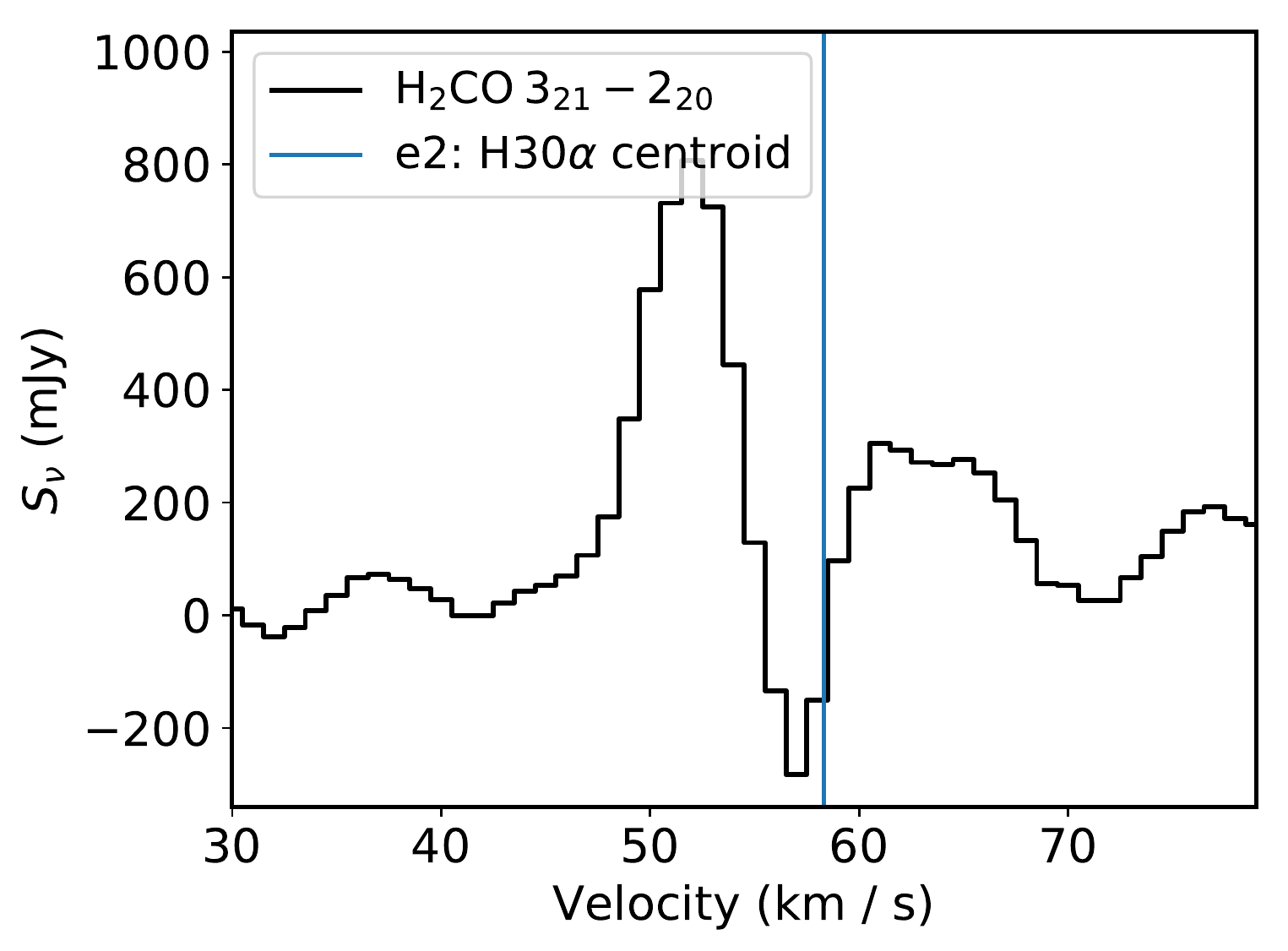}
   \includegraphics[scale=0.38]{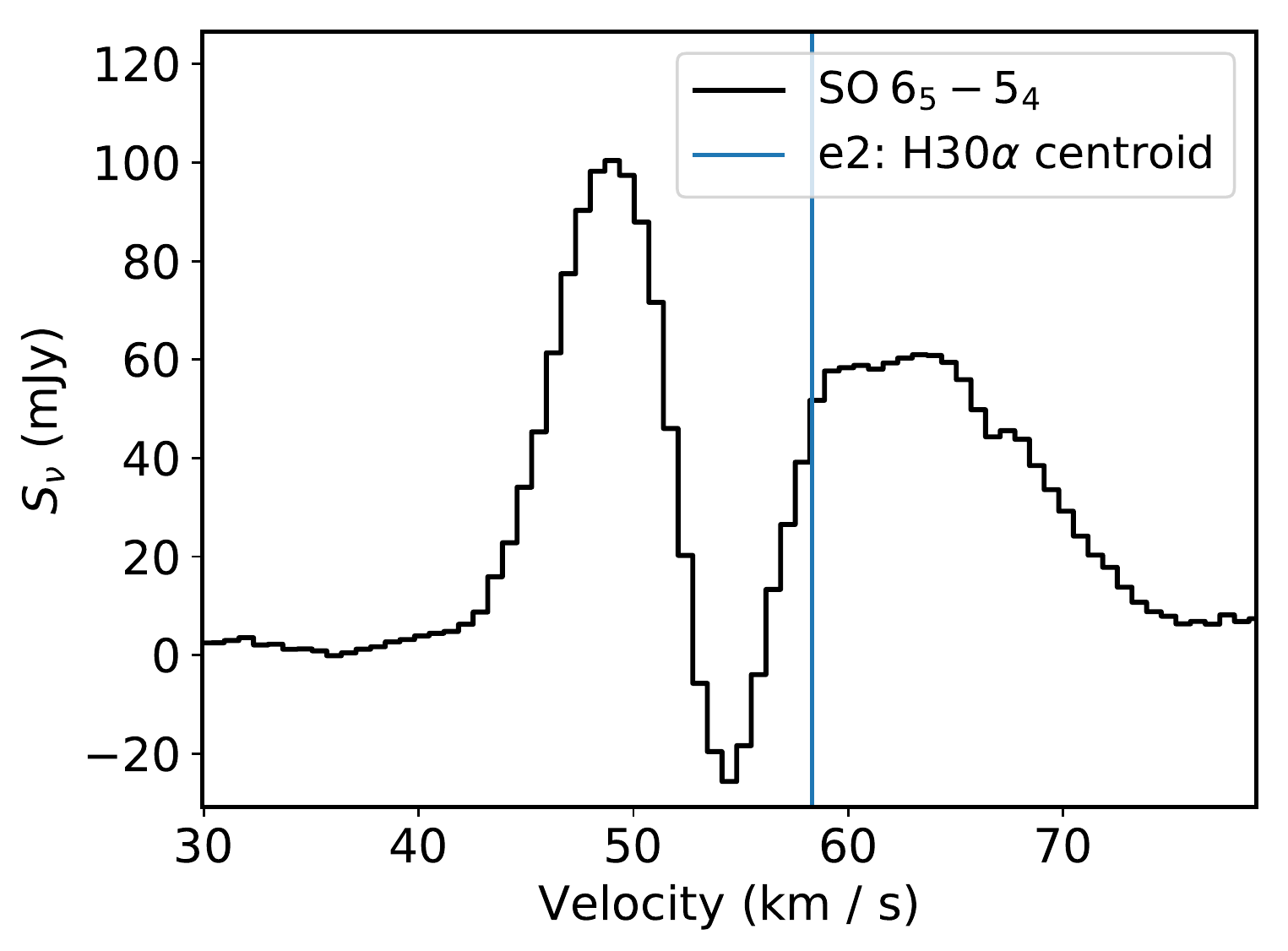}
   \includegraphics[scale=0.38]{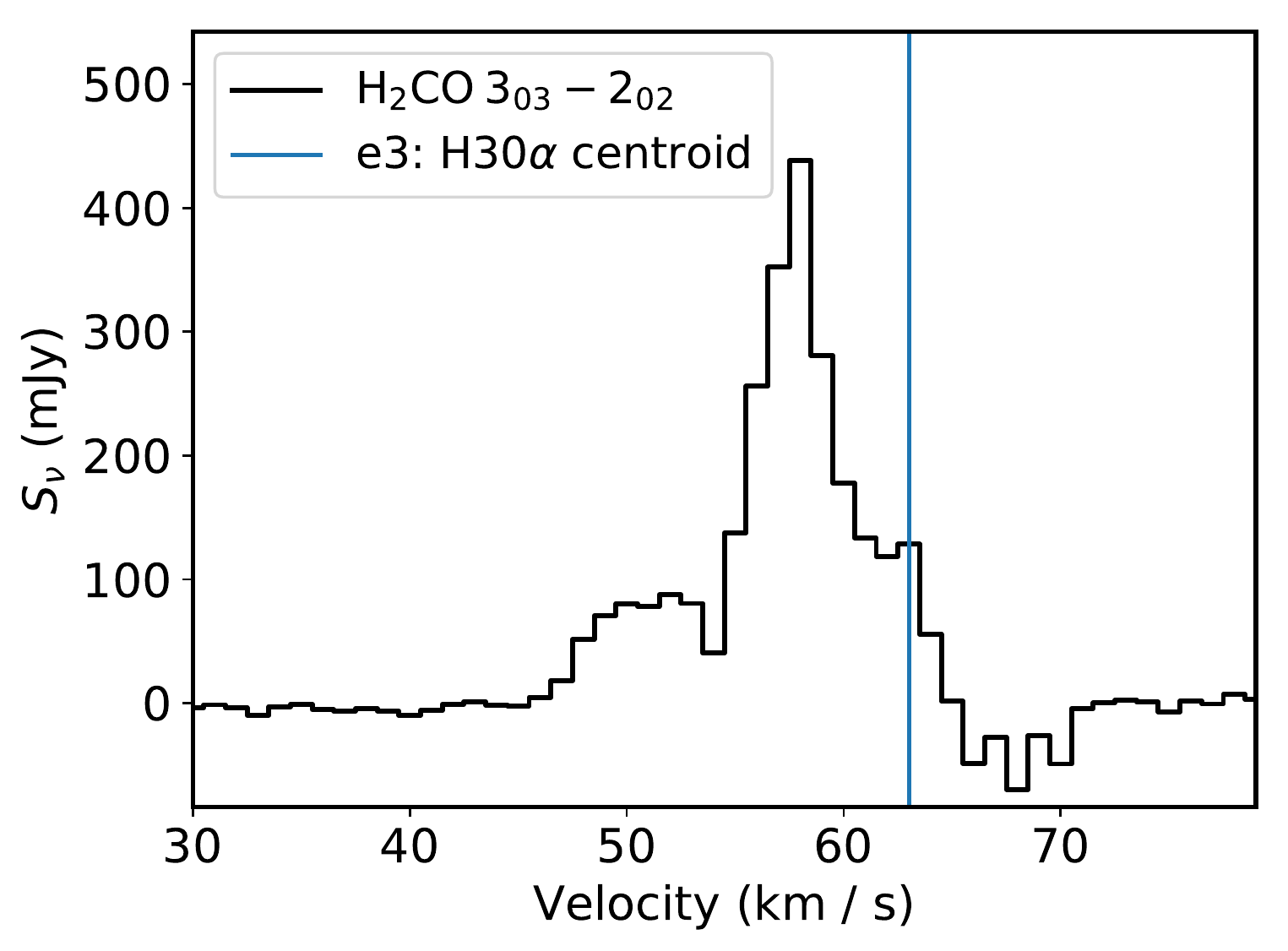}
   \includegraphics[scale=0.38]{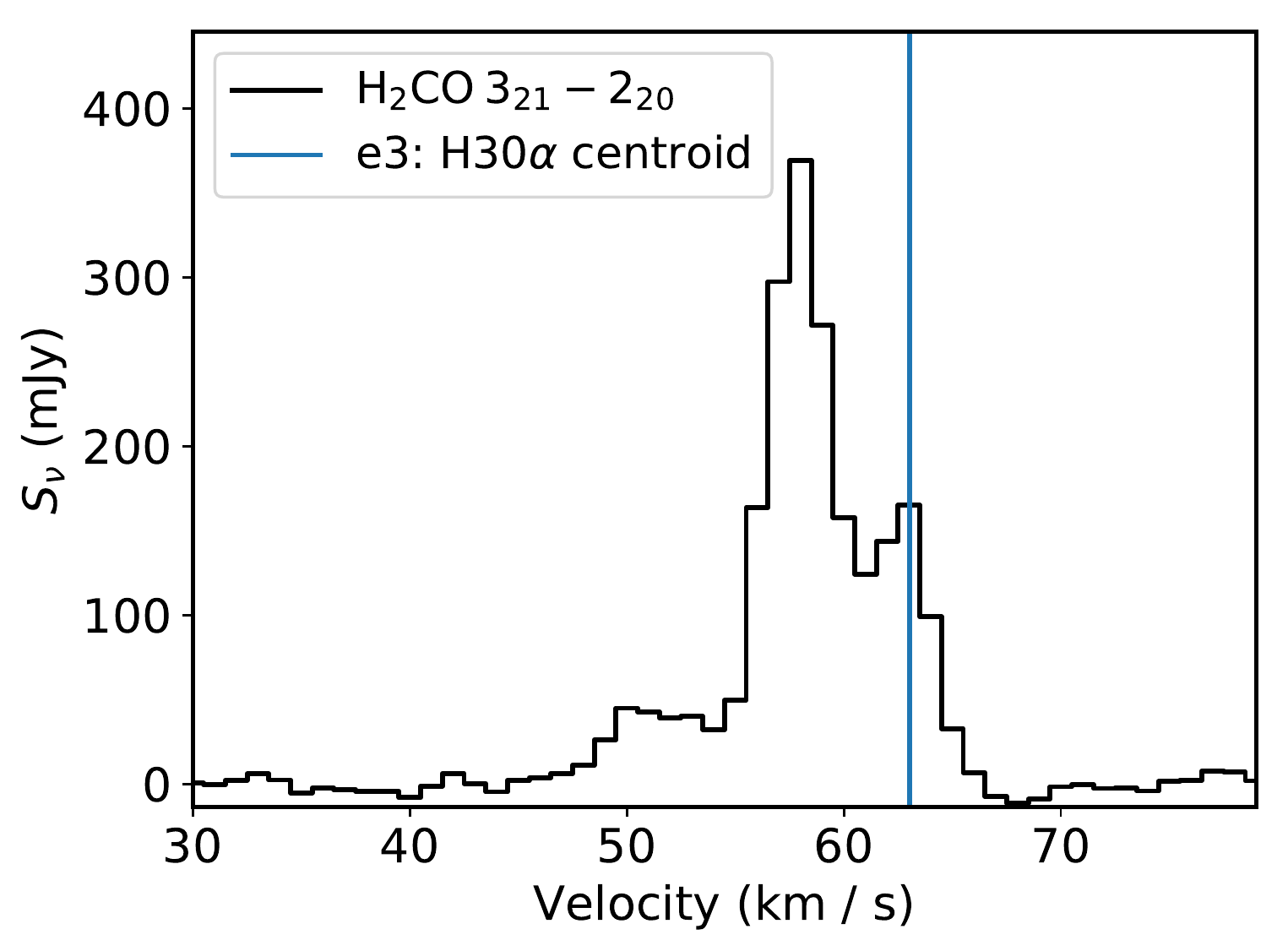}
   \includegraphics[scale=0.38]{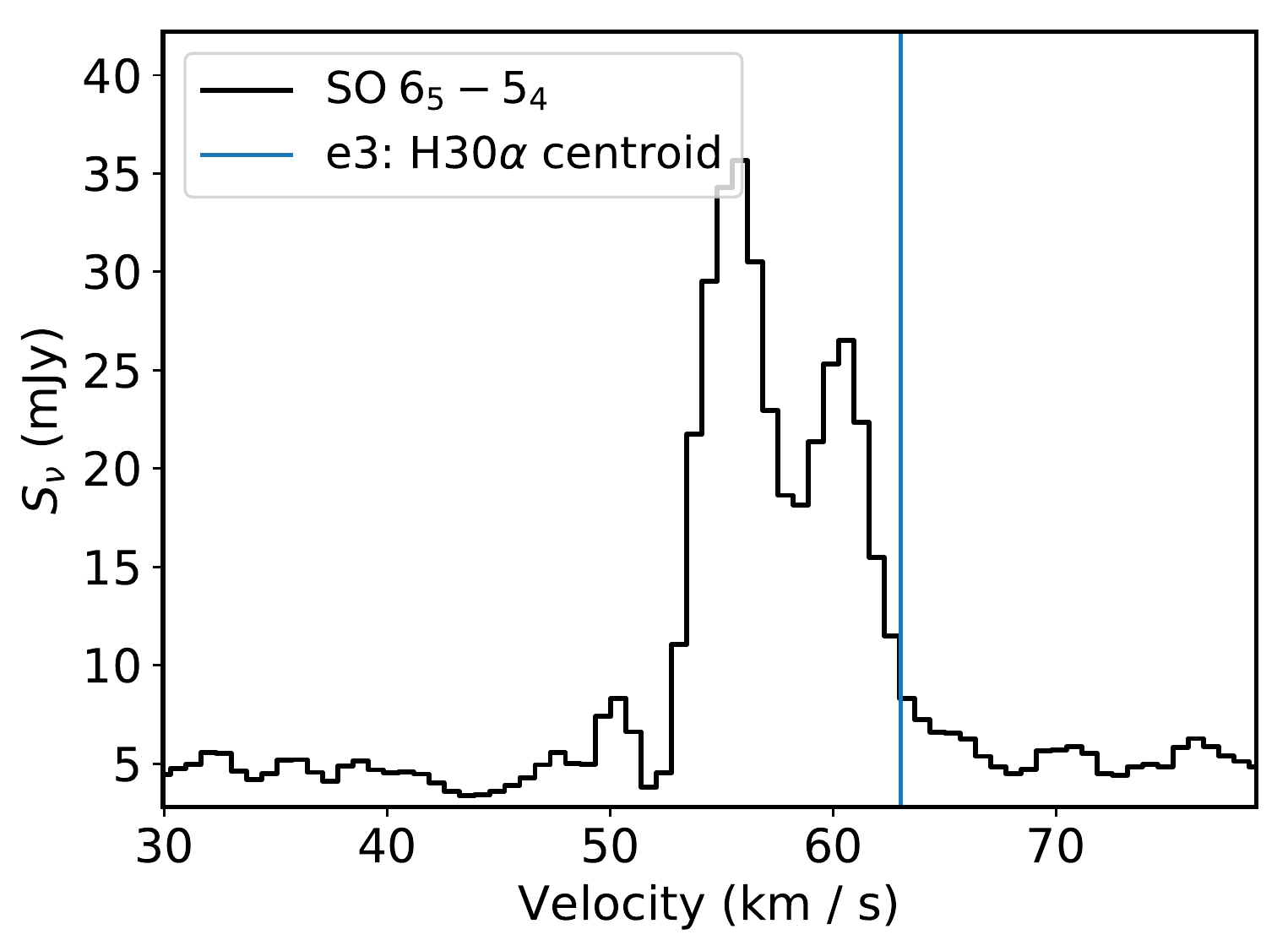}
    \caption{Spectral profiles of the molecular transitions \formai~ -- left --, \formaii~ -- middle --, \soi~ -- right --  for sources d2, e1, e2, and e3 from catalog \textit{B-H30}. The H30$\alpha$ velocity centroid  -- proxy for the stellar velocity -- is marked with a blue line.} \label{fig:StellarVelocities1}
\end{figure*}

\begin{figure*}[htb!]
   \ContinuedFloat
   \centering
   \includegraphics[scale=0.38]{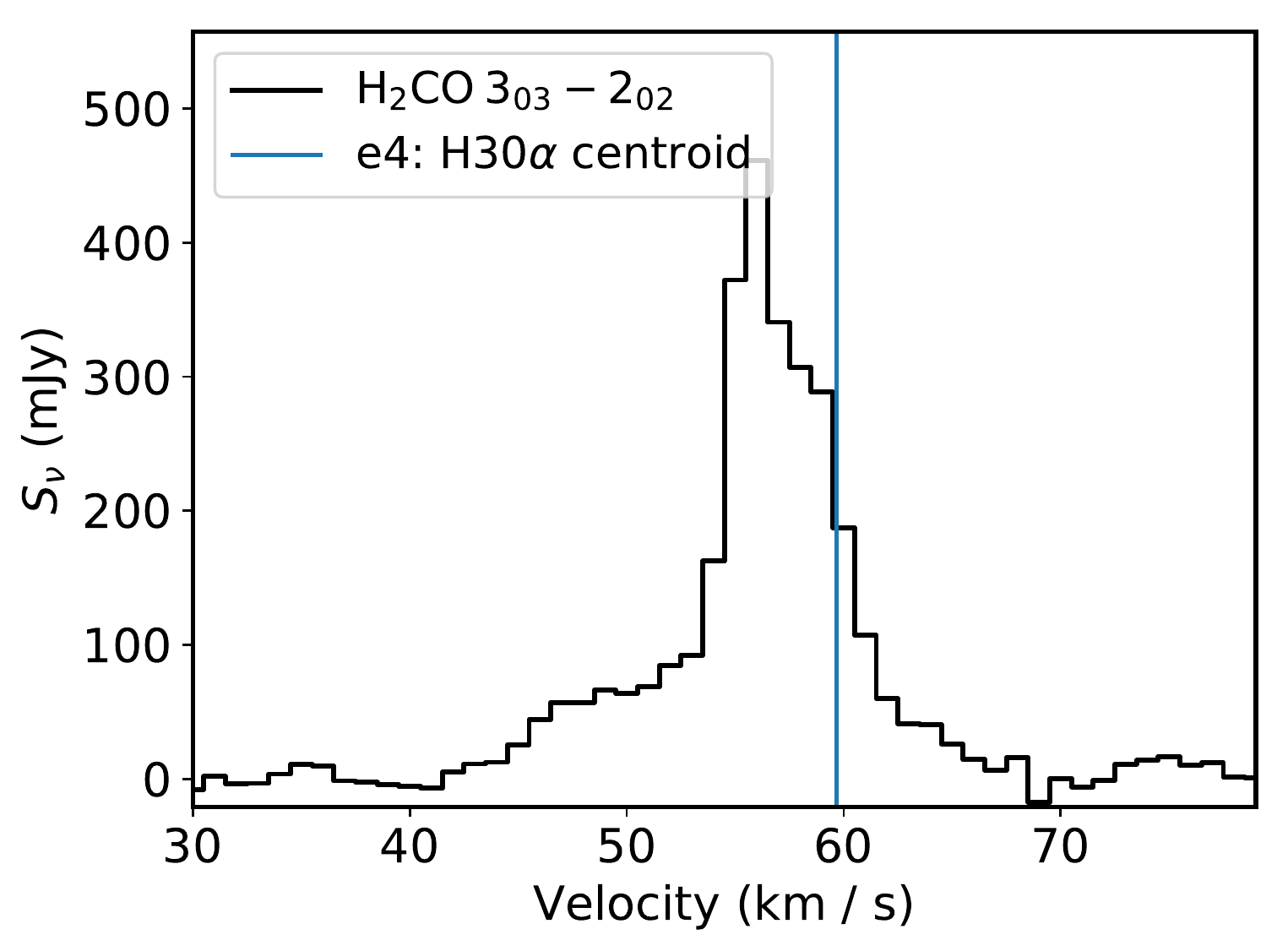}
   \includegraphics[scale=0.38]{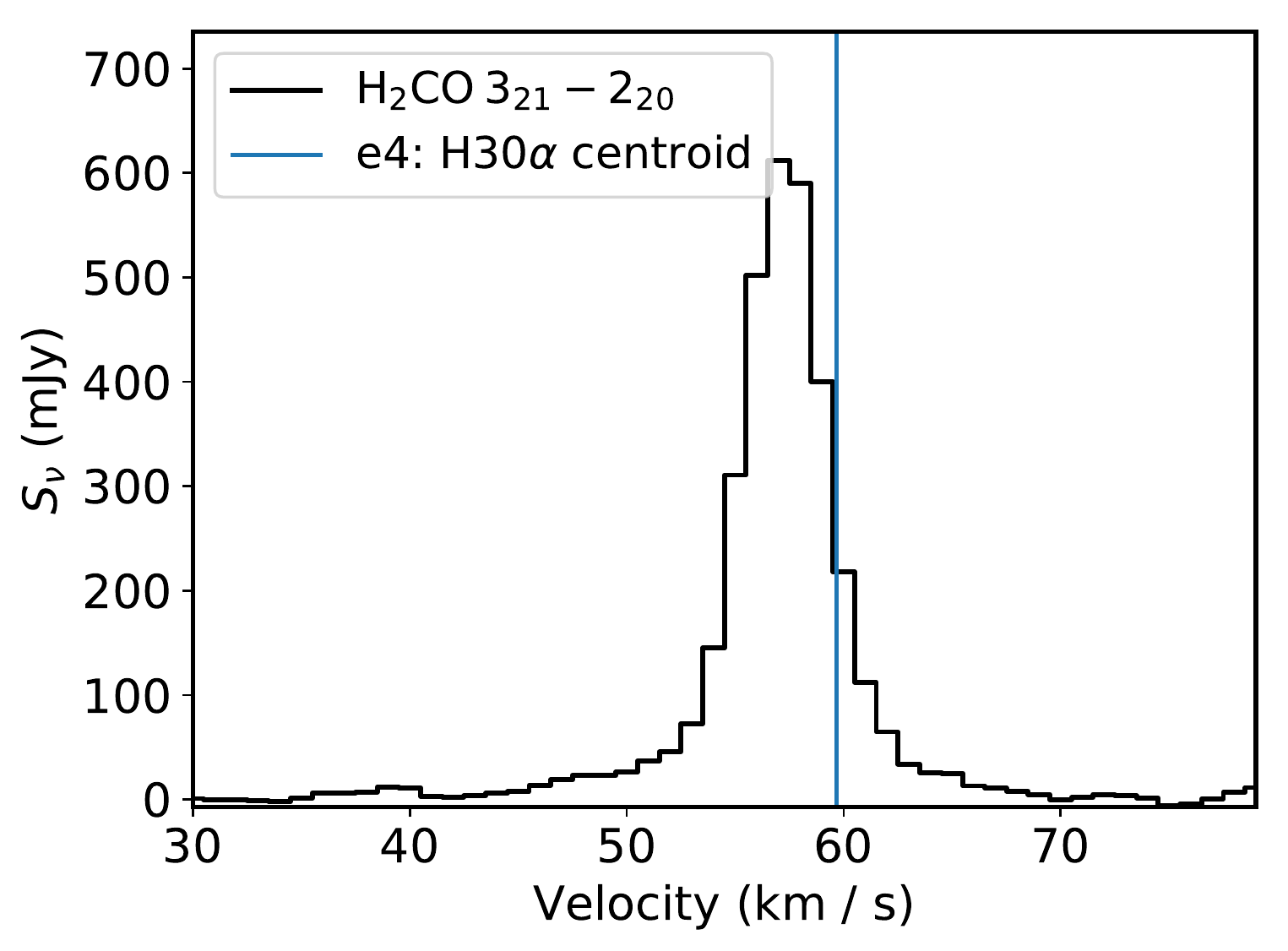}
   \includegraphics[scale=0.38]{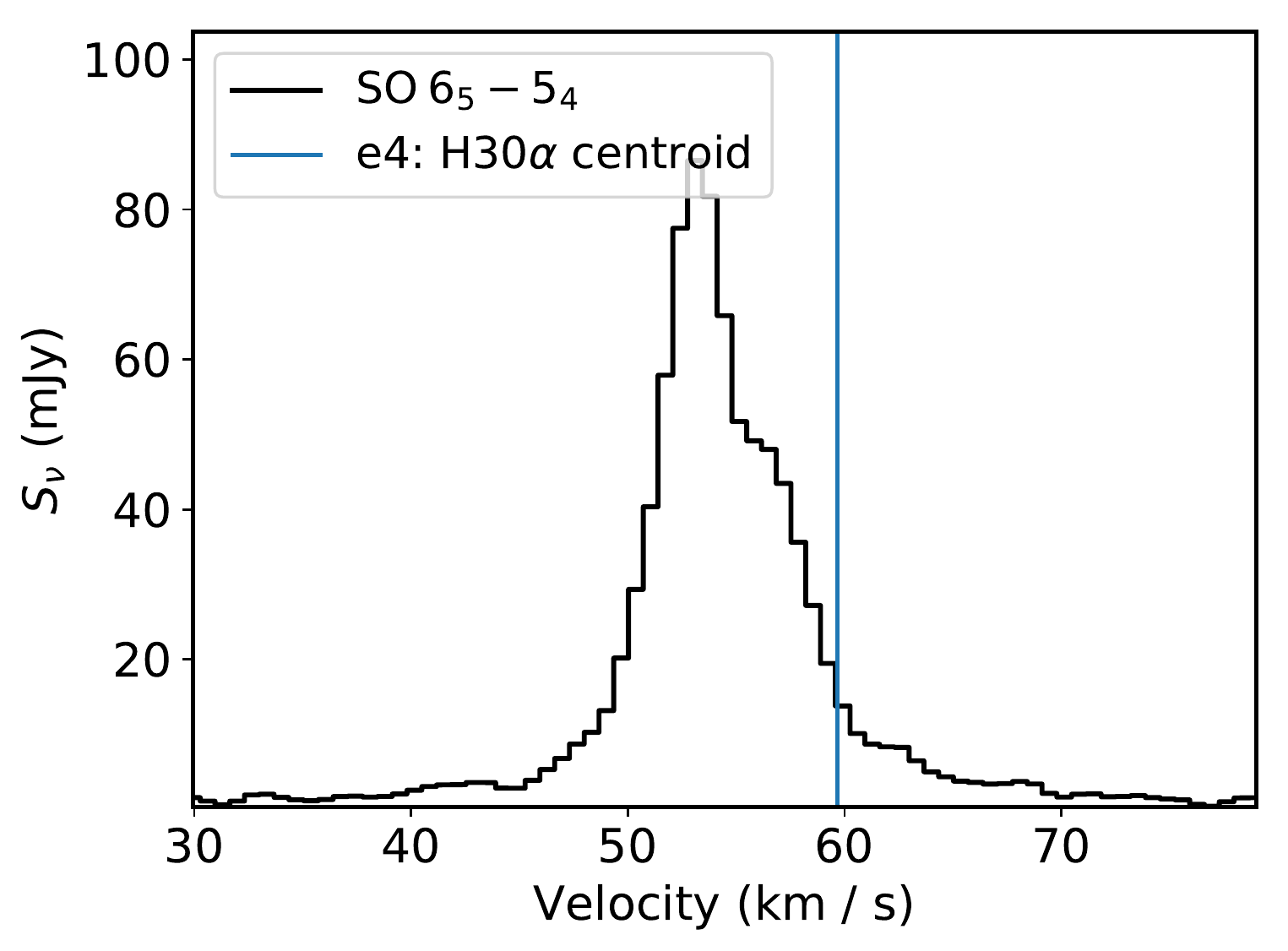}
   \includegraphics[scale=0.38]{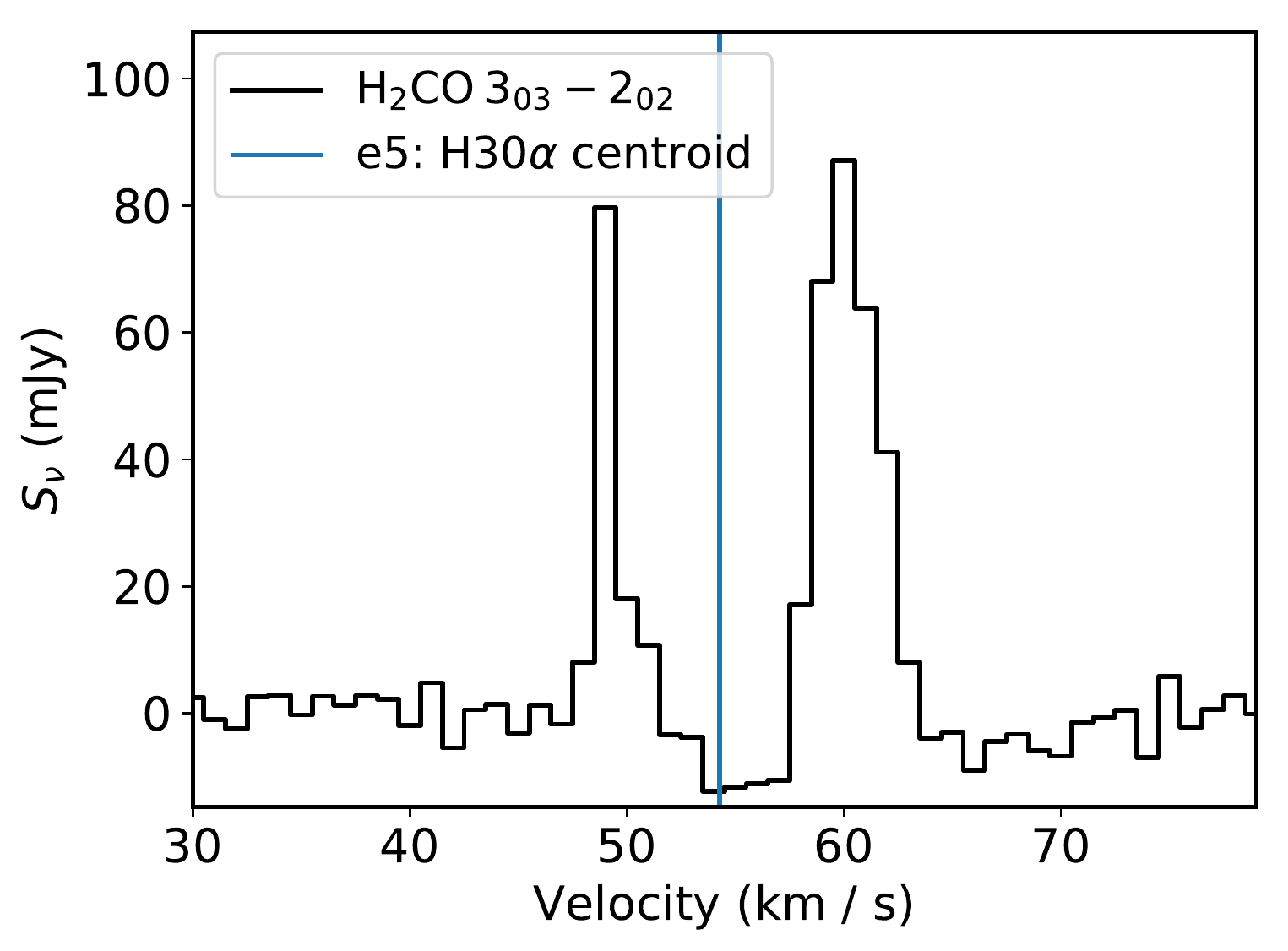}
   \includegraphics[scale=0.38]{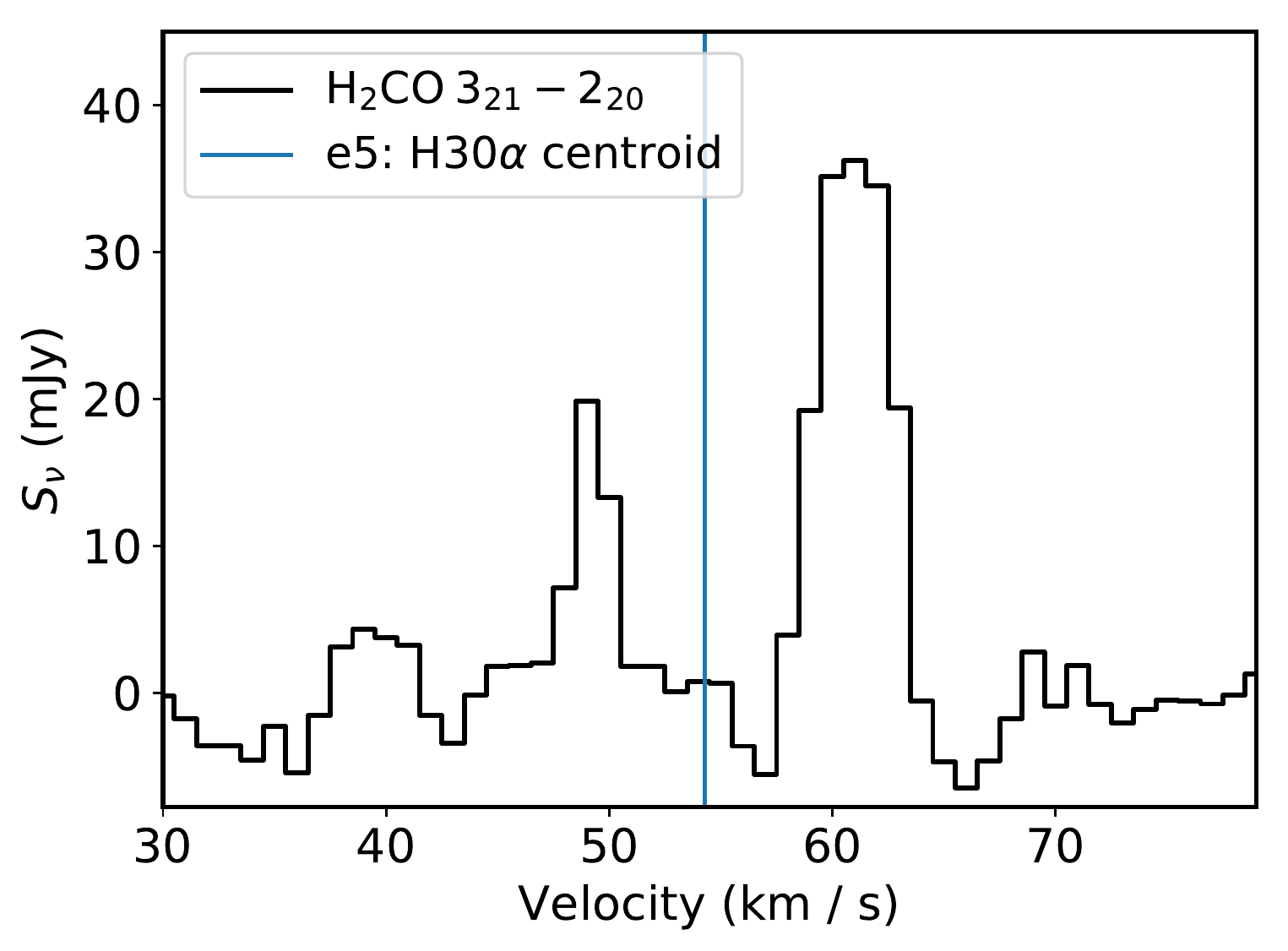}
   \includegraphics[scale=0.38]{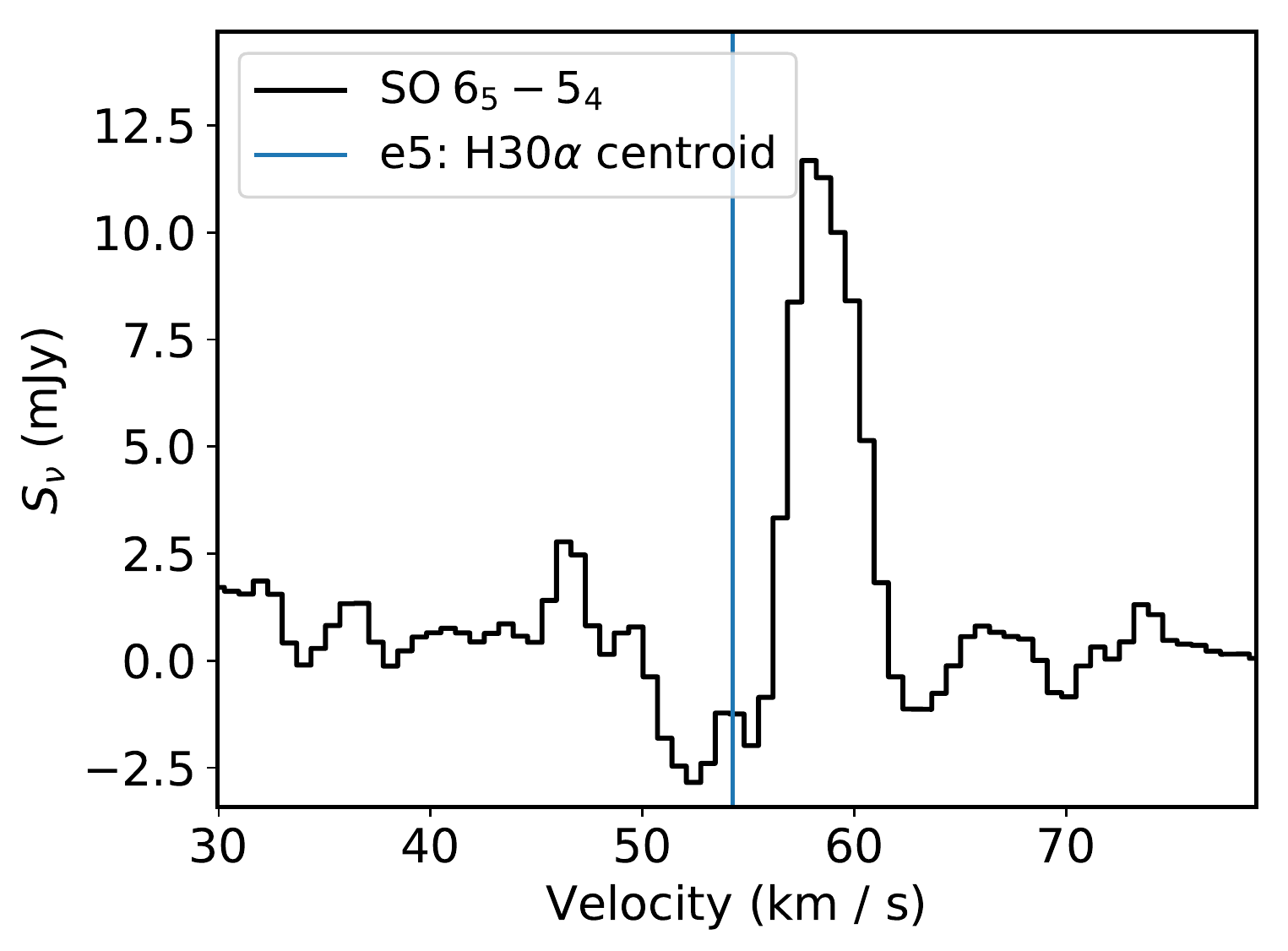}
   \includegraphics[scale=0.38]{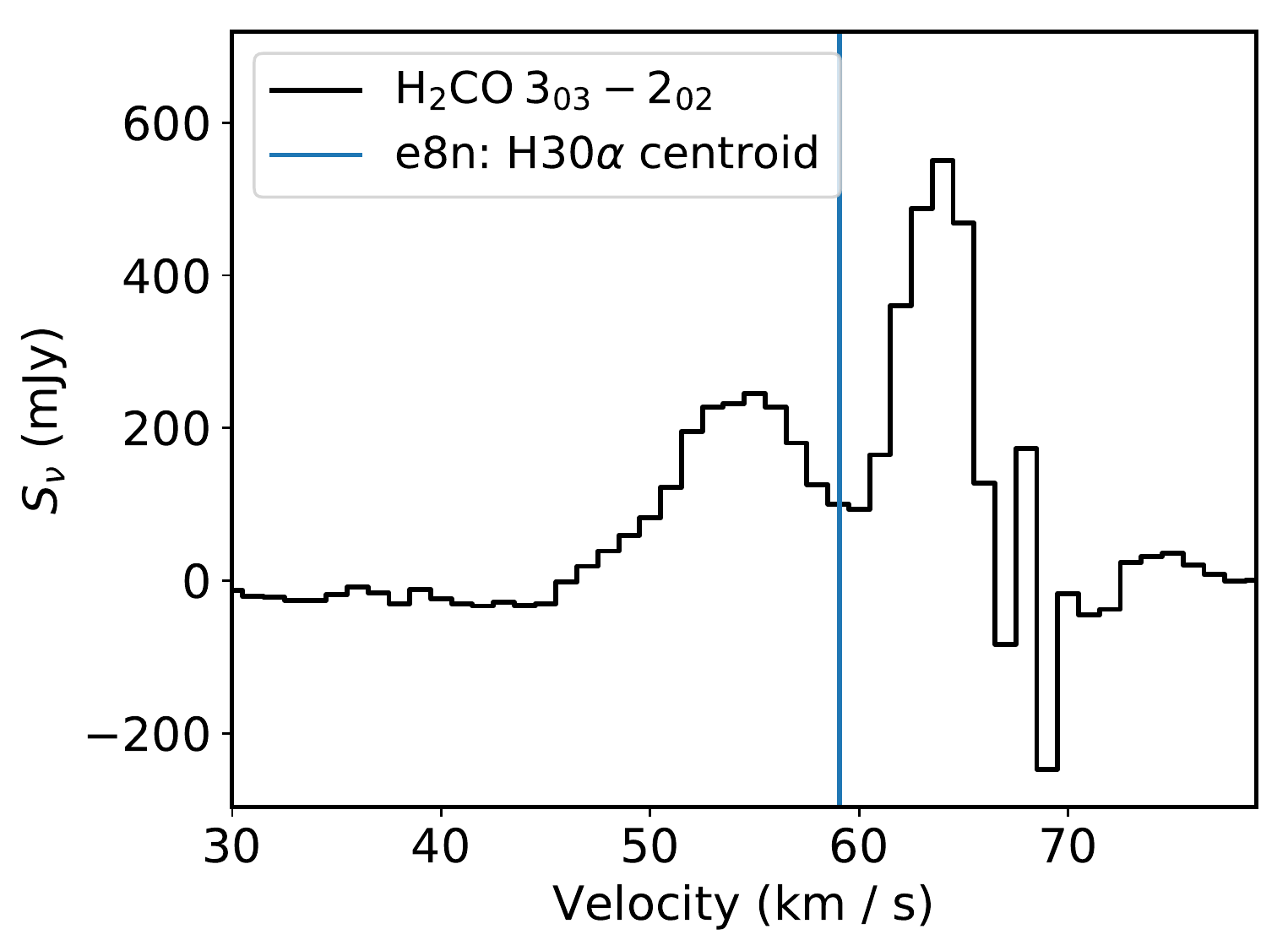}
   \includegraphics[scale=0.38]{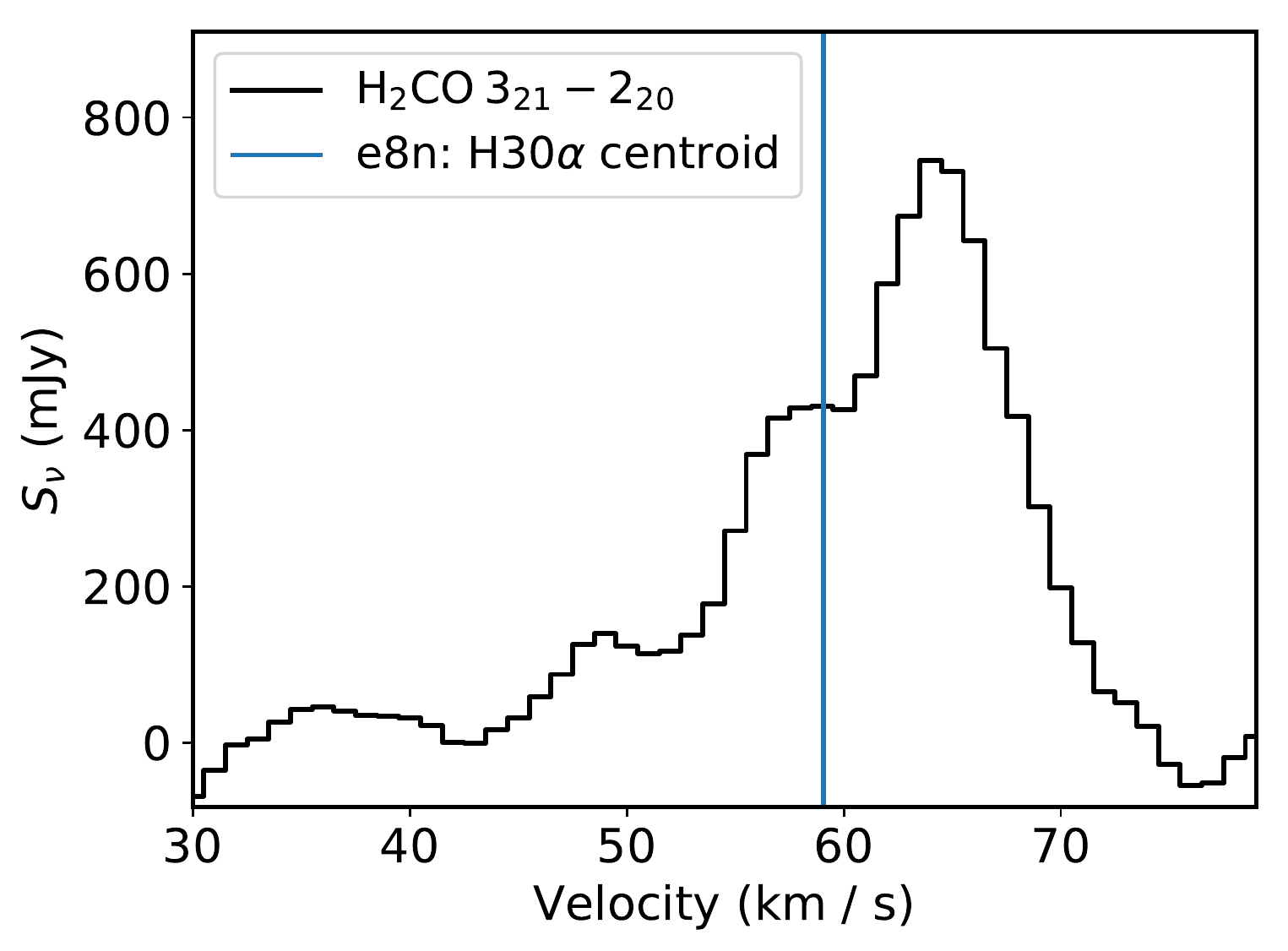}
   \includegraphics[scale=0.38]{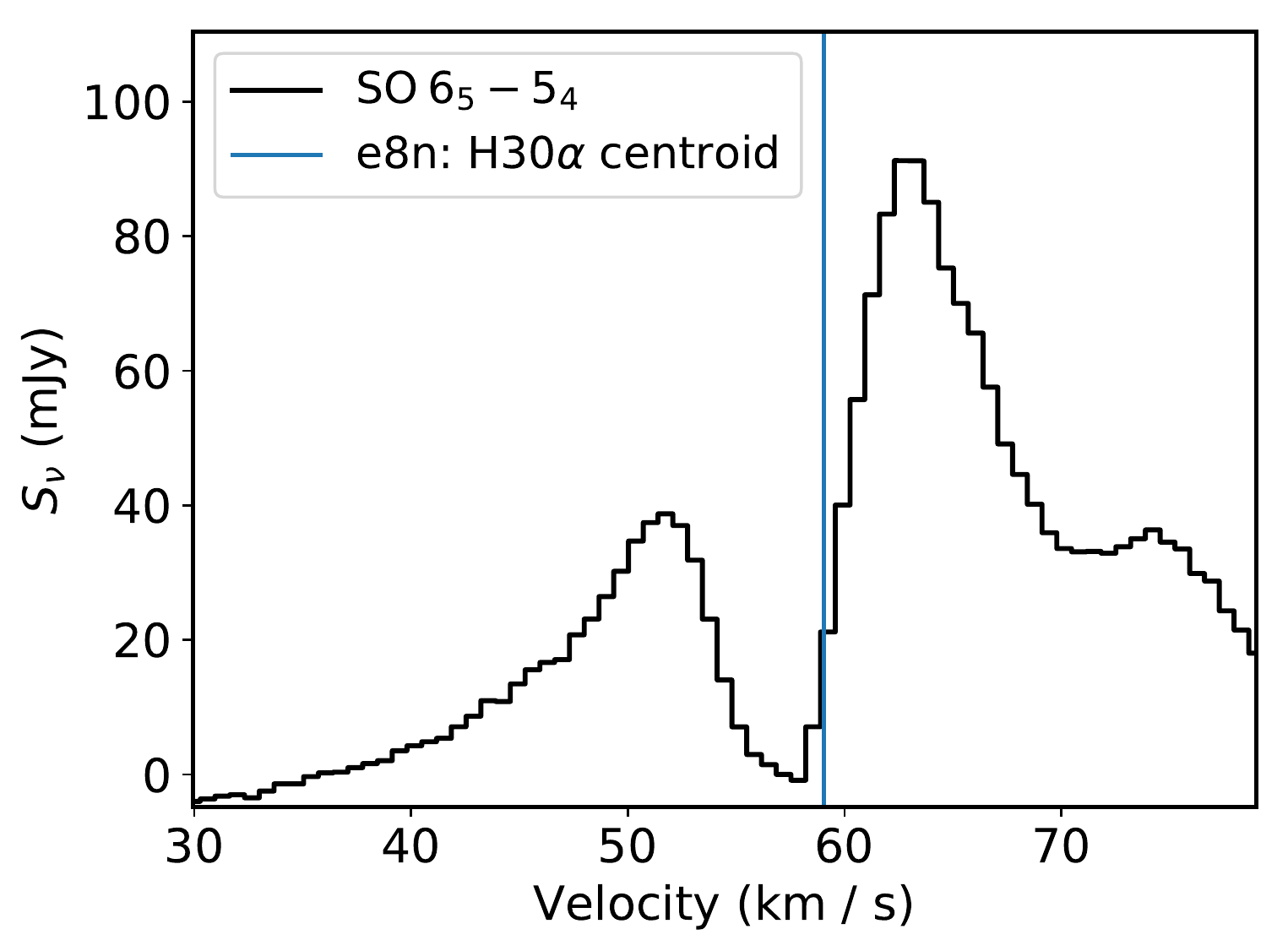}
    \caption{\textit{contd.}
    Spectral profiles of the molecular transitions \formai~ -- left --, \formaii~ -- middle --, \soi~ -- right --  for sources e4, e5, and e8n from catalog \textit{B-H30}. The H30$\alpha$ velocity centroid  -- proxy for the stellar velocity -- is marked with a blue line. 
    }
\end{figure*}

\section{Discussion} \label{sec:disc}

\subsection{The nature of hypercompact HII regions in W51 A}

We found bulk {\it outward} motions in the ionized gas for {\it all} the 7 sources for which we could infer their kinematics from a comparison of the H$30\alpha$ and H$77\alpha$ RLs. We also found that all 20 HC HII's in the common ALMA+VLA field are associated with dense molecular gas, although only 8 of them reside within compact molecular emission from a clearly defined HMC or envelope. For these 8 sources we determined that their molecular-gas kinematics is a mixture of infall and expansion motions. The ionized and molecular kinematics seem to be aligned within 5 \kms~.
We conclude that in our sample of HC HII's accretion is not dominant for the bulk of the ionized+molecular material at scales of a few $\times 10^3$ au. However, the presence of smaller-scale accretion streams has been observed with higher angular resolution for e8n and the younger sources W51 North and e2e (both without detected cm continuum and thus not in our catalog) by \citet{Goddi18}. 

\smallskip

We derived electron densities from the 2-cm continuum and from the RL analysis for a subset of the sources. 
Through both methods we find that the electron densities are larger than those of UC HII regions, yet smaller than typically defined for HC HII's \citep{Kurtz05}. Interestingly, we find that our sources follow the $n_e$ vs $D$ inverse relations previously found by \citet{GarayLizano99} and \citet{KimKoo} for samples of compact and UC HII regions. 
Our interpretation is that we are characterizing a population of HC HII regions that are more common than landmark objects. This, combined with the previously discussed expansion kinematics in ionized gas, suggests that the majority of these HC HII's are essentially smaller versions of expanding UC and compact HII regions.

The previous interpretation is also supported by the finding that most of our objects are not ionized by O-type stars, but by early B-type stars. Since the ionizing stars are close to or already in the ZAMS they should follow a mass distribution close to the IMF. Thus it is not surprising that there are more B-type than O-type HC HIIs. 

We also note that none of our sources fall in the category of ``broad  recombination line'' HII regions \citep[e.g.,][]{Sewilo04,DePree04}, 
which could be truly different objects where accretion is still ongoing.

We speculate that there could be two different physical objects worth receiving the  label of ``hypercompact'' HII regions because of their $D \sim 10^{-3} - 10^{-2}$ pc diameters: i) the most common would be stars with practically no leftover accretion producing tiny HII regions that are mostly expanding. The ionizing sources of these objects would be dominated by B-type stars, although a few O-stars are also expected  \citep[e.g., see discussion in][]{Kurtz02}; and  ii) rarer objects with extremely high densities and large dynamical recombination linewidths. These could be the ionized accretion flows expected to occur in the formation of stars more massive than about 20 to 30 $M_\odot$ \citep{Keto07,Peters10}.
Under the previously described scenario, not all HC HII's would evolve to become UC HII's ionized by ZAMS O-type stars. It would depend on the local gas reservoir and final accretion history \citep{Peters10b,GM11}. 

\subsection{Comparison to previous studies}

The compact radio continuum objects e1 and e2 were identified by \citet{Scott78}. 
Higher resolution observations at 3.6 and 1.3 cm by \citet{Gaume93} found d2, e3, e4, and e5. Those authors derived electron densities and emission measures in agreement with our values in most cases ($n_{\rm e} \sim 10^4$ to $10^6$ cm$^{-3}$, $EM \sim 10^7$ to $10^9$ pc cm$^{-6}$). Source e6 was first observed by \citet{Mehringer94}, who derived lower limit densities and emission measures for e1, e2, and e6 $n_{\rm e} \sim 10^3$ to $10^6$ cm$^{-3}$, $EM \sim 10^6$ $\rm pc ~ \rm cm^{-6}$, which are lower but still comparable with our calculations. Additionally, for e2 \citet{KZK08} found $n_{\rm e} \sim 2 \times 10^6$ cm$^{-3}$ derived from H$53\alpha$ and H$66\alpha$, which is about an order of magnitude larger than our H$77\alpha$ derived value but consistent with our continuum-derived lower limit (see tables \ref{tab:Cont_Params} and \ref{tab:Broadening_Components}). 
This difference can be explained if the RLs are partially optically-thick (the continuum of e2 has $\tau_\mathrm{c} >> 1$) and the lower quantum numbers are more sensitive to denser gas, or by our higher S/N data. Source e8n was first seen at 1.3 cm by \citet{ZhangQ97}, but not characterized. 

\smallskip

Earlier references found evidence for molecular infall toward e2 \citep[e.g.,][]{ZhangQ97}. Later it was found that e2 is 
resolved into three (sub)mm cores e2-E, e2-W, and e2-NW \citep{Shi10a}. 
e2-W corresponds to the cm e2 HC HII region. Accretion activity seems to be concentrated in e2-E \citep{Shi10b,Goddi2016}, so our line profiles could result from a mixture of emission and absorption from more than one (sub)mm core.

\section{Conclusions} \label{sec:concl}

$\bullet$ We derived deconvolved diameters for the 2-cm continuum of 10/20 sources in catalog \textit{B}, and found them to be within the regime of HC HII regions: $D \sim 10^{-3}$ to $10^{-2}$ pc.

$\bullet$ We calculated the electron densities of these HC HII's, finding $n_{\rm e, c} \sim 10^4$ to $10^5$ cm$^{-3}$. The respective emission measures are $EM \sim 10^7$ to $10^8$ pc cm$^{-6}$.  

$\bullet$ We analyzed the RL's of the 7 objects in catalog \textit{B-H30-H77} and calculated electron densities also of the order of $n_{\rm e, RL} \sim 10^4$ to $10^5$ cm$^{-3}$.

$\bullet$ The electron densities obtained from both methods are in the same range for the sample but with significant scatter for individual measurements. They are smaller than often  defined for HC HII regions ($n_{\rm e} \geq 10^6$ cm$^{-3}$, $EM \geq 10^{10}$ cm$^{-6}$ pc). 
However, they follow the relation between $n_e$ and $D$ previously found in the literature for samples UC and compact HII regions.

$\bullet$ We estimated the Lyman continuum photon rates of the ionizing stars and found that these HC HII's are ionized by early B-type stars in most cases (5 to 8 out of 10).

$\bullet$ From the analysis of the RL velocity centroids, we found that the bulk of the ionized gas in the 7 objects in catalog \textit{B-H30-H77} is going outwards, suggesting that accretion has mostly ceased within our sample.

$\bullet$ We found that all the 20 cm-continuum sources in catalog $B$ have detections in \forma, and 14 of them in \so. Eight of those sources still reside within a HMC or envelope. This suggests that, at least in a clustered environment such as in W51 A, the HC HII stage is always associated with the presence of some dense molecular gas. 

$\bullet$ The molecular-line profiles of the 8 HC HII's still embedded in compact molecular emission show that 5 of them have evidence of expansion motions, whereas 3 of them have evidence of infall. The three infall candidates (e2, e3, and e4) belong to the same (sub)cluster. 

$\bullet$ We compared the 
velocity centroids of the 6 H$30\alpha$ detections also embedded in compact cores to the molecular-line emission. The ionized and molecular tracers are always within 1 to 5 \kms~ from each other. For the neighbouring sources e2, e3, and e4 the H$30\alpha$ line is systematically redshifted, suggesting small relative motions between the star cluster and its surrounding  material. 

\smallskip

We have performed a thorough characterization of the HC HII population in W51 A, where \textit{hypercompact} is defined as having a diameter smaller than 0.05 pc. We find that these HC HII's behave as expected in some aspects (presence of dense molecular gas, infall and outflow molecular kinematics), but are surprising in some other (in most cases less dense than expected, purely outflows in ionized gas). The finding that they tend to behave like smaller UC HII's ionized by early B-type stars suggests that there could be two different types of very small HII regions: the more common one associated to the more abundant B-type stars, and a few truly \textit{hyperdense} objects associated with O-type stars in a specific evolutionary stage. 


\acknowledgments
The authors are grateful to the referee for providing thoughtful reviews which improved this paper. 
RRS, RGM, and AG acknowledge support from UNAM-PAPIIT project IN104319. The National Radio Astronomy Observatory is a facility of the National Science Foundation operated undercooperative agreement by Associated Universities, Inc. This paper makes use of the ALMA data set 2013.1.00308.S and VLA data set 13A-064. ALMA is a partnership of ESO (representing its member states), NSF (USA), and NINS (Japan), together with NRC (Canada), NSC, and ASIAA (Taiwan), and KASI (Republic of Korea), in cooperation with the Republic of Chile. The Joint ALMA Observatory is operated by ESO, AUI/NRAO, and NAOJ.

\facilities{ALMA, VLA} 
\software{NumPy \citep{Numpy}, PySpecKit \citep{GinsburgMirocha11}, APLpy \citep{AplPy}, CASA \citep{McMullin07}}

\smallskip

\bibliographystyle{yahapj}
\bibliography{references}

\newpage

\appendix

\begin{figure*}[!t]
    \centering
    \includegraphics[scale=0.46]{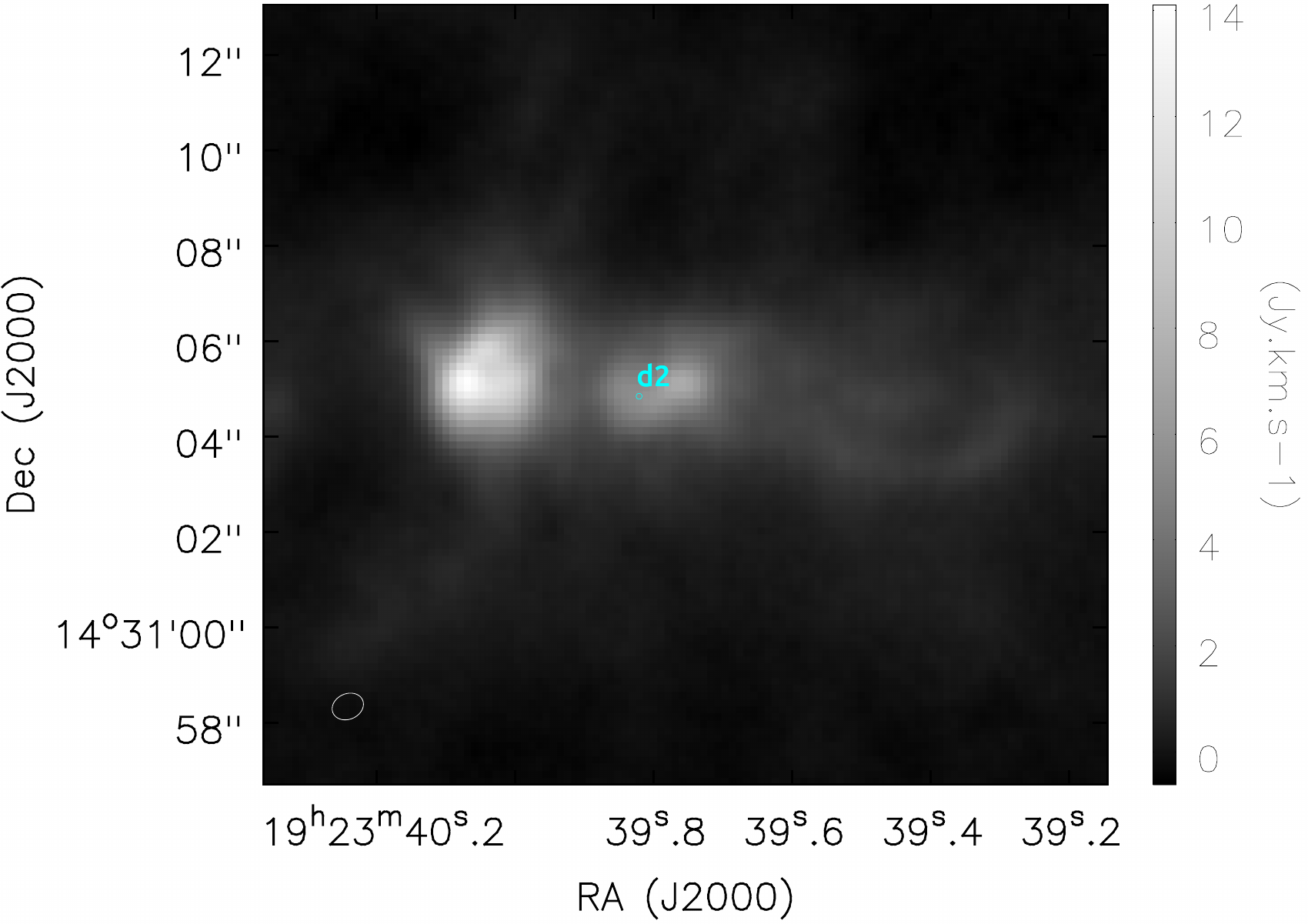}
    \includegraphics[scale=0.46]{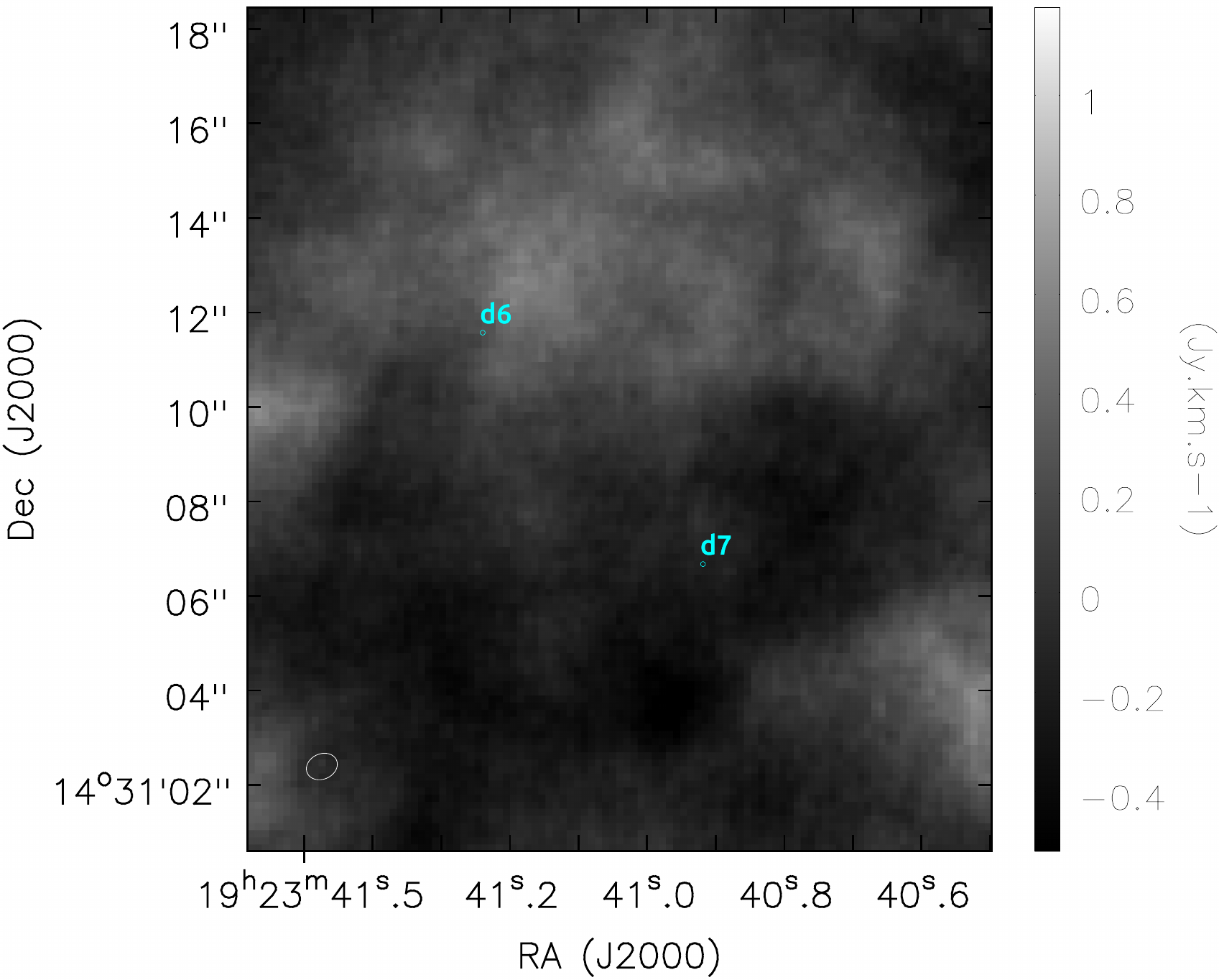}
    \includegraphics[scale=0.45]{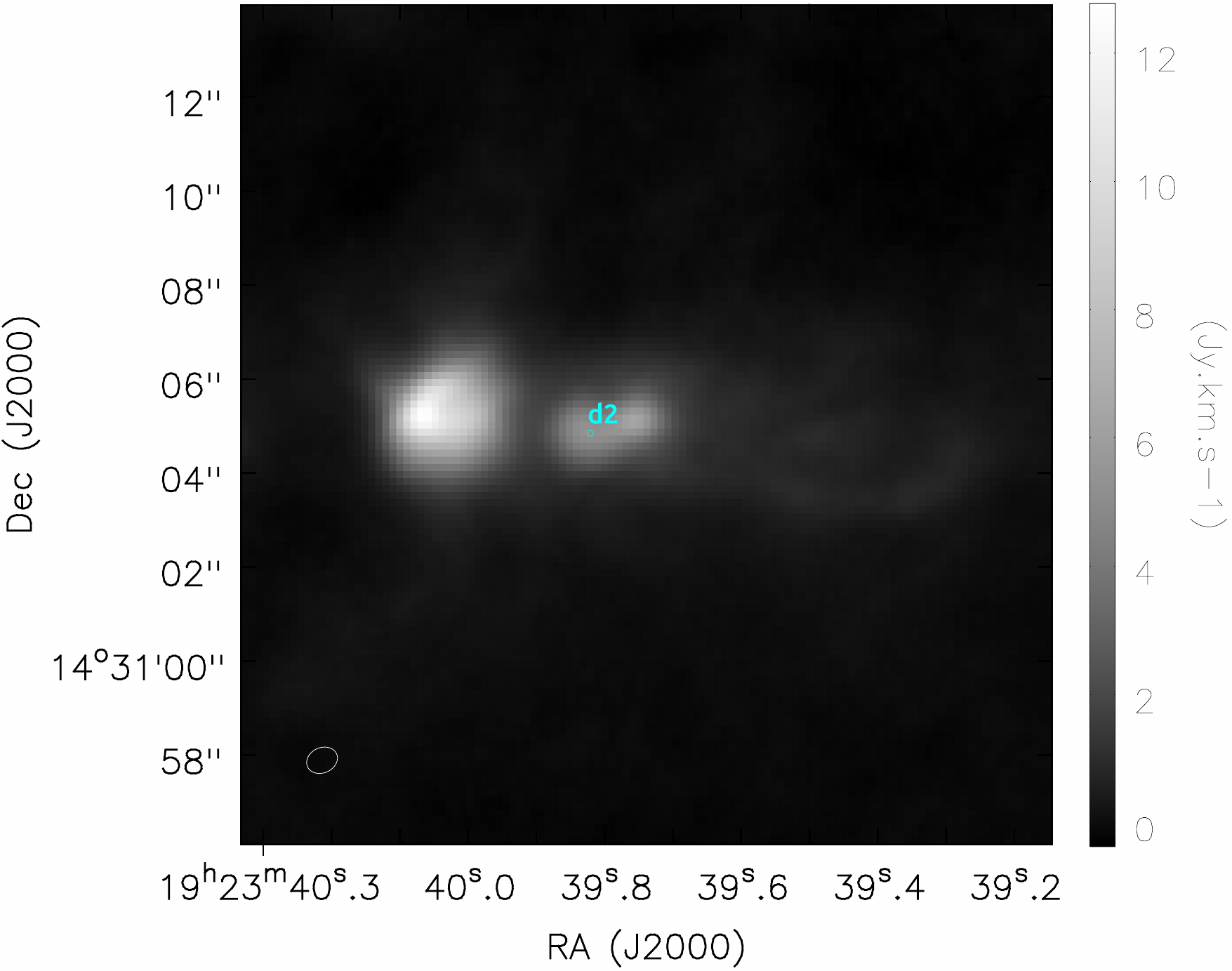}
    \includegraphics[scale=0.45]{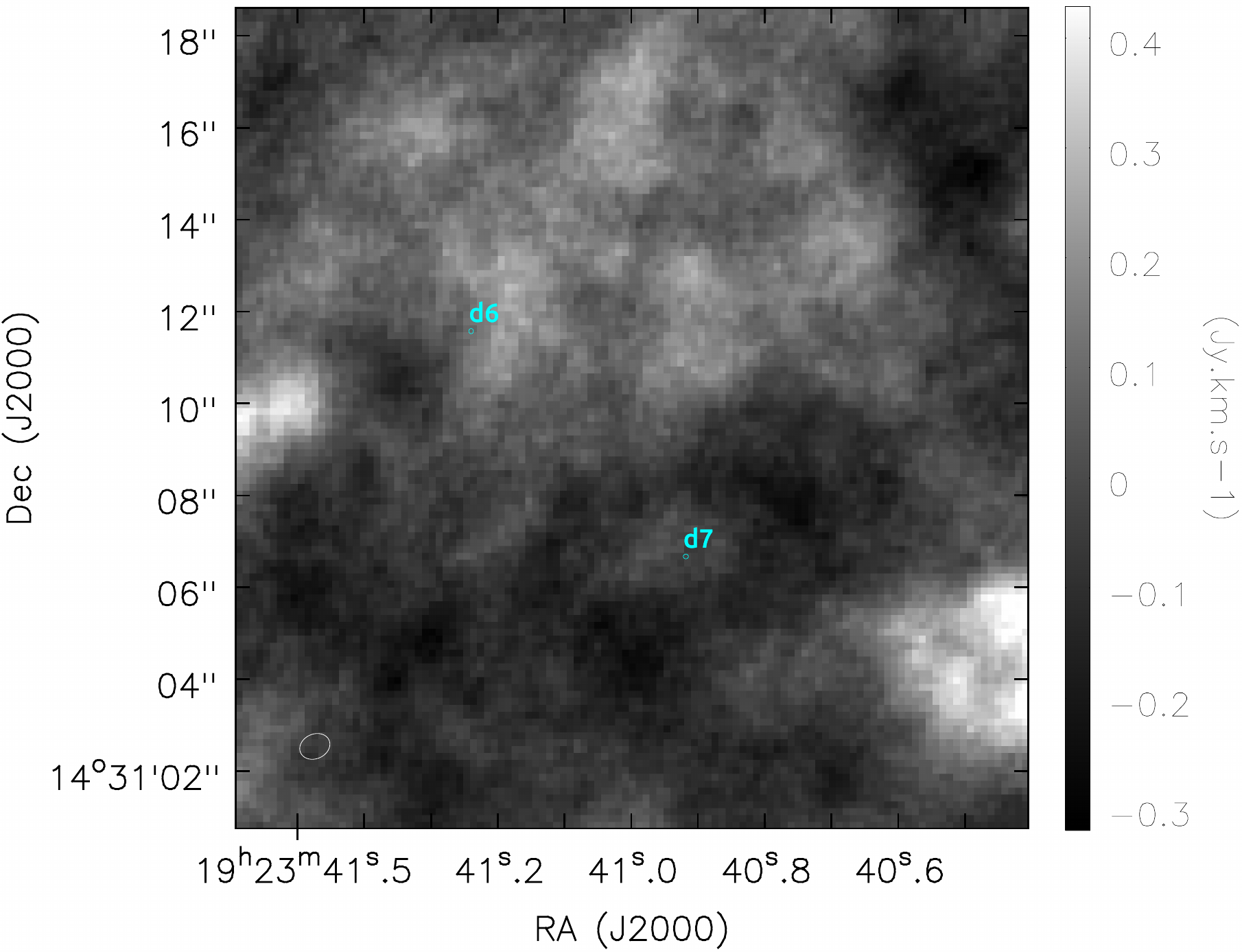}
    \includegraphics[scale=0.45]{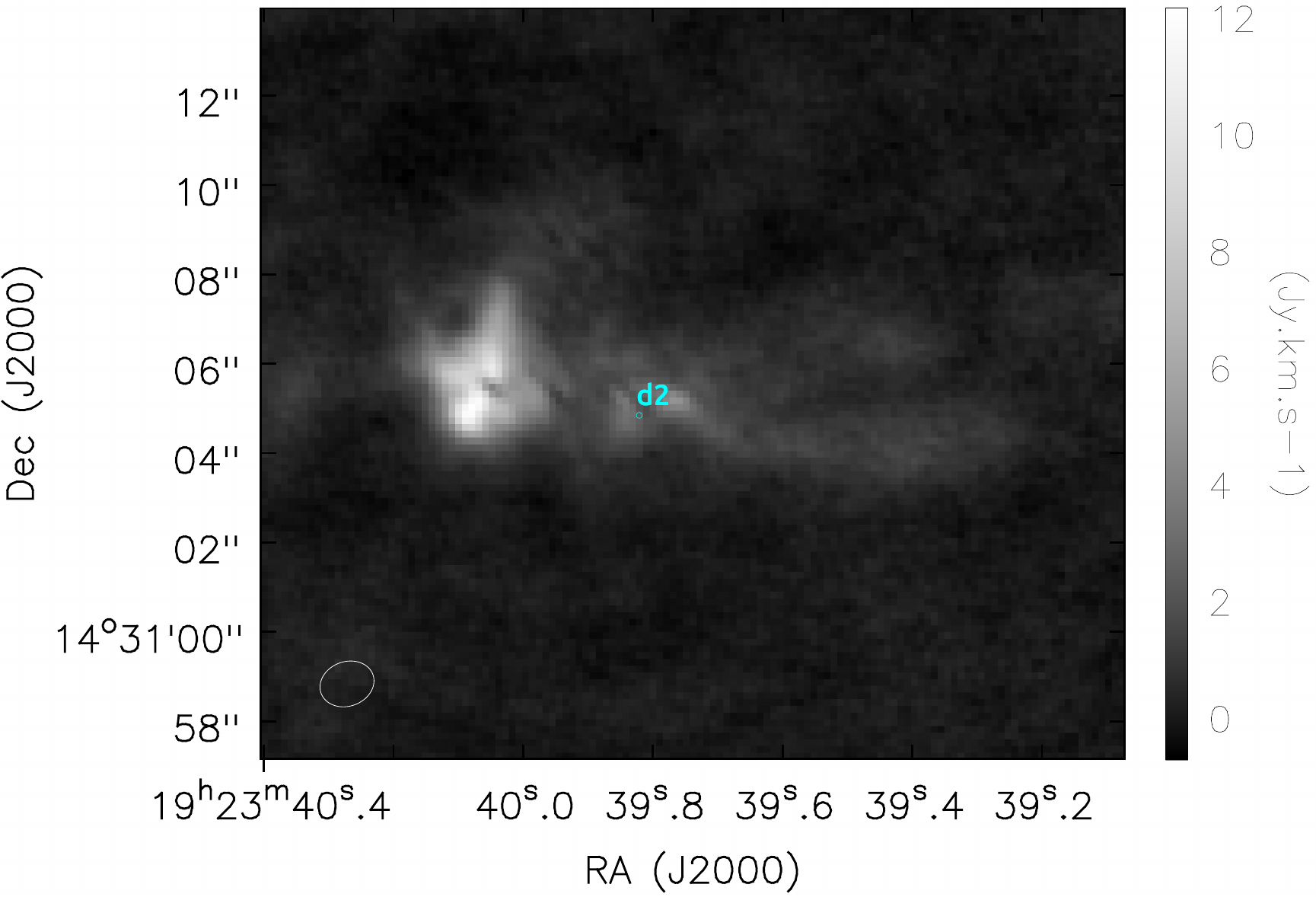}
    \includegraphics[scale=0.45]{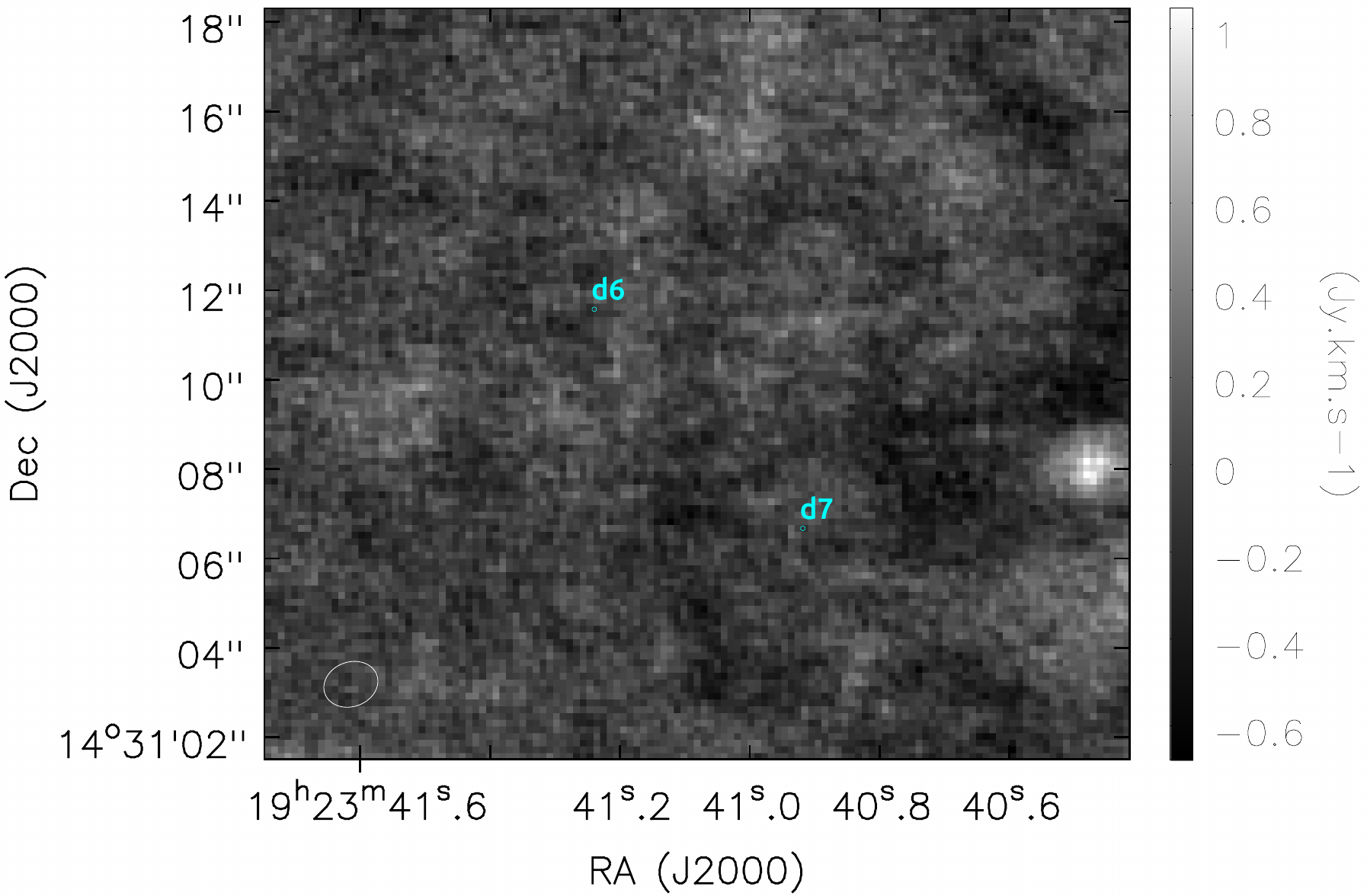}
    \label{fig:mom0_zoom}
    \caption{Moment 0 images around specific areas for d2 ({\it left} column) and d6/d7 ({\it right} column). \formai~ is in the {\it top} row, \formaii~ in the {\it middle}, and \soi~ in the {\it bottom}. Color stretch is square-root from local minimum to maximum. Velocity integration is from 35 \kms to 75 \kms. Cyan markers have the same size as for the spectral extraction.}
\end{figure*}

\begin{figure*}[!t]
    \ContinuedFloat
    \centering
    \includegraphics[scale=0.46]{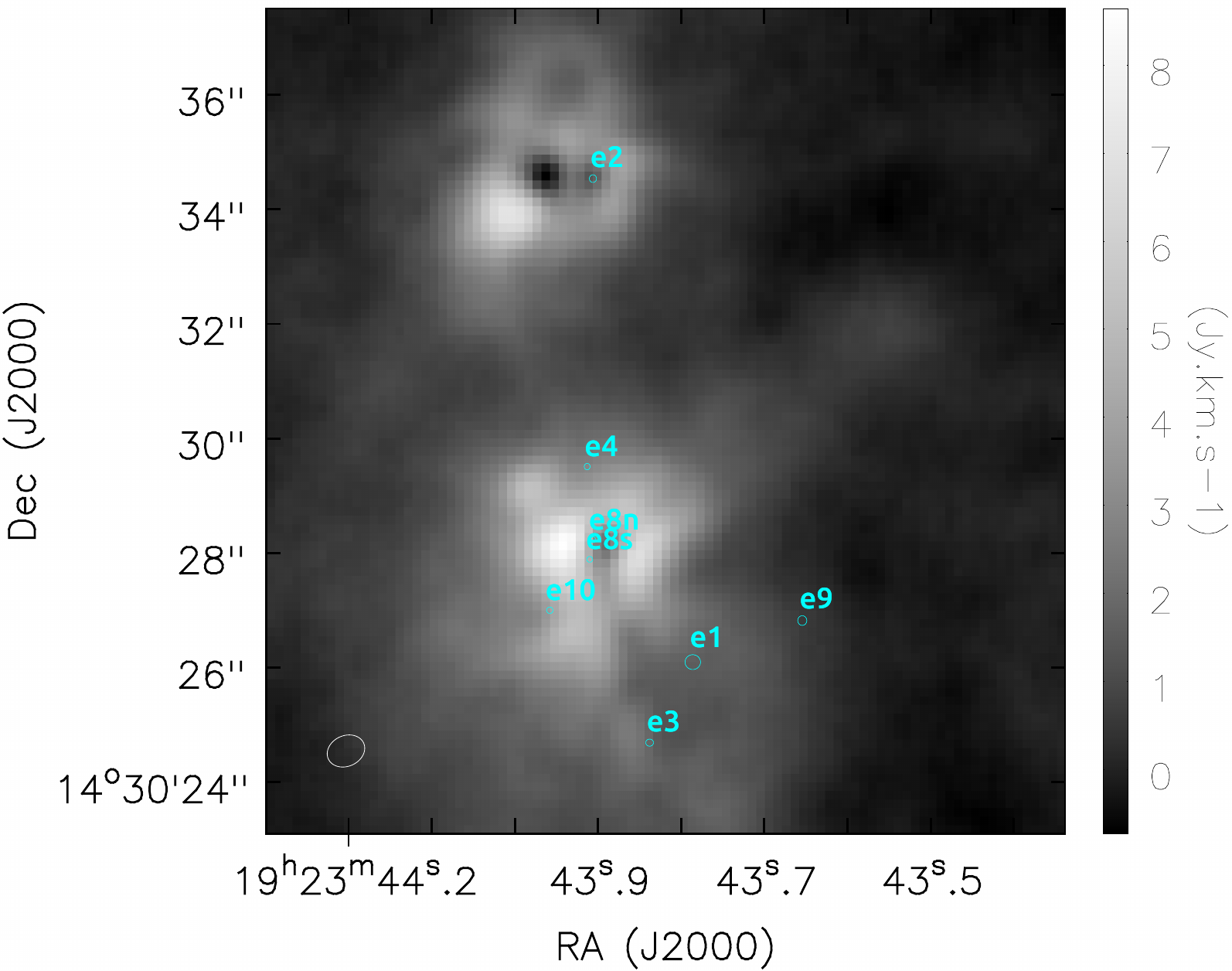}
    \includegraphics[scale=0.46]{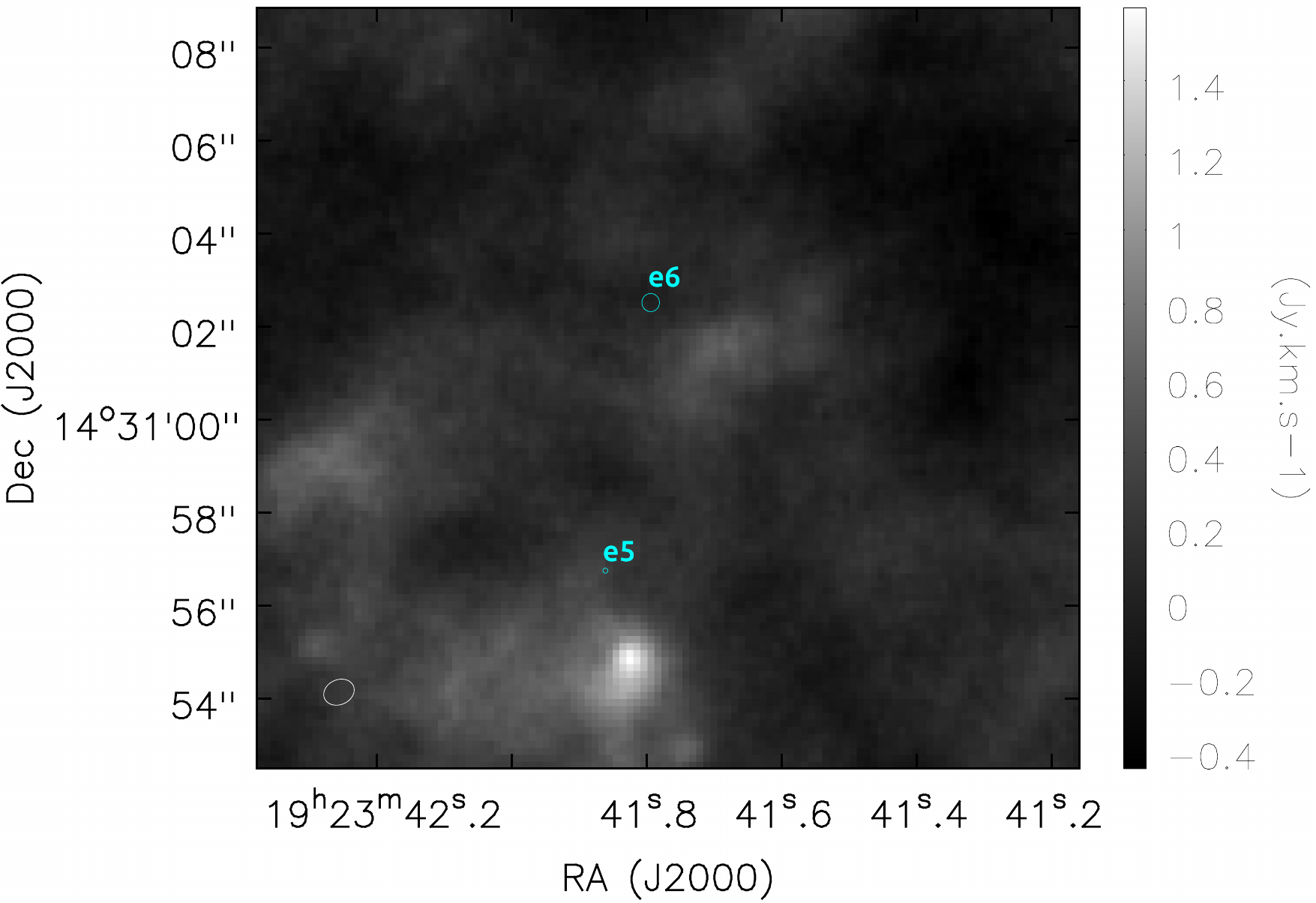}
    \includegraphics[scale=0.44]{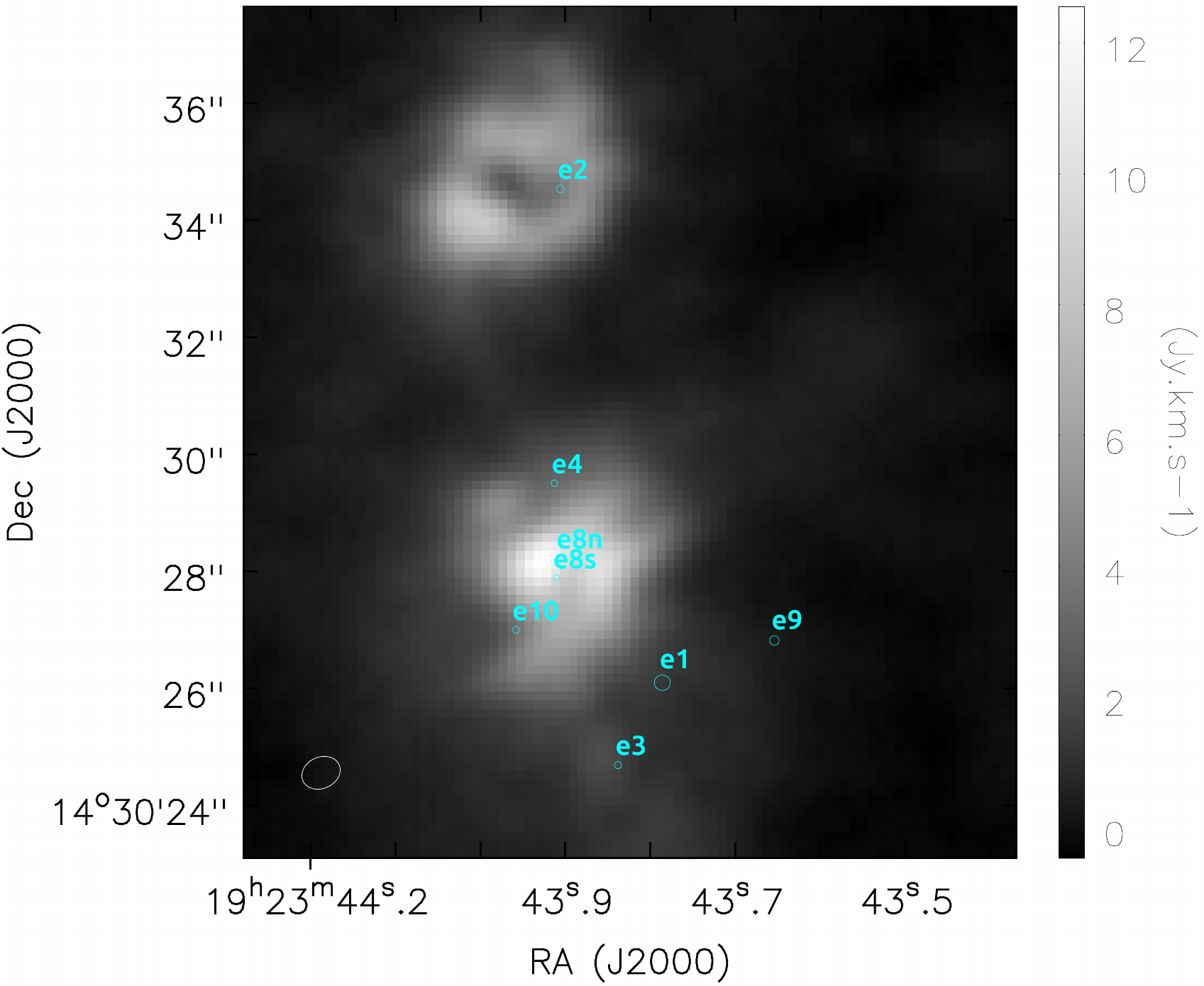}
    \includegraphics[scale=0.44]{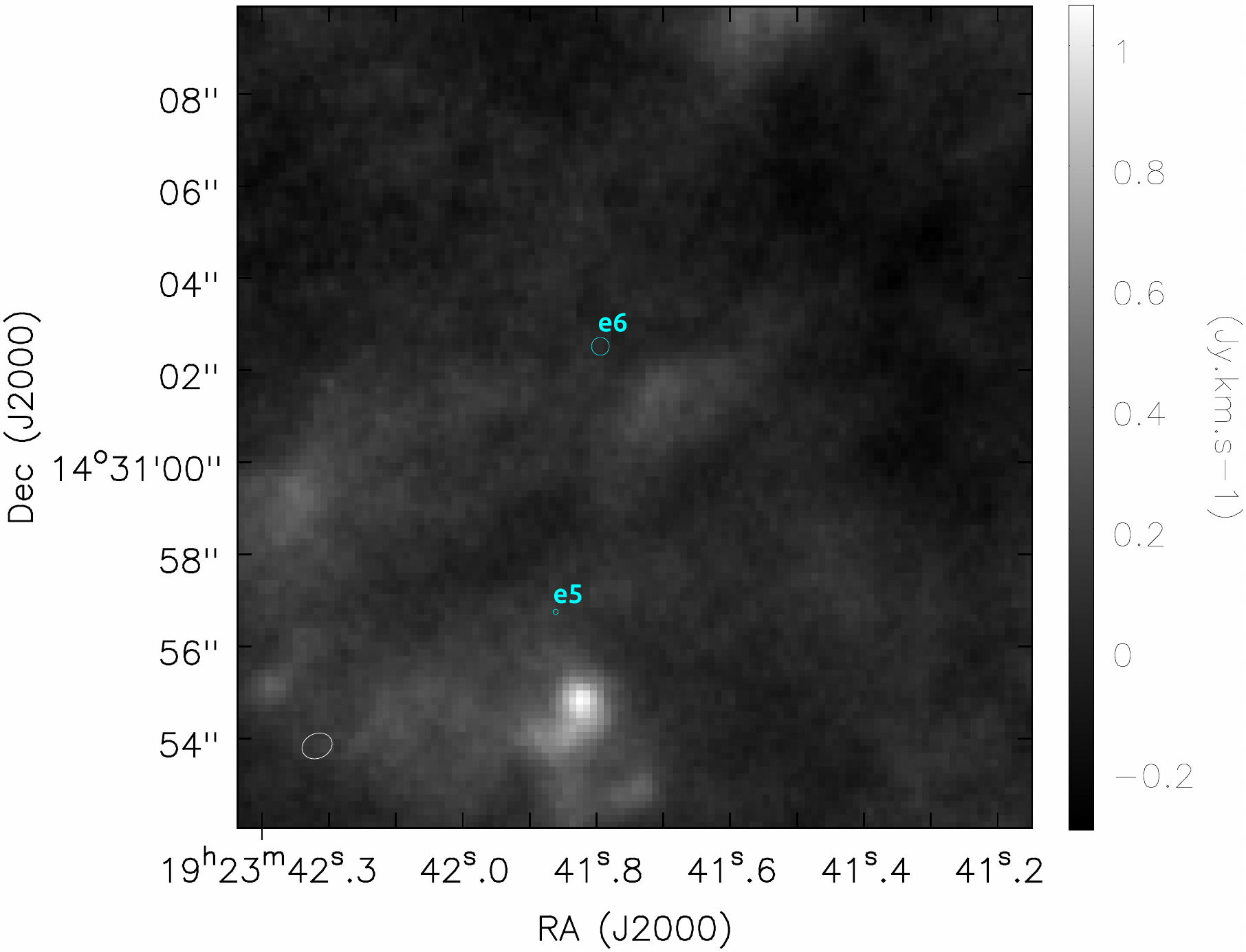}
    \includegraphics[scale=0.44]{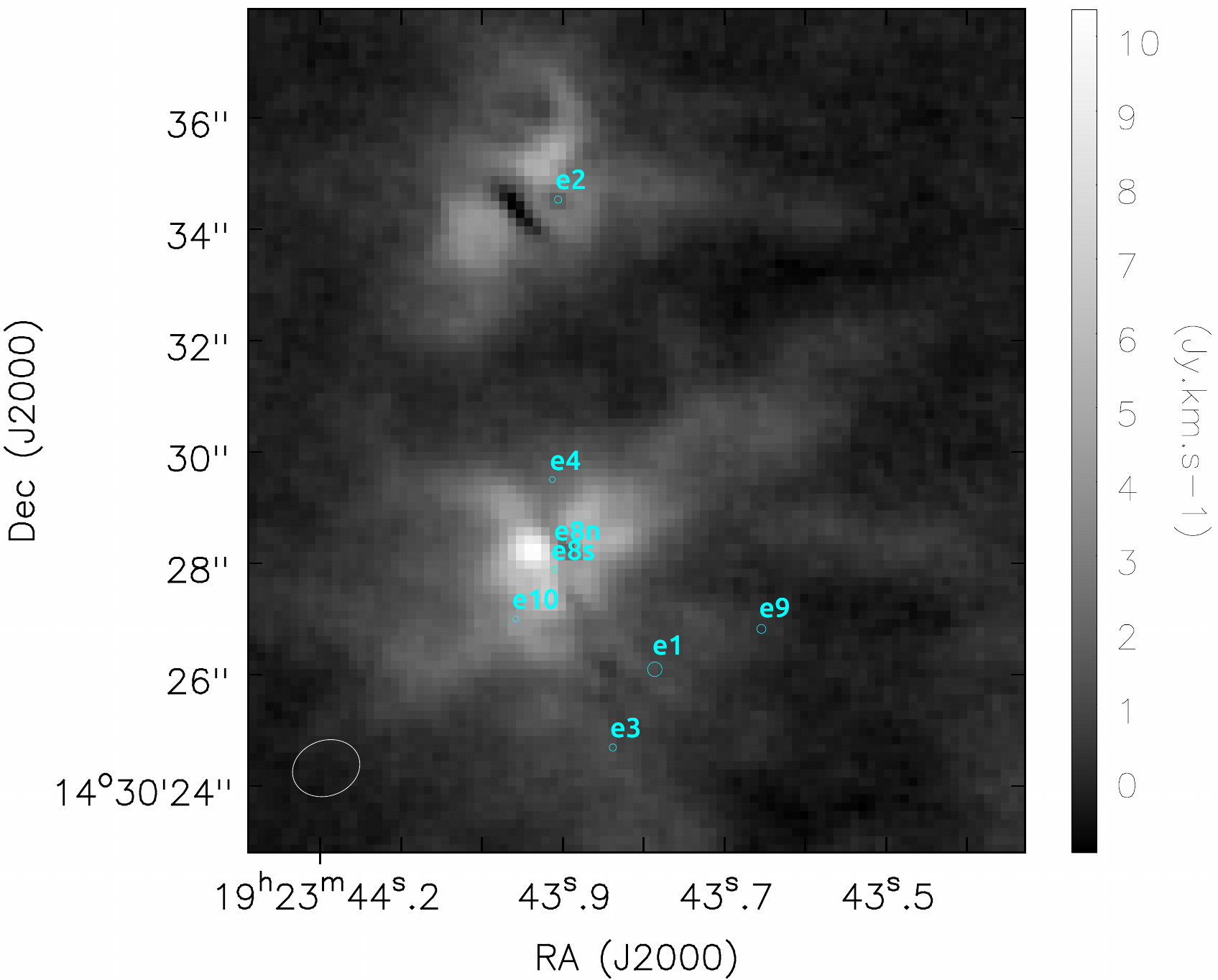}
    \includegraphics[scale=0.46]{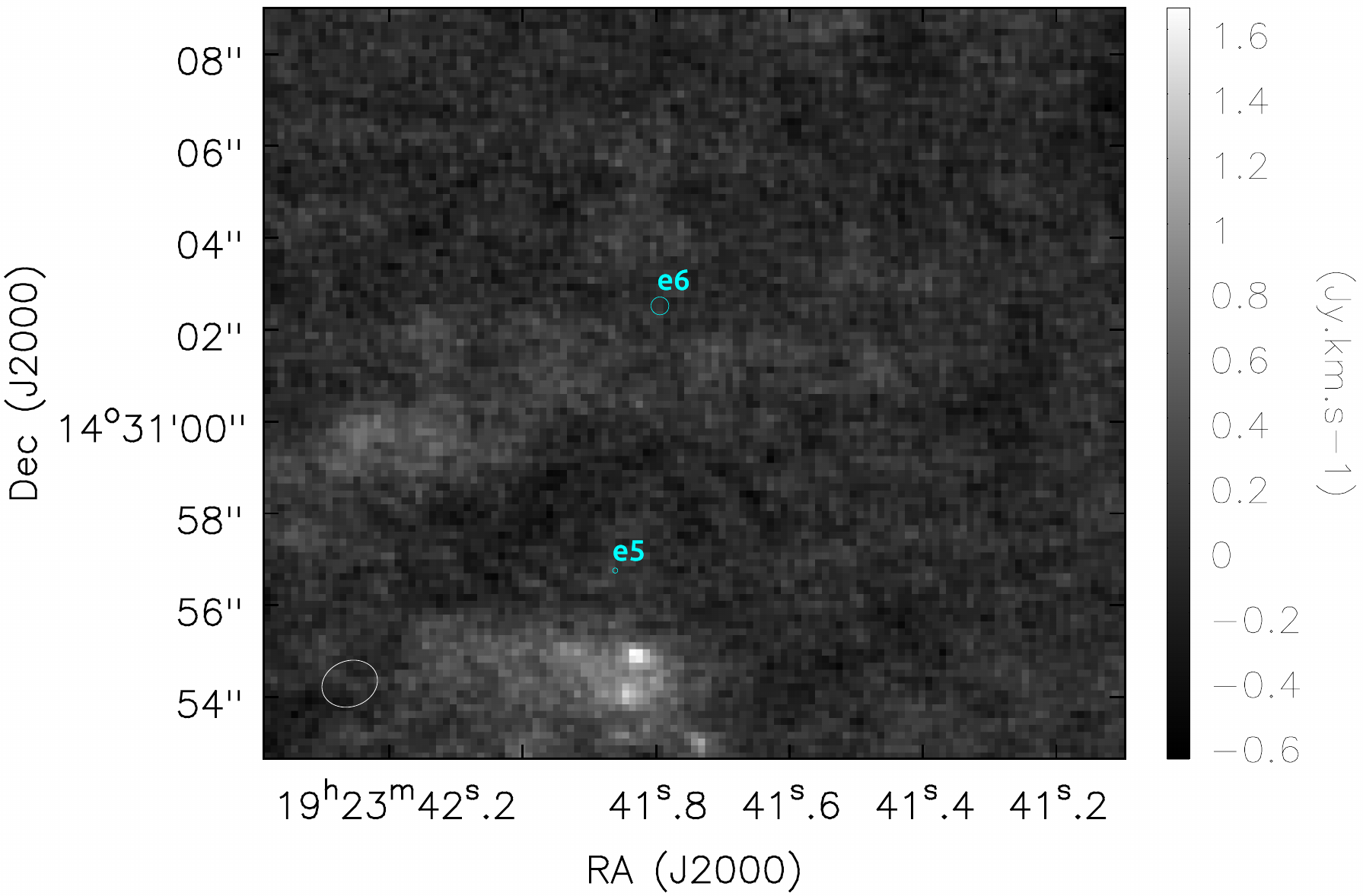}
    \label{fig:mom0_zoom}
    \caption{\textit{contd.} 
    Moment 0 images around specific areas for the e2/e8 cluster ({\it left} column) and e5/e6 ({\it right} column). \formai~ is in the {\it top} row, \formaii~ in the {\it middle}, and \soi~ in the {\it bottom}. Color stretch is square-root from local minimum to maximum. Velocity integration is from 35 \kms through 75 \kms. Cyan markers have same size as for the spectral extraction.}
\end{figure*}

\begin{figure*}[!t]
    \ContinuedFloat
    \centering
    \includegraphics[scale=0.45]{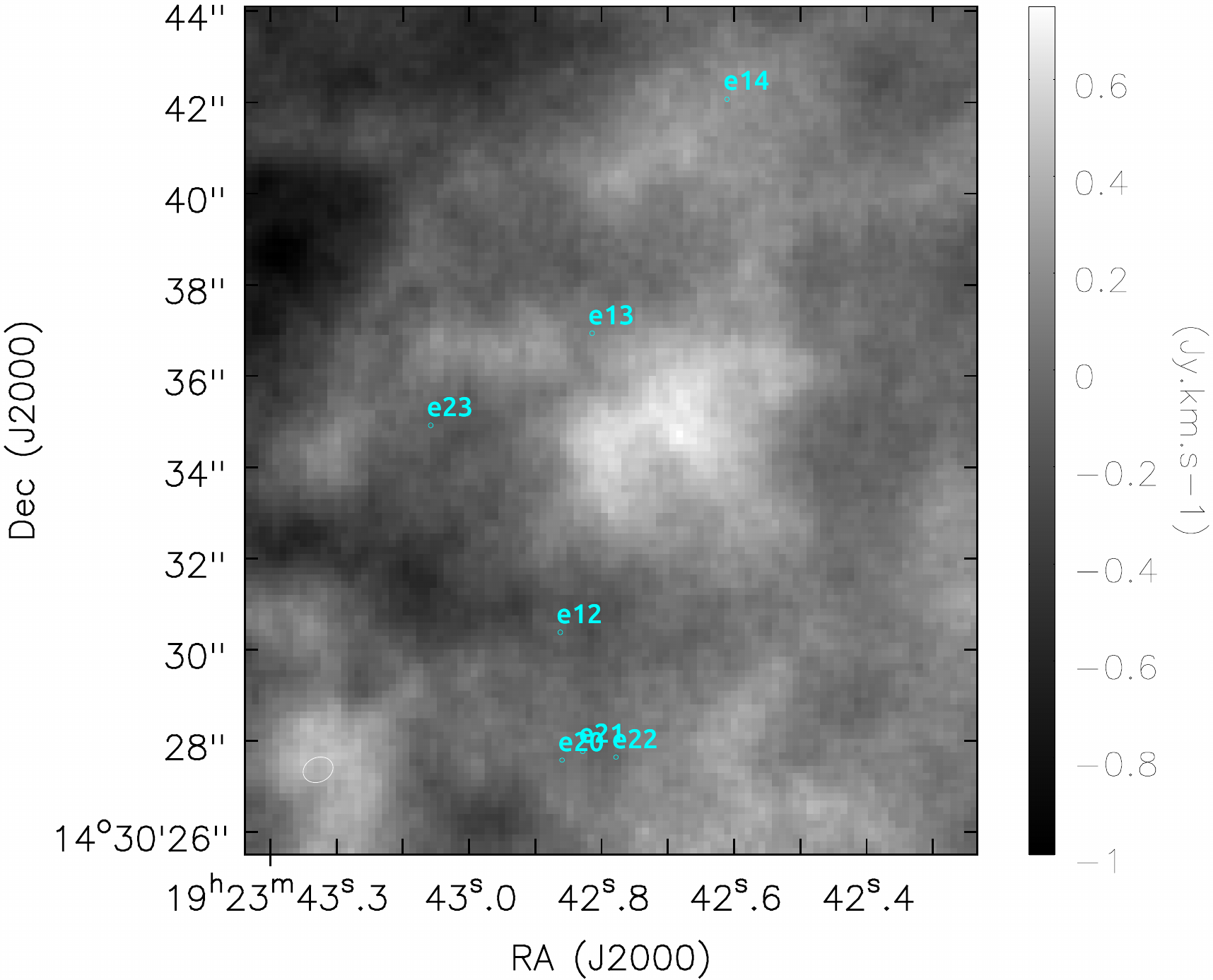}
    \includegraphics[scale=0.45]{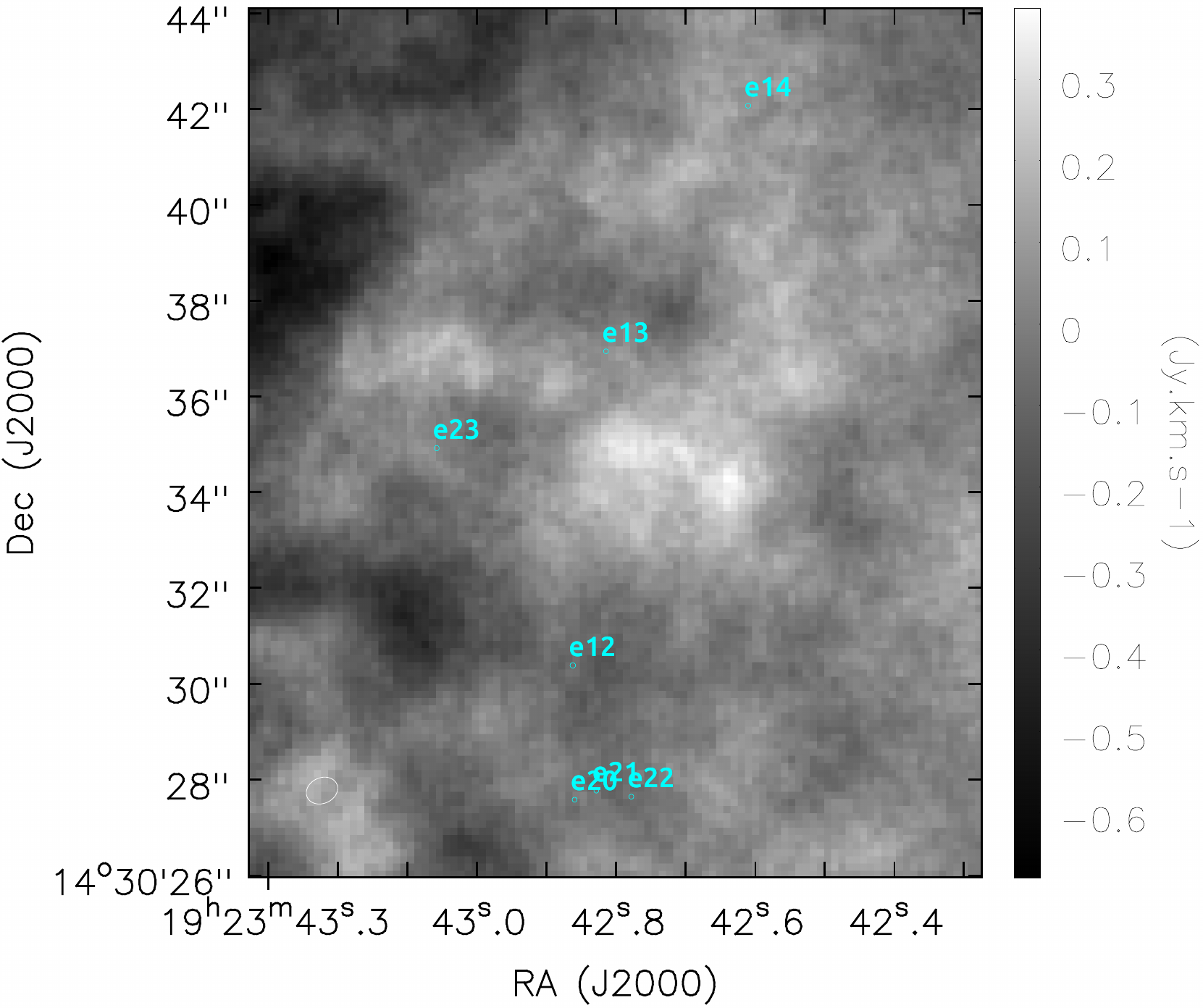}
    \includegraphics[scale=0.45]{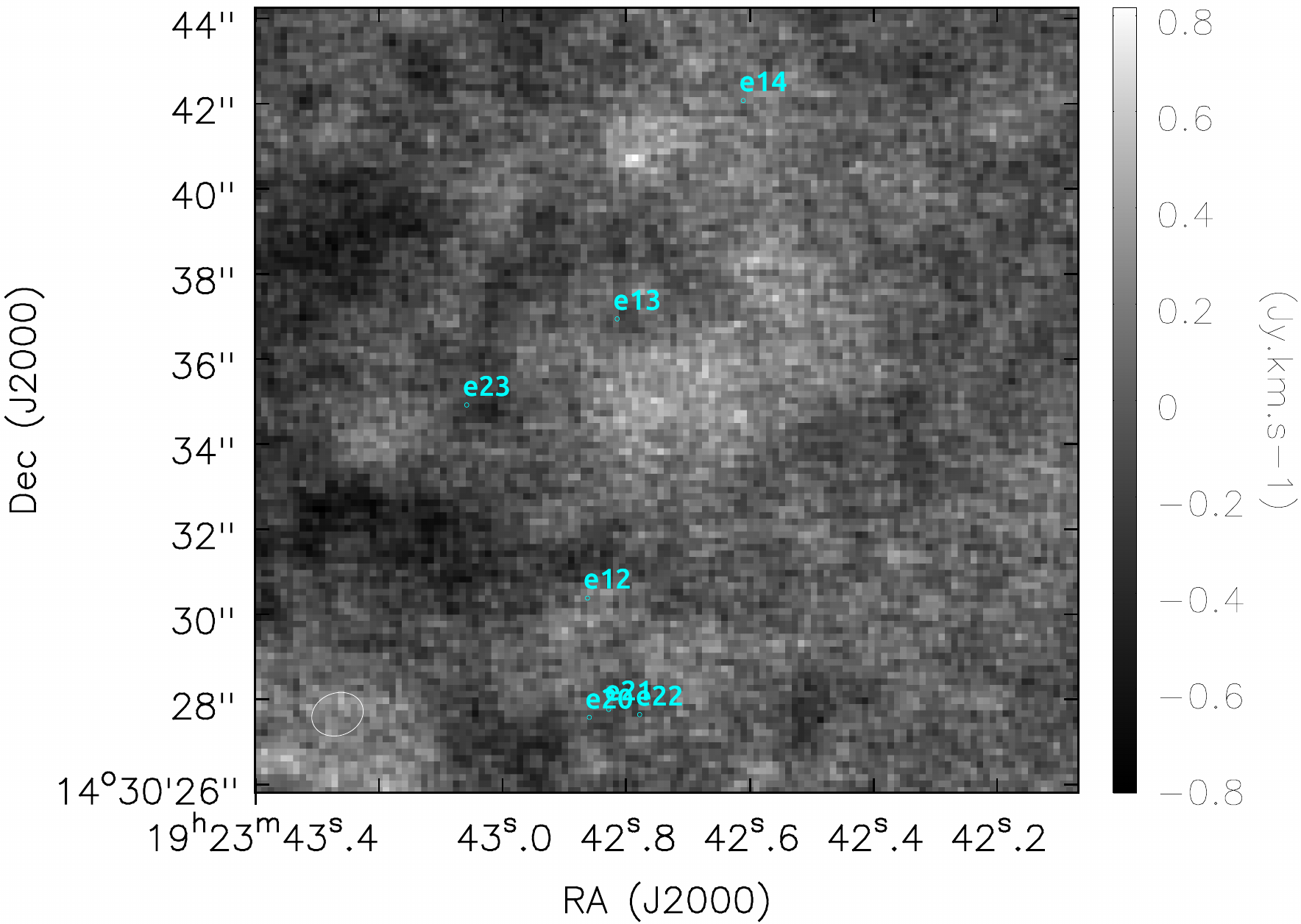}
    \label{fig:mom0_zoom}
    \caption{\textit{contd.} 
    Moment 0 images around the specific area of fainter e-sources. \formai~ ({\it left}) and \formaii~ ({\it right}) are in the {\it top} row and \soi~ is in the {\it bottom} row. Color stretch is square-root from local minimum to maximum. Velocity integration is from 35 \kms through 75 \kms. Cyan markers have same size as for the spectral extraction.}
\end{figure*}

\begin{figure*}[!h]
    \setlength{\lineskip}{0pt}
    \centering
    \includegraphics[scale=0.38]{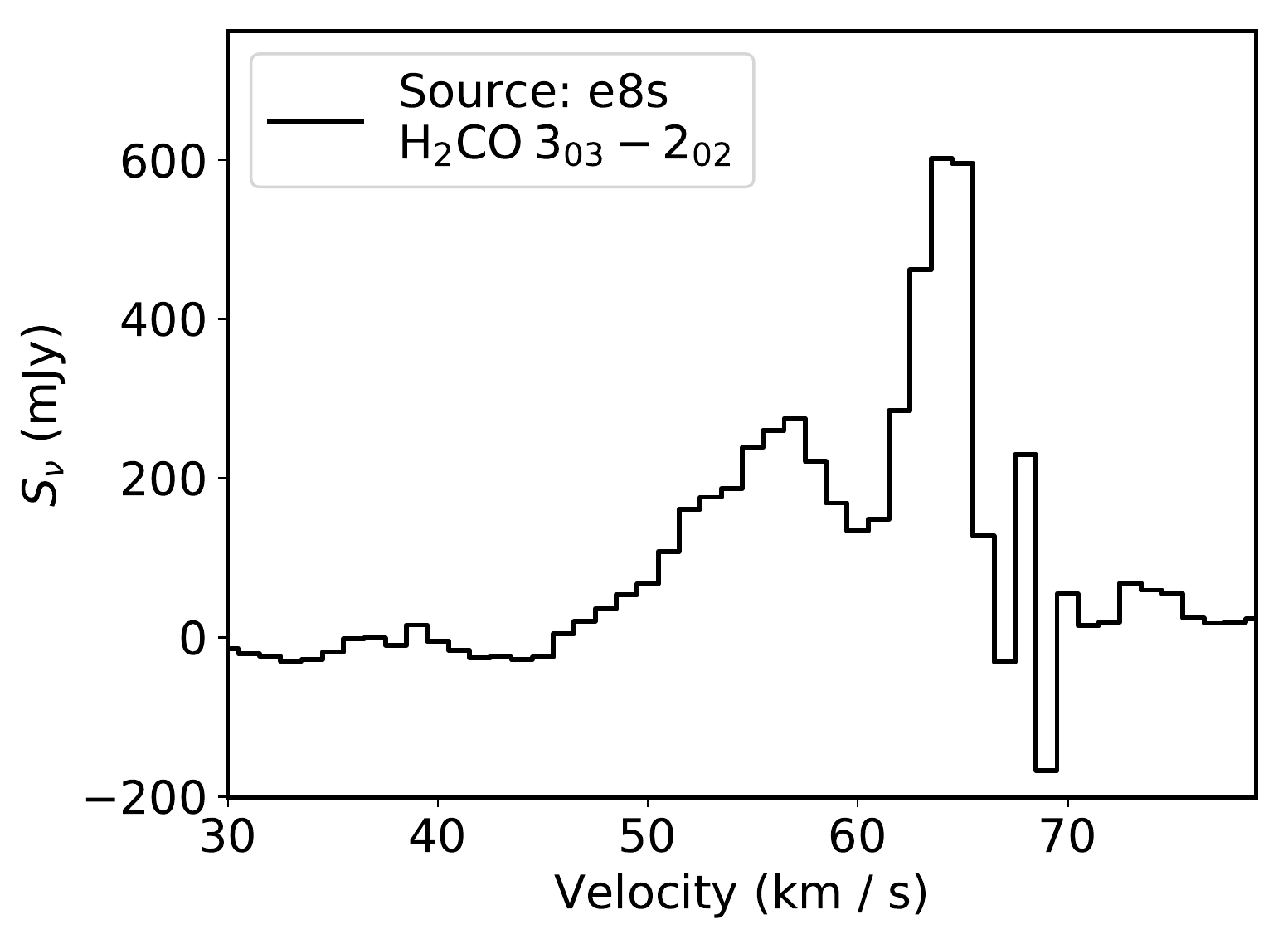}
    \includegraphics[scale=0.38]{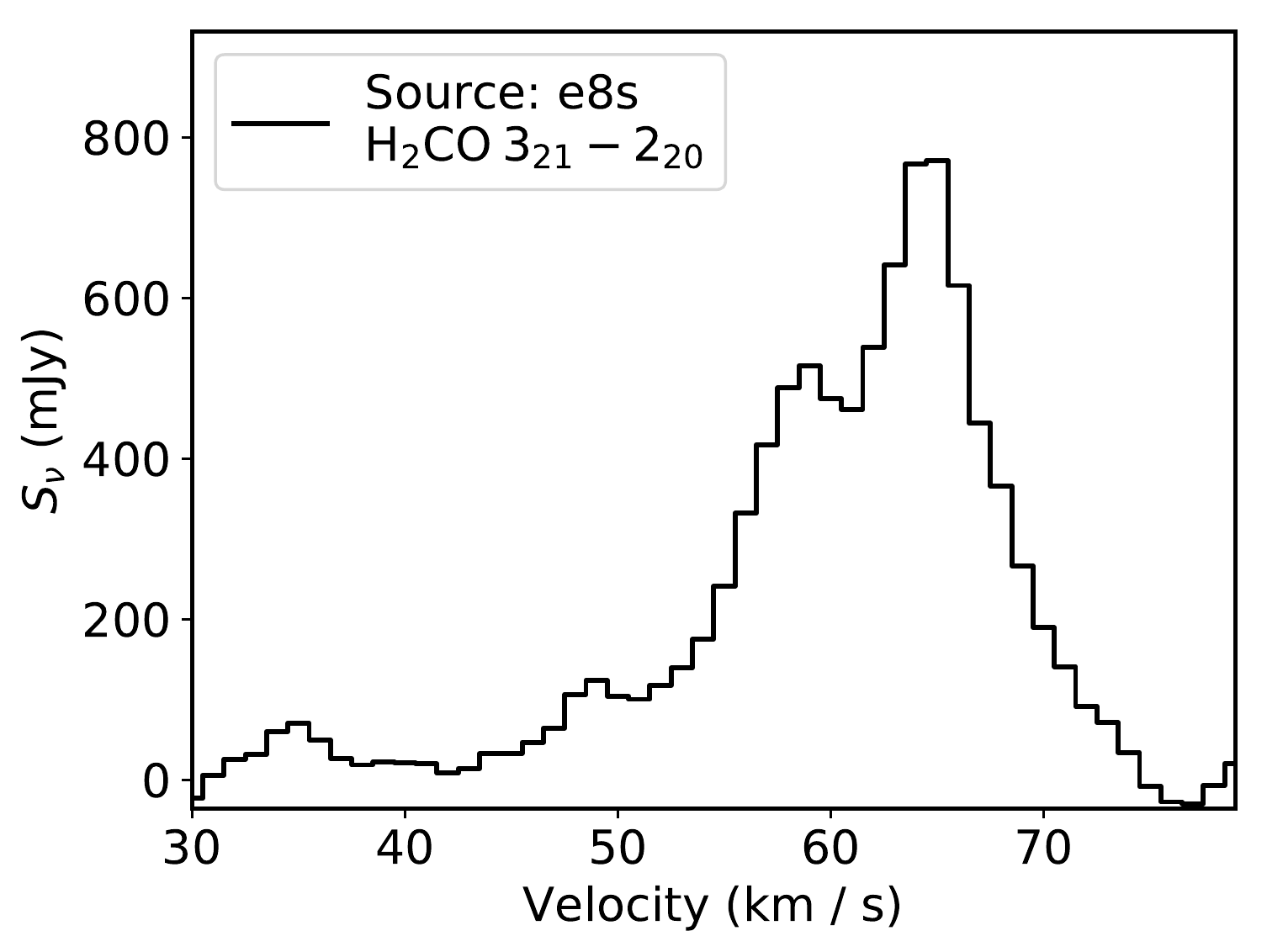}
    \includegraphics[scale=0.38]{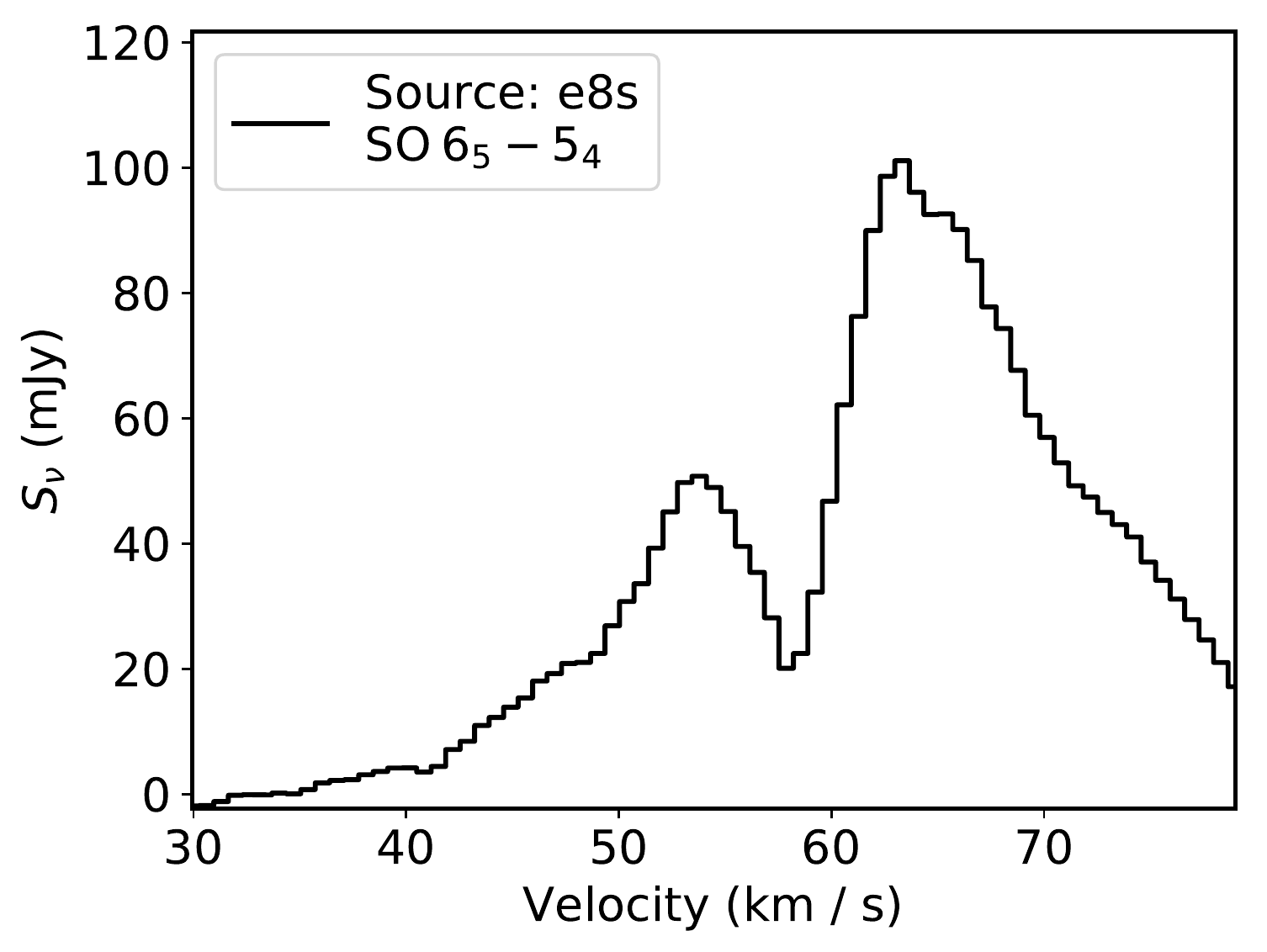}
    \includegraphics[scale=0.38]{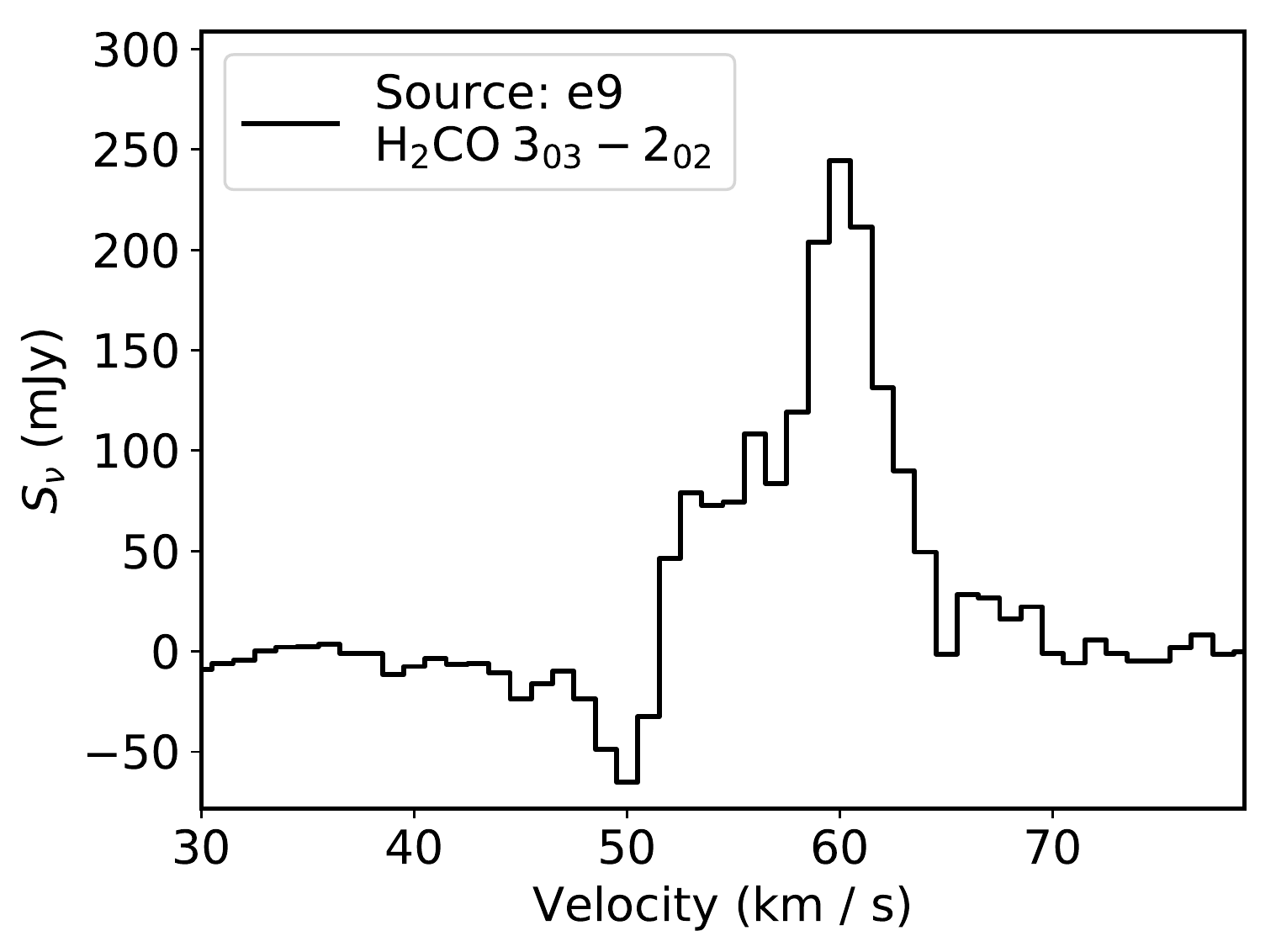}
    \includegraphics[scale=0.38]{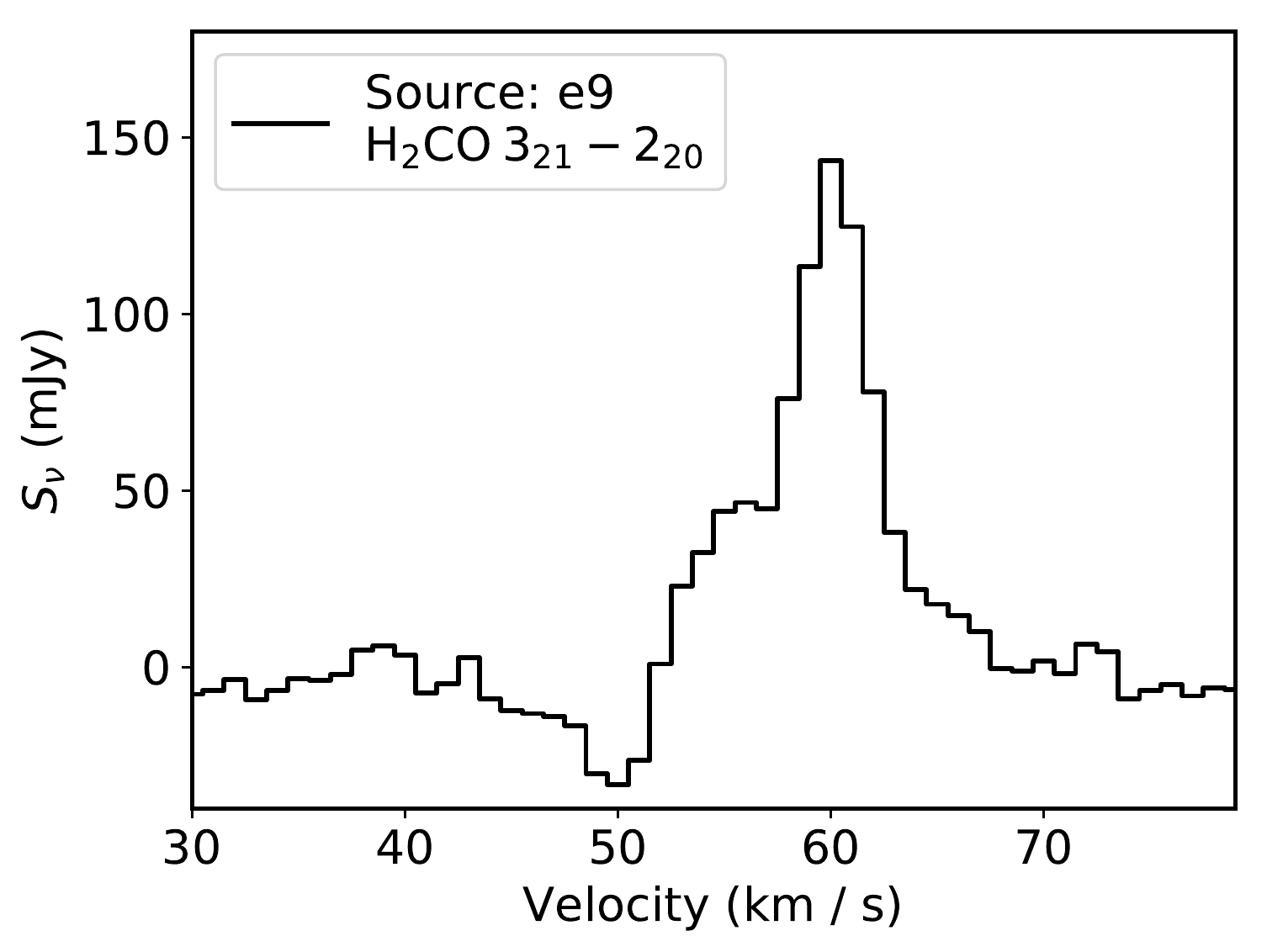}
    \includegraphics[scale=0.38]{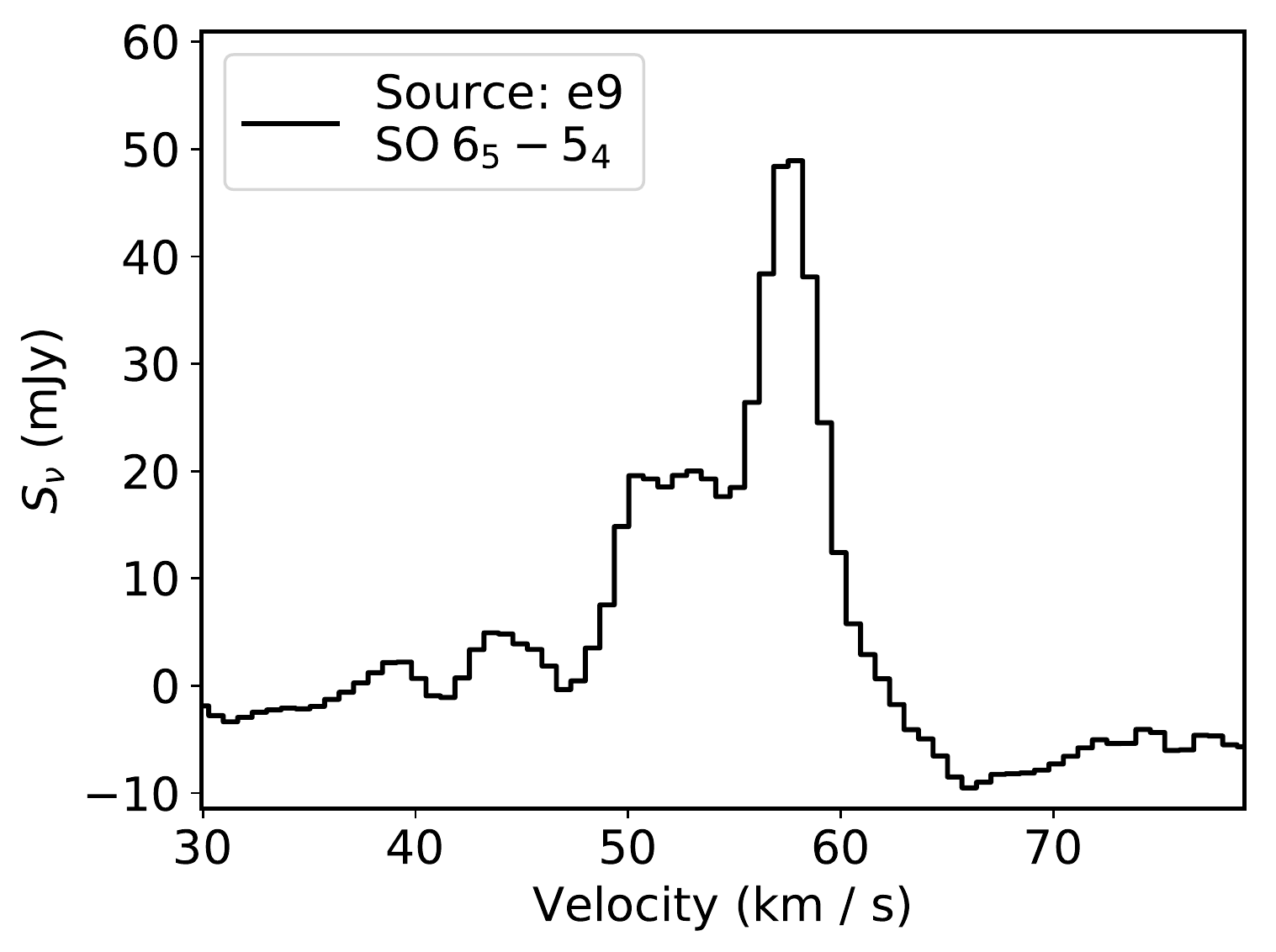}
    \includegraphics[scale=0.38]{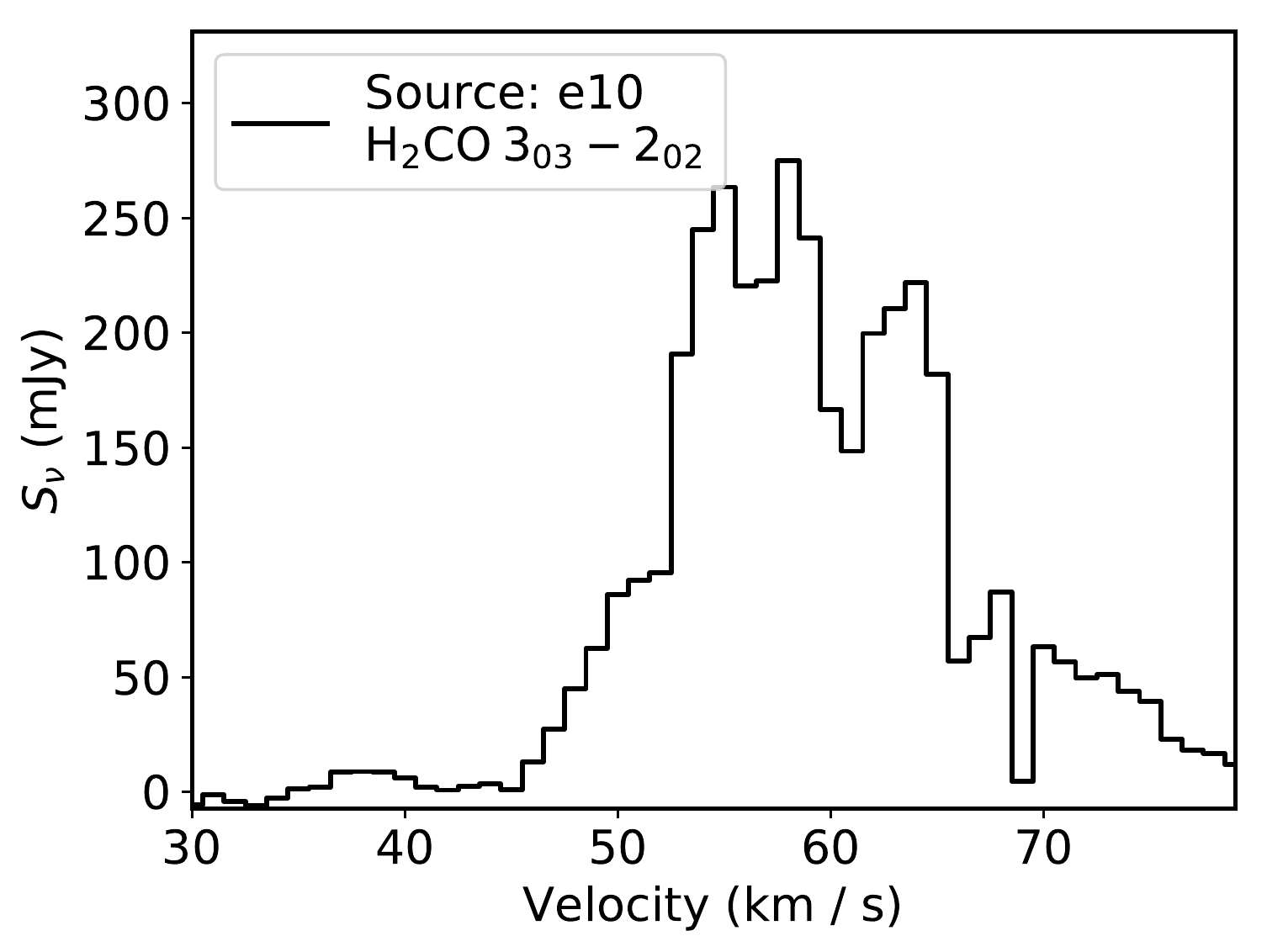}
    \includegraphics[scale=0.38]{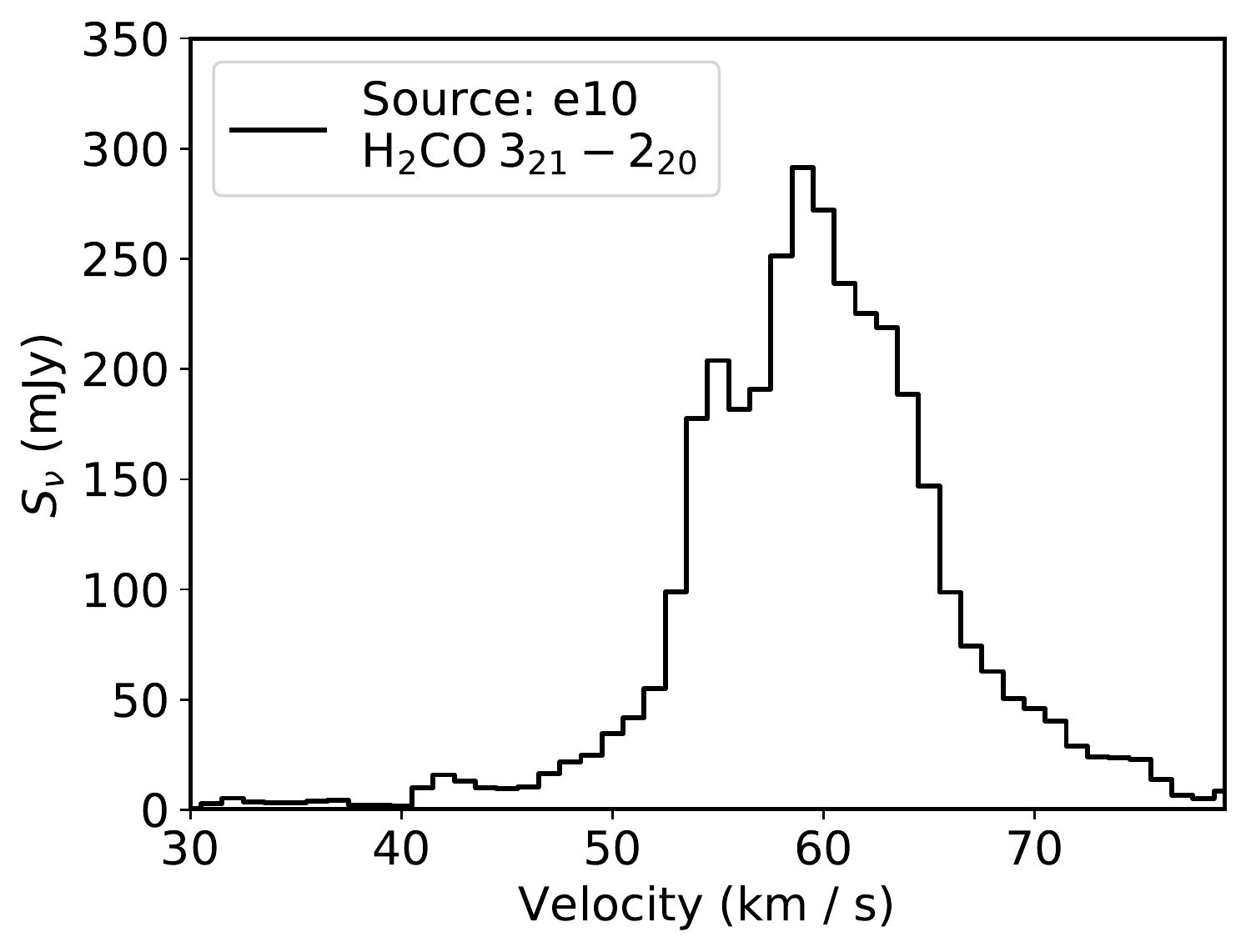}
    \includegraphics[scale=0.38]{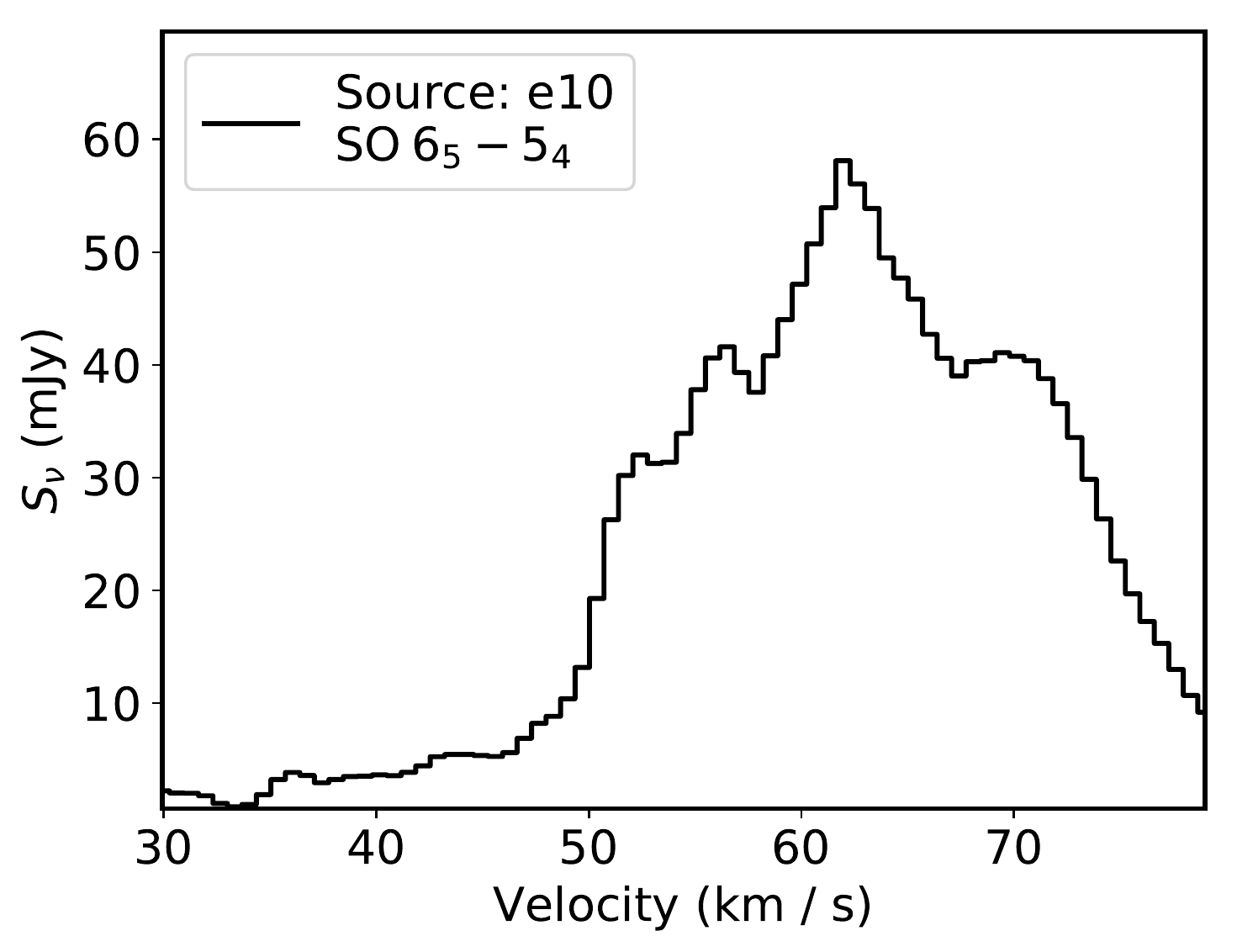}
    \includegraphics[scale=0.38]{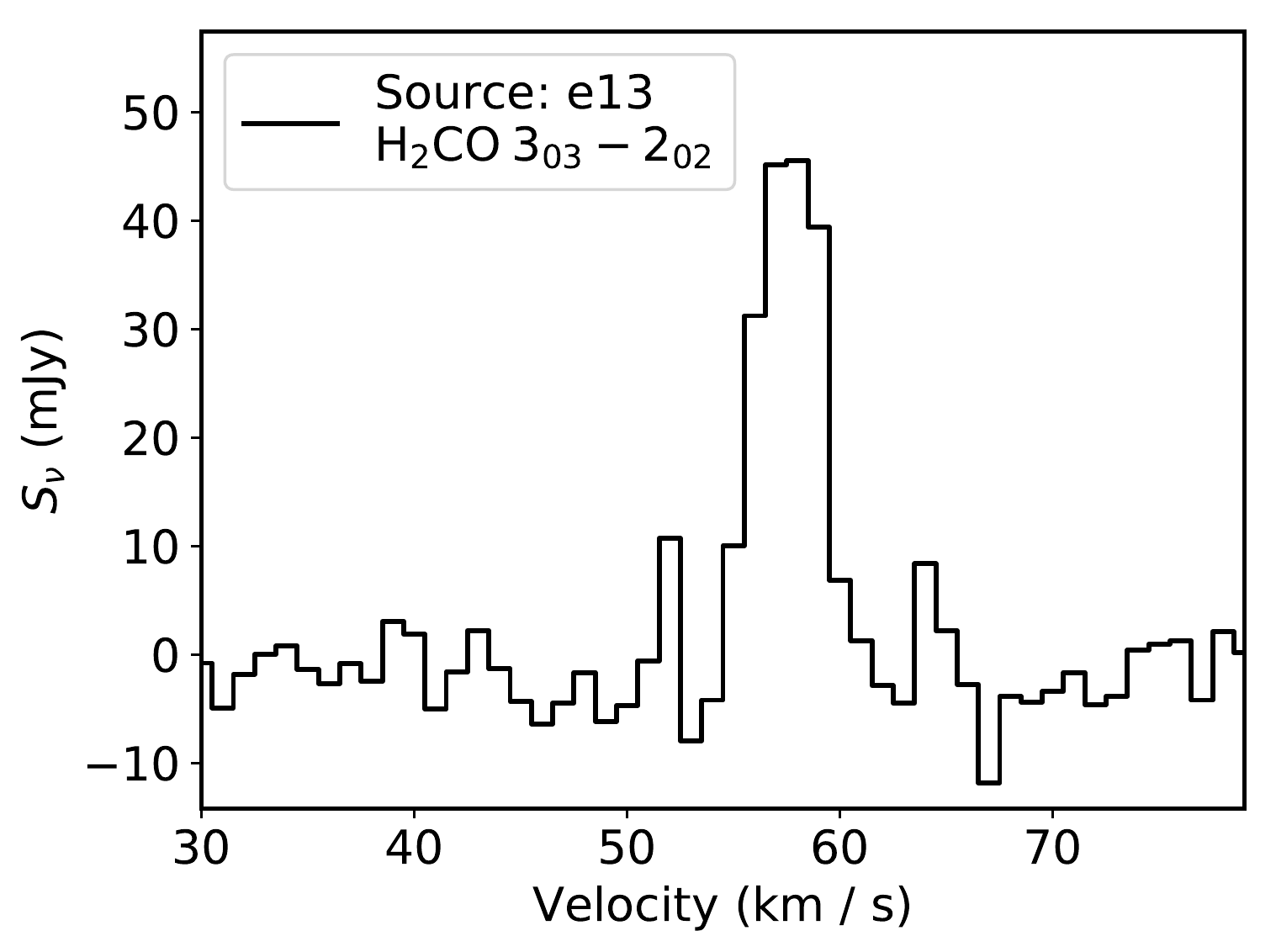}
    \includegraphics[scale=0.38]{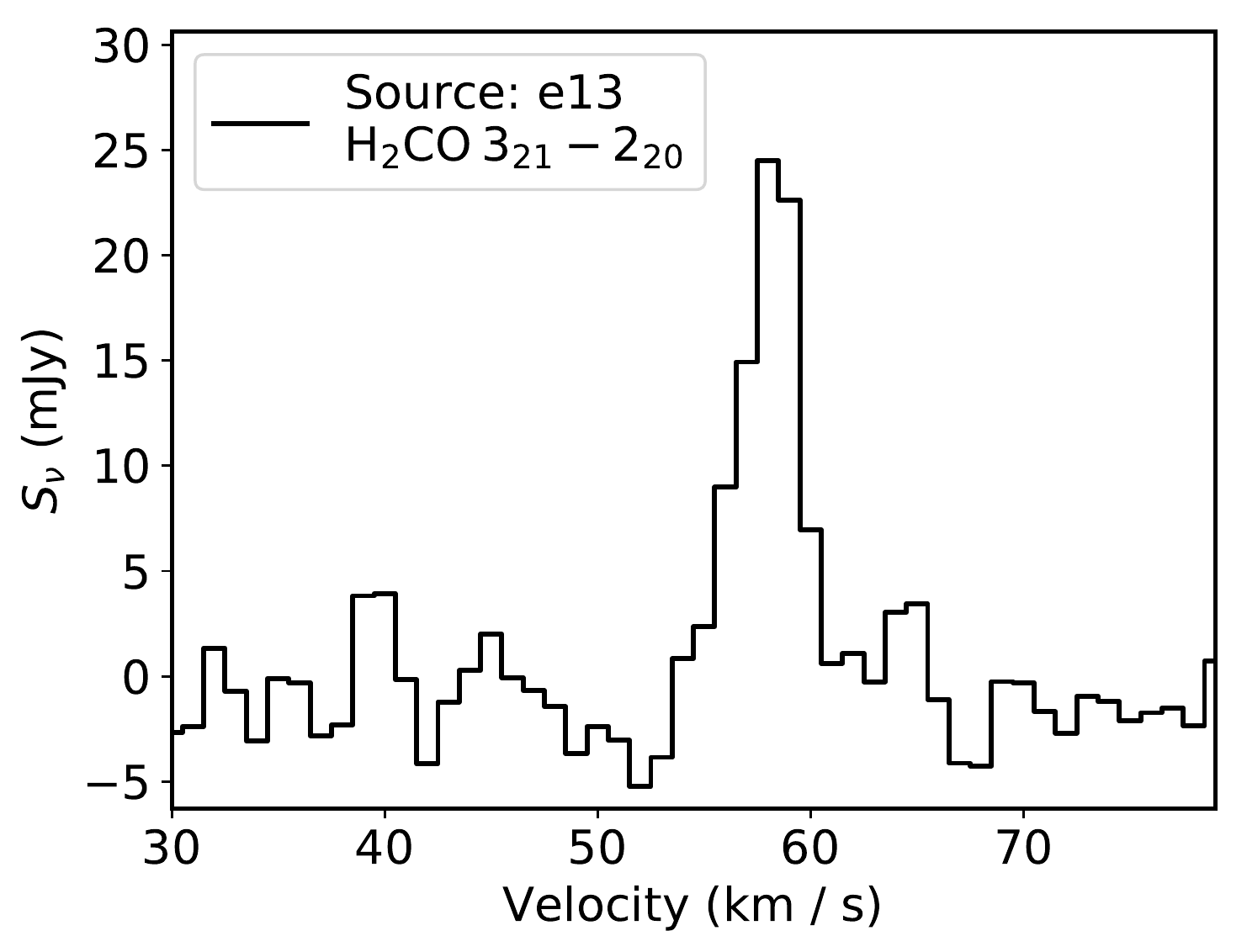}
    \includegraphics[scale=0.38]{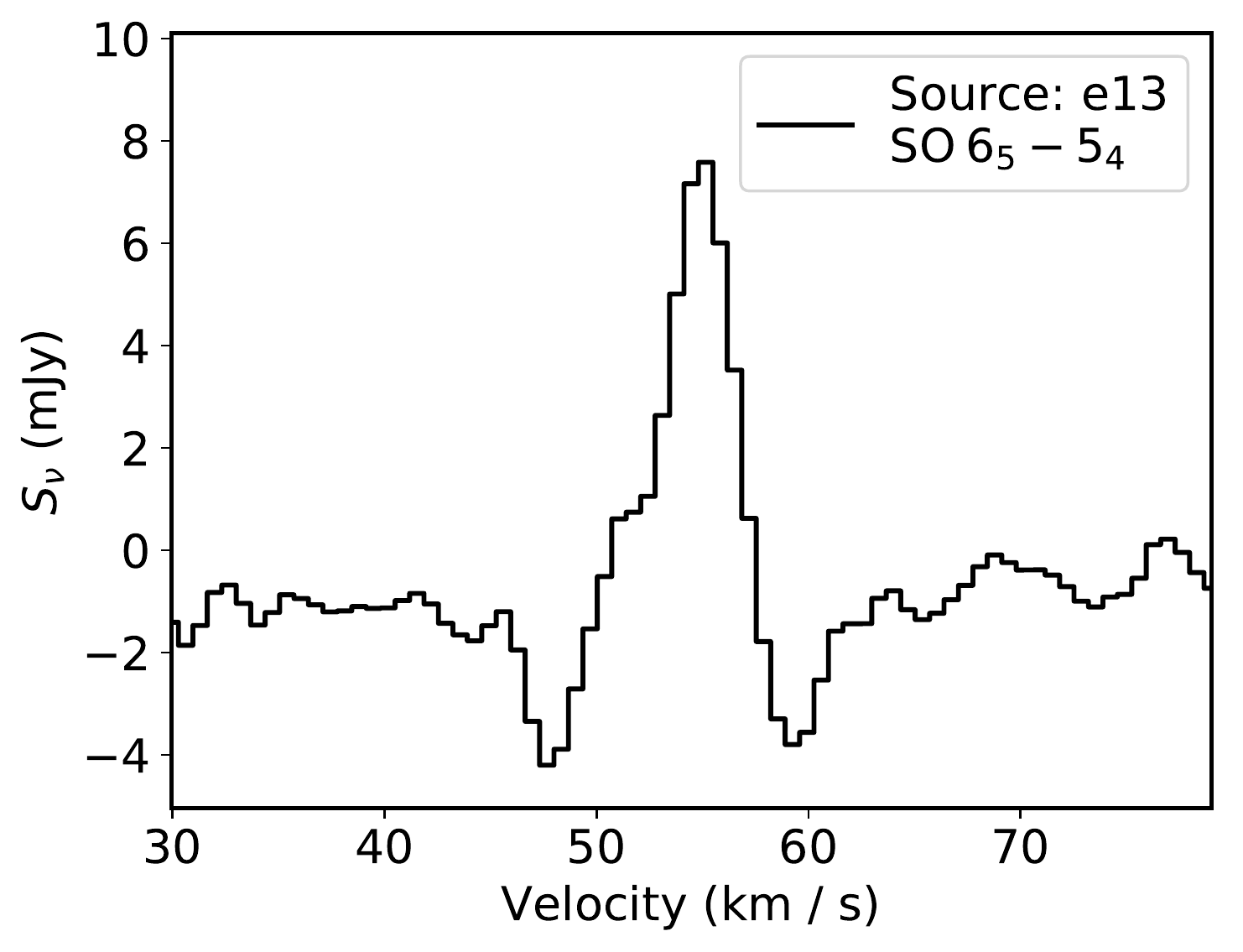}
    \includegraphics[scale=0.38]{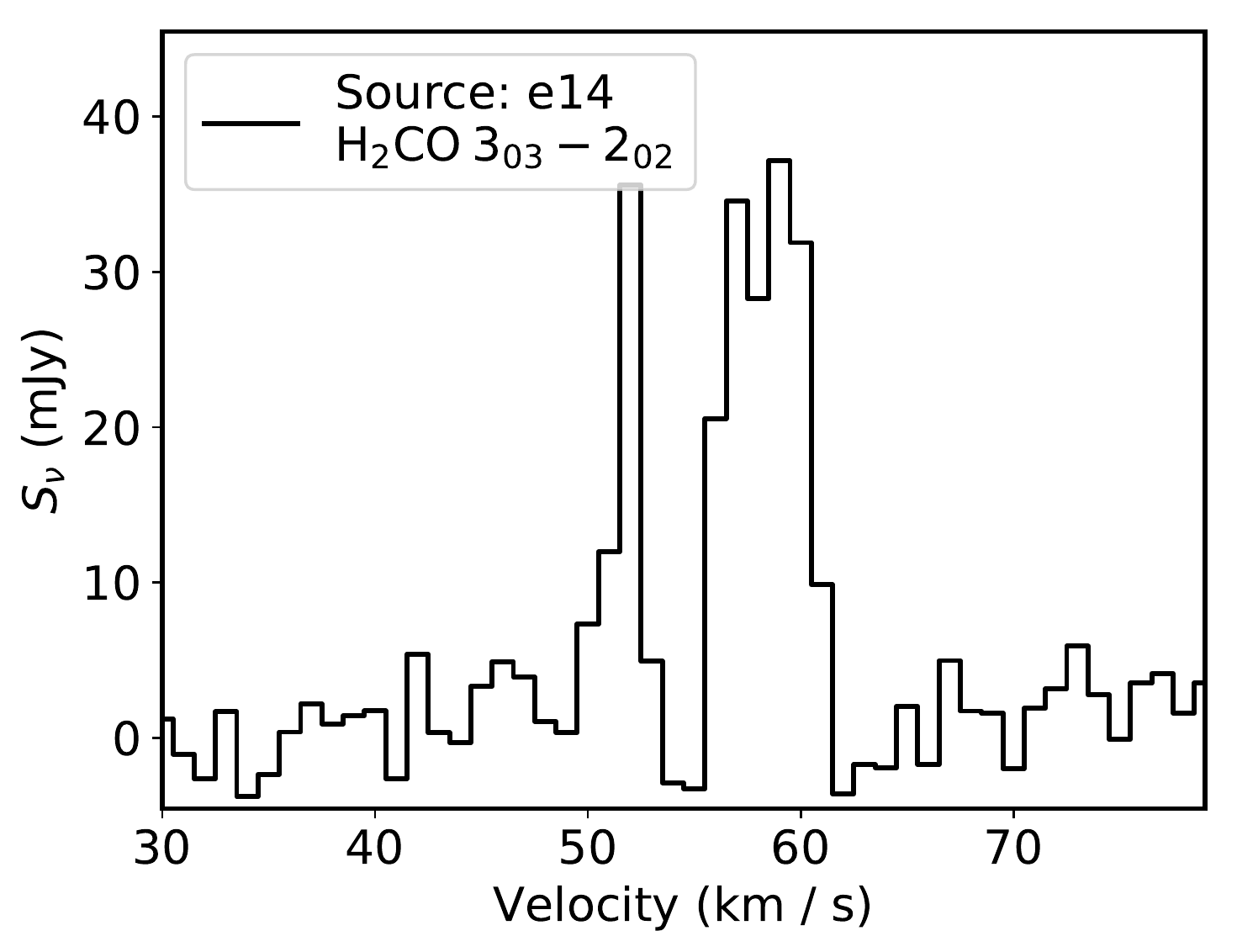}
    \includegraphics[scale=0.38]{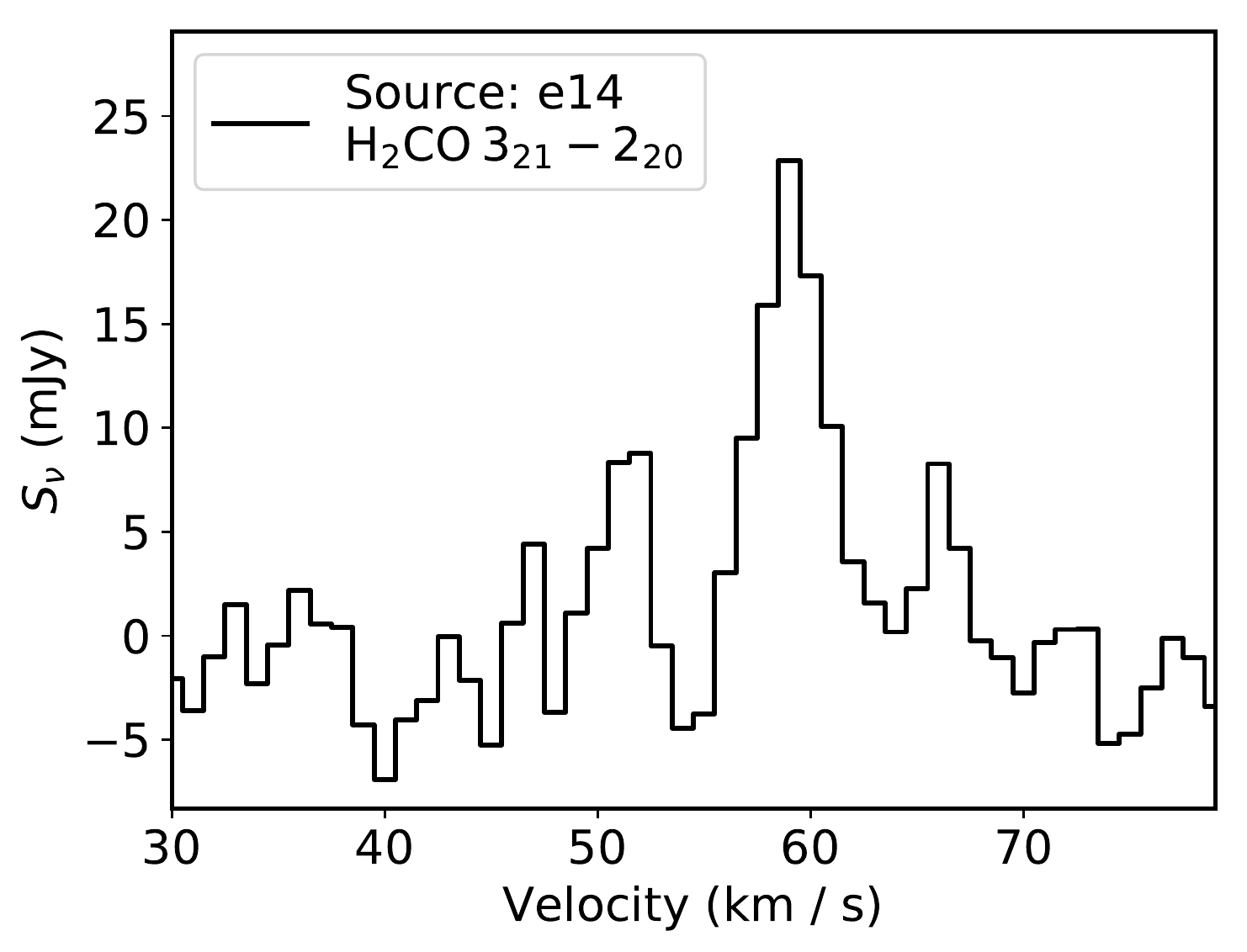}
    \includegraphics[scale=0.38]{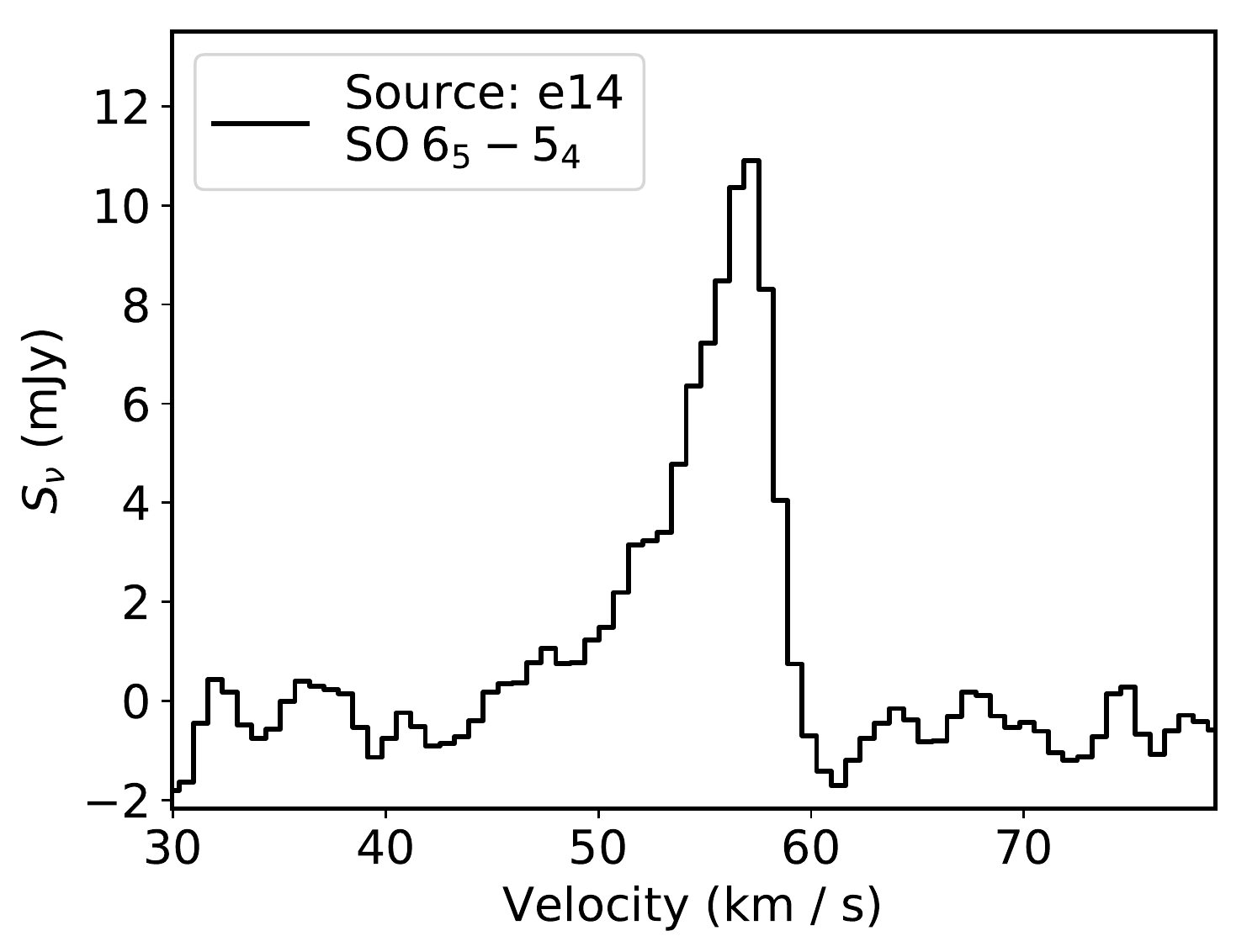}
    \caption{ 
    Spectral profiles of the molecular transitions \formai~ -- left --, \formaii~ -- middle --, \soi~ -- right --  for sources in catalog \textit{B} without H$30\alpha$ detection and not affected by image artefacts.} \label{fig:profiles_noH30a}
\end{figure*}

\begin{figure*}[!t]
    \setlength{\lineskip}{0pt}
    \centering
        \includegraphics[scale=0.38]{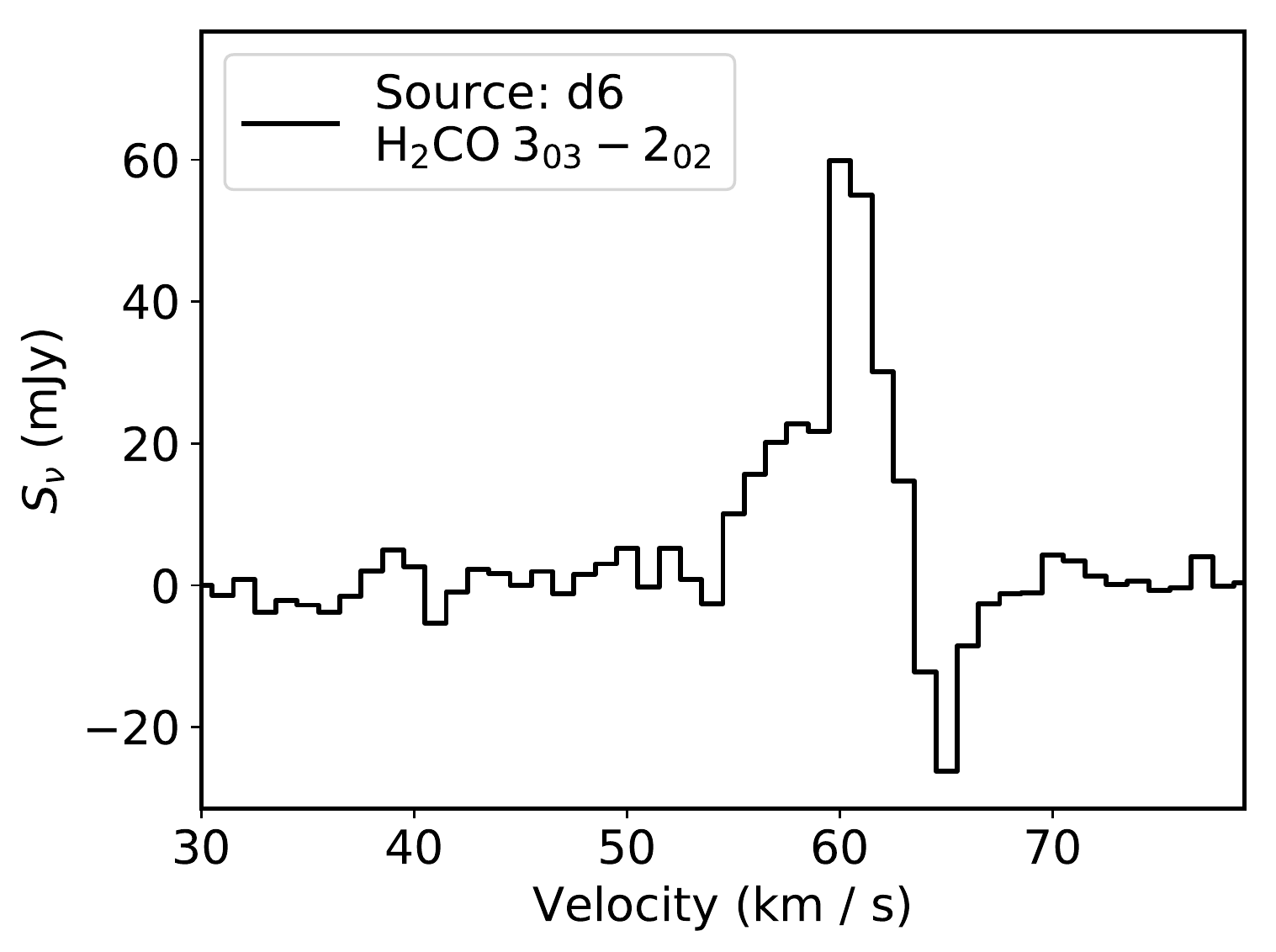}
        \includegraphics[scale=0.38]{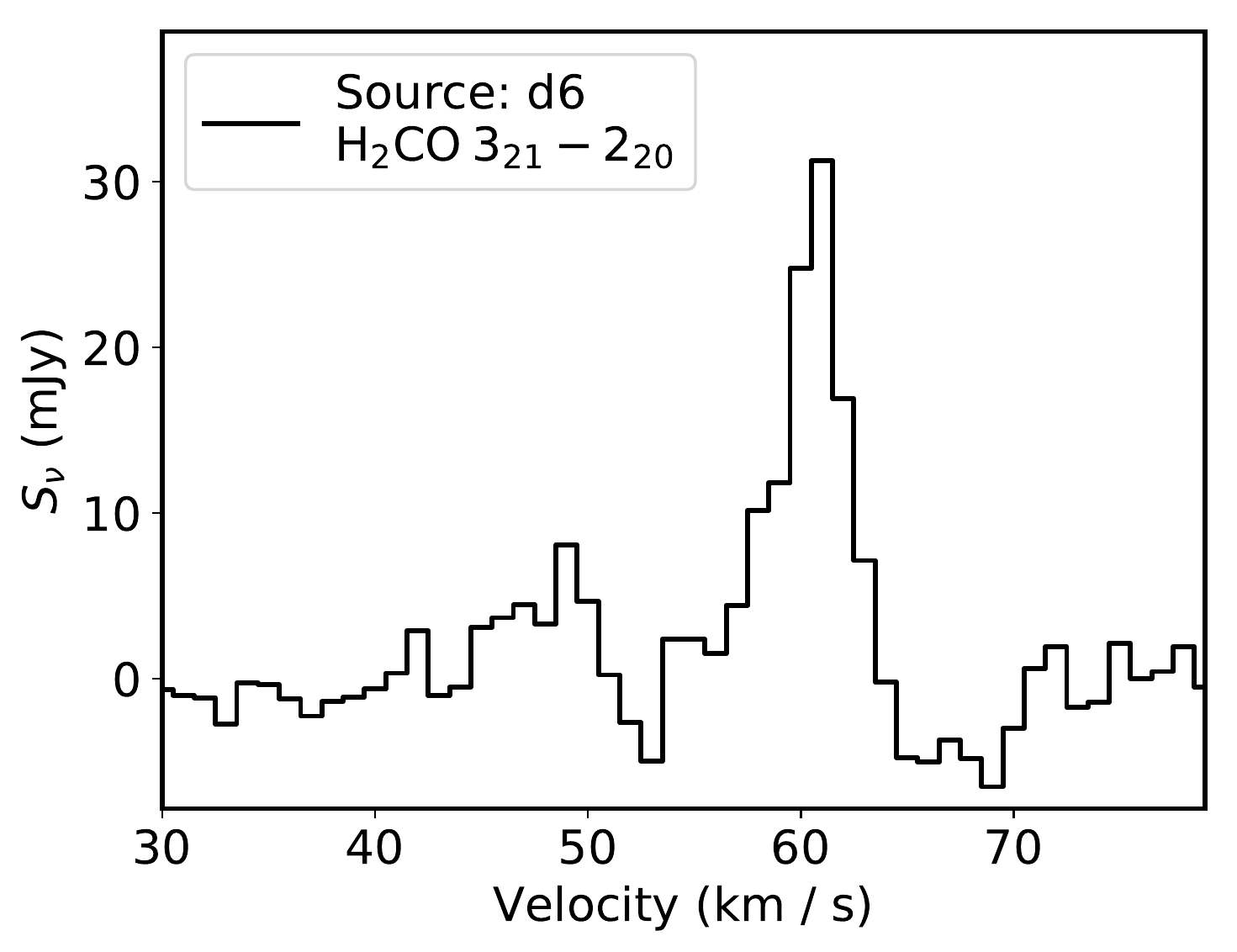}
        \includegraphics[scale=0.38]{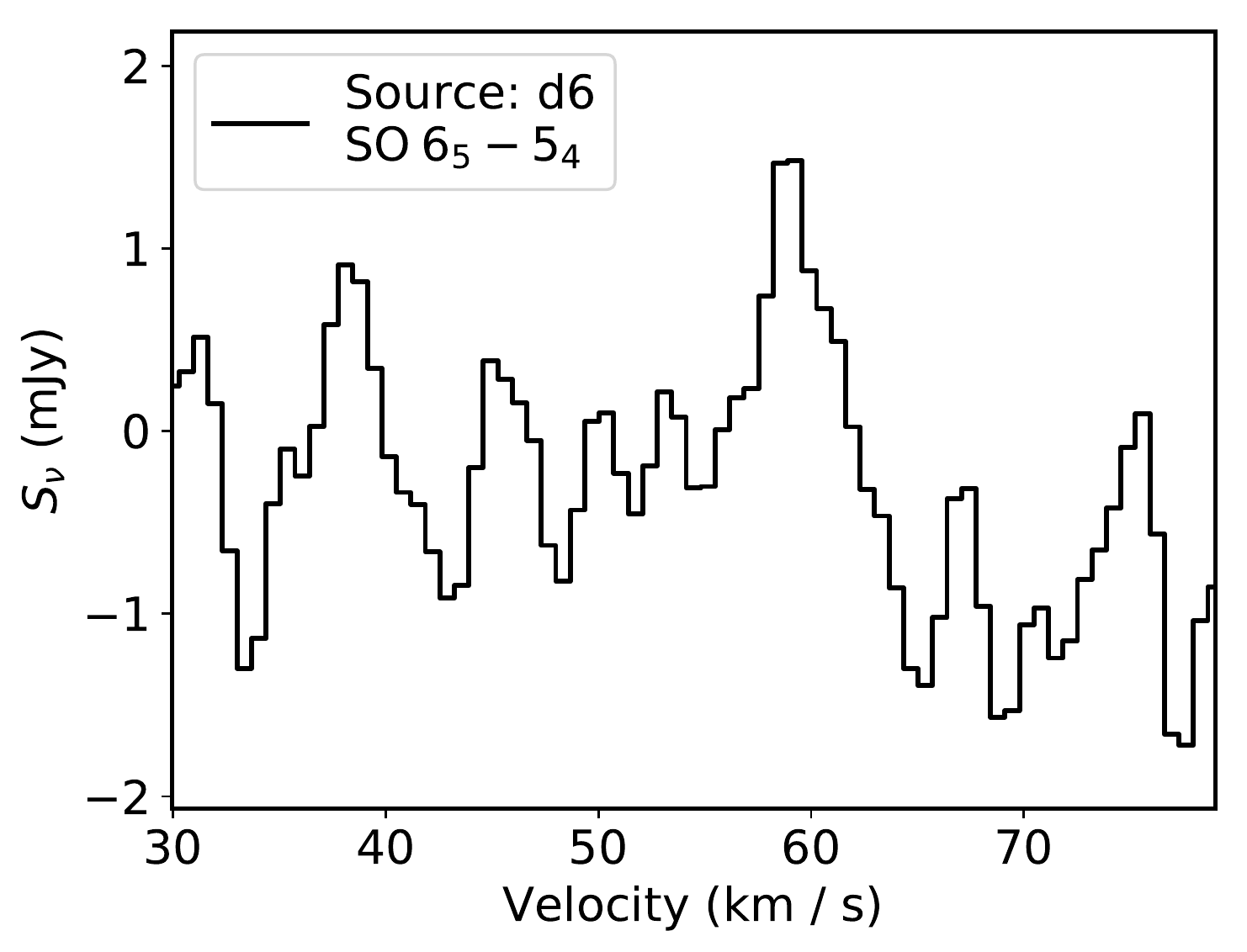}
        \includegraphics[scale=0.38]{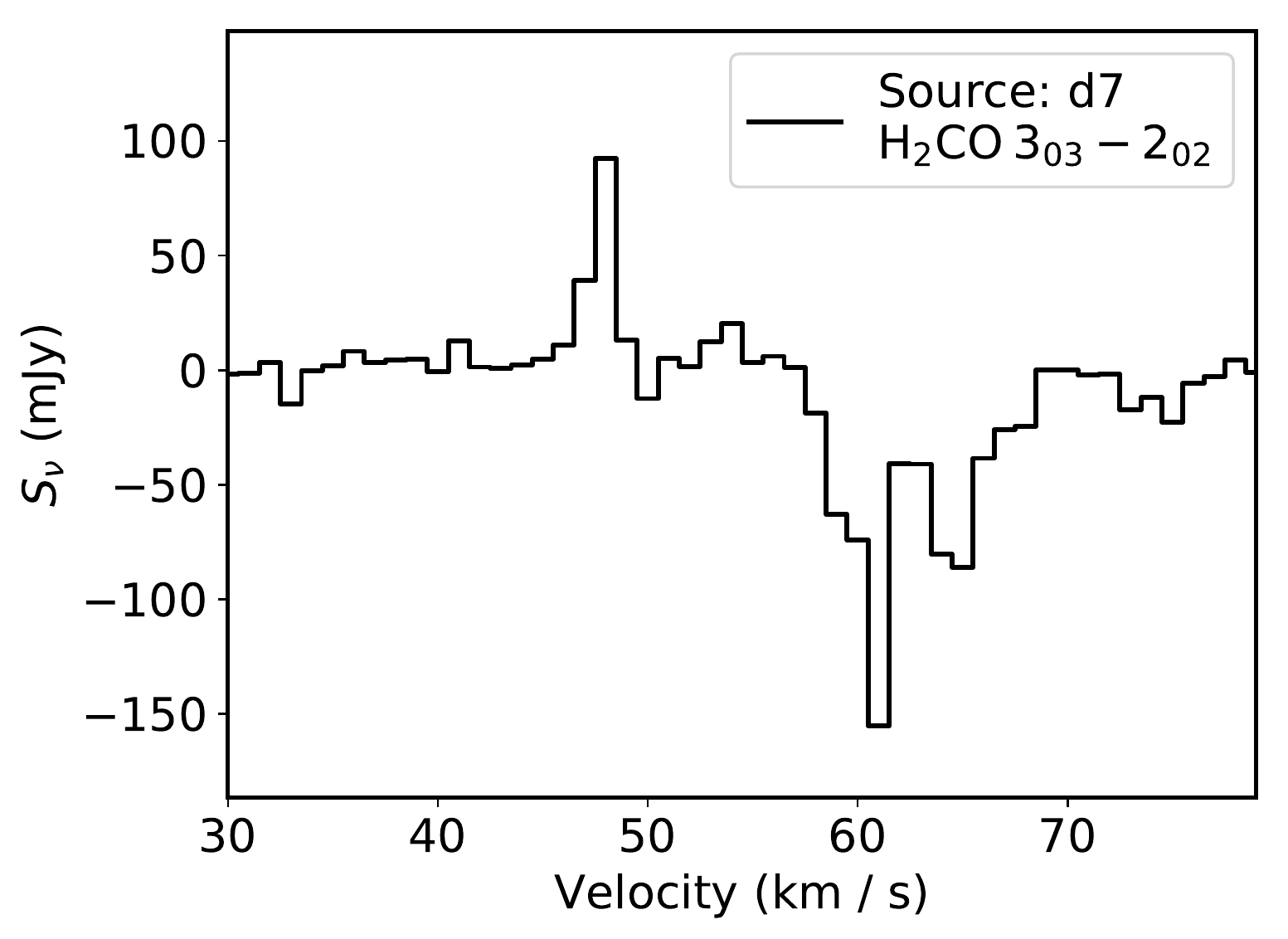}
        \includegraphics[scale=0.38]{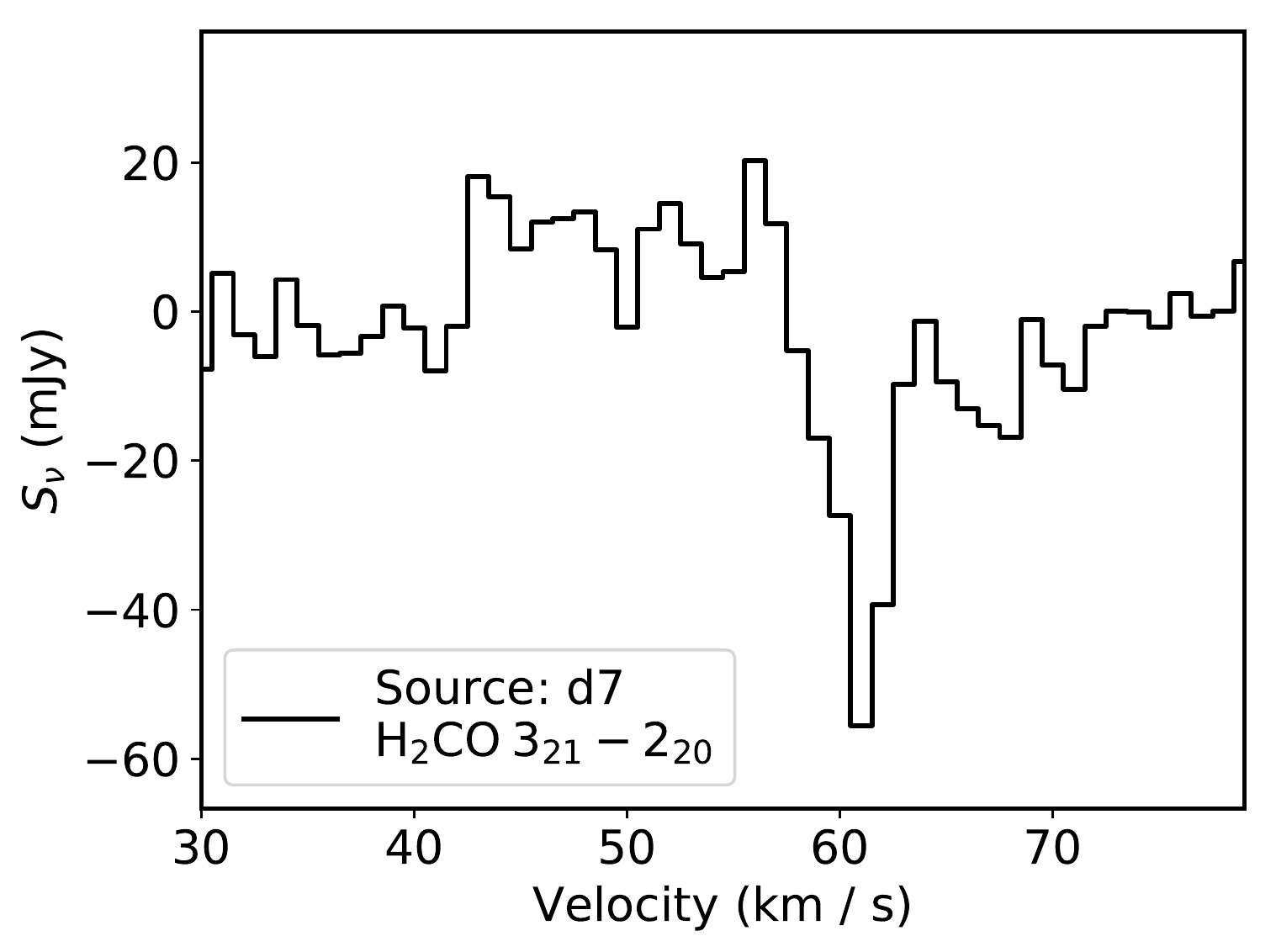}
        \includegraphics[scale=0.38]{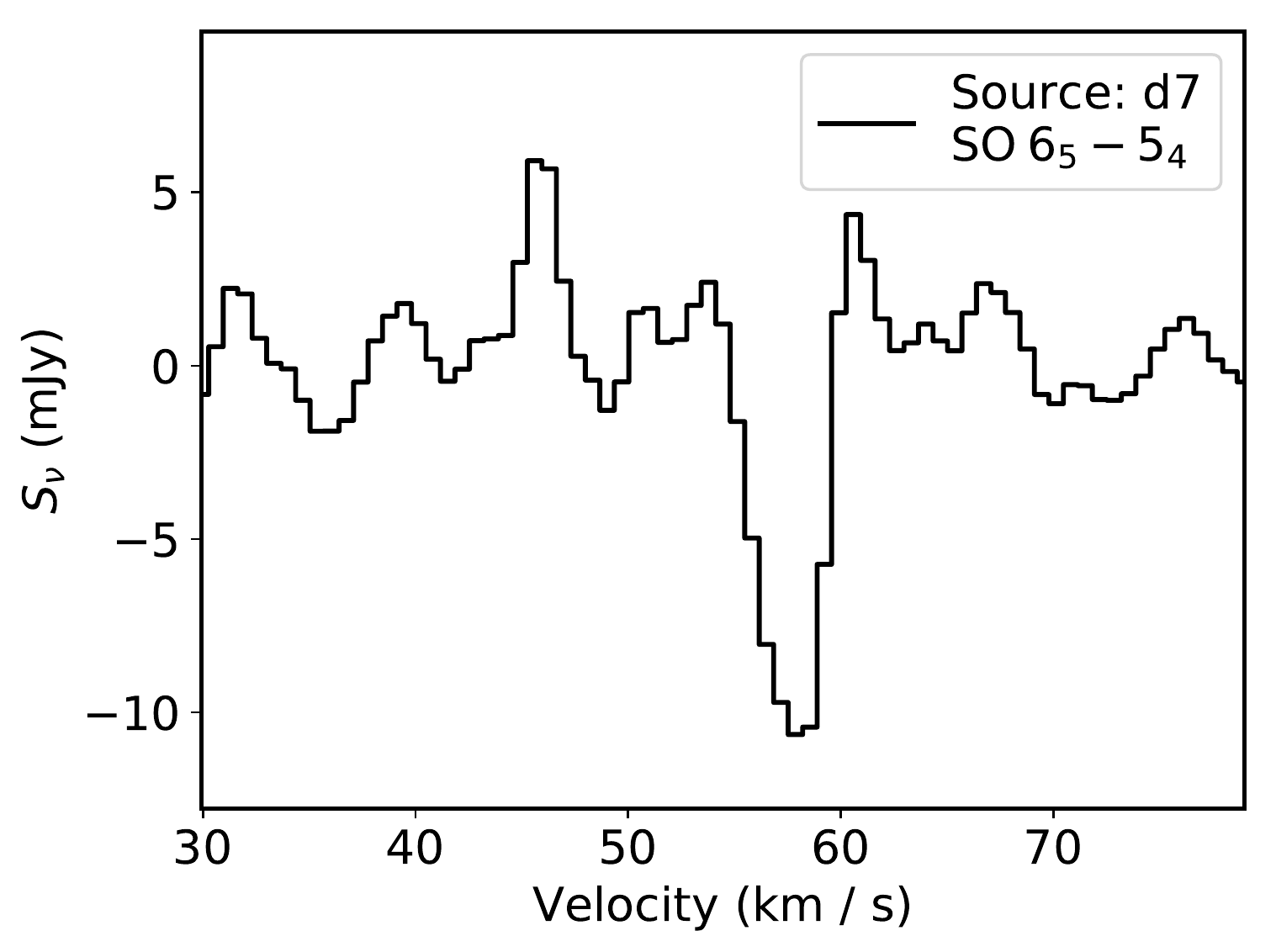}
        \includegraphics[scale=0.38]{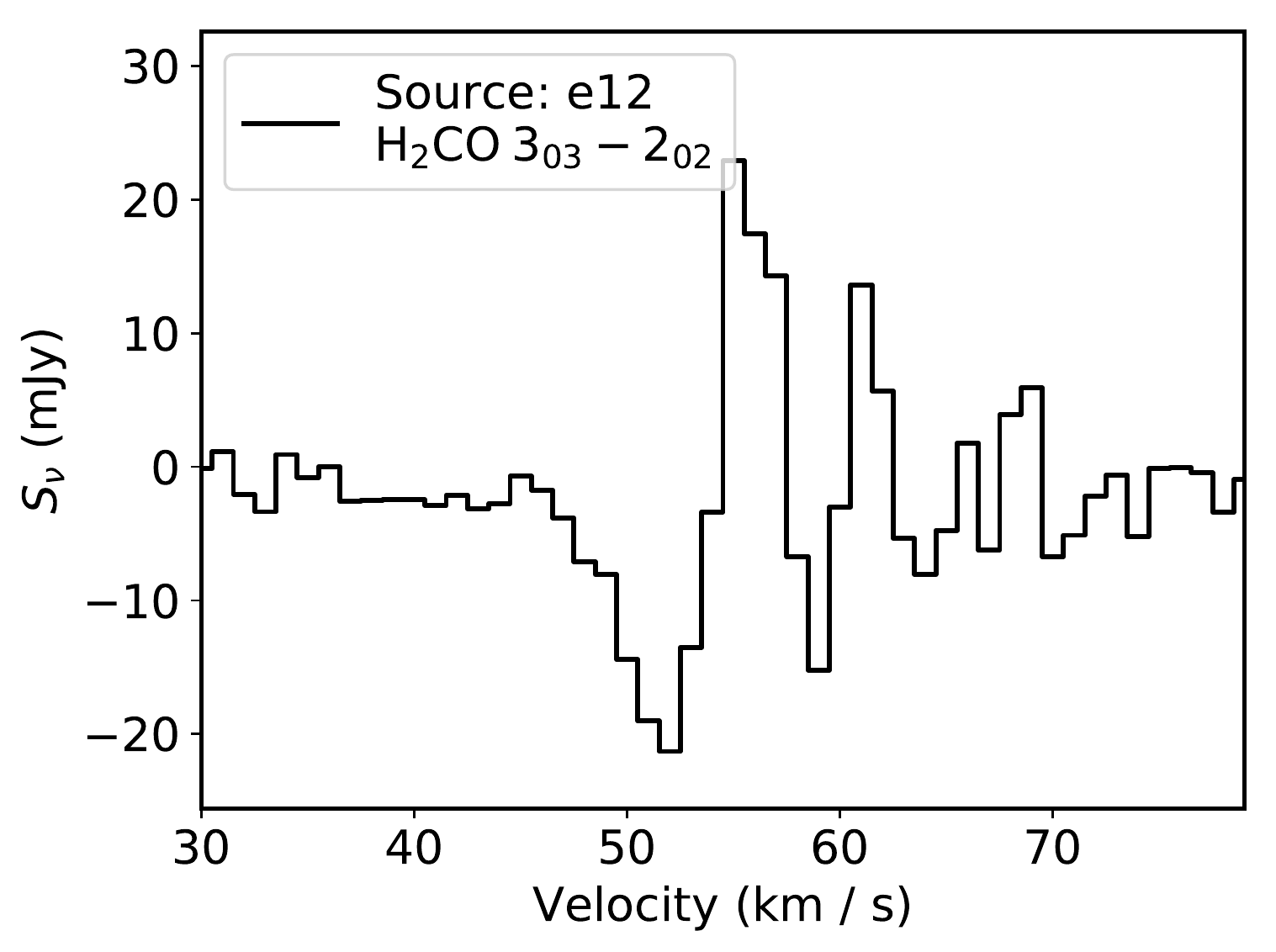}
        \includegraphics[scale=0.38]{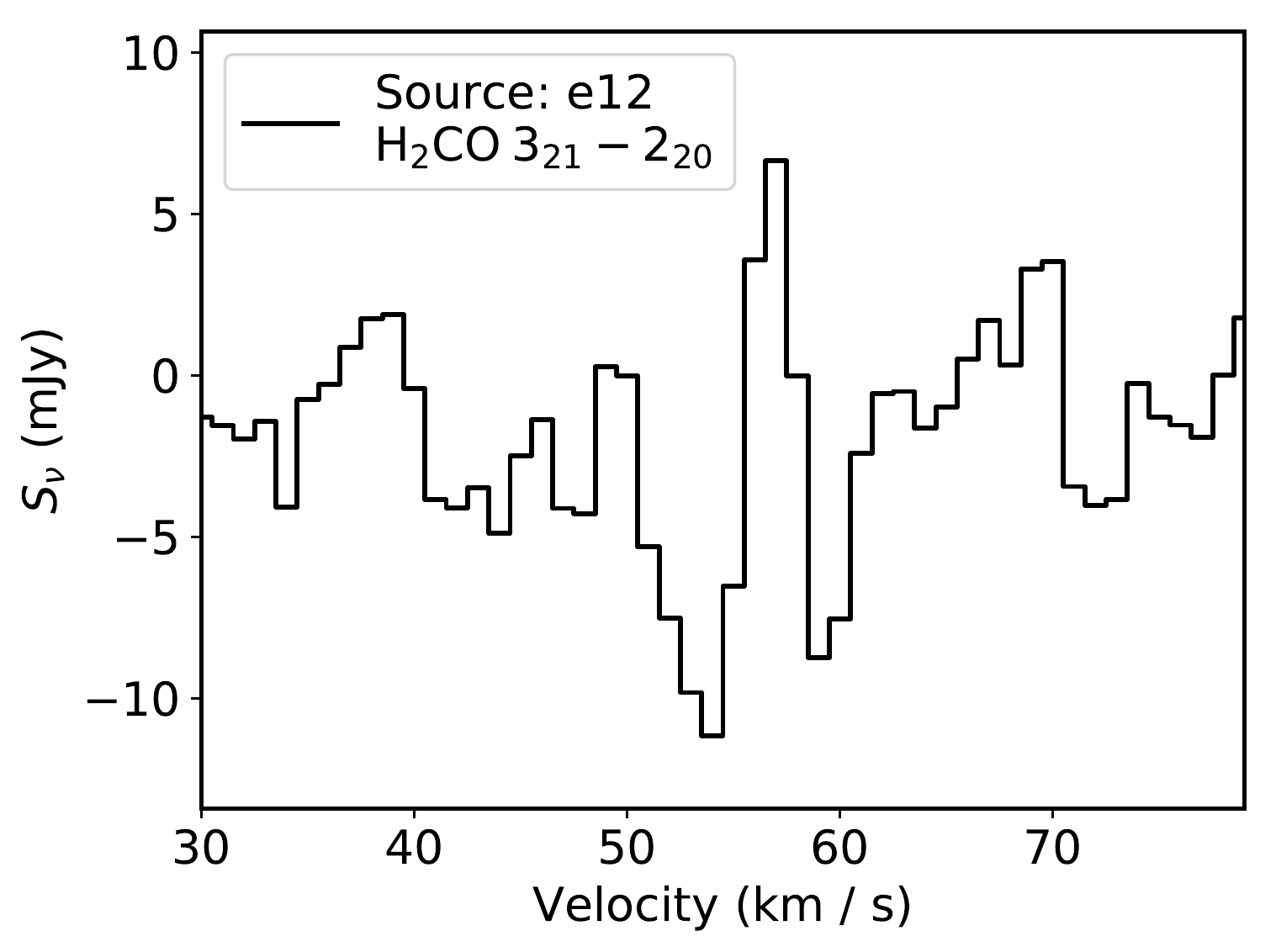}
        \includegraphics[scale=0.38]{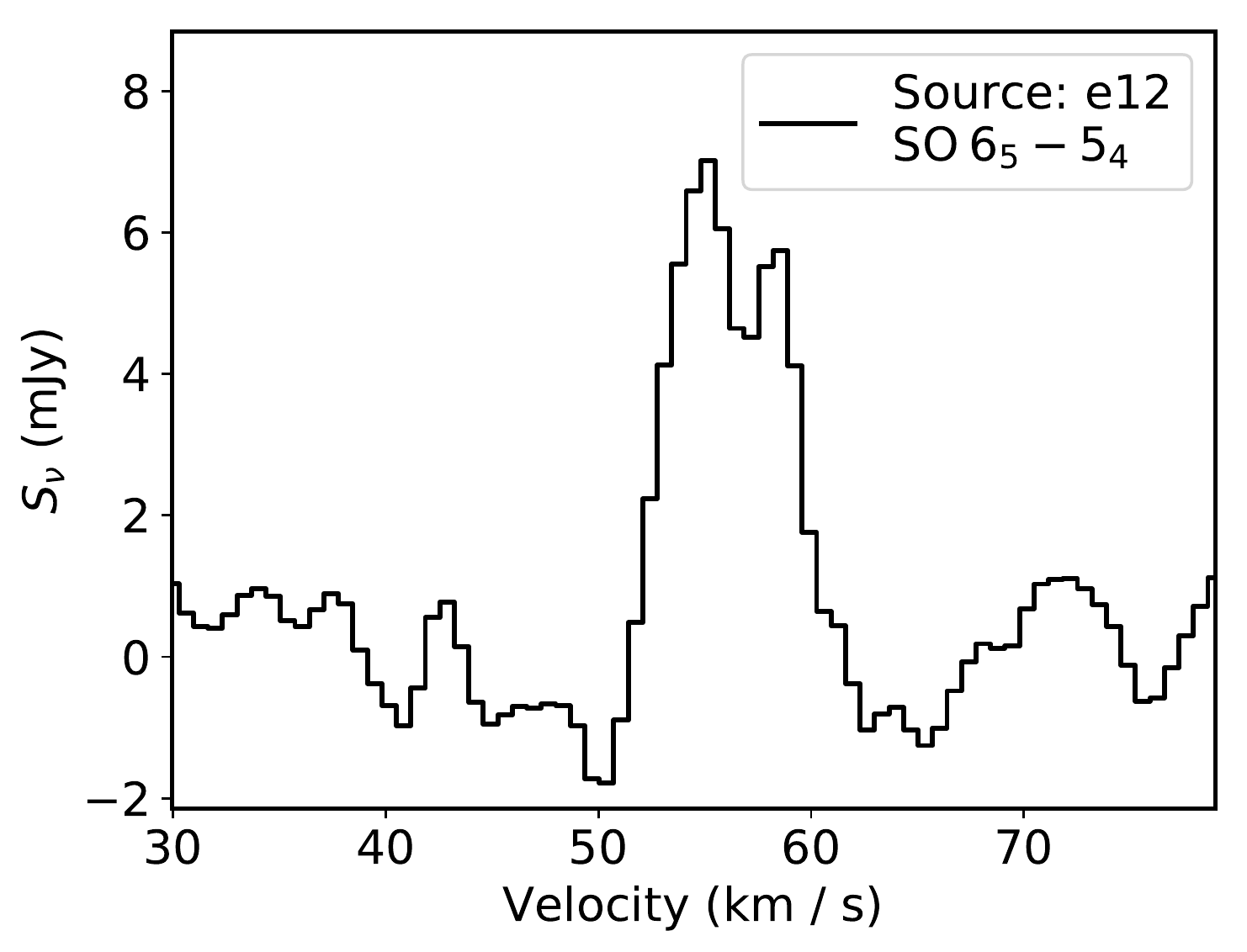}
        \includegraphics[scale=0.38]{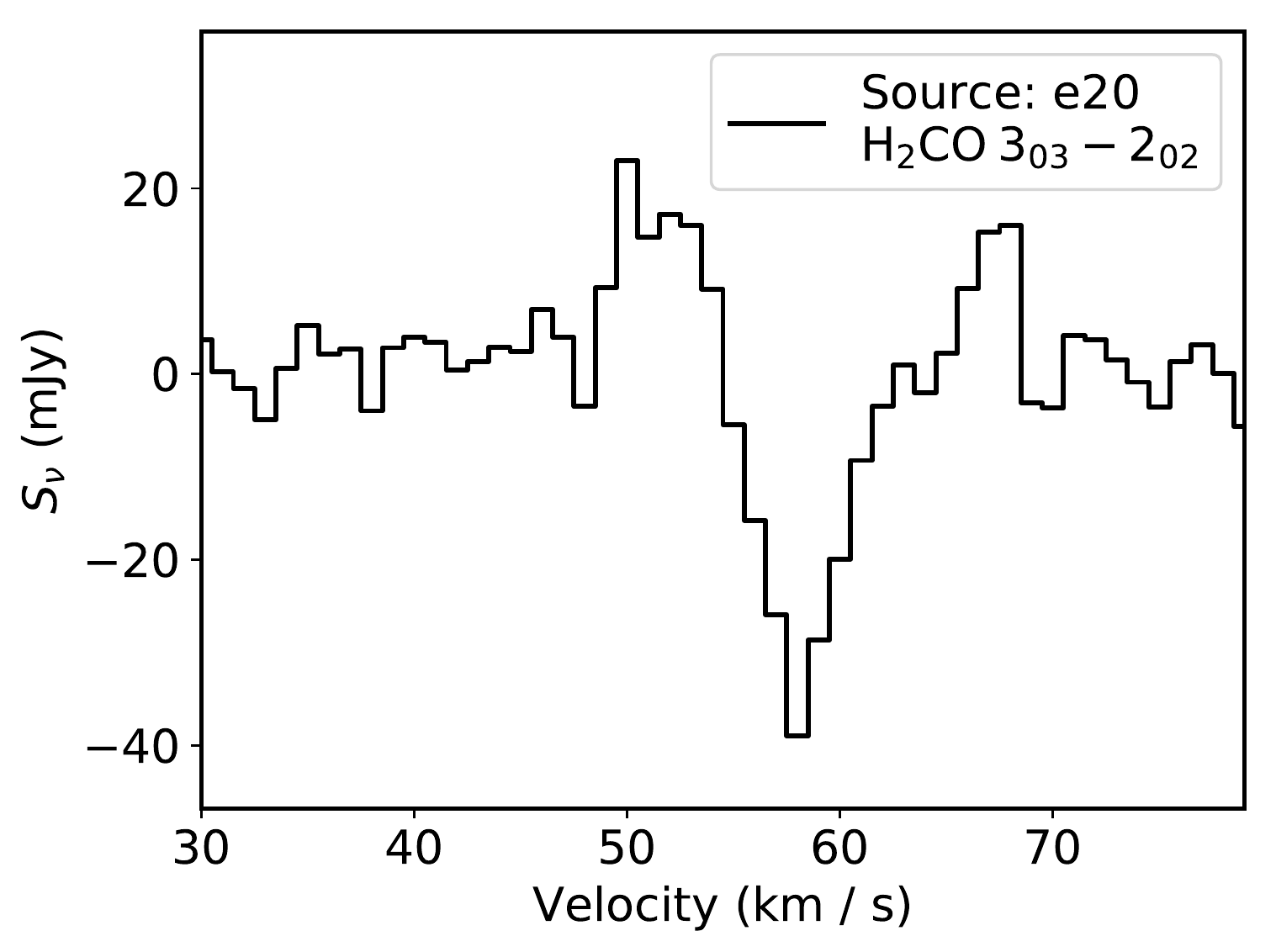}
        \includegraphics[scale=0.38]{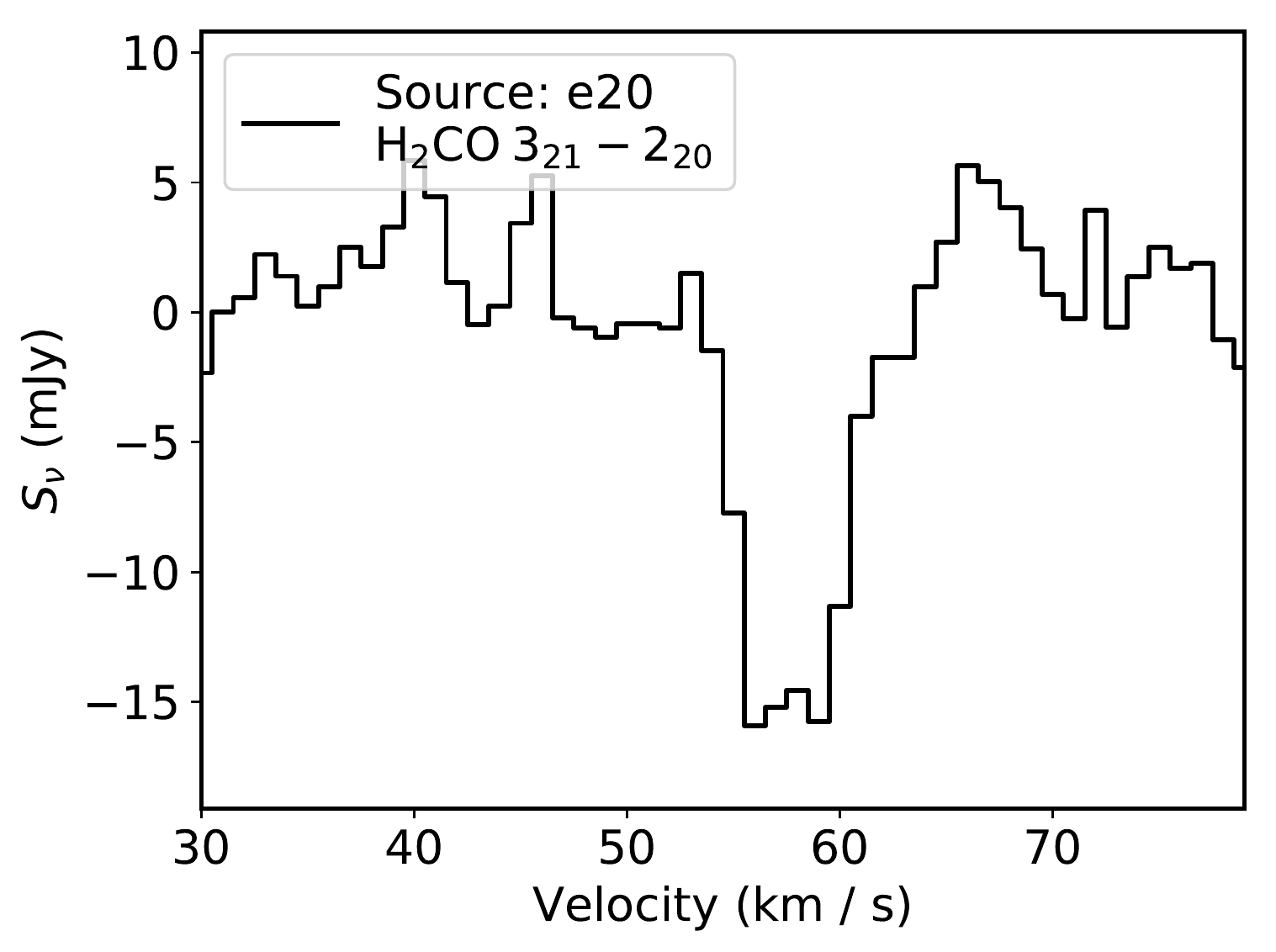}
        \includegraphics[scale=0.38]{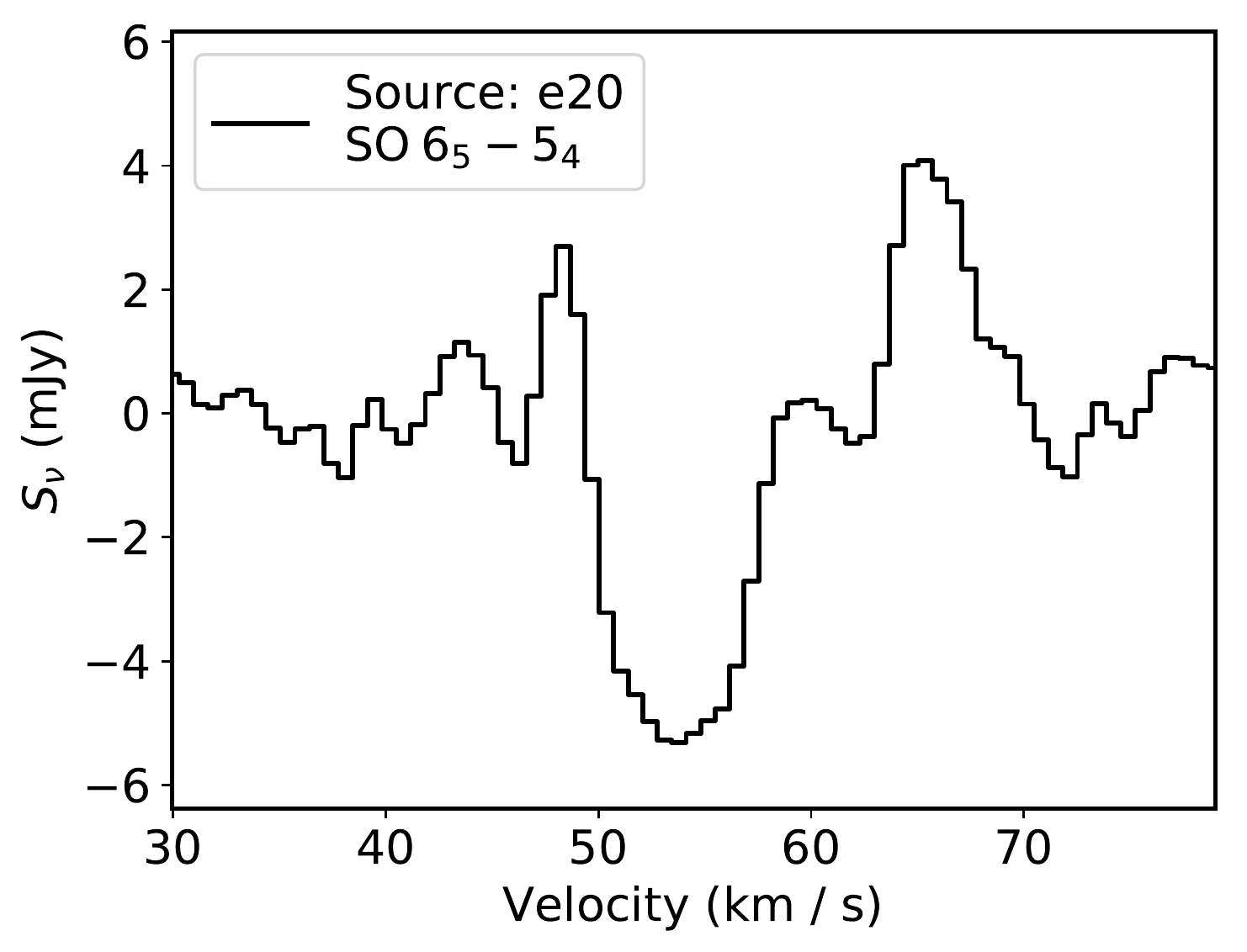}
        \caption{ 
        Spectral profiles of the molecular transitions \formai~ -- left --, \formaii~ -- middle --, \soi~ -- right --  for sources in catalog \textit{B} that are significantly affected by negative sidelobes from nearby, bright emission.} \label{fig:profiles_sidelobes}
\end{figure*}

\begin{figure*}[!t]
    \setlength{\lineskip}{0pt}
    \ContinuedFloat
    \centering
    \includegraphics[scale=0.38]{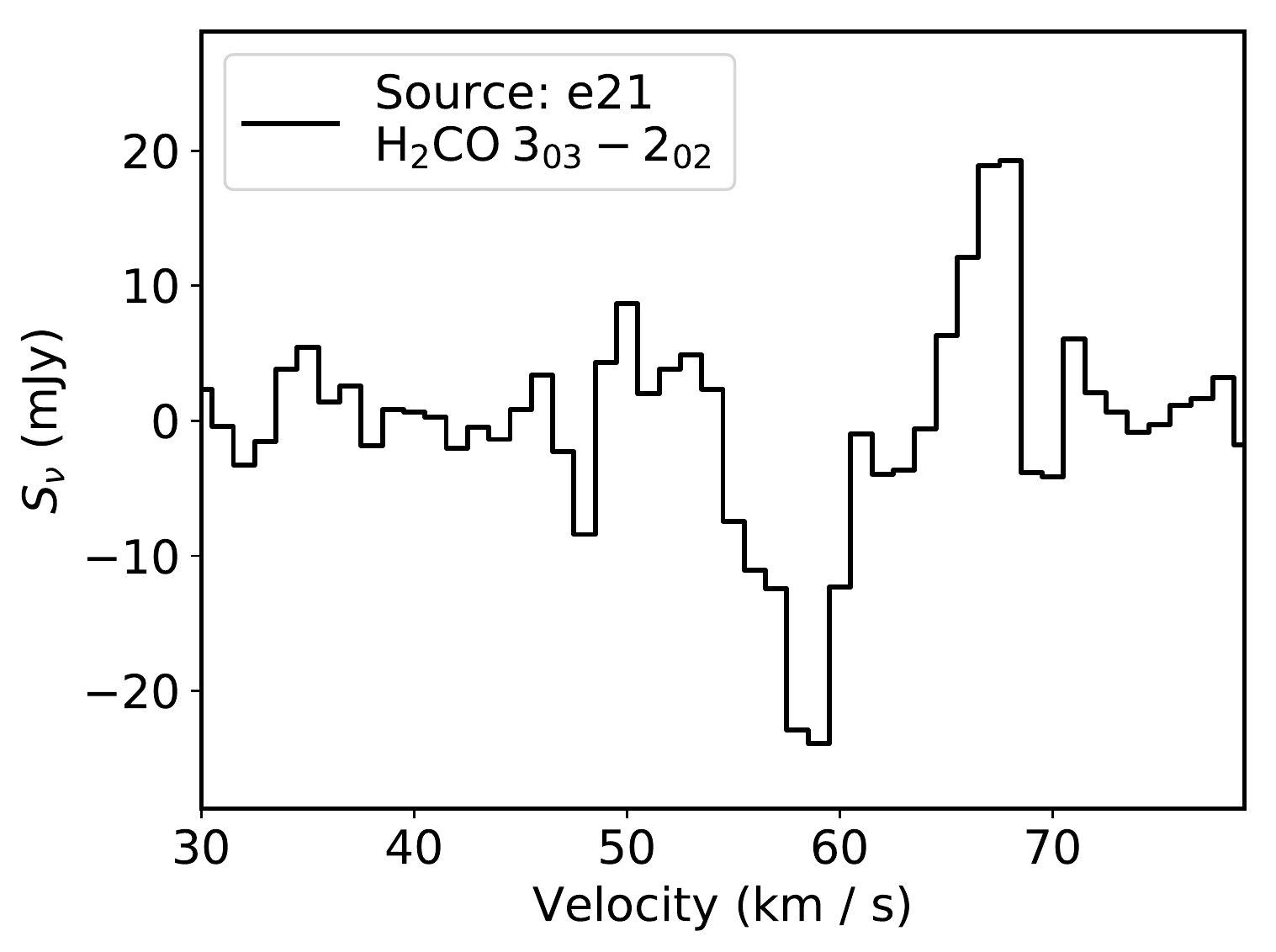}
    \includegraphics[scale=0.38]{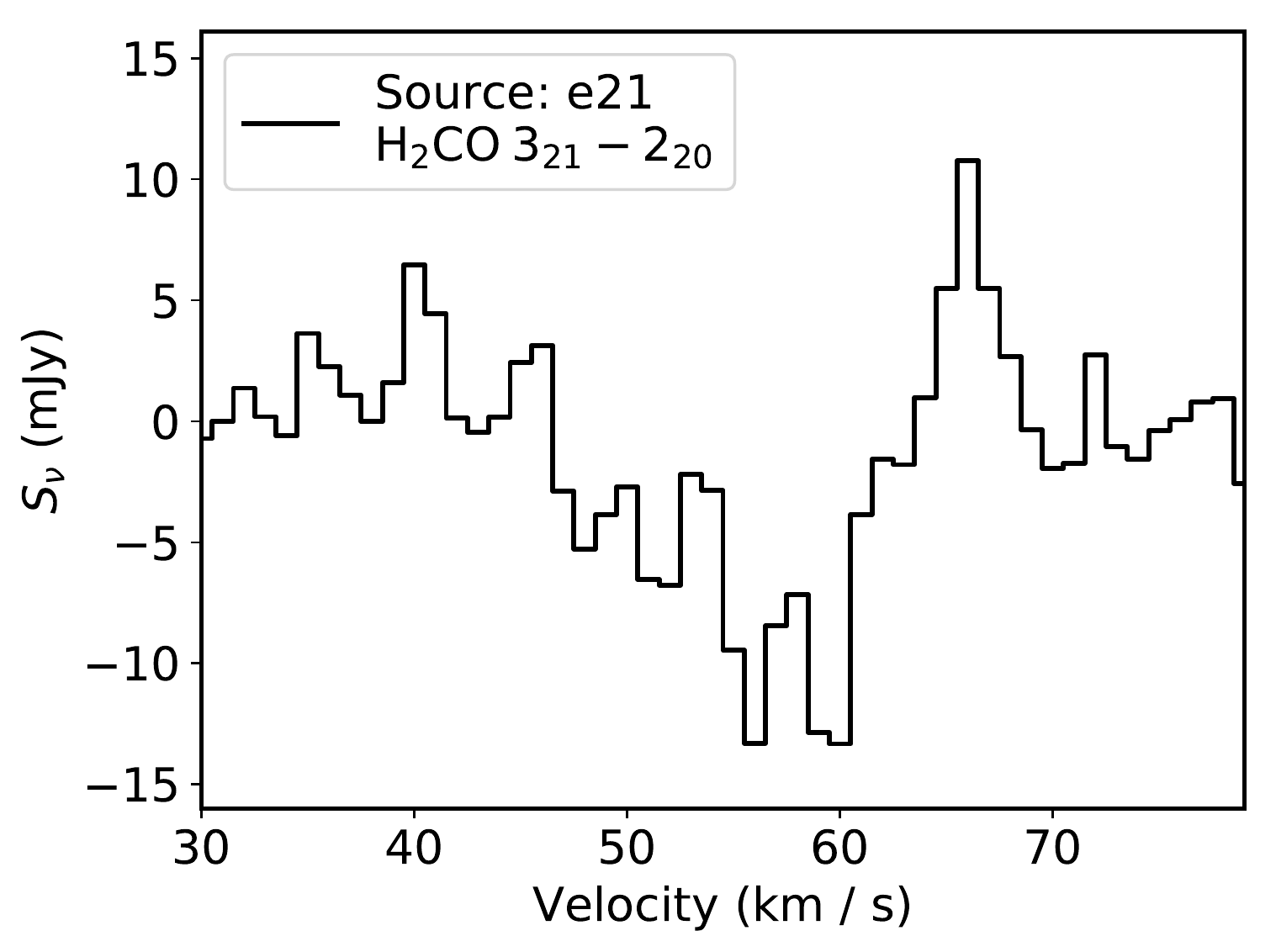}
    \includegraphics[scale=0.38]{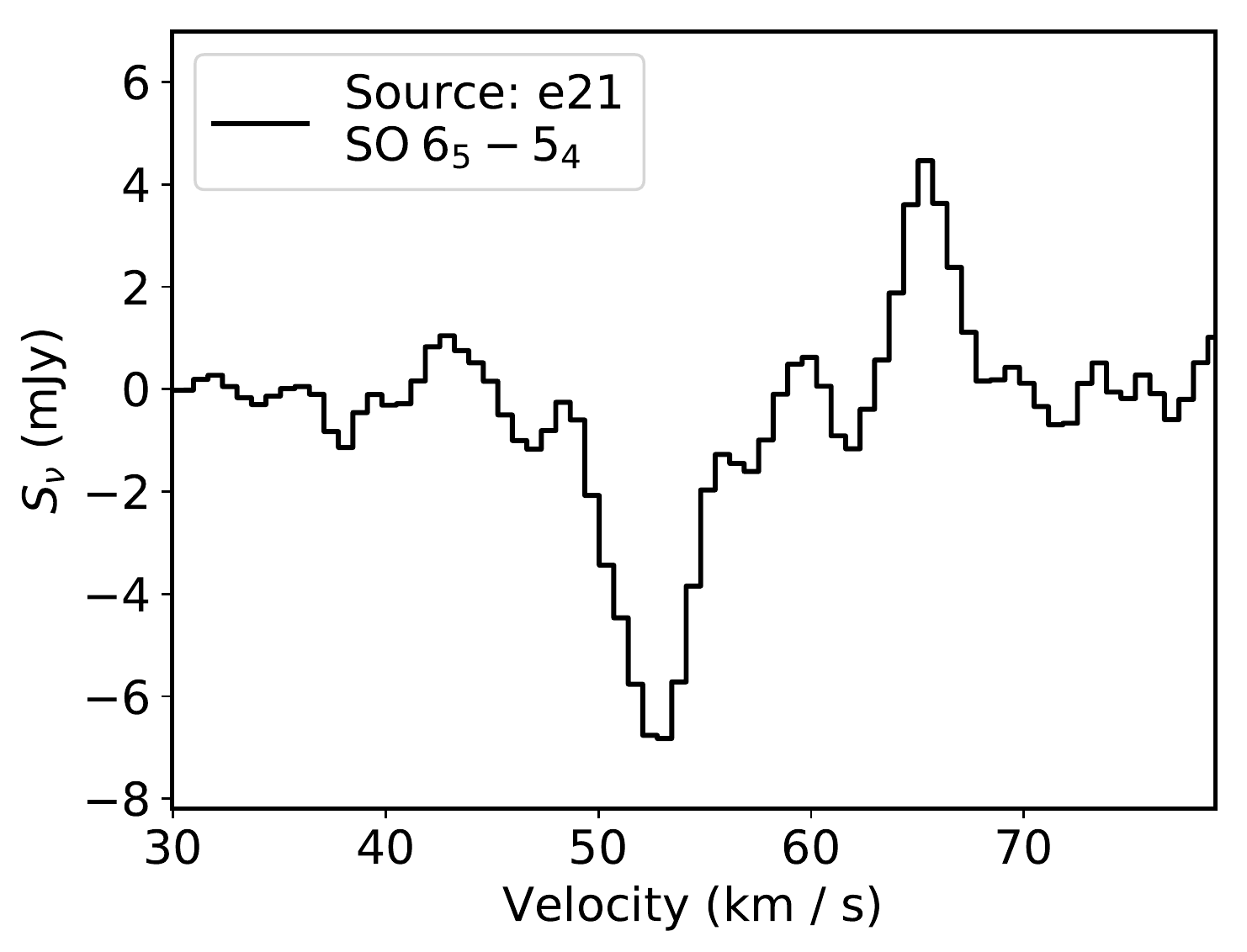}
    \includegraphics[scale=0.38]{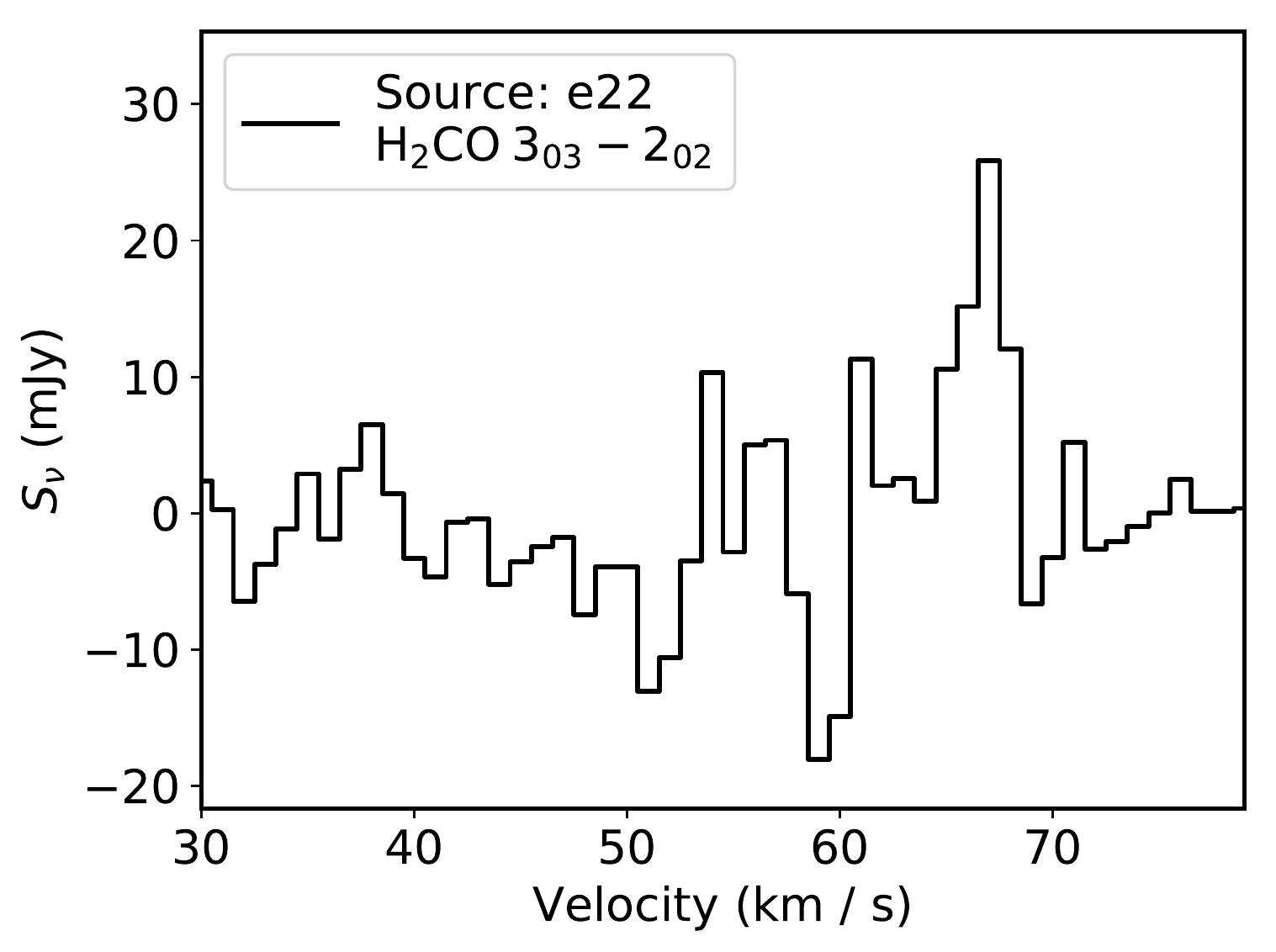}
    \includegraphics[scale=0.38]{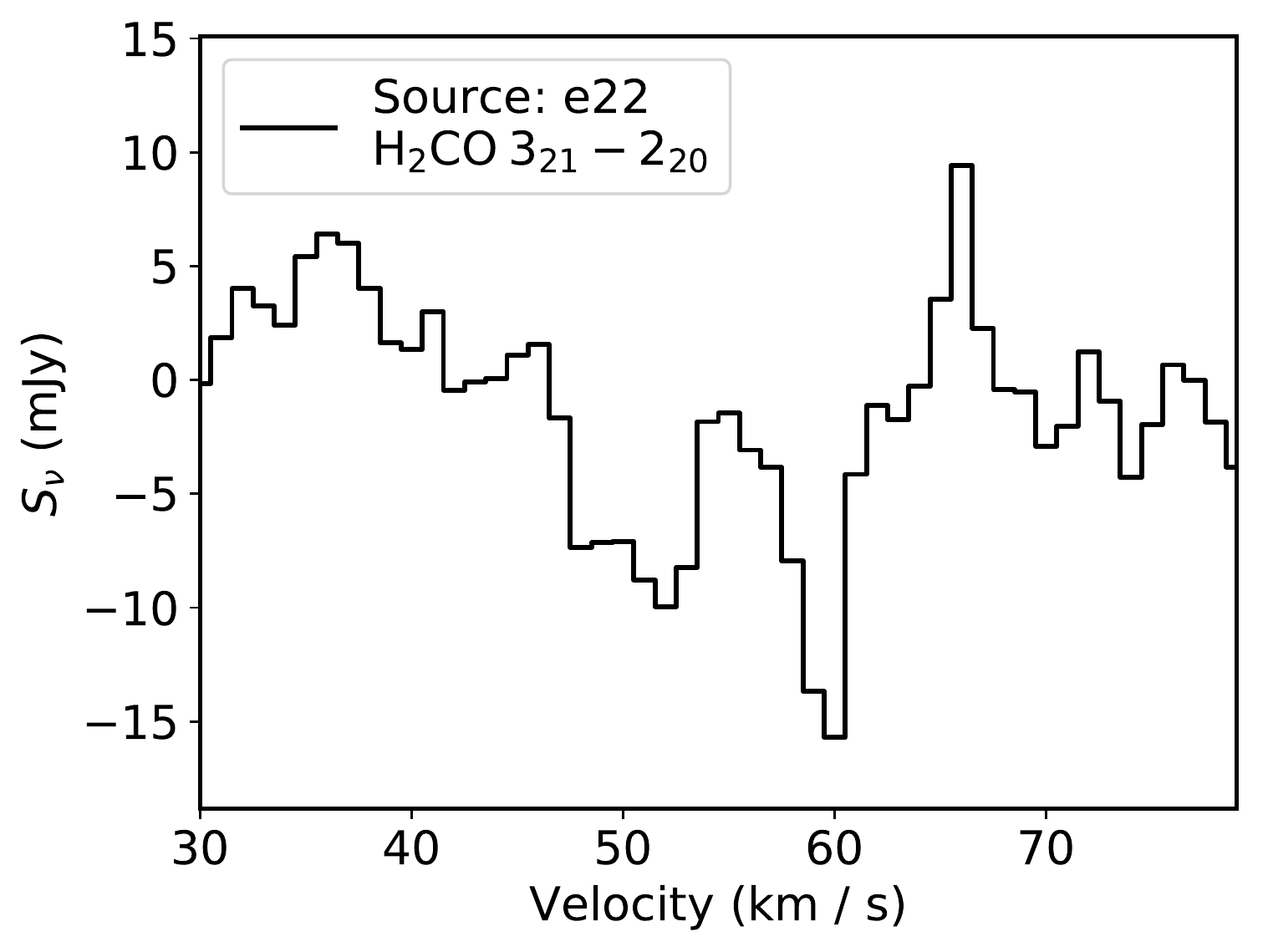}
    \includegraphics[scale=0.38]{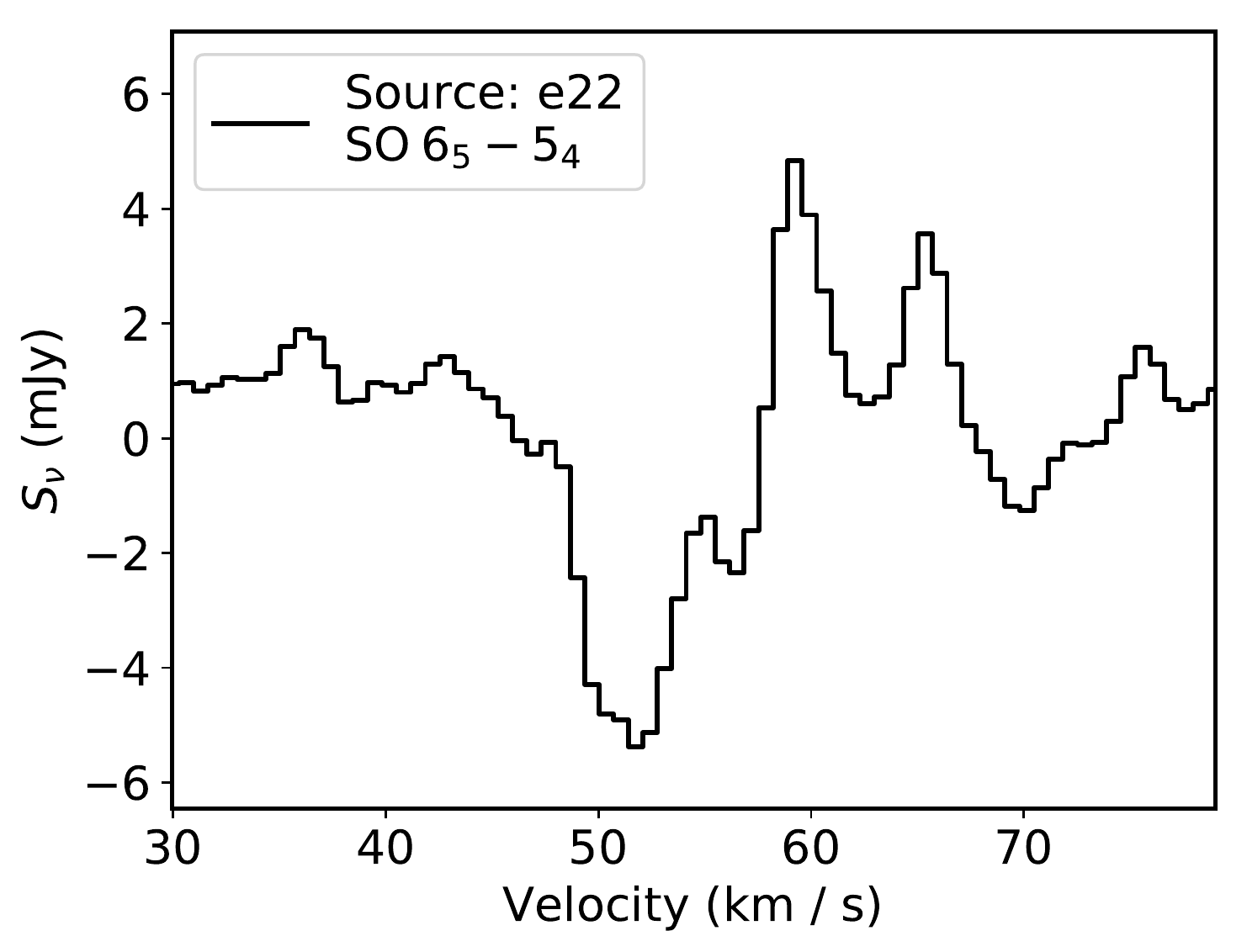}
    \includegraphics[scale=0.38]{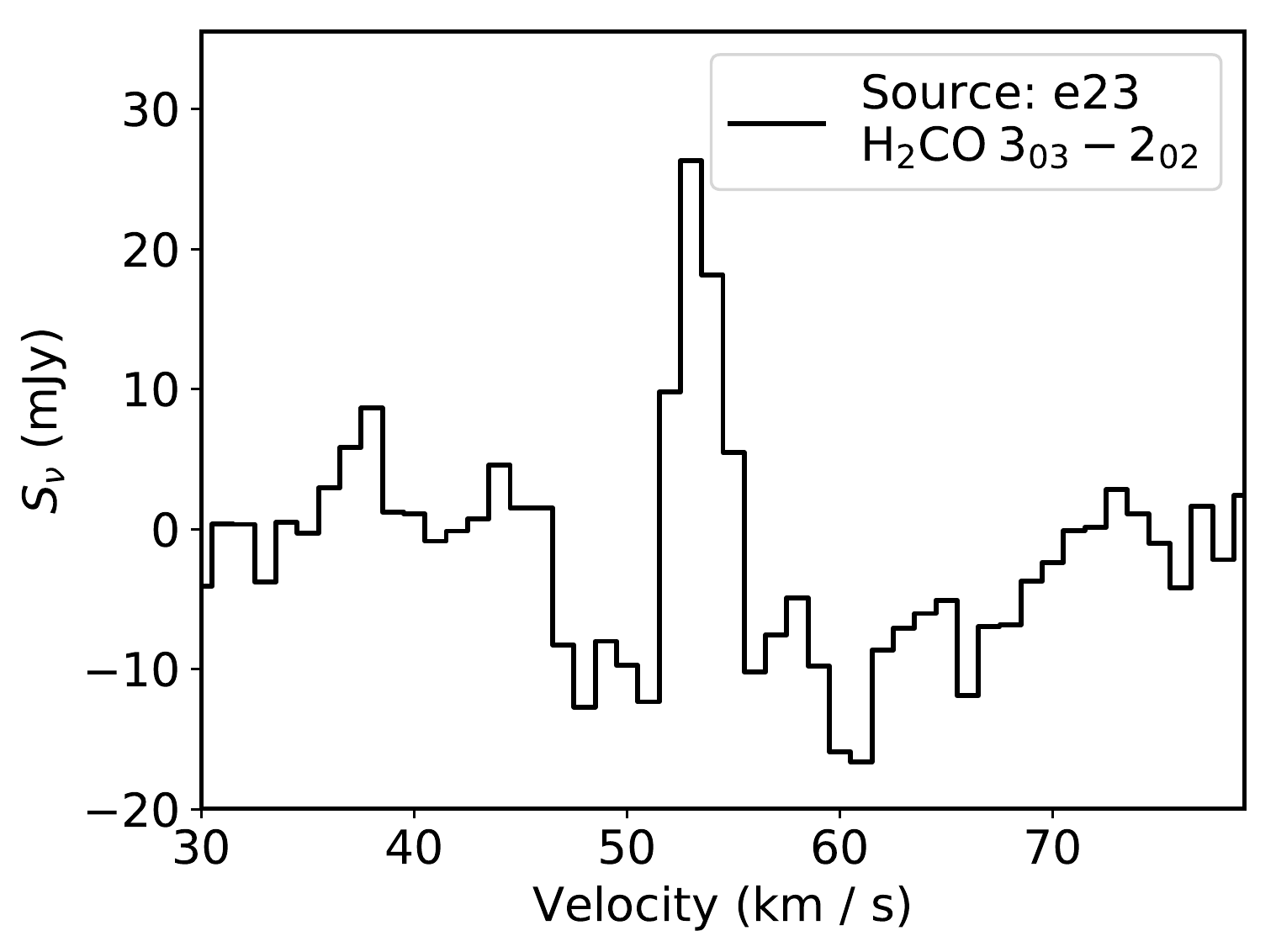}
    \includegraphics[scale=0.38]{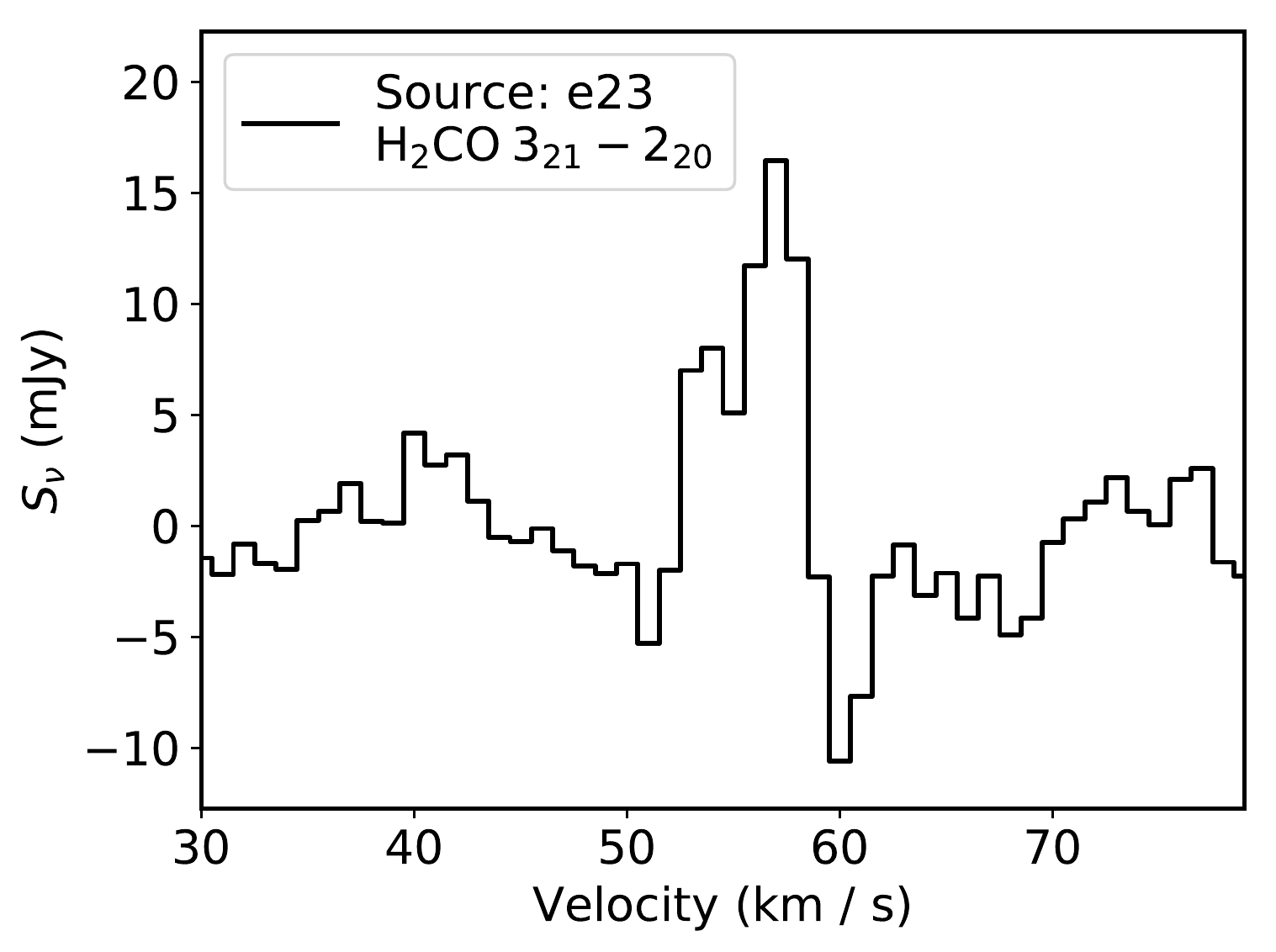}
    \includegraphics[scale=0.38]{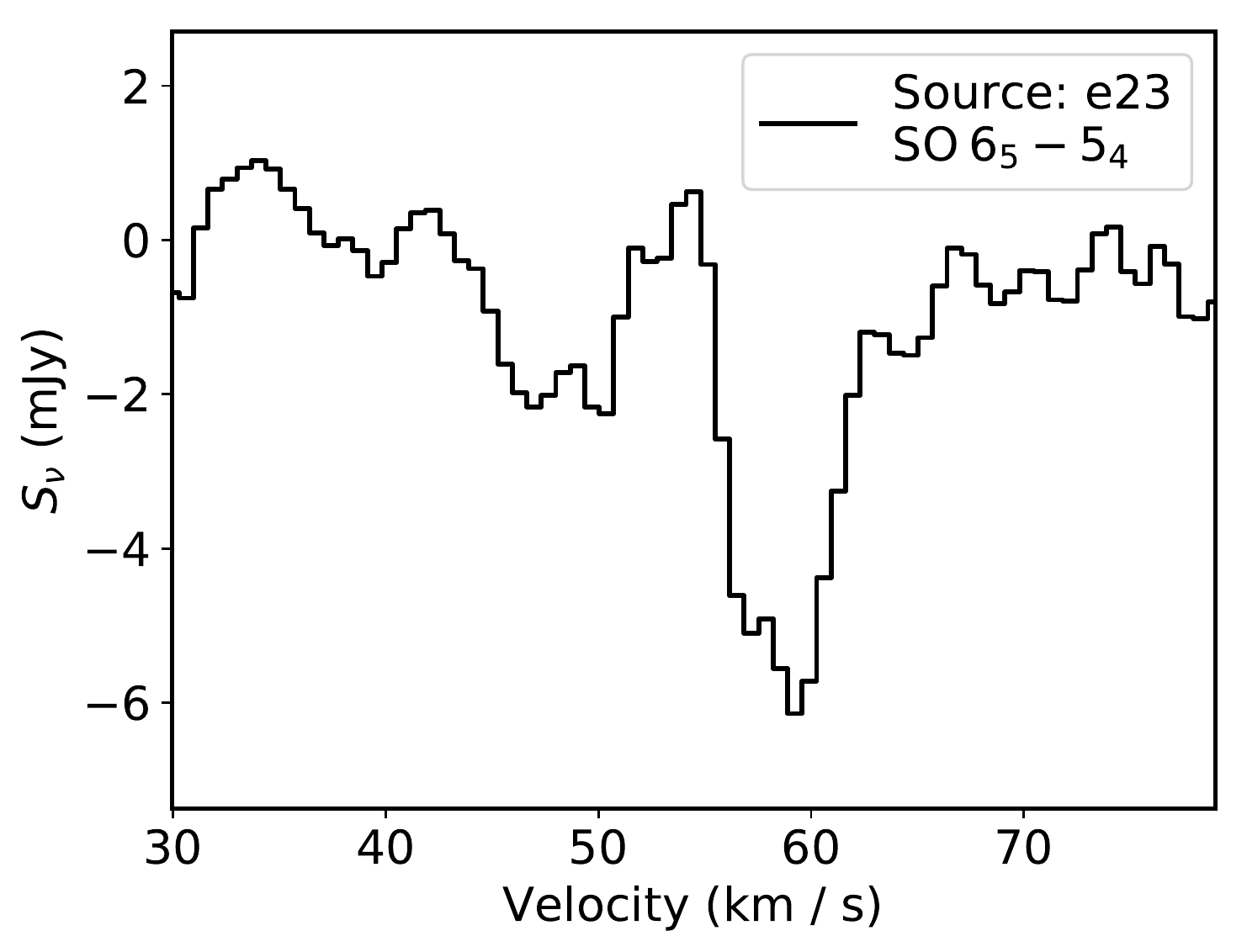} 
    \includegraphics[scale=0.38]{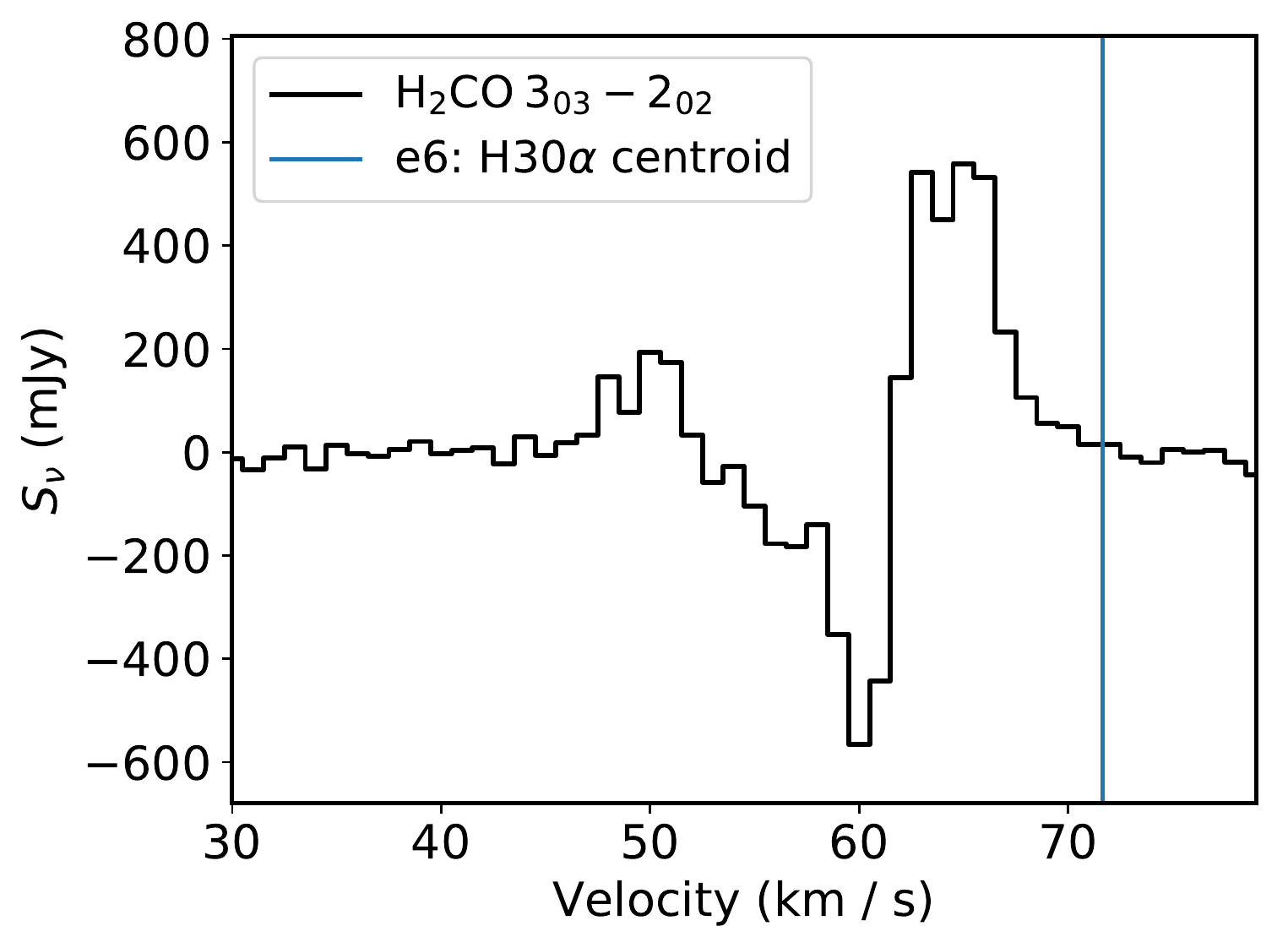}
    \includegraphics[scale=0.38]{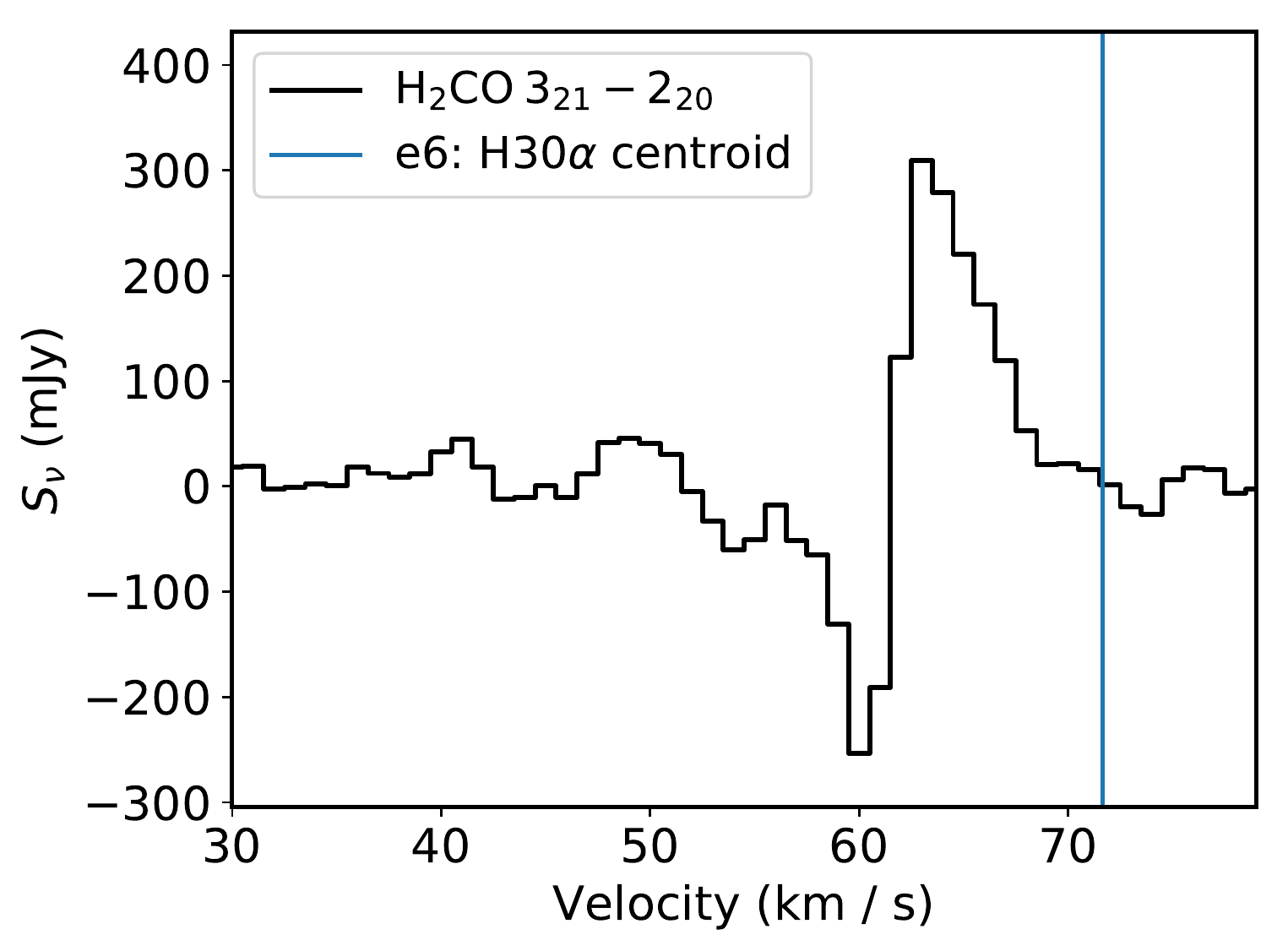}
    \includegraphics[scale=0.38]{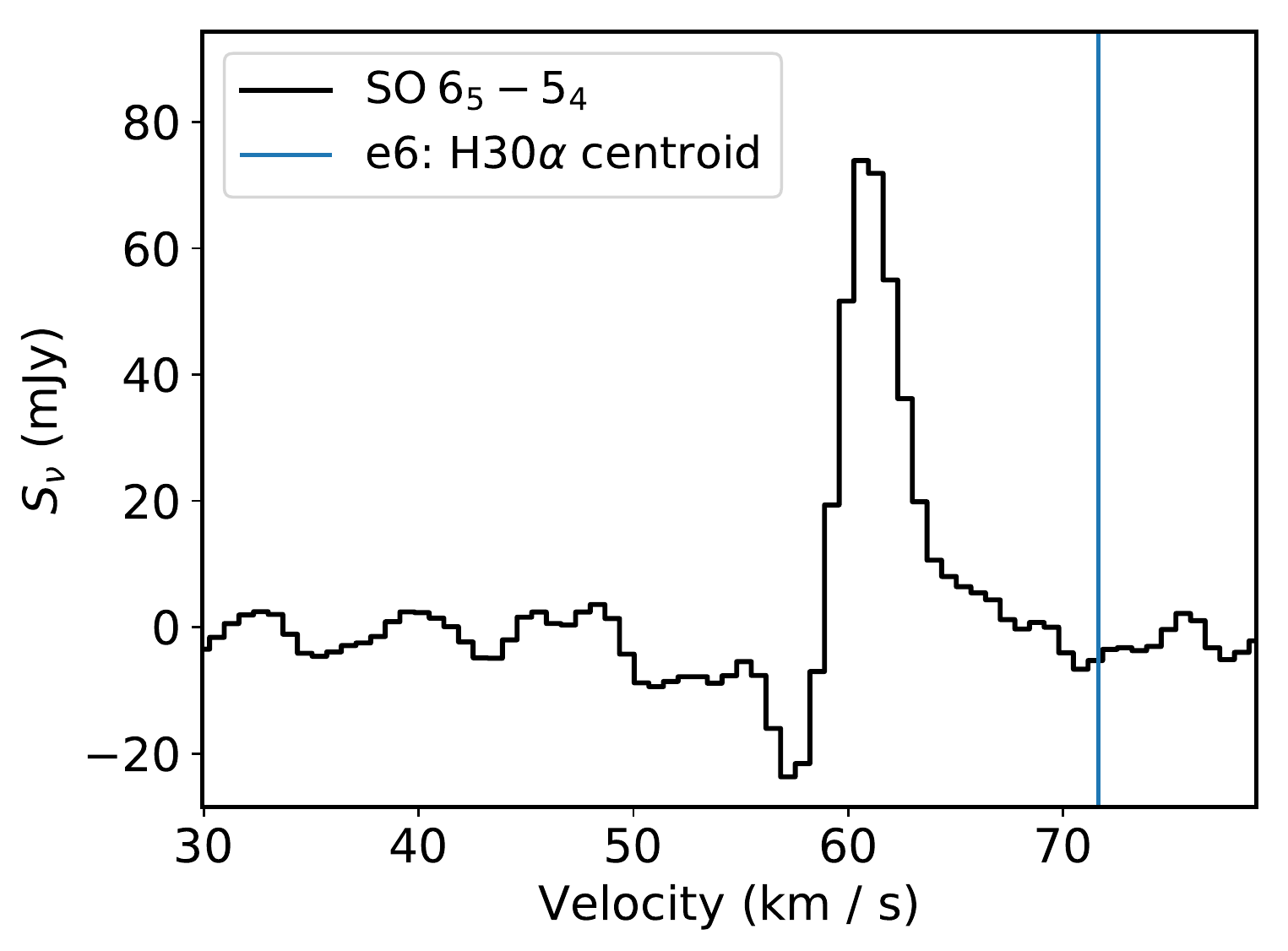}
    \caption{\textit{contd.}
    Spectral profiles of the molecular transitions \formai~ -- left --, \formaii~ -- middle --, \soi~ -- right --  for sources
    in catalog \textit{B} 
    that are significantly affected by negative sidelobes from nearby, bright  emission.}
\end{figure*}

\end{document}